\documentclass[reprint,
superscriptaddress,
 amsmath,amssymb,
 aps,
pra,
nofootinbib
]{revtex4-2}

\usepackage{amsmath,amssymb,times,graphicx,xcolor,physics}

\definecolor{linkColor}{RGB}{0,70,120}
\usepackage[colorlinks=true, allcolors=linkColor,pdfborder={0 0 0},pdfencoding = auto]{hyperref}

\usepackage{algorithm}
\usepackage{algpseudocode}

\usepackage{array}[=2016-10-06]
\usepackage{longtable}
\makeatletter
\g@addto@macro\@mkpream@relax{\let\@endpbox\relax}
\makeatother

\DeclareMathOperator{\argmax}{arg\,max}
\DeclareMathOperator{\argmin}{arg\,min}

\newcommand{\Wgrad}{\nabla_\mathrm{W}}
\newcommand{\BWgrad}{\nabla_\mathrm{BW}}
\newcommand{\FF}{\mathcal{F}}

\newcommand{\RR}{\mathbb{R}}
\newcommand{\dkl}{\mathrm{D}_\mathrm{KL}}
\newcommand{\VV}{\mathcal{V}}

\newcommand{\NN}{\mathcal{N}}
\newcommand{\bw}{\mathrm{BW}}

\newcommand{\EE}{\mathbb{E}}

\newcommand{\LL}{\mathcal{L}}

\definecolor{darkgreen}{RGB}{0,128,0}

\makeatletter \let\oldaddcontentsline\addcontentsline \newcommand{\StartHideFromToC}{\renewcommand{\addcontentsline}[3]{\def\tempA{##1}\def\tempB{toc}\ifx\tempA\tempB\else \oldaddcontentsline{##1}{##2}{##3}\fi }} \newcommand{\StopHideFromToC}{\let\addcontentsline\oldaddcontentsline }
\makeatother
\newcommand{\notationold}[1]{{\color{red}#1}}


\begin{document}

\title{Bridging Control, Inference, Transport, and Thermodynamics \\[1ex]
\normalsize 
From Theory to Applications in Learning }

\date{\today}

\author{Emmy Blumenthal} 
\email{eblu@princeton.edu}
\affiliation{Joseph Henry Laboratories of Physics, Princeton University, Princeton, New Jersey 08544, USA} 

\author{Nikolas Claussen} 
\email{nc1333@princeton.edu}
\affiliation{Joseph Henry Laboratories of Physics, Princeton University, Princeton, New Jersey 08544, USA} 
\affiliation{Princeton Center for Theoretical Science, Princeton University, Princeton, New Jersey 08544, USA}

\author{Benjamin Eysenbach} 
\email{eysenbach@princeton.edu}
\affiliation{Department of Computer Science, Princeton University, Princeton, New Jersey 08544, USA}

\author{Catherine Ji} 
\email{cj7280@princeton.edu}
\affiliation{Joseph Henry Laboratories of Physics, Princeton University, Princeton, New Jersey 08544, USA} 

\author{Gautam Reddy} 
\email{greddy@princeton.edu}
\affiliation{Joseph Henry Laboratories of Physics, Princeton University, Princeton, New Jersey 08544, USA}

\author{Colin Scheibner} 
\email{cs5096@princeton.edu}
\affiliation{Joseph Henry Laboratories of Physics, Princeton University, Princeton, New Jersey 08544, USA} 
\affiliation{Princeton Center for Theoretical Science, Princeton University, Princeton, New Jersey 08544, USA}

\author{Benjamin Sorkin} 
\email{bs4171@princeton.edu}
\affiliation{Joseph Henry Laboratories of Physics, Princeton University, Princeton, New Jersey 08544, USA} 
\affiliation{Princeton Center for Theoretical Science, Princeton University, Princeton, New Jersey 08544, USA}

\begin{abstract}
    The last decade has seen the development of powerful methods for learning complex structure from high-dimensional data. 
    These advances have brought to the foreground fundamental connections between subdisciplines of physics, applied mathematics, and machine learning. 
    In this review, we bring together some of these ideas, often expressed in different languages, to highlight a conceptual thread that links five distinct fields: control theory, optimal transport, probabilistic inference, non-equilibrium thermodynamics, and machine learning. 
    A common theme is the optimization of free-energy-like functionals under dynamical or statistical constraints. 
    We offer a guided tour through this thread and present selected applications in reinforcement learning, variational inference, and generative modeling. The review does not assume prior familiarity with these topics, and begins with principles originating from physics. 
\end{abstract}

\maketitle

\newpage

\tableofcontents

\section*{Introduction}

\begin{figure*}[t]
    \centering
    \begingroup\setlength{\unitlength}{\dimexpr\linewidth/1198\relax}\begin{picture}(0,0)\put(72.559,394.042){\hyperref[sec:mechanics_control]{\phantom{\rule{181.805\unitlength}{25.021\unitlength}}}}\put(916.878,394.042){\hyperref[sec:energy-based_modelling]{\phantom{\rule{195.548\unitlength}{25.021\unitlength}}}}\put(94.287,314.558){\hyperref[sec:mechanics_control]{\phantom{\rule{138.348\unitlength}{25.021\unitlength}}}}\put(450.417,314.558){\hyperref[sec:OTthermo]{\phantom{\rule{277.237\unitlength}{25.021\unitlength}}}}\put(916.690,314.558){\hyperref[sec:varational_basis_for_inference]{\phantom{\rule{195.944\unitlength}{25.021\unitlength}}}}\put(80.102,235.074){\hyperref[sec:pisc]{\phantom{\rule{166.679\unitlength}{25.021\unitlength}}}}\put(306.100,235.074){\hyperref[sec:path_inference]{\phantom{\rule{139.171\unitlength}{25.021\unitlength}}}}\put(513.351,235.074){\hyperref[sec:OT]{\phantom{\rule{151.367\unitlength}{25.021\unitlength}}}}\put(772.907,154.622){\hyperref[sec:sampling]{\phantom{\rule{98.536\unitlength}{25.021\unitlength}}}}\put(60.024,74.732){\hyperref[sec:RL]{\phantom{\rule{206.837\unitlength}{25.021\unitlength}}}}\put(472.732,74.732){\hyperref[sec:WGFApps]{\phantom{\rule{232.989\unitlength}{25.021\unitlength}}}}\put(772.907,74.732){\hyperref[sec:flows_diffusions]{\phantom{\rule{188.449\unitlength}{25.021\unitlength}}}}\end{picture}\includegraphics[width=\linewidth]{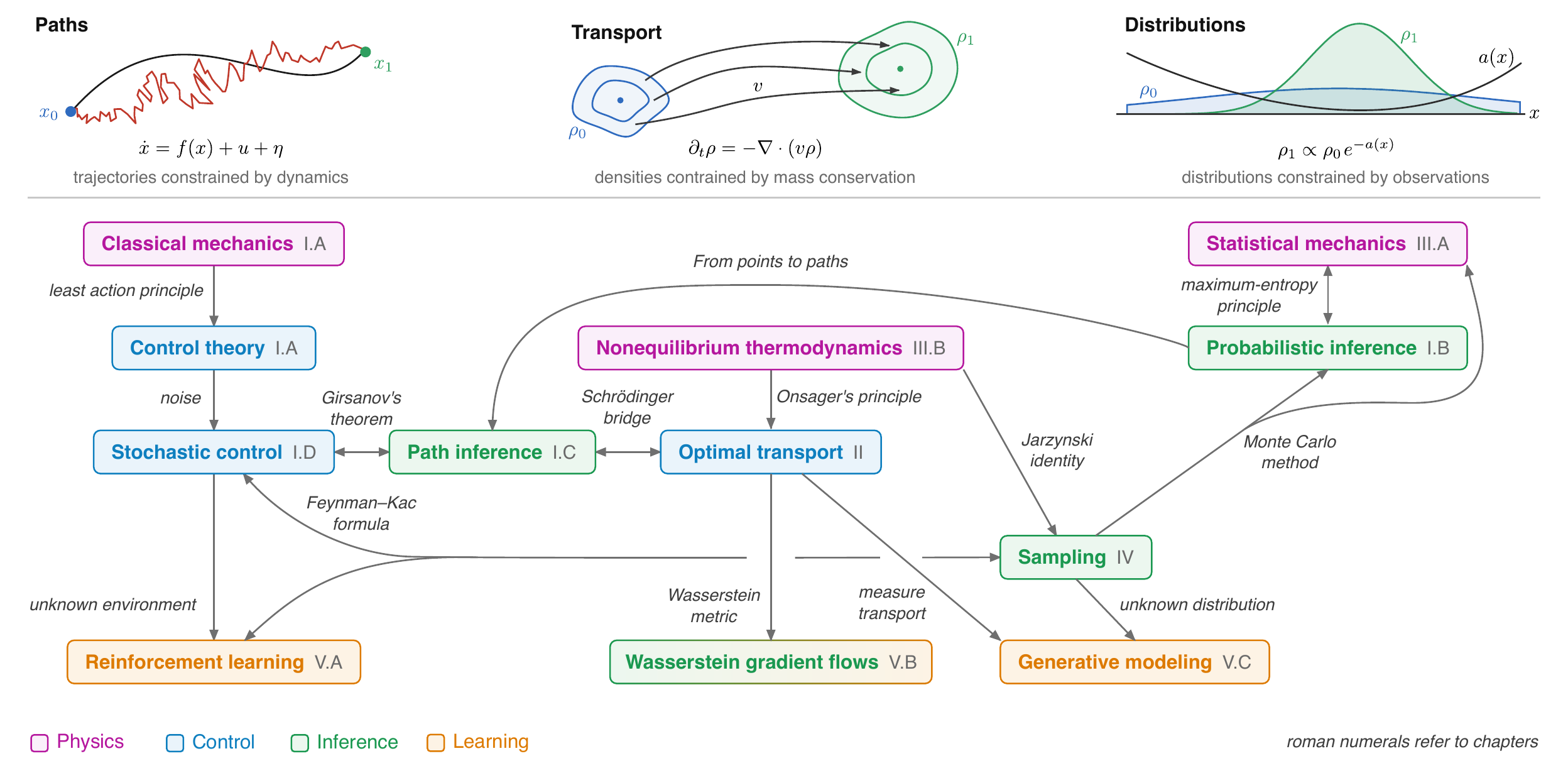}\endgroup
    \caption{\textbf{Graphical table of contents}: topics of the review and the connections between them.
    }
    \label{fig:visual_toc}
\end{figure*}
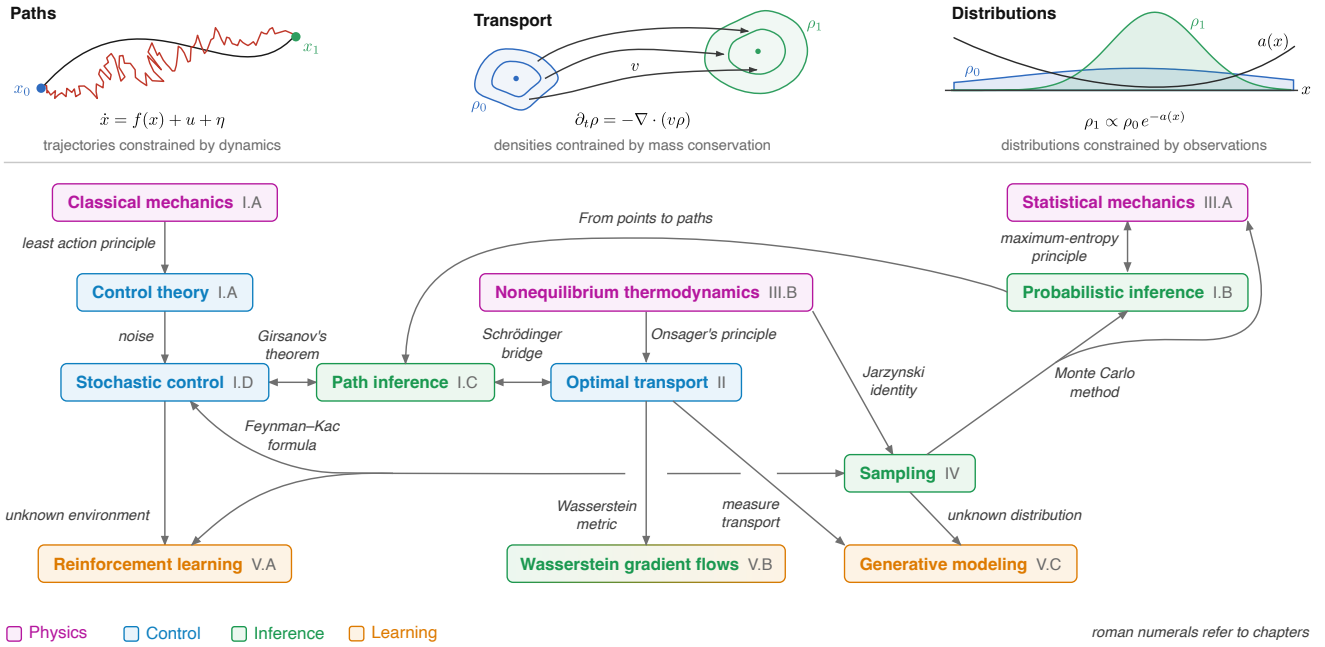

Recent decades have seen the development of remarkably powerful tools for learning complex structure from data. These methods have brought to the foreground certain key connections between subdisciplines of physics and applied mathematics. The conceptual ideas have appeared in different fields at different times, sometimes rediscovered and often expressed in different languages. In this review, we bring together some of these connections to highlight a fascinating conceptual thread that links five distinct fields: control, transport, inference, thermodynamics and learning. 
The visual table of contents, Fig.~\ref{fig:visual_toc}, lays out the topics we cover and the links between them.
These connections have proved fruitful for the development of powerful practical algorithms for generative modeling and control. 
We offer a guided tour through this thread and present selected applications in reinforcement learning, variational inference, and generative modeling. The review is written in the language of physicists and does not assume prior familiarity with these topics. Ideas from mechanics and thermodynamics are used as familiar anchors to introduce concepts in other areas.

The conceptual thread begins at the level of paths (left column in Fig.~\ref{fig:visual_toc}) in \hyperref[sec:variational_structure]{\emph{Variational structure of mechanics, control, and inference}}. This chapter develops a common variational language that is reused throughout the review. We introduce optimal control, which seeks the best way to steer a dynamical system---for example, guiding a rocket to a target while minimizing fuel consumption. We illustrate two complementary structural approaches for solving control problems. Introducing Lagrange multipliers for imposing a constraint on the dynamics leads to Hamiltonian equations. Another technique, known as dynamic programming, reformulates the optimization as a recursive equation. This reformulation leads to the Hamilton--Jacobi equation and, for stochastic control, its more general \emph{Hamilton--Jacobi--Bellman (HJB)} counterpart. Controls and rewards play roles analogous to velocities and the negative Lagrangian in classical mechanics. We then pass from optimizing individual paths to distributions (right column in Fig.~\ref{fig:visual_toc}). The \emph{Maximum Entropy Principle (MaxEnt)} selects a probability distribution that remains as close as possible to a reference distribution, as measured by the Kullback--Leibler (KL) divergence, while remaining consistent with data.
We then generalize these tools for probabilistic inference to distributions over paths.
For controlled drift-diffusive processes, Girsanov's theorem identifies this path-space KL divergence with a quadratic control cost, thereby connecting inference and control.

The next chapter, \hyperref[sec:OT]{\emph{Control and transport of densities}}, illustrates how the control of paths can be used to obtain a geometric description for the evolution of probability densities. 
The topic of transport brings together ideas from the study of (stochastic) paths and of distributions (center column in Fig.~\ref{fig:visual_toc}).
The first step is to impose constraints on both the initial and final densities of a stochastic control problem. These constraints lead to the \emph{Schr\"odinger bridge problem}, which finds the most likely ensemble of stochastic paths that connects the two densities. If one retains the endpoint constraints but describes only how initial and final positions are paired, the problem becomes \emph{entropic optimal transport}. In general, optimal transport asks how to move one density into another at minimum cost, much like reshaping a pile of earth while minimizing the cost of moving it. The entropic version adds a term that favors spreading probability among different pairings. For a quadratic cost, the zero-noise limit defines the $2$-Wasserstein distance, which has special mathematical properties. This distance equips the space of probability densities with a geometry and leads to \emph{Wasserstein gradient flows}: steepest-descent dynamics in which the evolving object is a probability distribution.

The chapter on \hyperref[sec:thermodynamics]{\emph{Thermodynamics}} grounds this geometry in physics. At equilibrium, statistical mechanics assigns probabilities to microscopic states while retaining only a small number of macroscopic constraints. Maximum-entropy inference reproduces the familiar Gibbs distributions from constraints on mean energy, making a heuristic connection between inference and equilibrium thermodynamics. Away from equilibrium, control and transport describe finite-time transformations between distributions. \textit{Onsager's least-dissipation principle} generates phenomenological field theories whose dissipation rate is bounded by the Wasserstein distance. For stochastic dynamics farther from equilibrium, the Wasserstein transport cost provides \emph{thermodynamic speed limits} on the entropy produced during a finite-time transformation. Nonequilibrium \emph{work fluctuation relations} provide a complementary link between fluctuations, free-energy differences, and the energetic cost of driving. Thermodynamics therefore supplies physical meaning for the variational structures introduced earlier, which in turn furnish quantitative limits on thermodynamic processes.

The fourth conceptual chapter, \hyperref[sec:sampling]{\emph{Sampling and control}}, presents sampling as a general approach to probabilistic inference, optimization, and control. Monte Carlo methods estimate averages from samples, while simulated annealing searches for low-energy states. Exploring possible trajectories and weighting them according to their rewards also links sampling to control and reinforcement learning; the Feynman--Kac formula makes this connection precise for a class of stochastic control problems. Sampling faces several distinct challenges, including high dimensionality, separated modes that impede efficient exploration, and observables whose averages are dominated by rare trajectories. When an unnormalized target density is available, \emph{Annealed Importance Sampling (AIS)} replaces direct sampling by a sequence of intermediate distributions and corrects the resulting nonequilibrium process through trajectory-dependent weights. The control perspective shows how guiding the dynamics toward important trajectories can reduce fluctuations. Although this ideal control is generally unknown, it provides a target for practical algorithms that learn progressively better sampling dynamics. This chapter thus closes the conceptual thread by turning efficient sampling into a control problem.

These connections provide a framework and common language for \hyperref[sec:application]{Applications} (bottom row in Fig.~\ref{fig:visual_toc}). The first application is \hyperref[sec:RL]{\emph{Reinforcement learning}}. Stochastic optimal control assumes that the dynamics of the environment are known. Reinforcement learning instead seeks good controls---called actions---by interacting with an environment and observing the resulting states and rewards. A policy gives the probability of taking each action in a given state and therefore defines an ensemble of trajectories. The connection between inference and control leads to \emph{Maximum Entropy Reinforcement Learning (MaxEnt RL)}, in which high-reward trajectories are favored while the policy retains some randomness. Applying dynamic programming to this objective leads to soft Q-learning and \emph{Soft Actor-Critic (SAC)}, practical algorithms for learning stochastically controlled paths when the dynamics are not known in advance.

The second application concerns \hyperref[sec:WGFApps]{\emph{Wasserstein gradient flows}}. The geometry of probability distributions provides a variational method for integrating Fokker--Planck equations. The same viewpoint helps describe large systems of interacting particles. In a suitable mean-field limit, the individual particles are replaced by an evolving density, which can obey a Wasserstein gradient flow. This description has also been used for wide neural networks, where many parameters are treated as interacting particles. Finally, in variational inference, a difficult target distribution is approximated within a simpler family of distributions. The KL objective has the form of a free energy, and projecting its Wasserstein gradient flow onto that family yields a practical approximation scheme.

The final application,
\hyperref[sec:flows_diffusions]{\emph{Generative modeling with flows and diffusion}}, 
treats targets represented by data rather than by a known density. Normalizing flows and diffusion models learn deterministic maps or stochastic dynamics that transform samples from a simple base distribution into samples from the target distribution. In this sense, generative modeling is a learned transport problem. Flow matching learns a velocity field that carries the base density toward the data, while diffusion models learn to reverse a gradual noising process. These methods need not find a minimum-cost transport; their goal is to learn a transformation that can be estimated from finite data and used efficiently for sampling. The earlier connections explain why ideas from transport, inference, control, and nonequilibrium physics repeatedly appear in their construction, and why additional guidance can be understood as a control that favors selected outcomes.
\\

\noindent\textbf{A brief note on notation.}
We have sought to unify notation across the chapters to the extent possible. The \hyperref[sec:notation]{notation tables} at the end of this review are organized by the
chapter in which a symbol is first introduced.

\section{Variational structure of mechanics, control and inference}\label{sec:variational_structure}

In this chapter, we develop a common variational language for mechanics, control, and inference. We introduce two complementary approaches for solving control problems. We then turn to inferring probability distributions from observations, and show how inference over stochastic paths leads back to a particular class of control problems.

\subsection{Finding optimal paths in classical mechanics and control theory}\label{sec:mechanics_control}

We begin with a gentle introduction to control theory. Consider an illustrative example\,---\,suppose we would like to guide a rocket towards a target in a certain amount of time by controlling its thrust. A large thrust accelerates the rocket but consumes more fuel, whereas insufficient thrust may not be enough to reach the target on time. How does one choose a thrust protocol so that the rocket reaches the target while consuming the least amount of fuel? Different control strategies generate different paths. In optimal control, we assume that there is a reward function that assigns a value or utility to each path. This reward may favor reaching a target, penalize control costs, or balance competing goals. Control theory offers mathematical tools for finding a control strategy that maximizes the total reward along the path. Although its variational roots are much older, modern optimal control took shape in the 1950s. Bellman's dynamic programming and the maximum principle of Pontryagin offer two powerful and complementary approaches for solving control problems~\cite{Bellman1954,Pontryagin1962,Dreyfus2002}. We now describe these two solution principles, which will 
reoccur throughout the
rest of the review. 
The principles are best illustrated by demonstrating their relationship with standard formulations of classical mechanics.

\paragraph*{The optimal control problem.}
In optimal control, a system is described by a (possibly multi-dimensional) state $x_t$ whose evolution is modulated by a control $u_t$. For the rocket, the state includes its position, velocity, and remaining fuel, while the control is the thrust. For convenience, we rescale time to the interval $0\leq t\leq 1$. 

In stochastic optimal control, the state obeys
\footnote{
    Eq.~\eqref{eq:soc-dynamics} is the physicists' notation for the stochastic process $\dd  x_t = g_t(x_t, u_t) \dd t + \sqrt{\varepsilon}  \dd W_t$, where $W_t$ is a standard Wiener process. 
    Throughout the review, we use the It\^o convention for stochastic calculus, in which $\eta_t$ is independent of the current state $x_t$. 
}
\begin{align}
   \dot{x}_t = g_t(x_t,u_t) + \sqrt{\varepsilon}\,\eta_t, \label{eq:soc-dynamics}
\end{align}
where $\varepsilon$ sets the noise scale and $\eta_t$ is standard Gaussian white noise, with $\langle\eta_{t,i}\eta_{s,j}\rangle=\delta_{ij}\delta(t-s)$. A reward $r_t(x_t,u_t)$ assigns a value per unit time to each state-control pair, while a terminal reward $r_1(x_1)$ assigns value to the final state. Starting from a fixed $x_0$, the optimized objective is
\begin{align}
    \begin{aligned}
    V_0(x_0) &= \max_{u_{0 \le s < 1} }
    \left\langle r_1(x_1) +\int_0^1 r_s(x_s,u_s)\,ds \;\middle|\; x_0
    \right\rangle,
    \end{aligned}\label{eq:soc-objective}
\end{align}
where the expectation is taken over the noise histories $\{\eta_t\}$ conditioned on the initial state $x_0$~\cite{evans2024control,oksendal2003stochastic,fleming1975deterministic}. The controls may depend on the observed state, but not on future noise. A fuel cost, for example, enters with a minus sign in $r_t$, while a terminal reward favors arrival near the target.

Perhaps intuitively, classical mechanics can be described as a special case of this optimization problem. Consider a particle with coordinates $x_t$ and velocity $\dot{x}_t$ whose motion is described by a Lagrangian $L_t(x,\dot{x})$. Its action is

\begin{align}
    S[x(\cdot)] = \int_0^1 L_t(x_t,\dot{x}_t)\,dt, \label{eq:action}
\end{align}
where $x(\cdot)$ denotes the full path. The principle of stationary action states that the physical path $\{x_t^*\}$ makes $S$ stationary under variations that leave the endpoints $x_0$ and $x_1$ fixed. Imposing stationarity leads to the Euler--Lagrange equations. We now map action minimization onto a control problem by choosing
\begin{align}
    g_t(x,u) &= u, \qquad \varepsilon=0, \\
    r_t(x,u) &= -L_t(x,u),
\end{align}
and imposing that the terminal endpoint is $x_1$. Here the control plays the role of velocity, $u_t=\dot{x}_t$. The endpoint constraint can equivalently be imposed through a terminal reward that is zero at $x_1$ and $-\infty$ elsewhere. Maximizing the reward then minimizes the action. Note that classical mechanics only requires stationarity, that is, physical trajectories need not be minima and can instead be saddles. The optimization therefore selects action-minimizing paths, while the variational equations also describe other stationary paths.

There are two structural approaches for solving the control problem. The first treats the dynamics in Eq.~\eqref{eq:soc-dynamics} as a constraint and introduces Lagrange multipliers, leading to a Hamiltonian formulation. The second recursively expresses the optimal solution in terms of optimal solutions to shorter remaining problems, leading to the Hamilton--Jacobi--Bellman equation. Relating the two will identify the Lagrange multiplier as the gradient of a \emph{value function}.

\paragraph*{Lagrange multipliers and Hamiltonian dynamics.}
A Lagrange multiplier converts a constrained optimization problem into a stationary problem in an enlarged set of variables. Variation with respect to the multiplier enforces the constraint, while variation with respect to the original variables supplies the optimality equations. To see this, let's first take the deterministic limit $\varepsilon=0$. Introducing a time-dependent multiplier $p_t$ for the constraint $\dot{x}_t=g_t(x_t,u_t)$ gives the augmented objective
\begin{align}
    \begin{aligned}
    \widetilde{V}(x,u,p)
    &= r_1(x_1)
    + \int_0^1 \left[
        r_t(x_t,u_t)\right.
    \\
    &\qquad\left.
        + p_t\cdot\bigl(g_t(x_t,u_t)-\dot{x}_t\bigr)
    \right]dt,
    \end{aligned}\label{eq:aug_obj}
\end{align}
with the initial endpoint $x_0$ held fixed. The sign of $p_t$ is conventional. Define the control Hamiltonian and its optimized form by
\begin{align}
    \mathcal{H}_t(x,p,u) &= p\cdot g_t(x,u)+r_t(x,u), \\
    H_t(x,p) &= \max_u \mathcal{H}_t(x,p,u).
\end{align}
Variation with respect to $p_t$ and $x_t$ (after integrating by parts the term involving $\dot{x}_t$ in Eq.~\eqref{eq:aug_obj}), together with maximization over $u$, gives the necessary optimality conditions~\cite{Pontryagin1962,evans2024control}. When $H_t$ is differentiable, these take the form
\begin{align}
    \dot{x}_t &= \nabla_p H_t(x_t,p_t), \\
    \dot{p}_t &= -\nabla_x H_t(x_t,p_t), \\
    u_t^* &= \argmax_u \mathcal{H}_t(x_t,p_t,u).
\end{align}
If the terminal state is free, the endpoint variation gives $p_1=\nabla r_1(x_1)$, whereas if it is fixed, the terminal variation vanishes. For a free terminal state, the multiplier equation is thus evolved backward from the terminal condition along the trajectory. We will see below that $p_t$ measures how the optimal future reward changes under a displacement of the state.

For the specialization $g_t(x,u)=u$ and $r_t(x,u)=-L_t(x,u)$, the optimized Hamiltonian becomes
\begin{align}
    H_t(x,p)
    = \max_u\left\{p\cdot u-L_t(x,u)\right\},
\end{align}
which is the Legendre transform of the Lagrangian. The stationarity condition with respect to $u$ gives
\begin{align}
    p_t=\nabla_u L_t(x_t,u_t)=\nabla_{\dot{x}}L_t(x_t,\dot{x}_t),
\end{align}
so the multiplier is the canonical momentum, and the remaining conditions are Hamilton's equations.

\paragraph*{Dynamic programming and the Hamilton--Jacobi equation.}
The second approach involves keeping track of the maximum reward that remains available from every possible intermediate state (`reward-to-go'). For the rocket, suppose we already knew the maximum total future reward attainable from every state $x_{t+dt}$ at time $t+dt$. The value of choosing a particular thrust at time $t$ is its immediate reward, including the fuel cost, plus the expected best future reward from the state it produces. The optimal thrust is the one that maximizes this sum. In other words, if the expected best future reward is known, solving for the optimal thrust involves optimizing over a single step rather than over a full trajectory. 

This method for solving a problem by breaking it into shorter subproblems and reusing their solutions is called \emph{dynamic programming}~\cite{Bellman1954}.  In this setting, the key observation is that the control at the next state for an optimal strategy must itself be optimal. This gives a recursive relation between the best reward now and the best reward at the next time step, which we can solve backward from the terminal reward. The reuse of partial results is analogous to the transfer-matrix method in statistical mechanics, where a partial sum over configurations is updated one site at a time~\cite{Mezard.Montanari2009}.

The central object in dynamic programming is the value function
\begin{align}
    \begin{aligned}
    V_t(x)
    &= \max_{u_{t \le s < 1}}
    \left\langle
        \int_t^1 r_s(x_s,u_s)\,ds +r_1(x_1)
        \;\middle|\; x_t=x
    \right\rangle,
    \end{aligned}\label{eq:value-function}
\end{align}
with terminal condition $V_1(x)=r_1(x)$. Thus, $V_t(x)$ is the largest expected total reward that can still be collected from time $t$ onward, starting at $x$, including the terminal reward. It does not include rewards already received. This is the optimal \emph{reward-to-go}, and $V_0(x_0)$ is the objective in Eq.~\eqref{eq:soc-objective}.

The controls after time $t+dt$ have already been optimized inside $V_{t+dt}$, so only the control applied during the first short interval remains to be chosen. Splitting the total reward into the immediate and future contributions gives Bellman's recursion, to first order in $dt$,
\begin{align}
    \begin{aligned}
    V_t(x)
    &= \max_u \left\{
        r_t(x,u)dt\right.
    \\
    &\quad\left.
        + \left\langle
            V_{t+dt}\!\left(
                x+g_t(x,u)dt+\sqrt{\varepsilon}\,dW_t
            \right)
        \right\rangle
    \right\}
    \\
    &\quad+o(dt).
    \end{aligned}\label{eq:bellman-recursion}
\end{align}
Here $dW_t$ is a Gaussian noise increment with zero mean and $\langle dW_{t,i}dW_{t,j}\rangle=\delta_{ij}dt$, and the expectation is over this increment. Assuming the value function is smooth, expanding $V_{t+dt}$ to first order in time and second order in the state increment, averaging over the noise, and retaining terms of order $dt$ gives the Hamilton--Jacobi--Bellman (HJB) equation~\cite{oksendal2003stochastic,fleming1975deterministic}
\begin{align}
    \begin{aligned}
    &\partial_t V_t(x)
    + \frac{\varepsilon}{2}\nabla^2V_t(x)
    \\
    &\quad+ \max_u\left\{
        g_t(x,u)\cdot\nabla V_t(x)+r_t(x,u)
    \right\}
    =0.
    \end{aligned}\label{eq:hjb-stoch}
\end{align}
The control chosen at state $x$ is the maximizer of the local expression,
\begin{align}
    u_t^*(x)
    = \argmax_u\left\{
        g_t(x,u)\cdot\nabla V_t(x)+r_t(x,u)
    \right\}. \label{eq:u_maximizer}
\end{align}
Thus, when $g_t$ and $r_t$ are known, the difficult optimization over an entire control history can be replaced by a backward equation for $V_t$ followed by a local maximization at each state and time. Finding this function over a high-dimensional state space can nevertheless be difficult.

In the deterministic limit, the Hamiltonian $H_t(x,p)$ introduced above allows us to write the HJB equation as
\begin{align}
    \partial_t V_t(x)
    + H_t\bigl(x,\nabla V_t(x)\bigr)=0. \label{eq:hjb_det}
\end{align}
This equation also makes the connection between dynamic programming and the Lagrange-multiplier formulation explicit. Where $V_t$ and $H_t$ are smooth, set $p_t=\nabla V_t(x_t)$ along an optimal trajectory. The state evolves as
\begin{align}
    \dot{x}_t= g_t(x_t,u_t^*) = \nabla_p H_t(x_t,p_t).
\end{align}
The total time derivative of $p_t$ along this trajectory is
\begin{align}
    \dot{p}_t = \partial_t\nabla V_t(x_t)
    + (\dot{x}_t\cdot\nabla)\nabla V_t(x_t).
\end{align}
Taking the spatial gradient of Eq.~\eqref{eq:hjb_det} gives
\begin{align}
    \partial_t\nabla V_t + \nabla_x H_t
    + (\nabla_p H_t\cdot\nabla)\nabla V_t=0,
\end{align}
where $\nabla_x H_t$ denotes differentiation with respect to the explicit state argument of $H_t(x,p)$ while holding $p$ fixed, and the Hamiltonian derivatives are evaluated at $p=\nabla V_t(x)$. Using $\dot{x}_t=\nabla_p H_t$ and combining the above two equations, we get
\begin{align}
    \dot{p}_t=-\nabla_x H_t(x_t,p_t).
\end{align}
Thus, the trajectories along which the Hamilton--Jacobi equation is solved, its \emph{characteristics}, obey precisely the Hamilton equations obtained from the Lagrange-multiplier approach. The multiplier $p_t$ is the gradient of the value function because it measures the sensitivity of the optimal future reward to a displacement of the current state.

For action-minimizing trajectories in classical mechanics, the value function is the negative of the minimum action remaining to the fixed endpoint $x_1$,
\begin{align}
    V_t(x)
    = -\min_{\substack{\{x_s\}:x_t=x,}}
    \int_t^1 L_s(x_s,\dot{x}_s)\,ds.
    \label{eq:hj-value}
\end{align}
On a minimizing trajectory, variation of the initial endpoint gives $\delta V_t=p_t\cdot\delta x$, while variation with respect to the velocity gives $p_t=\nabla_{\dot{x}}L_t$. Hence
\begin{align}
    p_t=\nabla V_t(x_t)=\nabla_{\dot{x}}L_t,
\end{align}
so the multiplier of optimal control becomes the canonical momentum of mechanics.

The two approaches thus give complementary descriptions: the \emph{Lagrangian} viewpoint follows individual trajectories, while the \emph{Eulerian} viewpoint of dynamic programming finds a field $V_t(x)$ over all states. Note that we have assumed that the reward $r_t$ and dynamics $g_t$ are known. Reinforcement learning (RL) addresses the challenging problem of finding the optimal control when this information is not available in advance. In RL, one interacts with the environment to learn which controls, or \emph{actions}, produce large cumulative reward from observed state transitions and rewards. The value function remains central because it provides an estimate of the long-term consequence of acting from each state. We return to the RL framework in Sec.~\ref{sec:RL} of the \hyperref[sec:application]{Applications} chapter.

\subsection{Variational basis for probabilistic inference and control}\label{sec:varational_basis_for_inference}

We now turn to a different but related problem, wherein measurements constrain probabilities of random degrees of freedom, instead of control functions that steer individual paths. Since measurements rarely provide every microscopic detail of a system, we describe the system by a probability distribution that we seek to infer from the available observations. Bayesian inference is an example of such a capability: what is the distribution of $x$, given observations $A$ and a prior $p^0(x)$ (which encodes existing knowledge about the system~\cite{Gelman.etal2013})? By Bayes' rule, the conditional distribution is $p(x|A) = p(A|x) p^0(x)/p^0(A)$, thereby recasting the problem as specifying a model for $A$ given $x$. More generally, the central task of probabilistic inference is to determine conditional probabilities. In this chapter, we show that this task can be formulated as constrained optimization\,---\,a variational problem under constraints arising from observations. This formulation highlights a conceptual connection to the optimization problems of mechanics and control (Sec.~\ref{sec:mechanics_control}), and provides the theoretical basis for path-integral stochastic control in Sec.~\ref{sec:pisc}. Later in the review, variational probabilistic inference will serve as a foundation for practical inference algorithms:  Chapter~\ref{sec:sampling} uses control theory to address inference problems, while Sec.~\ref{sec:variational_inference} considers variational inference, where the system distribution is approximated using a parametrized \textit{ansatz}.

Here, we will introduce maximum-entropy inference, a method for finding a probability distribution for the system without prior knowledge, constrained to be consistent with observations. The observations take the form of measurements of moments like the mean or variance, which, however, do not fully determine the distribution. There is thus a many-to-one correspondence between microscopic models and the measurements. How to choose one model out of the many?  One well-motivated approach is to construct a model of behavior $p^A(x)$ that is closest to a reference distribution $p^0(x)$, but remains consistent with the observations $A$. As we will show below, a natural candidate for measuring closeness is the Kullback-Leibler divergence,

\begin{equation}
    \dkl(p\Vert q)=\int\mathrm{d}x\,p(x)\log\frac{p(x)}{q(x)}.
\end{equation}
measuring how different (``the excess surprise'') the distribution $p$ is from $q$~\cite{cover1999elements}. We will furthermore see that using Lagrange multipliers that enforce consistency with observed paths has a natural interpretation as rewards. This choice will establish a conceptually and practically relevant connection between probabilistic inference and control.

\subsubsection{Sanov's theorem and the inference method}

We first make a plausibility argument for the inference method through Sanov's theorem. It equips $\dkl$ with a probabilistic interpretation as a measure of how a distribution estimated from samples deviates from the underlying true distribution.
Namely, suppose that $N$ independent and identically distributed (IID) samples $\{x_n\}_{1,\ldots,N}$ are drawn from a base measure, $p^0(x)$. Having drawn these samples, to what extent does the empirical measure, 
\begin{equation}
    \hat{p}(x)=\frac1N\sum_{n=1}^N\delta(x-x_n),
\end{equation}
depart from $p^0(x)$? Sanov's theorem~\cite{Sanov1957,TouchettePR2009} states that the probability of the empirical measure, $\mathbb P[\hat p]$, deviating from the ground truth $p^0(x)$ is exponentially suppressed with an increasing number of samples $N$:

\begin{equation}
    \mathbb{P}[\hat p]\sim e^{-N\dkl(\hat p\Vert p^0)},\label{eq:MeasureEmpiP}
\end{equation}
Eq.~\eqref{eq:MeasureEmpiP} is achieved using large-deviations theory, which concerns finding the asymptotic decline in probability of a rare event, wherein the probability typically takes the form $\sim e^{- N I}$ with $I$ called the rate function. The rate function here is the $\dkl$, which possesses two appealing properties: $\dkl=0$ and minimal if and only if $\hat p=p^0$, and the spread itself gets smaller as $N\to\infty$: $\hat p-p^0\sim N^{-1/2}$. The rarer the event, the larger the KL divergence, and the more the probability is suppressed.

We will prove Eq.~\eqref{eq:MeasureEmpiP} via a path-integral calculation, exemplifying the path-integral machinery we will encounter later in the review. We compute the distribution of empirical measures $\mathbb{P}[\hat p]$ by definition, from the expectation of the functional delta function over the drawn samples,

\begin{equation}
    \begin{aligned}
    \mathbb{P}[\hat p]&=\prod_{m=1}^N\left[\int\mathrm{d}x_mp^0(x_m)\right]
    \\
    &\quad\times\delta\left[\hat{p}(x)-\frac1N\sum_{n=1}^N\delta(x-x_n)\right].
    \end{aligned}\label{eq:MeasureEmpiPdef}
\end{equation}
Expressing the delta function via its Fourier transform, $\delta[f(x)]\propto\int\mathrm{D}[\mathrm{i}\mu(x)]e^{\int\mathrm{d}x\mu(x)f(x)}$, and noting the identical and independent integrals over samples, we rewrite

\begin{align}
    \label{eq:MeasureEmpiPdefFT}
    \mathbb{P}[\hat p]\propto\int\mathrm{D}[\mathrm{i}\mu(x)]\exp&\left\{N\int\mathrm{d}x\frac{\mu(x)}N\hat{p}(x)\right.
    \\
    &\left.+N\log\left[\int\mathrm{d}x'p^0(x')e^{-\mu(x')/N}\right]\right\} \nonumber 
\end{align}
Since $N\to\infty$, we note that only $\mu(x)\sim N$ contributes to the Fourier integral, overall implying that the argument of the exponential is $\sim N$. This justifies a saddle-point evaluation of the integral\,---\,the saddle point $\mu^*(x)$ satisfies
\begin{equation}
    \frac{p^0(x)e^{-\mu^*(x)/N}}{\int \mathrm{d}x'p^0(x')e^{-\mu^*(x')/N}}=\hat p(x),\label{eq:SPmu}
\end{equation}
Inserting Eq.~\eqref{eq:SPmu} in Eq.~\eqref{eq:MeasureEmpiPdefFT}, we obtain Eq.~\eqref{eq:MeasureEmpiP}.

Sanov's theorem ensures that the deviations of the estimated measure from the ground truth are minimized with more samples. However, during probabilistic inference, the choice of a reference distribution is often that which can be sampled easily, meaning that there is no inherent reason why the underlying model is that of the base $p^0$. This suggests that if we observe data inconsistent with $p^0$ and seek the most likely alternative explanation, we should find the distribution $p^A$ that is closest to $p^0$ in KL divergence while remaining consistent with the observations $A$. By observations we mean, \textit{e.g.}, measurable moments or state occupancies of the system, dictated by the unknown underlying distribution.
For instance, later we will be interested in inferring and controlling the distributions over paths. Via a large-deviation principle, we may motivate the fact that the distribution $p^A$ closest to $p^0$, consistent with observations $A$, satisfies
\begin{equation}
    \begin{gathered}
    p^A=\arg\min_{p}
    \dkl(
        p
        \Vert
        p^0
    ),
    \\
    \text{s.t.}\quad\int \mathrm{d}{x} p(x)=1\quad\text{and}\quad
    \int\mathrm{d}{x} p(x) a(x)=A.
    \end{gathered}
\label{eq:large-deviations-heuristic-form}
\end{equation}
This will serve as the basis for probabilistic inference throughout this review.

To motivate this inference, we recall that not all $\hat p(x)$s are possible if the value of a certain observable, $A$, is known and thus imposed as a constraint. The space of empirical distributions is thus reduced to those which satisfy $A=N^{-1}\sum_{n=1}^Na(x_n)$. Therefore, Eq.~\eqref{eq:MeasureEmpiPdef} should be rewritten instead as 

\begin{equation}
    \begin{aligned}
    \mathbb{P}[\hat p| A]&=\prod_{m=1}^N\left[\int\mathrm{d}x_mp^0(x_m)\right]
    \\
    &\quad\times\delta\left[\hat{p}(x)-\frac1N\sum_{n=1}^N\delta(x-x_n)\right]
    \\
    &\quad\times\delta\left[A-\frac1N\sum_{n=1}^Na(x_n)\right].
    \end{aligned}
\end{equation}
Even more so, note that Eq.~\eqref{eq:SPmu} implies additionally that $\hat p(x)$ must be normalized ($\int\mathrm{d}x\hat p(x)=1$). The rest of the procedure to obtain Eq.~\eqref{eq:MeasureEmpiP} remains unchanged other than the additionally imposed observables. Following the above saddle-point calculation, Eq.~\eqref{eq:MeasureEmpiP} can be rewritten as

\begin{equation}
    \begin{aligned}
    \mathbb{P}[\hat p | A]\sim\exp&\biggl\{-N\dkl(\hat p\Vert p^0)
    \\
    &+\eta^*\left[\int \mathrm{d}x\hat p(x)-1\right]
    \\
    &+\lambda^*\left[\int \mathrm{d}x\hat p(x)a(x)-A\right]\biggr\},
    \end{aligned}\label{eq:MeasureEmpiPconstr}
\end{equation}
where $\eta^*$ and $\lambda^*$ are such that the overall rate function (the argument of the exponential) is minimized. (This calculation should be regarded as qualitative, as the saddle point in $\eta$ and $\lambda$ may often not be justified.)
Therefore, we conclude that Sanov's theorem motivates turning the inference problem of $p^A(x)$ into a constrained optimization problem, Eq.~\eqref{eq:large-deviations-heuristic-form}. 

\subsubsection{Maximum-entropy inference}\label{sec:MEM}

Sanov's theorem is closely related to maximum-entropy inference. Suppose, for simplicity, that the prior is the uniform distribution, $p^0(x)=1/\int \mathrm dx'$. In this case, the KL divergence simplifies to the negative of the Shannon entropy of a distribution $p(x)$~\cite{cover1999elements},

\begin{equation}
    H[p]=-\int \dd{x}p(x) \log p(x).\label{eq:ShannonEnt}
\end{equation}
Shannon argued that Eq.~\eqref{eq:ShannonEnt} is a functional that can quantify information~\cite{Shannon1948}. When the only available information is a set of measured expectation values, the ``least-committal'' distribution consistent with them is the one that maximizes entropy~\cite{JaynesPR1957,JaynesIEEE1982}.

Mathematically put, suppose a set of observables was measured, $A=\int \mathrm{d}x p(x) a(x)$. Combined with normalization, according to Eq.~\eqref{eq:large-deviations-heuristic-form} (with entropy replacing the KL divergence) the inferred distribution is that which satisfies
\begin{equation}
    \begin{gathered}
    p^A=\arg\max_pH[p],\quad\text{s.t.}\quad\int \mathrm{d}{x} p(x)=1\quad\text{and}
    \\
    \left\{\int\mathrm{d}{x} p(x) a_m(x)=A_m\right\}_{m=1,\ldots,M}.
    \end{gathered}\label{eq:MaxEnt}
\end{equation}
Using the method of Lagrange multipliers, one finds the solution
\begin{equation}
    p^A(x)=\frac1Z\exp\left[-\sum_{m=1}^M\lambda_ma_m(x)\right],\label{eq:MaxEnt_q}
\end{equation}
where
\begin{equation}
    Z=\int \dd{x} \exp\left[-\sum_{m=1}^M\lambda_ma_m(x)\right]\label{eq:MaxEnt_Z}
\end{equation}
arose from the Lagrange multiplier enforcing normalization. $\lambda_m$ are the remaining Lagrange multipliers, determined from the implicit relation

\begin{equation}
    A_m=-\frac{\partial\log Z}{\partial\lambda_m}.\label{eq:MaxEnt_lambda}
\end{equation}
The maximized entropy reads

\begin{equation}
    H[p^A]=\log Z-\sum_{m=1}^M\lambda_mA_m.\label{eq:MaxEnt_H}
\end{equation}
The above is the general statement for an inference scheme based on the maximum-entropy principle. It can be readily generalized to the case where the prior $p^0(x)$ is not uniform, \textit{e.g.}, the inferred distribution will become the ``exponential tilt'' of the base distribution

\begin{equation}
    p^A(x)\propto p^0(x)\exp\left[-\sum_{m=1}^M\lambda_ma_m(x)\right].
\end{equation}
In the above expressions, $a_m(x)$ can be polynomials, in which case $A$ gives the moments of the underlying distribution. However, a particularly relevant observable for later is that which checks if $x$ is in a region $\mathcal{A}$, $a(x) := \mathbf{1}_{\mathcal{A}}(x)$, so $A = \int a(x) p(x)dx =1$. The constraint is then $p^A(x\in\mathcal{A}) := p^A(\mathcal{A}) = 1$. The maximum-entropy solution Eq.~\eqref{eq:MaxEnt} is exactly the conditional probability distribution. At the same time, Bayes' rule means $p(x|\mathcal{A}) = \mathbf{1}_{\mathcal{A}}(x)p^0(x) /  p^0(\mathcal{A})$. Thus, the constraint $p^A(\mathcal{A}) = 1$ implies:

\begin{align}
    \begin{aligned}
    &\left.\dkl(p^A\Vert p^0)\right|_{p^A(\mathcal{A})=1}
    =\dkl(\mathbf{1}_{\mathcal{A}}\cdot p^A\Vert\mathbf{1}_{\mathcal{A}}\cdot p^0)
    \\
    &=\dkl(p^A\Vert p(\cdot|\mathcal{A}))-\log p^0(\mathcal{A})
    \end{aligned}
\end{align}
whose minimum is $p^A(x) = p(x|\mathcal{A})$. In this setting, maximum-entropy inference exactly reproduces Bayesian inference. 

The minimum-KL-divergence formulation is closely related to Bayesian inference more generally. The ``update'' ratio $p^A(x)/p^0(x)=e^{\cdots}$ can be interpreted as the likelihood ratio, $p(A| x)$ by Bayes' rule\,---\,the exponential factor reweighing the prior according to the information encoded by the imposed constraints. In this sense, the minimum-KL solution is thus a Bayesian-style update of the reference distribution, with the exponential tilt playing the role of an (unnormalized) likelihood. These general functional forms will appear repeatedly in the upcoming sections for specialized observables. We will furthermore see that the Lagrange multipliers will take the form of rewards, thereby generalizing ideas from Sec.~\ref{sec:mechanics_control} to path-integral stochastic control. Furthermore, the $\dkl$ objective arises naturally in the context of Variational Inference (VI), the problem of approximating a complex distribution by a simpler\,---\,typically parametric\,---\,one, which will be discussed in Sec.~\ref{sec:variational_inference}. 

Another noteworthy connection of maximum-entropy models is to exponential-family models~\cite{ZhuNC1997,WainwrightBOOK2008}. Maximum-entropy distributions subject to linear expectation constraints take the exponential-family form of Eq.~\eqref{eq:MaxEnt_q}, in which the Lagrange multipliers $\lambda_m$ enter linearly in the exponent and determine the partition function $Z$. Beyond their conceptual connection to maximum entropy, exponential-family models are useful because their parameters can be estimated through convex optimization, making them particularly attractive for statistical inference and applications such as image processing.

In addition to Sanov's theorem and the Bayesian-inference connections, there is a subtle connection of maximum-entropy inference to equilibrium statistical mechanics. Sanov's theorem provides concrete grounds for the maximum-entropy approach, as large deviations are exponentially suppressed as the number of samples increases. In the thermodynamic limit the dimensionality of $x$ (the number of particles), reaching as much as $\sim10^{23}$, far exceeds any feasible number of samples that one may obtain. Instead, the second law of thermodynamics and statistical mechanics provide the theoretical basis for finding the microscopic statistical model of materials. Nevertheless, both mechanisms\,---\,equilibrium statistical mechanics and maximum-entropy modeling\,---\,provide the same mathematical structures~\cite{TouchettePR2009}. We shall discuss this correspondence in Sec.~\ref{sec:energy-based_modelling}.

\subsection{Probabilistic inference over paths}\label{sec:path_inference}
In Sec.~\ref{sec:mechanics_control}, we considered how to control a system with known dynamics to maximize a prescribed reward.
In Sec.~\ref{sec:varational_basis_for_inference}, we saw how inference problems can be formulated as optimization problems in which we seek a distribution that is consistent with observed data while remaining as close as possible to a prior distribution.
We now turn to the application of the large-deviation principle to the inference of distributions over paths. We will exploit the tools of stochastic optimal control to solve these path-distribution inference problems.

In a stochastic system, repeated realizations of the noise generate different trajectories, so in the context of stochastic optimal control, a given control determines a probability distribution over paths.
When investigating the properties of a distribution over paths, we will exploit that, in generic settings, there is a correspondence between stochastic differential equations (SDEs) and distributions over paths.
In particular, an SDE determines a distribution over paths in which each realization of the noise generates a different trajectory.
Under certain regularity conditions, there is a converse relationship where a distribution over paths can be represented by an SDE whose drift and diffusion coefficients are determined by the properties of the distribution; we will restrict the inference problems we consider to those for which this correspondence holds.
We use uppercase $P$ for distributions over complete paths, reserving lowercase $p$ for state densities. If we have a prior distribution over paths, $P^0[x(\cdot)]$, and we observe some data about the system, the large-deviation principle tells us how to find the distribution over paths, $P^u[x(\cdot)]$, that is most similar to the prior while remaining consistent with the observations.
Thus, because of the path distribution-SDE correspondence, the result of inference is a new SDE that describes the dynamics of the system under the inferred distribution over paths.
Importantly, the inferred SDE will contain \emph{additional forces} that will be determined by an optimization problem.
To see this, consider a simple example of such a path-conditioning problem, namely, the Brownian bridge.
In the Brownian bridge, we consider a particle known to start at a point $x_0$ at time $t=0$ and undergo Brownian motion, but we impose the condition that the particle is at a prescribed target $y$ at time $t=1$:

\begin{equation}
\begin{gathered}
    \dot x_t
    =
    \sqrt{\varepsilon}\,\eta_t,
    \\
    \text{conditioned on}
    \quad
    x_{t=0}=x_0,
    \quad
    x_{t=1}=y.
\end{gathered}
\label{eq:brownian_bridge}
\end{equation}
Without this conditioning, the particle would hit $y$ at time $t=1$ formally with probability zero.
Within the ensemble of such conditioned paths, there is apparently an additional force that drives the particle to $y$ at time $t=1$, which is not present in the unconditioned dynamics.
Here, we will consider similar problems in which we condition on more general observations, and we will see that the large-deviation principle provides a natural way to infer the additional forces that emerge from this conditioning.

\subsubsection{From inference to distributions over paths}

Consider the prior distribution over paths, $P^0[x(\cdot)]$, induced by the passive dynamics (i.e., the dynamics without any additional forces),
\begin{equation}
\begin{aligned}
    \dot{x}_t
    =
    f_t(x_t)
    +
    \sqrt{\varepsilon}\,\eta_t.
\end{aligned}
\end{equation}
We seek to infer how this ensemble changes when observations constrain its properties.
We will represent the inferred ensemble through an additional control,
\begin{equation}
\dot{x}_t=f_t(x_t)+u_t(x_t)+\sqrt{\varepsilon}\,\eta_t,
\end{equation}
and denote its path distribution by $P^u[x(\cdot)]$. The reference and controlled processes start at the same fixed $x_0$, as in Sec.~\ref{sec:mechanics_control}. The marginal density at a single time is
\[
    p_t^u(x)=\left\langle\delta(x-x_t)\right\rangle_{P^u}.
\]
We omit the superscript on a marginal when only one ensemble is under discussion. The large-deviation argument applies here with each sample being an entire trajectory.
Among empirical ensembles compatible with the observations, it identifies those most likely to arise from repeated sampling of the passive dynamics.
In the limit of many trajectories, the empirical path distribution concentrates on the corresponding inferred distribution.

Suppose the observations specify averages $A_m$ of trajectory observables $a_m[x(\cdot)]$, so that

\begin{equation}
    \int \mathcal{D}x\,P^u[x(\cdot)]a_m[x(\cdot)]=A_m.
\end{equation}
Eq.~\eqref{eq:large-deviations-heuristic-form} then states that the most likely path measure is found by minimizing over $P^u$ the KL divergence from the prior $P^0$~\cite{PresseRMP2013},

\begin{equation}
\begin{aligned}
    \dkl\qty(
        P^u\Vert P^0
    )
    =
    \int
    \mathcal{D}x
    \,
    P^u[x(\cdot)]
    \log
    \frac{
        P^u[x(\cdot)]
    }{
        P^0[x(\cdot)]
    }.
\end{aligned}
\end{equation}
while satisfying these constraints.
These constraints can be enforced with Lagrange multipliers $r_m$. We use the reward sign convention $r_m=-\lambda_m$, where $\lambda_m$ is the multiplier in the convention used in Sec.~\ref{sec:MEM}, so a positive reward appears with a positive sign in the exponential tilt. This gives the constrained objective:

\begin{equation}
\begin{aligned}
    \mathcal L
    &=
    \dkl\qty(
        P^u\Vert P^0
    )
    \\
    &\quad-
    \sum_m
    r_m
    \qty[
        \int
        \mathcal{D}x
        \,
        P^u[x(\cdot)]
        a_m[x(\cdot)]
        -
        A_m
    ].
\end{aligned}
\end{equation}
Taking the variation with respect to the path distribution and setting it to zero gives the minimizing distribution. When represented by the optimal control $u^*$, it reads

\begin{equation}
\begin{aligned}
    P^{u^*}[x(\cdot)]
    &\propto
    P^0[x(\cdot)] \exp\qty(
        \sum_m
        r_m
        a_m[x(\cdot)]
    ),
\end{aligned}
\label{eq:maxentrl_titled_path_measure}
\end{equation}
meaning the inferred path measure is an exponential tilt of $P^0$ where the Lagrange multipliers $r_m$ enforce consistency with the observations.
These Lagrange multipliers can be roughly interpreted as inferred rewards that the system, conditioned on the observations, is receiving.
Because of the correspondence between path distributions and SDEs, the `tilts' $r_m$ to the prior path distribution translate into apparent forces in the SDE describing paths drawn from $P^{u^*}$. Our goal now is to solve the constrained optimization problem introduced to get an expression for $u_t^*$ which is the apparent force that emerges when the system is conditioned on the observations.

\subsubsection{Girsanov's theorem}{\label{sec:Girsanov}}

To do this, we exploit Girsanov's theorem, which allows us to express the change of measure from $P^0$ to $P^u$ in terms of the control $u_t$~\cite{Girsanov1960}.
Girsanov's theorem is a fundamental result in stochastic calculus because it provides a way to relate probability measures of stochastic processes with different drift terms. In the context of inference over paths, Girsanov's theorem allows us to express the KL divergence between the prior and inferred path measures in terms of the added control $u_t$. 
This will allow us to formulate the inference problem as an optimization problem over the control, which can then be solved using techniques from stochastic optimal control.

Girsanov's theorem states that the ratio of the path measures of two SDEs which are driven by (potentially inhomogeneous) Gaussian noise and have differing drift terms can be expressed in terms of the difference in the drift terms. To see this, note that the controlled dynamics can be made to look like the passive dynamics by relabeling the noise:
\begin{equation}
\begin{aligned}
    \dot{x}_t
    &=
    f_t(x_t)
    +
    u_t(x_t)
    +
    \sqrt{\varepsilon}\,\eta_t^u
    \\
    &=
    f_t(x_t)
    +
    \sqrt{\varepsilon}
    \qty[
        \eta_t^u
        +
        \frac{1}{\sqrt{\varepsilon}}u_t(x_t)
    ].
\end{aligned}
\end{equation}
Here $\eta_t^u$ denotes the zero-mean noise under the controlled path measure. 
Thus, the same realized trajectory that requires the noise $\eta_t^u$ under the controlled dynamics would require the shifted noise
$
    \eta_t^0
    =
    \eta_t^u
    +
    \frac{1}{\sqrt{\varepsilon}}u_t(x_t)
$
under the passive dynamics.
The ratio between the two measures can then be written as:
\footnote{
Here, the integral $\int_0^1 \dd{t} u_t(x_t) \cdot \eta_t^u$ is formally an It\^o integral.
}

\begin{align}
\label{eq:girsanov-ratio}
    &\frac{P^u[x(\cdot)]}{P^0[x(\cdot)]}
    =
    \frac{
        \exp\qty[
            -\frac{1}{2}
            \int_0^1\dd{t}\,
            \norm{\eta_t^u}^2
        ]
    }{
        \exp\qty[
            -\frac{1}{2}
            \int_0^1\dd{t}\,
            \norm{\eta_t^0}^2
        ]
    }
    \\
    &=
    \exp\left[
        \frac{1}{2\varepsilon}
        \int_0^1\dd{t}\,
        \norm{u_t(x_t)}^2
        +
        \frac{1}{\sqrt{\varepsilon}}
        \int_0^1\dd{t}\,
        u_t(x_t)\cdot\eta_t^u
    \right]. \nonumber
\end{align}
The KL divergence is the expectation of the logarithm of this ratio under the controlled path measure:

\begin{equation}
\begin{aligned}
    \dkl\qty(P^u\Vert P^0)
    &=
    \frac{1}{2\varepsilon}
    \left\langle
        \int_0^1\dd{t}\,
        \norm{u_t(x_t)}^2
    \right\rangle_{P^u}
    \\
    &\quad+
    \frac{1}{\sqrt{\varepsilon}}
    \left\langle
        \int_0^1\dd{t}\,
        u_t(x_t)\cdot\eta_t^u
    \right\rangle_{P^u}
    \\
    &=
    \frac{1}{2\varepsilon}
    \left\langle
        \int_0^1\dd{t}\,
        \norm{u_t(x_t)}^2
    \right\rangle_{P^u}.
    \label{eq:girsanov-kl-section2}
\end{aligned}
\end{equation}
In the final equality, the term linear in the controlled noise has vanished because under the controlled path measure, the noise $\eta_t^u$ has mean zero.
Eq.~\eqref{eq:girsanov-ratio} is the standard statement of Girsanov's theorem~\cite{oksendal2003stochastic}.

\subsubsection{Probabilistic inference as a control problem}\label{sec:inference_as_control}

With Girsanov's theorem, this means that we can write Eq.~\eqref{eq:large-deviations-heuristic-form} more formally as an optimization problem:

\begin{equation}
\begin{aligned}
    u^*
    &=
    \argmax_{u}
    \left\langle
        - \frac{1}{2\varepsilon}
        \int_0^1 \dd{t}
        \norm{u_t(x_t)}^2
    \right.
    \\
    &\qquad\left.
        + \int_0^1
        \dd{t}
        r_t(x_t)
        + r_1(x_1)
    \right\rangle_{P^u}
\end{aligned}
\label{eq:pre-pisc}
\end{equation}
where $r_t$ and $r_1$ are again Lagrange multipliers chosen to enforce the constraint of consistency with observations. Here the trajectory reward is additive in time, with a separate terminal term. In this special case, the running reward of Sec.~\ref{sec:mechanics_control} is
\[
    \underbrace{r_t(x,u)}_{\text{total running reward}}
    =\underbrace{r_t(x)}_{\text{state reward}}
    -\frac{\|u\|^2}{2\varepsilon}.
\]
This now has the form of a control problem in which we are solving to find the optimal control $u_t^*(x_t)$ which maximizes the reward but has a cost from applying too strong a control~\cite{Theodorou2012,levine2018reinforcement,BoueDupuis1998}.
We will see that the problem in Eq.~\eqref{eq:pre-pisc} can be solved by converting it into a linear PDE for which Monte Carlo methods can be used to sample the optimal control.

In this context, the quadratic control cost has a probabilistic interpretation in that it measures how strongly the controlled ensemble departs from the passive dynamics.
The same relation between noise and control cost that makes this inference interpretation possible also gives the associated HJB equation a particularly simple mathematical structure.

\subsection{Mathematical structure of path integral stochastic control}\label{sec:pisc}

\begin{figure*}[t]
    \centering
    \includegraphics[width=0.8\linewidth]{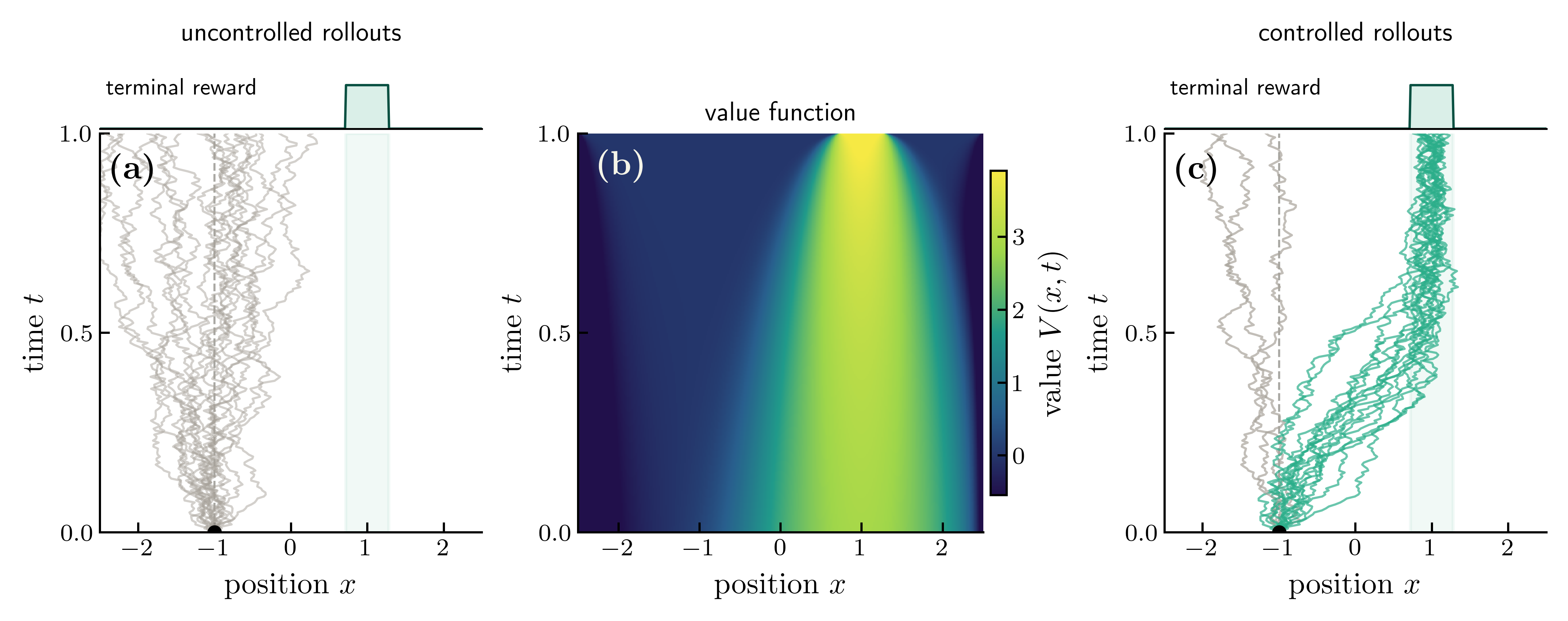}
    \caption{
        Path integral stochastic control (PISC) for an ensemble of freely diffusing particles.
        Left: an ensemble of particles begins at a common initial condition and freely diffuses, failing to achieve some distant terminal reward.
        Middle: the value function which solves the PISC problem of the ensemble of particles achieving the  maximum terminal reward subject to some control cost.
        Right: the resulting PISC policy leading a significant fraction of the particles to efficiently achieve the terminal reward.
    }
    \label{fig:pisc}
\end{figure*}

Problems with objectives such as Eq.~\eqref{eq:pre-pisc} are called path integral stochastic control (PISC) problems~\cite{Kappen2005PRL,Kappen2005JSTAT,todorov2006linearly,Todorov2009, Theodorou2010}.
PISC problems are a special class of stochastic control problems in which the cost of control is quadratic in the control and the dynamics are linear in the control (i.e., in Eq.~\eqref{eq:soc-dynamics}, $g_t(x,u)=f_t(x)+u$).
Fig.~\ref{fig:pisc} illustrates a solution to a PISC problem in which an ensemble of particles is controlled to achieve a distant terminal reward.
These problems can be solved using a path integral perspective like in Eq.~\eqref{eq:maxentrl_titled_path_measure}, but are most naturally solved using dynamic programming and the Hamilton--Jacobi--Bellman (HJB) equation as introduced in Sec.~\ref{sec:mechanics_control}.
PISC is a special class of stochastic control problems because the HJB equation takes a particularly simple form that can be transformed into a linear PDE.

\subsubsection{The Cole--Hopf transformation of the HJB equation}

To solve the inference problem, expressed as Eq.~\eqref{eq:pre-pisc}, we first define the value function to be the expected reward received if the optimal policy is taken starting at point $x$ at time $t$:

\begin{equation}
\begin{aligned}
    V_t(x)
    &=
    \max_{\substack{\{u_s\}\\ t \leq s < 1}}
    \left\langle
        -
        \frac{1}{2\varepsilon}
        \int_t^1
        \dd{s}
        \norm{u_s(x_s)}^2
    \right.
    \\
    &\qquad\left.
        + \int_t^1 r_s(x_s) \dd{s}
        + r_1(x_1)
        \,\middle|\, x_t=x
    \right\rangle_{P^u}
\end{aligned}
\label{eq:value-function-pisc}
\end{equation}
which using dynamic programming (Eq.~\eqref{eq:hjb-stoch}) must satisfy

\begin{equation}
\begin{aligned}
    0
    &=
    \max_u
    \left\{
        r_t(x)
        + \partial_t V_t
        + \nabla V_t \cdot
        \qty[f_t(x)+u]
    \right.
    \\
    &\qquad\left.
        + \frac{\varepsilon}{2}
        \nabla^2 V_t
        - \frac{1}{2\varepsilon}
        \norm{u}^2
    \right\}
\end{aligned}
\end{equation}
with the optimal control given by
\begin{equation}\label{eq:PISC_optimal_control}
\begin{aligned}
    u_t^*(x)
    =
    \varepsilon
    \nabla V_t(x).
\end{aligned}
\end{equation}
For PISC problems, the HJB equation is then the nonlinear PDE~\cite{Fleming2006}

\begin{equation}
\begin{aligned}
    &r_t(x)
    +
    \partial_t V_t
    +
    \nabla V_t \cdot
    f_t(x)
    \\
    &\quad+
    \frac{\varepsilon}{2}
    \nabla^2 V_t
    +
    \frac{\varepsilon}{2}
    \norm{
        \nabla V_t
    }^2
    =
    0
\end{aligned}
\end{equation}
with the boundary condition $V_1(x) = r_1(x)$.
Remarkably, this nonlinear PDE can be converted into a linear equation by the Cole--Hopf transformation~\cite{Kappen2005PRL,Kappen2005JSTAT},
\begin{equation}
\begin{aligned}
    V_t(x) = \log \psi_t(x)
    ,
\end{aligned}
\label{eq:cole-hopf-transformation}
\end{equation}
which upon substitution gives:

\begin{equation}
\begin{aligned}
    \qty[
        r_t(x)
        +
        \partial_t
        +
        f_t(x)\cdot
        \nabla
        +
        \frac{\varepsilon}{2}
        \nabla^2
    ]
    \psi_t(x)
    =
    0
\end{aligned}
\label{eq:cole-hopf-hjb-psi}
\end{equation}
which is a backward Kolmogorov equation with a potential term.
Notably, this is also the Schr\"odinger equation in imaginary time (i.e., with $t \to i t$) when $f_t=0$ with $\psi$ being the wavefunction, $\varepsilon$ playing the role of $\hbar/m$, and $-r_t(x)$ playing the role of a potential energy term.
In the small-$\varepsilon$ limit, this parallels the passage from wave optics to geometrical optics, where the Hamilton--Jacobi equation becomes the eikonal equation.
With this change of variables, the optimal control can be expressed as:
\begin{equation}
\begin{aligned}
    \label{eq:optimal-control-psi}
    u_t^*(x)
    =
    \varepsilon
    \nabla \log \psi_t(x).
\end{aligned}
\end{equation}
\subsubsection{Feynman--Kac formula and Doob's $h$-transform}
\label{sec:feynman-kac-formula-doobs-h-transform}

The transformed equation is linear, but solving it over the full state space can still be difficult in high dimensions.
An alternative is to express its solution as an average over trajectories of the passive dynamics, weighted by their accumulated rewards.
This is the stochastic counterpart of the imaginary-time path-integral representation from quantum mechanics and provides a starting point for the sampling methods discussed in Chapter~\ref{sec:sampling}.
From the same reasoning as in Eq.~\eqref{eq:maxentrl_titled_path_measure}, the optimal path measure conditional on starting at $x$ at time $t$ is given below. In these conditional path laws, $x(\cdot)$ denotes the remaining trajectory on $t\leq s\leq1$:

\begin{equation}
\begin{aligned}
    P^{u^*}[x(\cdot)| x_t=x]
    &=
    \frac{
        1
    }{
        Z_t(x)
    }
    P^0[x(\cdot)| x_t=x]
    e^{
        R_t[x(\cdot)]
    }
\end{aligned}
\label{eq:optimal-conditional-path-measure}
\end{equation}
where $R_t[x(\cdot)] = \int_t^1 r_s(x_s) \dd{s} + r_1(x_1)$ is the cumulative state and terminal reward from time $t$ to the final time $1$, excluding control cost. We temporarily denote the normalization factor by $Z_t(x)$, which is a conditional partition function.
Observe from Eq.~\eqref{eq:value-function-pisc} and Girsanov's theorem that we can write the value function as

\begin{equation}
\begin{aligned}
    \hspace{-.2em}
    V_t(x)
    &=
    \left\langle
        R_t[x(\cdot)]
        -
        \frac{1}{2\varepsilon}
        \int_t^1
        \dd{s}
        \norm{
            u_s^*(x_s)
        }^2
    \;\middle|\;x_t=x\right\rangle_{P^{u^*}}
    \\
    &=
    \left\langle
        R_t[x(\cdot)]
        -
        \log
        \frac{
            P^{u^*}[x(\cdot)| x_t=x]
        }{
            P^0[x(\cdot)| x_t=x]
        }
    \;\middle|\;x_t=x\right\rangle_{P^{u^*}}
    \\
    &=
    \left\langle
        R_t[x(\cdot)]
        -
        \qty(
            R_t[x(\cdot)]
            -
            \log Z_t(x)
        )
    \;\middle|\;x_t=x\right\rangle_{P^{u^*}}
    \\
    &=
    \log Z_t(x),
\end{aligned}
\end{equation}
where the expectations above are all taken with respect to the optimal path measure $P^{u^*}[x(\cdot)| x_t=x]$.
From Eq.~\eqref{eq:cole-hopf-transformation}, we then have $\psi_t(x)=Z_t(x)$. We henceforth use $\psi_t$ for this conditional partition function. It is a positive function of the current state, not a normalized state density, and can be expressed as a path integral over the passive dynamics:

\begin{equation}
\begin{aligned}
    \psi_t(x)
    &=
    \int
    \mathcal{D}x
    \,
    P^0[x(\cdot)| x_t=x]
    \\
    &\quad\times
    \exp
    \qty(
        \int_t^1 r_s(x_s) \dd{s} + r_1(x_1)
    ).
\end{aligned}
\label{eq:psi-as-path-integral}
\end{equation}
This is generalized to a result that a function $\psi_t(x)$ satisfying Eq.~\eqref{eq:cole-hopf-hjb-psi} can be expressed as a conditional expectation over the passive dynamics, which is known as the Feynman--Kac formula~\cite{oksendal2003stochastic,Kappen2005JSTAT}.
Expressing $\psi_t$ in terms of a path integral is familiar from quantum mechanics, where the solution to the Schr\"odinger equation can be expressed as a path integral over all possible paths weighted by the exponential of the action, and the Feynman--Kac formula is the stochastic process analog of this result.

We can apply this result to the case where we are interested in inferring the dynamics of a system that is observed to be in a particular set $\mathcal A$ at time $t=1$.
For a path measure $P$, we write $P(\mathcal A)\equiv P(x_1\in\mathcal A)$ for the probability of this terminal event. In this conditioning example the running state reward is zero, $r_t(x)=0$.
We are therefore interested in studying the case $P^u(\mathcal A) = 1$, which requires $r_1(x_1)$ to be zero for $x_1 \in \mathcal A$ and $-\infty$ for $x_1 \notin \mathcal A$ (i.e., $\psi_1(x_1) = \mathbf{1}_{\mathcal A}(x_1)$ where $\mathbf{1}_{\mathcal A}$ is the indicator function for the set $\mathcal A$).
From Eq.~\eqref{eq:psi-as-path-integral}, we then have $\psi_t(x) = P^0(\mathcal A \vert x_t = x)$, which is the probability that the system is in the set $\mathcal A$ at time $t=1$ given that it is in state $x$ at time $t$ under the passive dynamics; this conditioning function is the function conventionally called $h$ in Doob's $h$-transform~\cite{ChetriteTouchette2015}. We retain the symbol $\psi_t(x)$ here.
Eq.~\eqref{eq:optimal-conditional-path-measure} is then directly Bayes' rule:

\begin{equation}
\begin{aligned}
    \hspace{-.2em}
    P^{u^*}[x(\cdot)| x_t=x]
    &=
    P^0[x(\cdot)| x_t=x] \frac{
        \mathbf{1}_{\mathcal A}(x_1)
    }{
        P^0(\mathcal A \vert x_t = x)
    }
\end{aligned}
\end{equation}
From Eq.~\eqref{eq:optimal-control-psi}, the inferred dynamics (known as Doob's $h$-transform) that lead to the system being in the set $\mathcal A$ at time $t=1$ are then given by

\begin{equation}
\begin{aligned}
    \dot{x}_t
    &=
    f_t(x_t)
    +
    \varepsilon
    \nabla\log\psi_t(x_t)
    +
    \sqrt{\varepsilon} \,\eta_t
    .
\end{aligned}
\label{eq:doobs-h-transform-dynamics}
\end{equation}
The logarithmic gradient $\nabla\log\psi_t(x)$ is the conditioning or guidance term. It is closely related to the score $\nabla\log p_t(x)$ used in generative modeling and diffusion models, but the two are generally distinct: $\psi_t(x)$ is a conditioning probability, whereas $p_t(x)$ is a marginal density. We return to scores and guidance in Sec.~\ref{sec:flows_diffusions}.

\subsubsection{Example: Brownian bridge}\label{sec:brownian_bridge}

The result in Eq.~\eqref{eq:doobs-h-transform-dynamics} allows us to now study the Brownian bridge (Eq.~\eqref{eq:brownian_bridge}) introduced in Sec.~\ref{sec:path_inference}.
In particular, we consider one-dimensional Brownian motion as the passive dynamics (i.e., $f_t(x) = 0$) and the system is observed to be in a particular state $x_1 = y$ at time $t=1$ while starting at $x_0 = 0$ at time $t=0$.
To use Eq.~\eqref{eq:doobs-h-transform-dynamics} to solve this problem, we treat the target set $\mathcal A$ as a small ball of radius $\delta \to 0^+$ around the target state $y$, so that $\mathcal A = \{x: \norm{x - y} < \delta\}$.
Let $k_{t| s}(z| x)$ denote the transition density of the passive process from state $x$ at time $s$ to state $z$ at time $t>s$. Then:

\begin{equation}
\begin{aligned}
    P^0
    &
    (\mathcal A \vert x_t = x)
    =
    \int_{\mathcal A}
    \dd{x_1}
    \,
    k_{1| t}(x_1| x)
    \\
    &=
    \int_{\mathcal A}
    \dd{x_1}
    \,
    \frac{1}{\sqrt{2\pi\varepsilon(1-t)}}
    \exp
    \qty(
        -\frac{\norm{x_1 - x}^2}{2\varepsilon(1-t)}
    )
    \\
    &\approx
    \frac{1}{\sqrt{2\pi\varepsilon(1-t)}}
    \exp
    \qty(
        -\frac{\norm{y - x}^2}{2\varepsilon(1-t)}
    )
    \int_{\mathcal A}
    \dd{x_1}
\end{aligned}
\end{equation}
The dynamics of a Brownian bridge are given by Eq.~\eqref{eq:doobs-h-transform-dynamics}:

\begin{equation}
\begin{aligned}
    \dot x_t
    =
    -
    \frac{x_t - y}{1-t}
    +
    \sqrt{\varepsilon}\,\eta_t
    .
\end{aligned}
\end{equation}
This means that in the conditioned dynamics, there is a drift term that pushes the system to move in a straight line from the starting point to the target point, with a strength that diverges as $t \to 1$.

Conditioning a stochastic system can therefore be understood dynamically. 
The information supplied by an observation is expressed as a control that makes the observed behavior typical.
The Brownian bridge makes this equivalence concrete, with an endpoint constraint appearing as a drift toward the target.

\section{Control and transport of densities}\label{sec:OT}

The variational viewpoint of Chapter~\ref{sec:variational_structure} compares trajectories through their action or control cost. We now consider the problem of moving an entire distribution between prescribed initial and final distributions, rather than a particle between prescribed locations. An intuitive picture is moving a pile of earth into another shape: the total amount of earth moved is unchanged, but there are many ways to rearrange it. Choosing a rearrangement that minimizes a specified cost is an \emph{optimal transport} problem. When the motion of mass is stochastic, one considers the most likely evolution consistent with the two distributions at the endpoints. We begin with a version of transport, called the Schr\"odinger bridge problem, formulated in terms of stochastic control. We then consider more general formulations of transport and the geometry it defines on probability distributions. This geometry allows us to describe the evolution of densities as steepest descent, connecting the present chapter to the discussions of nonequilibrium thermodynamics, sampling, and learning further below.

\subsection{The Schr\"odinger bridge problem}
\label{sec:dynamic_OT}

\begin{figure*}[t]
    \centering
    \includegraphics[width=0.85\linewidth]{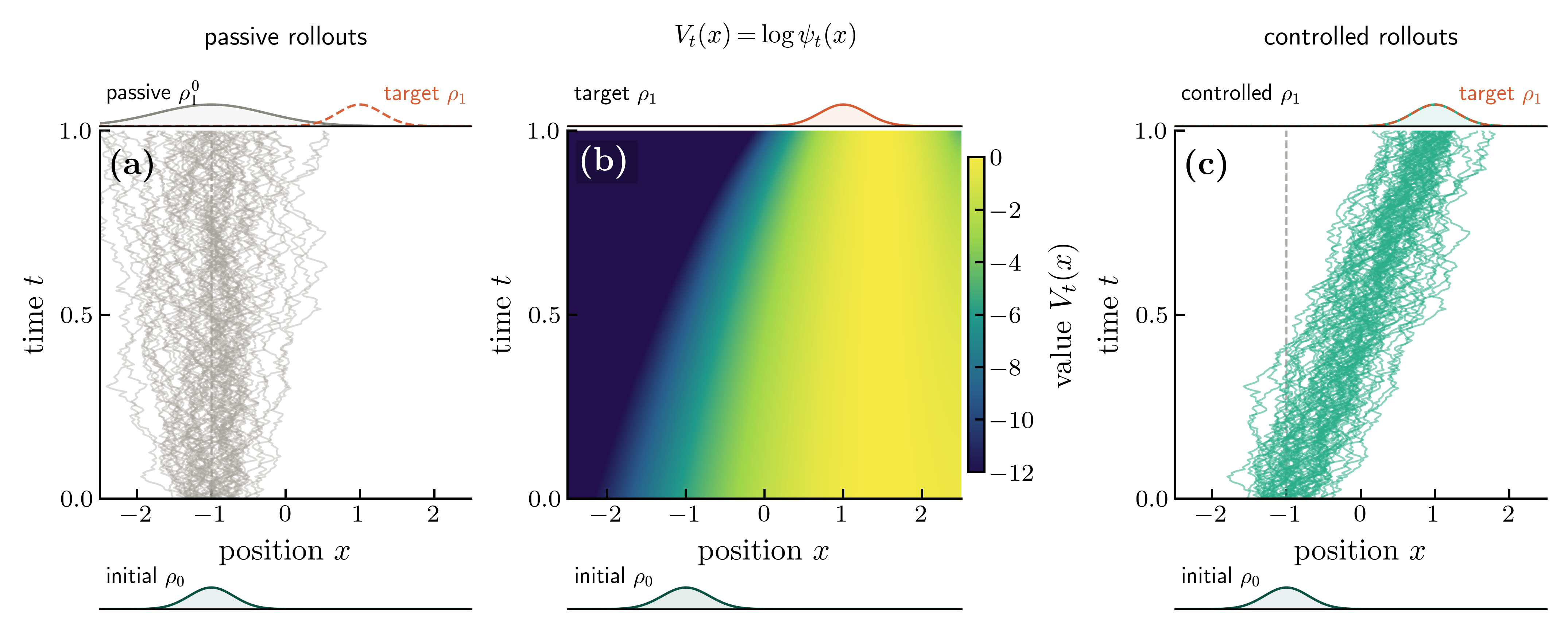}
 \caption{
        The Schr\"odinger bridge problem (SBP) for an ensemble of freely diffusing particles.
        Left: an ensemble of particles begins distributed according to an initial density $\rho_0$ and freely diffuses to $\rho_1^0$, failing to achieve the target terminal density $\rho_1$.
        Middle: the value function $V_t(x)=\log\psi_t(x)$ which solves the SBP of transporting the ensemble between the prescribed initial and terminal densities subject to a control cost. Right: the resulting control policy transports the particles so that the terminal density of the controlled ensemble matches the target density.}
    \label{fig:sbp}
\end{figure*}

The Schr\"odinger bridge problem (SBP) asks for the distribution over stochastic trajectories that matches two prescribed endpoint densities while differing as little as possible from a reference process~\cite{Schrodinger1931,Schrodinger1932,Leonard2014,Chen2021}. Unlike the fixed-endpoint bridge considered in Sec.~\ref{sec:brownian_bridge}, the initial and final positions are now drawn from distributions $\rho_0$ and $\rho_1$. The control generates a curve of densities $t\mapsto\rho_t$ connecting these endpoints while accounting for all possible particle trajectories between them. Fig.~\ref{fig:sbp} illustrates this passage from controlling trajectories to transporting densities.

\subsubsection{From trajectories to densities}

As in Sec.~\ref{sec:path_inference}, we consider non-interacting particles with controlled Langevin dynamics,
\begin{equation}
    \dot{x}_t=f_t(x_t)+u_t(x_t)+\sqrt{\varepsilon}\,\eta_t,
    \qquad 0\leq t\leq 1,
    \label{eq:sb_dynamics}
\end{equation}
which is Eq.~\eqref{eq:soc-dynamics} with $g_t(x,u)=f_t(x)+u$ and feedback $u=u_t(x)$. Here $\eta_t$ is standard Gaussian white noise and $\varepsilon>0$. The reference dynamics have $u_t=0$. Both the reference and controlled processes are initialized with $x_0\sim\rho_0$. We write $P^0[x(\cdot)]$ and $P^u[x(\cdot)]$ for the reference and controlled distributions over complete paths, respectively; these are the path distributions introduced in Sec.~\ref{sec:path_inference}. In contrast, $\rho_t(x)$ is the density of the position at a single time under $P^u$, i.e., it is the marginal $p_t^u(x)$ of Chapter~\ref{sec:variational_structure}. We use $\rho_t$ when emphasizing the transport of this density, and add a superscript when distinguishing different ensembles. For the independent particles considered here, this is both the single-particle marginal density and the density described by a large ensemble. 
The density generated by Eq.~\eqref{eq:sb_dynamics} obeys the Fokker--Planck equation, which expresses local conservation of probability:

\begin{equation}
\begin{aligned}
    \partial_t\rho_t
    &=-\nabla\cdot\left[(f_t+u_t)\rho_t\right]
    +\frac{\varepsilon}{2}\nabla^2\rho_t,\\
    &\qquad \rho_{t=0}=\rho_0,\quad \rho_{t=1}=\rho_1.
\end{aligned}
\label{eq:sb_fokker_planck}
\end{equation}
It is useful to distinguish the added control $u_t$ from the velocity $v_t$ that transports the density. Eq.~\eqref{eq:sb_fokker_planck} can equivalently be written as
\[
\begin{gathered}
    \partial_t\rho_t+\nabla\cdot(\rho_t v_t)=0,\\
    v_t=f_t+u_t-\frac{\varepsilon}{2}\nabla\log\rho_t.
\end{gathered}
\]
Here $\rho_t v_t$ is the probability current. In the deterministic, drift-free case, $v_t=u_t$.
Note that even when intermediate densities $\rho_t$ are specified, we cannot identify a unique velocity field; adding a field $w_t$ with $\nabla\cdot(\rho_t w_t)=0$ leaves the continuity equation unchanged. Specifying the two endpoint densities leaves still more freedom, since the intermediate densities are also unspecified. Thus, as in maximum-entropy inference, the constraints must be supplemented by a criterion for choosing among the compatible path distributions.

Suppose a large collection of independent particles evolving under $P^0$ is observed to begin with density $\rho_0$ and end with density $\rho_1$. The large-deviation argument of Eq.~\eqref{eq:large-deviations-heuristic-form}, now applied to paths, selects the distribution over paths that satisfies these observations and is closest to $P^0$ in KL divergence. This gives the SBP,
\begin{equation}
\begin{aligned}
    \min_{P^u}\;&\quad \dkl(P^u\Vert P^0),\\
    \text{s.t.}\;&\quad \rho_{t=0}=\rho_0,\qquad \rho_{t=1}=\rho_1,
\end{aligned}
\label{eq:sb_path}
\end{equation}
where $\rho_t$ is the marginal density under $P^u$. To relate this problem to control, we write $P^u[x(\cdot)]=\rho_0(x_0)P^u[x(\cdot)| x_0]$, and similarly for $P^0$. The chain rule for KL divergence gives
\begin{equation}
\begin{aligned}
    &\dkl(P^u\Vert P^0)\\
    &\quad=\int\dd{x_0}\,\rho_0(x_0)
    \dkl\!\left(P^u[\cdot| x_0]\Vert P^0[\cdot| x_0]\right)\\
    &\quad=\frac{1}{2\varepsilon}\int_0^1\dd{t}\int\dd{x}\,
    \rho_t(x)\norm{u_t(x)}^2,
\end{aligned}
\label{eq:sb_kl_decomposition}
\end{equation}
where the second equality follows from Girsanov's theorem, Eq.~\eqref{eq:girsanov-kl-section2}, applied to paths with the same initial position. Consequently, minimizing Eq.~\eqref{eq:sb_path} is equivalent to minimizing the expected quadratic control effort subject to Eq.~\eqref{eq:sb_fokker_planck} and its endpoint constraints. The SBP thus chooses the least surprising controlled ensemble relative to the reference process, or equivalently the one requiring the least expected control effort. This introduces the link between control, inference and transport, i.e., the control changes the path distribution and, through its single-time marginals, transports the density from $\rho_0$ to $\rho_1$.

\subsubsection{The coupled forward--backward construction}

We can impose the initial constraint directly by drawing $x_0$ from $\rho_0$, and impose the terminal constraint using a Lagrange multiplier function $r_1(x_1)$. As in Sec.~\ref{sec:pisc}, this multiplier acts as a terminal reward. For a fixed $r_1$, and after dropping terms independent of the control, the problem is

\begin{equation}
\begin{gathered}
    \min_u\;\left\langle
        \frac{1}{2\varepsilon}\int_0^1\dd{t}\,\norm{u_t(x_t)}^2
        -r_1(x_1)
    \right\rangle_{P^u},\\
    x_0\sim\rho_0.
\end{gathered}
\label{eq:sb_pisc}
\end{equation}
An initial reward $r_0(x_0)$ would have the fixed expectation $\int\dd{x}\,\rho_0(x)r_0(x)$ and would not change this minimization. Maximizing the negative of the objective in Eq.~\eqref{eq:sb_pisc} is analogous to solving the PISC problem considered in Sec.~\ref{sec:pisc}, with no additional running reward. Its value function therefore admits the Cole--Hopf transformation $V_t=\log\psi_t$, and Eq.~\eqref{eq:optimal-control-psi} gives

\begin{equation}
\begin{aligned}
    u_t^\ast(x)&=\varepsilon\nabla\log\psi_t(x),
    \qquad \psi_1(x)=e^{r_1(x)},\\
    \partial_t\psi_t&+f_t\cdot\nabla\psi_t
    +\frac{\varepsilon}{2}\nabla^2\psi_t=0.
\end{aligned}
\label{eq:psi-backward}
\end{equation}
Unlike a prescribed terminal reward in PISC, $r_1$ must now be chosen so that the terminal density is $\rho_1$. Since $\psi_t$ satisfies the backward Kolmogorov equation, it is  interpreted as a field that propagates information backwards from $t=1$.

To find the density under this control, define a second field $\psi_t^\dagger(x)=\rho_t(x)/\psi_t(x)$, so that
\begin{equation}
    \rho_t(x)=\psi_t(x)\psi_t^\dagger(x).
    \label{eq:sb_density}
\end{equation}
The two factors need not be normalized densities. Schr\"odinger's key insight was that $\psi_t^\dagger$ obeys the forward Kolmogorov equation of the reference process,

\begin{equation}
    \partial_t\psi_t^\dagger
    +\nabla\cdot(f_t\psi_t^\dagger)
    -\frac{\varepsilon}{2}\nabla^2\psi_t^\dagger=0.
    \label{eq:psi-hat-forward}
\end{equation}
To see this, substitute $\rho_t=\psi_t\psi_t^\dagger$ and $u_t^\ast=\varepsilon\nabla\log\psi_t$ into Eq.~\eqref{eq:sb_fokker_planck}. Expanding the derivatives gives

\begin{equation}
\begin{aligned}
    \partial_t(\psi_t\psi_t^\dagger)
    &=-\nabla\cdot(f_t\psi_t\psi_t^\dagger)\\
    &\quad-\varepsilon\nabla\cdot(\psi_t^\dagger\nabla\psi_t)
    +\frac{\varepsilon}{2}\nabla^2(\psi_t\psi_t^\dagger)\\
    &=\psi_t^\dagger\left[-f_t\cdot\nabla\psi_t
    -\frac{\varepsilon}{2}\nabla^2\psi_t\right]\\
    &\quad+\psi_t\left[-\nabla\cdot(f_t\psi_t^\dagger)
    +\frac{\varepsilon}{2}\nabla^2\psi_t^\dagger\right].
\end{aligned}
\label{eq:sb_factor_cancellation}
\end{equation}
The first bracket is $\partial_t\psi_t$ by Eq.~\eqref{eq:psi-backward}. Cancelling $\psi_t^\dagger\partial_t\psi_t$ from both sides and dividing by $\psi_t$ yields Eq.~\eqref{eq:psi-hat-forward}. Thus, $\psi_t$ propagates information backward from the final constraint, while $\psi_t^\dagger$ propagates information forward from the initial constraint.

Let $k_{t| s}(x| y)$ denote the transition density of the reference process, namely, the conditional density of being at $x$ at time $t$ given position $y$ at time $s<t$, with $u=0$ in Eq.~\eqref{eq:sb_dynamics}. It is the propagator of the reference Fokker--Planck equation. The solutions of the backward and forward equations can therefore be written as

\begin{equation}
\begin{aligned}
    \psi_t(x)&=\int\dd{y}\,k_{1| t}(y| x)\psi_1(y),\\
    \psi_t^\dagger(x)&=\int\dd{y}\,k_{t| 0}(x| y)\psi_0^\dagger(y).
\end{aligned}
\label{eq:sb_system2}
\end{equation}
The endpoint constraints couple these otherwise linear equations,
\begin{equation}
\begin{aligned}
    \psi_0(x)\psi_0^\dagger(x)&=\rho_0(x),\\
    \psi_1(x)\psi_1^\dagger(x)&=\rho_1(x).
\end{aligned}
\label{eq:sb_boundary}
\end{equation}
Note that neither endpoint factor is known in advance; specifying $\psi_1$ determines $\psi_0$ by backward propagation, while specifying $\psi_0^\dagger$ determines $\psi_1^\dagger$ by forward propagation. One may solve this Schr\"odinger system by alternating between the two constraints. Starting from a positive guess for $\psi_1$, propagate it backward using the first line of Eq.~\eqref{eq:sb_system2} to obtain $\psi_0$. Set $\psi_0^\dagger=\rho_0/\psi_0$, propagate this field forward using the second line to obtain $\psi_1^\dagger$, and update $\psi_1=\rho_1/\psi_1^\dagger$. Repeating these steps enforces the initial and final constraints in turn. This procedure is the so-called Sinkhorn algorithm~\cite{Sinkhorn1967,Chen2021}.

With the two fields determined, the optimal distribution over paths is an endpoint reweighting, or `tilt', of the reference distribution~\cite{Leonard2014,Chen2021},
\begin{equation}
\begin{aligned}
    P^{u^\ast}[x(\cdot)]
    &=P^0[x(\cdot)]\,
    \frac{\psi_1(x_1)}{\psi_0(x_0)}.
\end{aligned}
\label{eq:sb_path_tilt}
\end{equation}
Eq.~\eqref{eq:sb_path_tilt} has the same exponential-tilt structure encountered in maximum-entropy inference and extends the Doob $h$-transform to prescribed endpoint distributions. Instead of conditioning an endpoint to lie in a set, the endpoint factors are chosen together to reproduce two full densities. The central result is therefore not a new form of the optimal control, which remains $u_t^\ast=\varepsilon\nabla\log\psi_t$, but the coupled forward--backward construction that determines this control from the two distributional constraints. 

\subsection{Entropic optimal transport and the Wasserstein distance}
\label{sec:eot}

The Schr\"odinger bridge specifies a complete distribution over paths, and thus solves a \emph{dynamic transport} problem. More generally, \emph{optimal transport} considers the least expensive rearrangement of mass from one particular configuration to another, given a cost $c(x_0,x_1)$ for moving a unit mass from $x_0$ to $x_1$. In Monge's original formulation, one views  optimal transport as the most efficient strategy for moving a pile of earth from one location to another. Each point $x_0$ is assigned a destination $T(x_0)$, and one minimizes $\int\dd{x_0}\,\rho_0(x_0)c(x_0,T(x_0))$ while requiring $T(x_0)\sim\rho_1$ for $x_0\sim\rho_0$. However, a deterministic map cannot split mass, and therefore need not exist; for example, a point mass cannot be mapped deterministically into two separated point masses. Kantorovich's formulation instead optimizes over a joint endpoint distribution $\pi(x_0,x_1)$ satisfying
\begin{equation}
\begin{aligned}
    \int\dd{x_1}\,\pi(x_0,x_1)&=\rho_0(x_0),\\
    \int\dd{x_0}\,\pi(x_0,x_1)&=\rho_1(x_1).
\end{aligned}
\label{eq:couplings}
\end{equation}
Any such joint distribution is called a \emph{coupling} of $\rho_0$ and $\rho_1$. Minimizing the mean cost over couplings allows the probability at one $x_0$ to be distributed among several destinations~\cite{Villani2003topics}. 

To connect this static problem to the Schr\"odinger bridge, let $k(x_1| x_0)$ denote the transition density of the reference process from $t=0$ to $t=1$ (we have dropped the $1|0$ subscript from our notation in Sec.~\ref{sec:dynamic_OT}). Consider the following optimization over couplings~\cite{Leonard2014}:
\begin{equation}
    \pi^*=\argmin_{\pi}\left\{
        \varepsilon\dkl\bigl(\pi\Vert\rho_0 k\bigr)
    \right\},
    \label{eq:static_maxent}
\end{equation}
subject to Eq.~\eqref{eq:couplings}, where $(\rho_0 k)(x_0,x_1)=\rho_0(x_0)k(x_1| x_0)$. Now, let's say
$k(x_1| x_0)=\exp\left(-c(x_0,x_1)/\varepsilon\right)/Z(x_0)$. For example, if the reference process is Brownian motion, we have $c(x_0,x_1)=\norm{x_0-x_1}^2/2$.
Here the same $\varepsilon>0$ that sets the noise covariance in Eq.~\eqref{eq:sb_dynamics} sets the strength of the regularization (and is analogous to a temperature), and $Z(x_0)$ normalizes the kernel over $x_1$. Using the fixed marginal constraints, Eq.~\eqref{eq:static_maxent} is equivalent, up to constants, to
\begin{equation}
\begin{aligned}
    &\pi^*=\argmin_{\pi}\Bigg\{
        \int\dd{x_0}\dd{x_1}\,\pi(x_0,x_1)c(x_0,x_1)\\
    &\qquad+\varepsilon\int\dd{x_0}\dd{x_1}\,\pi(x_0,x_1)\log\pi(x_0,x_1)
    \Bigg\}.
\end{aligned}
\label{eq:static_maxent2}
\end{equation}
This objective corresponds to \emph{entropic optimal transport}~\cite{Cuturi2013}. The first term is the mean transport cost, while the second is minus the Shannon entropy of the coupling introduced in Sec.~\ref{sec:MEM}. Eq.~\eqref{eq:static_maxent2} therefore has the form of a constrained free-energy minimization: the cost favors efficient transport, whereas the entropy spreads probability among several possible pairings. As $\varepsilon\to0$, the entropic term disappears and one obtains the Kantorovich formulation of optimal transport~\cite{Villani2003topics,BenamouNUMATH2000}.

Introducing Lagrange multipliers $\varphi_0(x_0)$ and $\varphi_1(x_1)$ for imposing the constraints in Eq.~\eqref{eq:couplings}, and absorbing the overall normalization into them, gives

\begin{equation}
\begin{aligned}
    &\pi^*(x_0,x_1)=\exp\left(\frac{\varphi_0(x_0)+\varphi_1(x_1)-c(x_0,x_1)}{\varepsilon}\right)\\
    &\quad=Z(x_0)\exp\left(\frac{\varphi_0(x_0)}{\varepsilon}\right)
    k_{1| 0}(x_1| x_0)\\
    &\qquad \times \exp\left(\frac{\varphi_1(x_1)}{\varepsilon}\right).
\end{aligned}
\label{eq:static_sol}
\end{equation}
These multipliers depend on $\varepsilon$ and their zero-noise limits are known as the Kantorovich potentials. A connection to the Schr\"odinger potentials can be made by setting $\psi_0^\dagger=Z\exp(\varphi_0/\varepsilon)$ and $\psi_1=\exp(\varphi_1/\varepsilon)$, which gives

\begin{equation}
\begin{aligned}
    \pi^*(x_0,x_1)&=\psi_0^\dagger(x_0) k_{1| 0}(x_1| x_0)\psi_1(x_1),\\
    \psi_0(x_0)&=\int\dd{x_1}\, k_{1| 0}(x_1| x_0)\psi_1(x_1),\\
    \psi_1^\dagger(x_1)&=\int\dd{x_0}\,k_{1| 0}(x_1| x_0) \psi_0^\dagger(x_0),
\end{aligned}
\label{eq:sb_static_system}
\end{equation}
together with $\rho_0=\psi_0^\dagger\psi_0$ and $\rho_1=\psi_1^\dagger\psi_1$. These are precisely the endpoint versions of Eqs.~\eqref{eq:sb_system2} and~\eqref{eq:sb_boundary}. Sinkhorn iteration, analogous to the Sinkhorn algorithm introduced previously, is the alternating rescaling of $\psi_0^\dagger$ and $\psi_1$ needed to solve for these quantities, and thus the optimal coupling $\pi^*$~\cite{Cuturi2013}. 

The quadratic cost $c(x_0,x_1)=\norm{x_0-x_1}^2/2$ has special geometric structure. For the zero-noise argument, write $\pi_\varepsilon^*$ and $\varphi_1^\varepsilon$ to display their dependence on $\varepsilon$. From Eq.~\eqref{eq:static_sol}, the conditional density of $x_1$ given $x_0$ is

\begin{equation}
    \frac{\pi_\varepsilon^*(x_0,x_1)}{\rho_0(x_0)} \propto
    \exp\left[\frac{\varphi_1^\varepsilon(x_1)-\frac{1}{2}\norm{x_0-x_1}^2}{\varepsilon}\right],
    \label{eq:conditional_quadratic}
\end{equation}
where factors depending only on $x_0$ are absorbed into the normalization. Rewriting the exponent as

\begin{equation}
    -\frac{\norm{x_0}^2}{2\varepsilon}
    +\frac{1}{\varepsilon}\left[
        x_0\cdot x_1-\left(\frac{\norm{x_1}^2}{2}-\varphi_1^\varepsilon(x_1)\right)
    \right],
    \label{eq:laplace_rewrite}
\end{equation}
shows by the Laplace principle that, as $\varepsilon\to0$, the conditional density concentrates at the maximizing endpoints. Assuming the potentials converge with a fixed choice of additive constants, denote the limiting potential by $\varphi_1=\lim_{\varepsilon\to0}\varphi_1^\varepsilon$ and define the supremum
\begin{equation}
\begin{gathered}
    \zeta(x_0)=\sup_{x_1}\left\{x_0\cdot x_1-\bar\varphi_1(x_1)\right\},\\
    \text{where }\bar\varphi_1(x_1)=\frac{1}{2}\norm{x_1}^2-\varphi_1(x_1).
\end{gathered}
\label{eq:legendre_transform}
\end{equation}
If the maximizer $x_1^*(x_0)$ is unique, differentiating the supremum with respect to $x_0$ gives $\nabla\zeta(x_0)=x_1^*(x_0)$, so the limiting optimal coupling is supported on the map
\begin{equation}
    x_1=T(x_0)=\nabla\zeta(x_0).
    \label{eq:brenier_map}
\end{equation}
Because $\zeta$ is a convex conjugate, it is convex. This result is known as \emph{Brenier's theorem}, which states that for a quadratic cost and absolutely continuous $\rho_0$, the optimal transport map is the gradient of such a convex function~\cite{Brenier1991polar}. In one dimension, this reduces to the familiar method for mapping a uniform distribution to an arbitrary distribution using the inverse of its cumulative distribution. Specifically, defining the cumulative distribution functions $C_i(x)=\int_{-\infty}^x\dd{y}\,\rho_i(y)$, $C_0(x_0)$ is uniform on $(0,1)$ when $x_0\sim\rho_0$, and $x_1=C_1^{-1}(C_0(x_0))$ has density $\rho_1$, with $C_i^{-1}$ denoting the quantile function. Hence,
\begin{align}
    \label{eq:CDF_OT}
    T=C_1^{-1}\circ C_0
\end{align}
is the one-dimensional Brenier map.

\begin{figure*}[t]
    \includegraphics[width=0.8\linewidth]{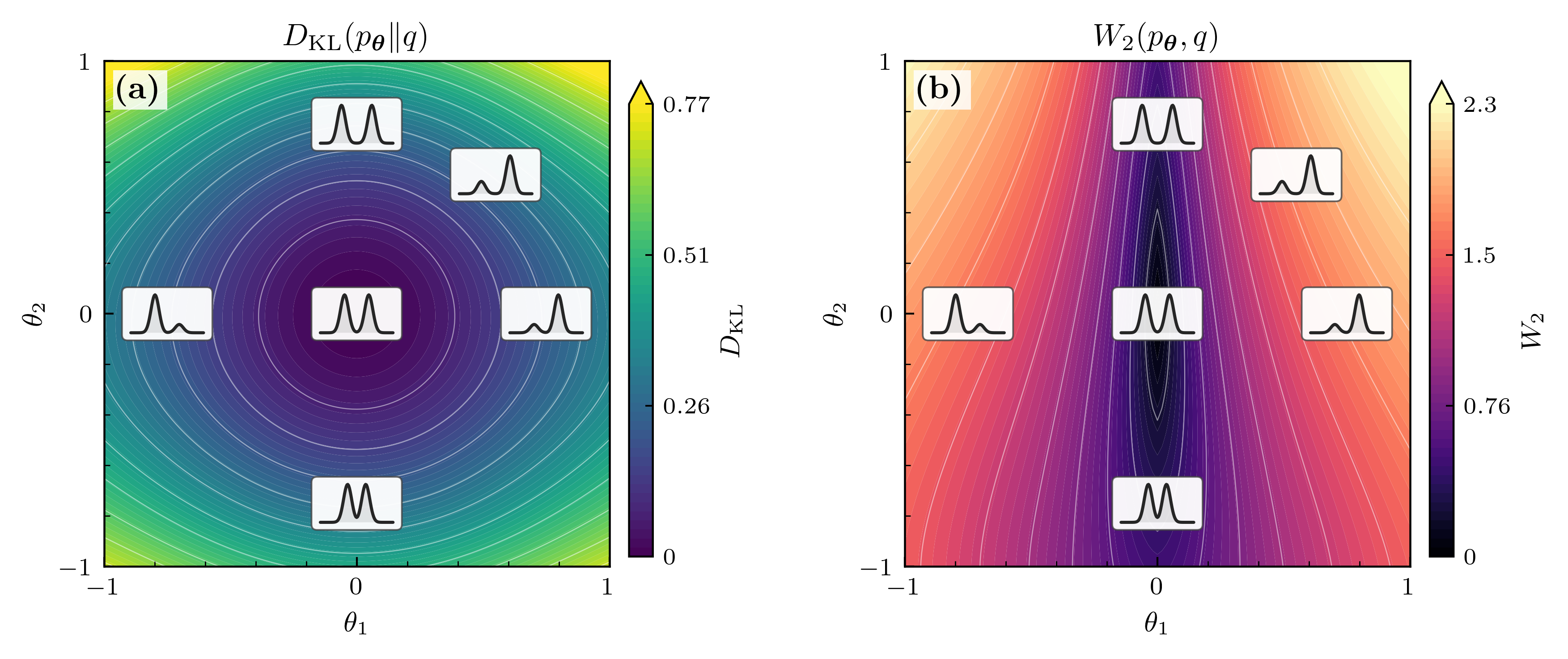}
 \caption{\label{fig:kl-wasserstein-geometry}
        A comparison of the geometries induced by the KL divergence and the $2$-Wasserstein distance on a two-parameter family of probability densities $p_\theta$ relative to the reference density $q$.
        The inset curves show representative densities at different points in parameter space.
        The $\theta_1$ parameter changes the balance of modes of a bi-modal distribution; the $\theta_2$ parameter shifts the distance between the modes.
        Left: the KL divergence assigns similar costs to changes along the two parameter directions, producing approximately isotropic contours.
        Right: the $2$-Wasserstein distance assigns a greater cost to changes that transport probability between the separated modes, producing strongly anisotropic contours.
        The figure is adapted from Ref.~\cite{blumenthal2026selforganized}, and the distribution's parameterization is described in the Supplementary Information of that work.
    } 
\end{figure*}

These special properties motivate the \emph{$2$-Wasserstein distance},

\begin{equation}
\begin{aligned}
    &W_2^2(\rho_0,\rho_1)\\
    &\quad=\min_{\pi}\left\{
        \int\dd{x_0}\dd{x_1}\,\norm{x_0-x_1}^2\pi(x_0,x_1)
    \right\},
\end{aligned}
\label{eq:w2_static}
\end{equation}
where the minimization is over the couplings in Eq.~\eqref{eq:couplings}. With the convention $c=\norm{x_0-x_1}^2/2$ used above, the unregularized minimum mean cost is $W_2^2/2$. In one dimension, the relationship to the cumulative distribution gives

\begin{equation}
    W_2^2(\rho_0,\rho_1)=\int_0^1\dd{z}\,
    \left|C_0^{-1}(z)-C_1^{-1}(z)\right|^2.
    \label{eq:w2_1d}
\end{equation}
The distinction between KL divergence and the 2-Wasserstein distance is illustrated in Fig.~\ref{fig:kl-wasserstein-geometry}. 
In the chosen family of bimodal densities, changes in the relative weights and separation of the modes produce comparable KL divergences, whereas Wasserstein distance distinguishes them strongly because changing the relative weights requires transporting probability between the modes.

\begin{figure*}[t]
    \centering
    \includegraphics[width=0.8\linewidth]{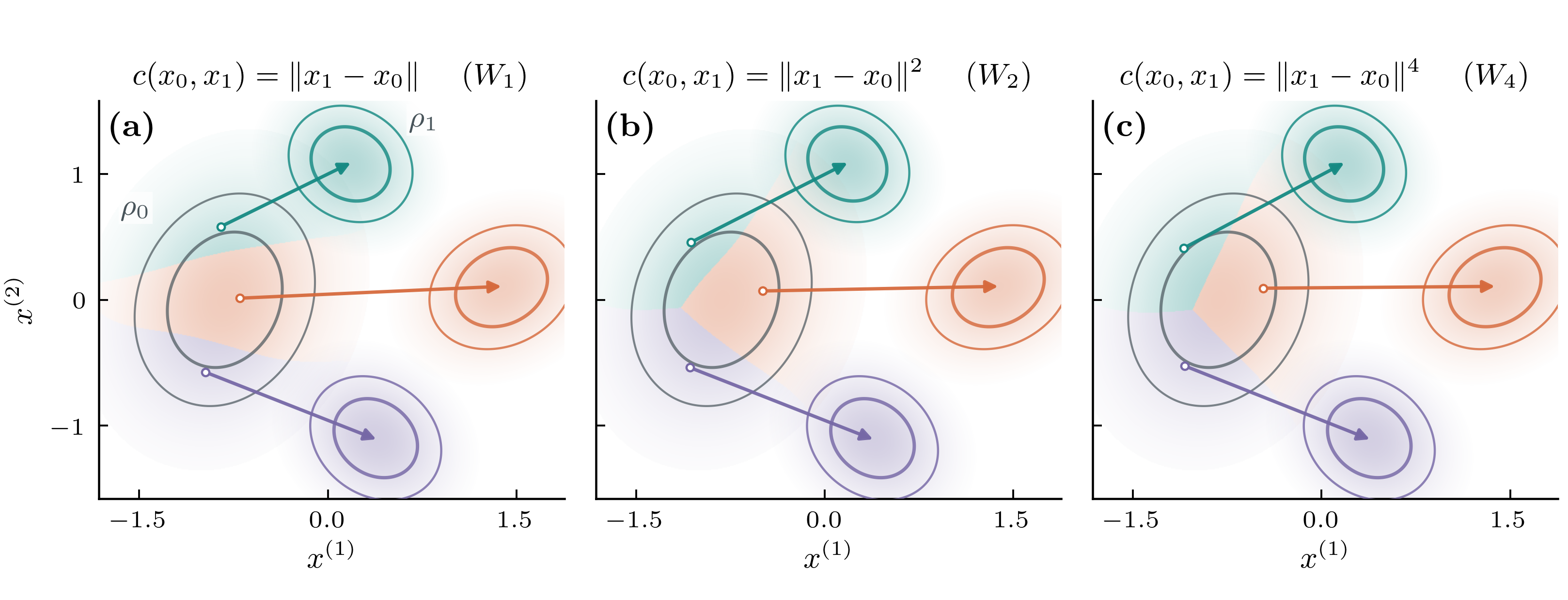}
 \caption{\label{fig:transport-cost-maps}
        Optimal transport maps between the same source density $\rho_0$ and three-mode target density $\rho_1$ under different choices of transport cost, $c$.
        In each panel, the three colored regions partition the initial density according to the correspondingly colored target mode, while one arrow shows a representative assignment from each region.
        Increasing the exponent penalizes long displacements more strongly and consequently changes how the initial density is partitioned among the target modes.
        The axes $x^{(1)},x^{(2)}$ denote spatial components, whereas the subscripts in $x_0,x_1$ denote initial and final endpoints.}
\end{figure*}

Note that the static formulation specifies which initial and final points are paired, but not how mass moves between them. 
In this deterministic, drift-free setting, the transport velocity equals the control, $v_t=u_t$. For a particle following $\dot{x}_t=v_t(x_t)$, the Cauchy--Schwarz inequality gives
$\norm{x_1-x_0}^2\leq\int_0^1\dd{t}\,\norm{v_t(x_t)}^2$, with equality for straight motion at constant speed. For an entire density, the velocity becomes a field $v_t(x)$ and valid paths must conserve probability locally. The remarkable content of the Benamou--Brenier theorem is that minimizing this kinetic action over all density paths and velocity fields gives exactly the same quantity as minimizing the static endpoint cost over couplings~\cite{BenamouNUMATH2000,Benamou2002monge}:

\begin{equation}
\begin{gathered}
    \begin{aligned}
    &W_2^2(\rho_0,\rho_1)\\
    &\quad=\min_{\rho_t,v_t}\left\{
        \int_0^1\dd{t}\int\dd{x}\,\rho_t(x)\norm{v_t(x)}^2
    \right\},
    \end{aligned}\\
    \partial_t\rho_t+\nabla\cdot(\rho_t v_t)=0,
\end{gathered}
\label{eq:bb}
\end{equation}
with endpoint constraints on $\rho_0$ and $\rho_1$. For a Brownian reference with $f_t=0$, multiplying the objective in Eq.~\eqref{eq:sb_path} by $2\varepsilon$ does not change its minimizer. In the limit $\varepsilon\to0$, the diffusion term in Eq.~\eqref{eq:sb_fokker_planck} vanishes, so Eq.~\eqref{eq:bb} can be interpreted as the zero-noise limit of the Schr\"odinger bridge, namely, the rescaled minimum $2\varepsilon\min_{P^u}\dkl(P^u\Vert P^0)$ approaches $W_2^2$~\cite{Chen2016}.

Once the optimal map $T$ is known, the minimizing density path is generated by linear interpolation of each transported particle,
\begin{equation}
    x_t= (1-t)x_0+tT(x_0),
    \qquad x_0\sim\rho_0.
    \label{eq:intermediate}
\end{equation}
The resulting $\rho_t$ is called displacement interpolation, which is analogous to a straight line in the geometry of probability distributions~\cite{McCann1997}. A useful feature of the 2-Wasserstein distance is that the static map $T$ also gives us the transport map for all intermediate times $t$, which we will return to later when we discuss flow matching in Sec.~\ref{sec:flow_matching}. As expected, this picture is consistent with the Brownian bridge discussed previously (Eq.~\eqref{eq:brownian_bridge}). For a fixed endpoint pair, the mean at an intermediate time $t$ follows $(1-t)x_0+tx_1$, with covariance $\varepsilon t(1-t)I$. As $\varepsilon\to0$, the fluctuations disappear and the bridge collapses onto that straight path. 

Although we focus on the quadratic cost here, other choices can produce different optimal assignments between the same endpoints. Increasing the exponent in $c(x_0,x_1)=\norm{x_0-x_1}^p$ penalizes long displacements more strongly and changes how the source is partitioned among those targets. The optimal pairing also depends on the choice of transport cost. 
Fig.~\ref{fig:transport-cost-maps} shows an example with three target locations and compares linear, quadratic, and quartic distance costs for the same initial and final densities, showing how this choice changes the regions of the initial density assigned to each target mode.

\subsection{Wasserstein geometry and Wasserstein gradient flows}
\label{sec:WGF}

As we've now seen, the $2$-Wasserstein distance $W_2$ in Eq.~\eqref{eq:bb} provides a notion of distance between probability densities in terms of the optimal flow field to transport one to another. This notion of distance provides the starting point for Wasserstein geometry, in which one treats the space of probability distributions like a Riemannian manifold in which the square root of the minimum in Eq.~\eqref{eq:bb} is the distance between points~\cite{Otto2001}. This will allow us to generalize the notion of gradient flow for particles in a potential $\dot x = - \nabla F(x)$ to that of densities flowing down gradients of a functional. 
This geometric picture will provide the basis for elegant algorithms presented in the Applications Sec.~\ref{sec:WGFApps}.

In Wasserstein geometry, a probability density $\rho$ is regarded as a point, a time-dependent density $\rho_t$ as a curve, and its instantaneous rate of change $s_t(x) = \partial_t \rho_t(x)$ as a tangent vector. Crucially, Wasserstein geometry only permits trajectories $\rho_t(x)$ that obey the continuity constraint, namely that they can be realized via a flow field $v_t$

\begin{align}
    \partial_t \rho_t = - \nabla \cdot (\rho_t v_t) \label{eq:contin}
\end{align}
Some trajectories are not allowed because they cannot be realized by a smooth flow field, for instance $\rho_t = \rho_1 t + (1-t ) \rho_0 $ when $\rho_0$ and $\rho_1$ have disjoint support.
This space of probability measures with finite second moment, equipped with $W_2$, is denoted $\mathcal{W}_2$~\cite{ChewiBOOK2025,Ambrosio2008gradient}.

Due to the continuity constraint in Eq.~\eqref{eq:contin}, it is possible to set up a duality between each tangent vector $s(x) = \partial_t \rho_t$ and the velocity field $v$ that generates it. The subtlety is that the velocity is not unique: adding a field whose probability current has zero divergence leaves $s$ unchanged. The canonical way to remove this redundancy is motivated by the metric $W_2$. Namely, given the tangent vector $s$ in the tangent space of the distribution $\rho$, choose the velocity field $v$ with minimum kinetic energy that generates $s$
\begin{equation}
\begin{aligned}
    v=\arg \min_{v'}\Bigl\{&\int\dd{x}\,\rho\norm{v'}^2 \\
    &\text{s.t.}\quad s+\nabla\cdot(\rho v')=0 \Bigr\}
\end{aligned}
\label{eq:w2_min_velocity}
\end{equation}
This minimization problem can be solved with a Lagrange multiplier for the continuity constraint. Let's say we choose its sign and normalization by adding $-2\int\dd x\,\chi(x)[s(x)+\nabla\cdot(\rho v)]$ to the kinetic energy. Variation with respect to $v$ then gives $v(x)=-\nabla\chi(x)$. Thus, the minimum energy velocity is a potential gradient.

Having fixed $v$ in this way, the minimum $\|s\|_\rho^2=\int\dd x\,\rho\|v\|^2$ in Eq.~\eqref{eq:w2_min_velocity} defines the squared norm of the tangent at $\rho$. Along a regular curve, this is the squared speed measured by $W_2$:
\begin{align}
    \|s_t\|_{\rho_t}=\lim_{\Delta t\to0}
    \frac{W_2(\rho_t,\rho_{t+\Delta t})}{|\Delta t|},
    \qquad s_t=\partial_t\rho_t.
    \label{eq:II}
\end{align}
Equivalently, $W_2^2(\rho_t,\rho_{t+\Delta t})=(\Delta t)^2\|s_t\|_{\rho_t}^2+o((\Delta t)^2)$. The inner product of two tangents $s,s'$ follows from the polarization identity $\langle s,s'\rangle_\rho=\tfrac12[\|s+s'\|_\rho^2-\|s\|_\rho^2-\|s'\|_\rho^2]$, giving

\begin{equation}
    \langle s,s'\rangle_\rho
    =\int\dd{x}\,\rho(x)v(x) \cdot v'(x).
    \label{eq:w2_tangent_inner_product}
\end{equation}
where $v(x)$ and $v'(x)$ are the minimum-energy velocity fields associated with $s$ and $s'$, respectively.

With the ingredients of Wasserstein geometry in hand, we can now define gradient descent for a functional $\mathcal{F}[\rho]$. Recall that for a function $F(x)$ in Euclidean space, its gradient is defined in such a way as to enforce the chain rule: $\nabla F$ is the vector with the property $\dd F(x_t)/\dd t=\dot x_t\cdot\nabla F(x_t)$ for all trajectories $x_t$. 
In complete analogy, the Wasserstein gradient  is defined to be the velocity field $\Wgrad\FF[\rho]$, with associated tangent vector $s=-\nabla\cdot(\rho\Wgrad\FF[\rho])$, such that $\frac{\dd\FF[\rho_t]}{\dd t} = \langle  s , \partial_t \rho_t \rangle_{\rho_t}$ for all trajectories $\rho_t$. To derive an explicit expression for $\Wgrad\FF[\rho]$, consider $\dd \FF/\dd t$ over a flow  generated by a potential $v=-\nabla\chi$. Writing $\delta\mathcal F/\delta\rho$ for the functional derivative, evaluated at the current density $\rho_t$, one obtains

\begin{equation}
\begin{aligned}
    \frac{\dd\mathcal{F}[\rho_t]}{\dd t}
    &=\int\dd{x}\,\frac{\delta\mathcal{F}}{\delta\rho}\partial_t\rho_t =\int\dd{x}\,\frac{\delta\mathcal{F}}{\delta\rho}\nabla\cdot(\rho\nabla\chi)\\
    &=-\int\dd{x}\,\rho\,\nabla\frac{\delta\mathcal{F}}{\delta\rho}\cdot\nabla\chi.
\end{aligned}
\label{eq:w2_functional_derivative}
\end{equation}
Comparing Eq.~\eqref{eq:w2_functional_derivative} with the inner product in Eq.~\eqref{eq:w2_tangent_inner_product}, the Wasserstein gradient velocity field is given by $\Wgrad\mathcal{F}[\rho]=\nabla(\delta\mathcal{F}/\delta\rho)$. This is the gradient in its velocity representation, whereas the associated density tangent is $-\nabla\cdot(\rho\Wgrad\mathcal F[\rho])$.
A \emph{Wasserstein gradient flow (WGF)} follows the negative gradient, with descent velocity $v_t=-\Wgrad\mathcal F[\rho_t]$. The resulting density evolution is

\begin{equation}
\begin{aligned}
    \partial_t\rho_t&=\nabla\cdot\left(\rho_t\Wgrad\mathcal{F}[\rho_t]\right),\\
    \frac{\dd\mathcal{F}[\rho_t]}{\dd t}
    &=-\int\dd{x}\,\rho_t\norm{\Wgrad\mathcal{F}[\rho_t]}^2\leq0.
\end{aligned}
\label{eq:gradflow}
\end{equation}
The WGF is a distributional analog of gradient descent $\dot{x}=-\nabla F(x)$ in Euclidean space. For example, consider the free energy
\begin{equation}
    \mathcal{F}[\rho]=\int\dd{x}\,\rho(x)\Phi(x)
    +\int\dd{x}\,\rho(x)\log\rho(x).
    \label{eq:wgf_free_energy_example}
\end{equation}
Eq.~\eqref{eq:gradflow} becomes
$\partial_t\rho_t=\nabla\cdot(\rho_t\nabla\Phi)+\nabla^2\rho_t$, the familiar drift-diffusion Fokker--Planck equation~\cite{JKOSIAM1998}. The potential energy term drives probability down $\Phi$, while the entropy term produces diffusion. Thus, the relaxation of a diffusive system can be read as steepest descent of a free energy in Wasserstein geometry, a connection that will later be discussed in the context of non-equilibrium thermodynamics (Sec.~\ref{sec:OTthermo}) and applications (Sec.~\ref{sec:WGFApps}). 

Once steepest descent has been defined, one can ask familiar questions from optimization, now in the space of densities. For example, it is useful to define a notion of convexity that rules out competing local minima and controls the convergence rate---these ideas will come up again in Sec.~\ref{sec:WGFApps} on applications of WGF, particularly in the setting of variational inference.  Recall that in Euclidean space, a function $F$ is \emph{$\alpha$-convex} if

\begin{equation}
\begin{aligned}
    &F((1-t)x_0+tx_1)\\
    &\quad\leq(1-t)F(x_0)+tF(x_1)\\
    &\qquad-\frac{\alpha}{2}t(1-t)\norm{x_0-x_1}^2.
\end{aligned}
\label{eq:alpha_convex}
\end{equation}
Ordinary convexity corresponds to $\alpha=0$. Analogously, a functional $\mathcal{F}:\mathcal{W}_2\to\RR$ is \emph{$\alpha$-geodesically convex} if, along the constant-speed Wasserstein geodesic $\rho_t$~\cite{McCann1997},

\begin{equation}
\begin{aligned}
    \mathcal{F}[\rho_t]
    &\leq(1-t)\mathcal{F}[\rho_0]+t\mathcal{F}[\rho_1]\\
    &\quad-\frac{\alpha}{2}t(1-t)W_2^2(\rho_0,\rho_1).
\end{aligned}
\label{eq:alpha_geodesic_convex}
\end{equation}
We emphasize that the comparison in Eq.~\eqref{eq:alpha_geodesic_convex} is made along geodesic interpolation, rather than along the point-wise mixture of two densities.
For $\alpha>0$, under standard regularity assumptions, if $\rho_t$ follows the Wasserstein gradient flow of an $\alpha$-geodesically convex $\mathcal{F}$, then

\begin{equation}
    \mathcal{F}[\rho_t]-\inf\mathcal{F}
    \leq e^{-2\alpha t}\left[\mathcal{F}[\rho_0]-\inf\mathcal{F}\right].
    \label{eq:convrate}
\end{equation}
assuming $\FF$ is bounded from below~\cite{ChewiBOOK2025}. 
Eq.~\eqref{eq:convrate} states that the functional gap decays exponentially. If a minimizer $\rho_*$ exists, positive geodesic convexity makes it unique and gives $\mathcal F[\rho]-\mathcal F[\rho_*]\geq\alpha W_2^2(\rho,\rho_*)/2$, so the same estimate also implies exponential convergence in $W_2$. In addition to defining convex functionals, we can define convex sets. We say that a set of probability distributions is \emph{geodesically convex} if it contains the Wasserstein geodesic between any two of its densities. Functionals that are $\alpha$-geodesically convex over $\mathcal{W}_2$ retain their convexity properties and convergence guarantees when restricted to geodesically convex subsets. 
In later sections, we will use this framework to describe irreversible thermodynamic relaxation, numerical integration of Fokker--Planck equations, variational inference, and the mean-field dynamics of interacting particles.

\section{Thermodynamics}\label{sec:thermodynamics}

As we elucidate in this section, the previously explored ideas in control and inference have a deep connection to the time-honored subject of thermodynamics. 
Upon specifying a state-space geometry, statistical inference aims to select probability distributions consistent with partial information, whereas control theory seeks protocols that transport systems between prescribed configurations while minimizing a cost. Thermodynamics simultaneously determines admissible equilibrium macroscopic states and identifies the least costly transformations among them, compatible with the imposed constraints such as fixed temperature, fixed volume, or adiabaticity. Here, we aim to make these connections concrete. In Sec.~\ref{sec:energy-based_modelling}, we discuss the maximum-entropy interpretation of equilibrium statistical mechanics. Subsequently, in Sec.~\ref{sec:OTthermo} we demonstrate that the Onsager least-dissipation principle and its extension to stochastic processes amount to an optimal transport problem.

\subsection{Equilibrium thermodynamics as maximum-entropy inference}\label{sec:energy-based_modelling}

Here, we discuss the extent to which statistical mechanics of materials in equilibrium can be interpreted as a maximum-entropy inference. As materials are composed of $\sim10^{23}$ degrees of freedom, it is infeasible to keep track of their individual dynamics. Instead, equilibrium statistical mechanics provides a means of describing microscopic particle configurations probabilistically. Using thermodynamic axioms (such as the thermodynamic entropy $S$ reaching a maximum in an isolated system at equilibrium, and the ergodic hypothesis wherein all states are equiprobable within a fixed energy shell), Boltzmann, Gibbs, and Maxwell obtained the expressions for the microstate distribution at equilibrium~\cite{KardarBOOK2007}. 

Boltzmann argued that by the ergodic hypothesis, all microstates $x$ with the same energy are equally likely in an isolated equilibrium system. He postulated that the microstate multiplicity $W$ compatible with the conserved quantities is maximized at equilibrium (by a reasoning akin to modern saddle-point approaches), and thus sets the thermodynamic entropy via $S=k_\mathrm{B}\log W$ (where the logarithm guarantees that entropy is additive). This is called the microcanonical ensemble and can be refined to accommodate additional conserved quantities, like particle number.

Boltzmann then considered a system whose energy fluctuates due to energy exchange with a large thermal bath, constituting the canonical ensemble. Suppose that the system has energy $E(x)$ and the bath $E_\mathrm{tot}-E$, so that $E_\mathrm{tot}$ is the conserved energy of the joint system (system and bath). The equilibrium probability distribution of the joint configuration $x_\mathrm{tot}=(x,x_\mathrm{bath})$ is uniform, $p^\mathrm{eq}(x_\mathrm{tot})=1/W_\mathrm{tot}(E_\mathrm{tot})$. However, integrating the bath degrees of freedom for a fixed system configuration $x$ yields
\begin{equation}
\begin{aligned}
p^\mathrm{eq}(x) &=\frac{p^\mathrm{eq}(x, x_\mathrm{bath})}{p^\mathrm{eq}(x_\mathrm{bath} | x)} = \frac{W_\mathrm{bath}(E_\mathrm{tot}-E(x))}{W_\mathrm{tot}(E_\mathrm{tot})}\\
&\propto e^{-k_\mathrm{B}^{-1}S_\mathrm{bath}(E_\mathrm{tot}-E(x))}
\end{aligned}
\label{eq:distribution_bath_entropy}
\end{equation}
(since $W_\mathrm{tot}(E_\mathrm{tot})=\mathrm{const}$). Therefore, physics naturally equipped us with a probability following a large-deviation principle, as $\log p^{\mathrm{eq}}$ is an extensive function. Since the bath is much larger than the subsystem $E(x)\ll E_\mathrm{tot}$, we Taylor expand $S_\mathrm{bath}(E_\mathrm{tot}-E(x))\simeq S_\mathrm{bath}(E_\mathrm{tot})-E(x)/T$, where $1/T = \mathrm{d} S_\mathrm{bath}(E)/\mathrm{d}E$ is the bath temperature. We write $\beta=(k_{\mathrm B}T)^{-1}$ for the inverse thermal energy. Thus, the equilibrium distribution of the subsystem microstates $x$ reads:

\begin{equation}
p^\mathrm{eq}(x)=\frac1Ze^{-\beta E(x)},\label{eq:Bfactor}
\end{equation}
We thus arrive at the Boltzmann distribution from \emph{physical} principles: energy conservation, the ergodic hypothesis, and a large-deviation principle arising from the thermodynamic limit. 
In the above, $Z$ is the partition function, and it is related to the free energy via $\mathcal F^{\mathrm{eq}} = -k_\mathrm{B}T\log Z$.  
Similar fluctuation principles upon coupling to different reservoirs give rise to all other statistical-mechanical ensembles.
Notably, the entropy across all ensembles can be written as

\begin{equation}
S=-k_{\mathrm B}\int\dd x\,p^\mathrm{eq}(x)\log p^\mathrm{eq}(x).\label{eq:TDentropy}
\end{equation}
Note that the Boltzmann factor (Eq.~\eqref{eq:Bfactor}) belongs to the exponential family (Eq.~\eqref{eq:MaxEnt_q}), and the thermodynamic entropy (Eq.~\eqref{eq:TDentropy}) takes the form of a Shannon entropy (Eq.~\eqref{eq:ShannonEnt}). This suggests a link between equilibrium statistical mechanics and maximum-entropy modelling (discussed in Sec.~\ref{sec:MEM}). Indeed, Jaynes proposed interpreting thermodynamics as enforcing a maximum-entropy inference~\cite{JaynesPR1957} wherein, for example, the average energy of a system, $\mathcal{E}=\int dxp^\mathrm{eq}(x)E(x)$, when coupled to a thermal bath, acts as a constraint. Upon maximization, the thermodynamic entropy is given by the Shannon entropy of the entropy-maximizing distribution,
\begin{equation}
S=k_\mathrm{B} H[p^\mathrm{eq}].\label{eq:EntEntRelation}
\end{equation}
It is important to stress, however, that the conceptual justification behind maximum-entropy modelling and Boltzmann statistics is fundamentally different. The former is motivated by the principle that, given limited information, the least-committal distribution is the one that maximizes entropy, while the latter follows from physical laws governing equilibrium statistical mechanics. The connection between maximum-entropy modelling and statistical mechanics thus arises from the coincidence between the functional forms obtained in the two frameworks. In that regard, while Eq.~\eqref{eq:EntEntRelation} is conceptually fascinating, there is no indication that either this relation or the information-theoretic interpretation of statistical mechanics should hold away from equilibrium.

While thermodynamic entropy (like other thermodynamic state functions such as temperature~\cite{CasasVazquez2003} and pressure~\cite{solon2015pressure}) is an equilibrium property of   matter, is it possible to extend thermodynamic notions far from equilibrium? Since Shannon entropy is well-defined for a system out of equilibrium given its instantaneous distribution $p(x)$, can it serve as an extension of entropy outside equilibrium, $S\to k_\mathrm{B}H[p]$, even though this quantity is not generally maximized? The answer lies in whether the emergent relations obtained by presuming a particular entropy functional are consistent with thermodynamic theorems and whether they provide useful predictions, \textit{e.g.}, for the extractable work during thermodynamic transformations.

For example, in certain systems (most often those uncoupled from an external bath), Shannon entropy is explicitly an inadequate measure of thermodynamic entropy~\cite{LebowitzPHYSA1999,LevinPR2014}. For instance, it is a Lyapunov function of the Liouville equation describing isolated systems, indicating that it does not increase as the system approaches equilibrium (despite the system potentially being ergodic). Therefore, not only may there be no physical reason to invoke maximum-entropy inference in such systems, but Shannon entropy may not be an adequate functional in the first place.

On the other hand, in other cases (most often stochastic systems away from the thermodynamic limit, where the number of particles is small), Shannon entropy provides a useful generalization of thermodynamic entropy out of equilibrium. For example, as we show in Sec.~\ref{sec:thermodynamic_fluctuation_theorems}, it recovers the second law of thermodynamics as well as additional useful fluctuation theorems concerning the extractable work from stochastic systems~\cite{SeifertROPP2012,PelitiBOOK2021}. Furthermore, returning to Shannon's construction, Shannon entropy is a functional that adequately quantifies uncertainty~\cite{Shannon1948}. Therefore, in the absence of any knowledge about a problem's structure beyond a few known expectation values, maximizing entropy under constraints is not an unreasonable inference methodology~\cite{JaynesIEEE1982}. Indeed, maximum-entropy inference, both static and dynamical (where $P$ denotes a distribution over trajectories), has proved useful across scales, including predicting protein structure~\cite{WeigtPNAS2009}, modeling neural communication~\cite{MeshulanRMP2025}, identifying bacterial swarming transitions~\cite{SorkinSM2023}, and characterizing alignment interactions during flocking~\cite{CavagnaARCMP2014}. Since thermodynamic variables are often uninformative in these systems, the imposed constraints are no longer temperature, pressure, and related quantities, but rather static and dynamic correlation functions~\cite{CavagnaPRE2014,SorkinPRE2023}, compressed snapshot data~\cite{AvineryPRL19,MartinianiPRX19}, transport coefficients~\cite{DyreJCP18,SorkinPRL2023}, and more, which nonetheless carry substantial information about the microstructure of the system. The success of such inference schemes depends not only on the adequacy of the maximum-entropy principle itself, but also on the judicious choice of the moments used to constrain the distribution~\cite{SorkinSM2023}: moments that capture relevant aspects of the underlying dynamics can yield accurate predictions, whereas constraints on physically irrelevant observables may provide little useful information.

To conclude, there is a deep but subtle connection between equilibrium thermodynamics and information theory. Upon accepting the axioms of thermodynamics, Gibbsian statistical mechanics can be formulated as an inference problem on rigorous grounds. Since entropy is constructed to provide a consistent quantification of statistical uncertainty, exponential-family models can also be useful for inferring the steady-state microstructure of athermal systems, albeit with important limitations. The thermodynamic meaning of these models, however, remains subject to debate.

\subsection{Nonequilibrium thermodynamics as a control problem}\label{sec:OTthermo}

In Sec.~\ref{sec:energy-based_modelling}, we recalled the statistical-mechanical characterization of systems at thermodynamic equilibrium, and discussed its information-theoretic interpretation. 
In this section, we will frame thermodynamic processes close to and far from equilibrium in the language of optimal transport. 
Thermodynamics provides a concrete notion of efficiency grounded in the second law,
\begin{equation}
    T\Sigma=T\Delta S-Q=W-\Delta \FF\geq0,\label{eq:2nd_law}
\end{equation}
where $W$ is the mechanical work along a process and $\Delta \FF$ is the resulting change in free energy $\FF$. Here $\Sigma$ is the entropy production, so that $T\Sigma$ is the dissipated heat\,---\,the energy irreversibly lost during the transformation.
(Note that, confusingly, $\Sigma$ is often called entropy production; more often than not, entropy production $\Sigma$ and change in entropy $\Delta S$ are different objects.) 
In particular, an efficient process between an initial and final state is one which minimizes the entropy production $\Sigma$, in turn allowing the extractable work $-W$ to reach the thermodynamic upper bound $-\Delta \FF$. 

In this section, we will discuss useful approaches for estimating the extractable work $W$ and different computational strategies to minimize entropy production $\Sigma$ using optimal-transport and path-integral methods. We begin by studying thermodynamic relaxation for small departures from equilibrium, described by the Onsager least-dissipation theory. We then extend these ideas to stochastic processes far from the thermodynamic limit (few degrees of freedom instead of $N_{\mathrm p}\sim10^{23}$ particles), for which we will derive useful thermodynamics-inspired bounds, a geometric interpretation for minimizing the entropy production, and nontrivial equalities for the extractable-work fluctuations.

\subsubsection{Optimal transport bounds efficiency via Onsager's least-dissipation principle}\label{sec:onsager}

Let's start with a modest question\,---\,what are the dynamics when we depart from equilibrium but remain close to it?
In this setting, Onsager formulated an approach that is phenomenological in nature~\cite{OnsagerPR1931i,OnsagerPR1931ii}, which has since permitted the calculation of bounds on dissipation in terms of the Wasserstein metric.   
The connection to Wasserstein geometry ultimately occurs in the context of field theories, but Onsager's governing principles are perhaps easiest to state in a discrete setting.  In brief, consider an isolated system described by the generalized variables $\{X_m\}_{m=1,\ldots,M}$\,---\,for instance, concentrations of various compounds in a chemical mixture\,---\,that determine the system free energy, $\FF(X_1,\ldots,X_M)$. 
Onsager postulated that, around equilibrium, the system responds linearly to  gradients in the free energy:
\begin{equation}
    \frac{\dd X_m}{\dd t}= - \sum_{n=1}^M L_{mn}\frac{\partial \FF}{\partial X_n}.\label{eq:LITdisc}
\end{equation}
where $L_{mn}$ is the Onsager matrix of phenomenological transport coefficients.  
Physically, the coefficients $L_{mn}$, which can be measured experimentally as well as predicted theoretically in some cases~\cite{ZwanzigBOOK2001}, describe how changing one variable affects all other variables in response. For instance, in the thermoelectric effect, the corresponding $L_{mn}$ is known as the thermoelectric coefficient and describes how changes in temperature drive changes in charge density. Eq.~\eqref{eq:LITdisc} serves as the basis for modern linear-irreversible thermodynamics~\cite{deGrootBOOK84}, wherein a system relaxes to equilibrium linearly in the thermodynamic forces ($\partial \FF /\partial X_m$), but the forces may be arbitrarily nonlinear in the $X$'s. 

It is worth noting that Onsager proved various nontrivial properties of $L_{mn}$~\cite{OnsagerPR1931ii}. First, to satisfy the second law of thermodynamics (the nonnegativity of heat dissipation, Eq.~\eqref{eq:2nd_law}), $L_{mn}$ must be positive-definite. Namely, close to equilibrium, very generally~\cite{OnsagerPR1931i} the dissipation functional is given by the thermodynamic dissipative force $-\sum_m(L^{-1})_{nm}\dot X_m$ (friction coefficient times velocity) along a displacement $\dot X_n$,
\begin{equation}
    T\dot\Sigma=\sum_{n,m}(L^{-1})_{nm}\dot X_n\dot X_m,\label{eq:onsagerEPR}
\end{equation}
which therefore requires a positive-definite $L_{mn}$ to satisfy $\dot\Sigma>0$.
Moreover, the so-called Onsager reciprocity theorem states that $L_{mn}$ must be symmetric, $L_{mn}=L_{nm}$, when the microscopic equations of motion are time-reversal symmetric.  

Useful for our purposes is that, for systems obeying Onsager reciprocity~\cite{OnsagerPR1931ii}, Eq.~\eqref{eq:LITdisc} can be obtained from a variational principle. Namely, near equilibrium, the velocities $\{\dot X_m\}$ extremize the following function:

\begin{align}
\begin{aligned}
\mathcal R(\dot X_1, \dots , \dot X_M ) &= \frac12 T\dot \Sigma + \frac{\dd \FF}{\dd t}\\
&= \frac12 T\dot \Sigma + \sum_m \frac{\partial \FF}{\partial X_m} \dot X_m
\end{aligned}
\label{eq:rate}
\end{align}
where $\mathcal R$ is known as the Rayleighian, and in the second equality we used the chain rule for the free energy. Solving $\partial \mathcal R/\partial \dot X_m =0$ is equivalent to Eq.~\eqref{eq:LITdisc} with the heat dissipation of Eq.~\eqref{eq:onsagerEPR}. In this formulation, it is natural to think of gradients of $\FF$ as driving forces, and $L^{-1}$ as yielding a metric that defines steepest descent. 

Onsager's theory can be connected with optimal transport when Eq.~\eqref{eq:rate} is extended to field theories. We follow the lines of Ref.~\cite{DoiJPCM2011}. In a continuum (hydrodynamic) approach, one models a fluid via a collection of conserved density fields (\textit{e.g.},  mass, momentum, or energy density fields~\cite{HansenBOOK2003}). For simplicity, we will concentrate on a single scalar field $n_t(x)$ representing the density of a conserved quantity, such as particle density. We denote the conserved particle number by $N_{\mathrm p}=\int n_t(x)\,\dd x$; unlike a probability density, $n_t$ need not integrate to one. Here $n_t(x)$ plays the role of $X_m(t)$ in Eqs.~\eqref{eq:LITdisc}--\eqref{eq:rate}, and the spatial coordinate $x$ plays the role of the index $m$. The field-theoretical generalization of Eq.~\eqref{eq:rate} can be used as a prescription to derive the equations of motion for $n_t(x)$. First, because the field is conserved, it evolves according to the continuum equation

\begin{equation}
    \frac{\partial n_t(x)}{\partial t}+\nabla\cdot[n_t(x)v_t(x)]=0,\label{eq:cons}
\end{equation}
where $v$ is the velocity field. Next, we assume that the free energy may be expressed as a local functional of $n$, namely $\FF[n]$. 
Finally, we consider phenomenologically that the dissipation $\dot \Sigma$ is quadratic in $v_t(x)$ as in Eq.~\eqref{eq:onsagerEPR},

\begin{equation}
    T\dot{\Sigma}=\int\dd{x} n_t(x)\Gamma(n_t(x))|v_t(x)|^2.\label{eq:EPR}
\end{equation}
where the phenomenological dissipation coefficient  $\Gamma(n)$ may be an arbitrary function of density $n$. 

As a concrete example, one can have in mind a system of very dilute colloidal particles in a viscous fluid. In this case $n_t(x)$ represents the number density of the particles; neglecting hydrodynamic interactions, the drag is given by the Stokes drag $\Gamma=6\pi \eta  r$ (where $r$ is the radius of the suspended particles, and $\eta$ is the viscosity); and, assuming the particles are non-interacting, the free energy is the negative of the entropy $\FF = - TS = k_\mathrm{B}T\int n_t(x) \log n_t(x) \dd x$.  

Upon extending the Onsager Rayleighian Eq.~\eqref{eq:rate} to fields, we obtain

\begin{equation}
\begin{aligned}
    \mathcal R&=\frac12T\dot\Sigma+\frac{\dd\mathcal{F}}{\dd t}\\
    &=\frac12T\dot\Sigma+\int\dd x n_t(x)v_t(x)\cdot\nabla\frac{\delta \mathcal{F}}{\delta n_t(x)}.
\end{aligned}
\label{eq:Rayleighian}
\end{equation}
with the entropy production $\dot\Sigma$ of Eq.~\eqref{eq:EPR}, and we used the functional chain rule and Eq.~\eqref{eq:cons} for evaluating the change in free energy. According to the Onsager variational principle, the equation of motion for $n_t(x)$ is determined by choosing the flow $v_t(x)$ that minimizes $\mathcal R$. Taking $\delta \mathcal R/\delta v =0$ yields 

\begin{equation}
    \Gamma(n_t(x))v_t(x)=-\nabla\frac{\delta \mathcal{F}}{\delta n_t(x)}.\label{eq:LIT}
\end{equation}
We write $\mu_t(x)=\delta\mathcal F/\delta n_t(x)$ for the local chemical potential. Eq.~\eqref{eq:LIT}, when combined with Eq.~\eqref{eq:cons}, yields a closed dynamical equation for $n$:

\begin{align}
 \partial_t  n = \nabla\cdot \left(\frac{n}{\Gamma (n)} \nabla \frac{\delta \mathcal F}{\delta n} \right)\label{eq:modelB}
\end{align}
Returning to our example of a dilute colloidal suspension, one obtains

\begin{equation}
    \frac{\partial n_t(x)}{\partial t}=D\nabla^2n_t(x),
\end{equation}
where $D=k_\mathrm{B}T/(6\pi\eta r)$ is the Stokes-Einstein diffusion constant.
Indeed, many of the known dynamical theories in soft matter, ranging from diffusion to phase separation and viscoelasticity, can be derived from Onsager's variational principle~\cite{DoiJPCM2011}. 

Onsager's least-dissipation theorem is closely related to optimal transport\,---\,as we show, the dissipation associated with the WGF serves as a lower bound on the true dissipated heat. Note the close resemblance of Eq.~\eqref{eq:modelB} to Eq.~\eqref{eq:gradflow}. The Onsager Rayleighian, Eq.~\eqref{eq:Rayleighian}, can be compared per particle with probability transport. Defining $\rho_t(x)=n_t(x)/N_{\mathrm p}$ gives

\begin{align}
\label{eq:W2R}
    \frac{2}{\Gamma N_{\mathrm p}}\int_0^1\dd t\,\mathcal R_t
    &=\int_0^1\dd t\int\dd x\,\rho_t(x)\|v_t(x)\|^2\\
    &+\int_0^1\dd t\int\dd x\,\rho_t(x)
      \left[\frac{2}{\Gamma}\nabla\mu_t(x)\right]\cdot v_t(x).
    \nonumber
\end{align}
where, for simplicity, we have taken $\Gamma=\mathrm{const}$. At the same time, return to Eq.~\eqref{eq:bb}. Using the method of Lagrange multipliers, the minimum of the 2-Wasserstein distance subject to the conservation law and initial and final distributions is found from the saddle point of the following Lagrangian,

\begin{equation}
\begin{aligned}
    &\int_0^1\dd t\int\dd x\,\rho_t(x)\|v_t(x)\|^2\\
    &+\int_0^1\dd t\int\dd x\,\rho_t(x)
       [2\nabla\chi_t(x)]\cdot v_t(x)\\
    &+(\text{terms independent of }v).
\end{aligned}
\label{eq:W2L}
\end{equation}
Here $\chi_t(x)$ is the transport potential of Chapter~\ref{sec:OT}, introduced by adding $-2\chi_t[\partial_t\rho_t+\nabla\cdot(\rho_t v_t)]$ to the integrand before integrating by parts. Variation with respect to $v_t$ gives $v_t=-\nabla\chi_t$, where terms independent of $v_t$ have been ignored.

Eqs.~\eqref{eq:W2R} and~\eqref{eq:W2L} have the same quadratic velocity term. The former describes the particle density $n_t=N_{\mathrm p}\rho_t$, with dissipation expressed per particle, whereas the latter is the probability transport variational problem. Thermodynamics fixes the potential to $\chi_t=\mu_t/\Gamma$, where $\mu_t=\delta\mathcal F/\delta n_t$ is the chemical potential, instead of optimizing over $\chi_t$. Thus, the Benamou-Brenier solution is a more efficient transportation mechanism;\footnote{The optimized $\chi_t$ satisfies the Hamilton--Jacobi equation $\partial_t\chi_t=\|\nabla\chi_t\|^2/2$, up to a time-dependent additive gauge; its gradient gives the corresponding Burgers equation.} linear-irreversible thermodynamics is a Benamou-Brenier-type ``partially'' optimized transport, as the chemical potential is specified. We will return to the far-reaching consequence of this realization\,---\,a bound, tighter than Eq.~\eqref{eq:2nd_law}, for finite-time transformations\,---\,shortly. But until then\,---\,may a similar conclusion apply far from equilibrium?

\subsubsection{Dynamics far from equilibrium}\label{sec:far_from_equilibrium}

In constructing Onsager's least-dissipation principle, we considered a coarse-grained hydrodynamic theory close to equilibrium. Yet, the same mathematical structures arise in many other types of dynamics, notably in stochastic processes far from equilibrium. In the following, we will be working within the framework of stochastic thermodynamics~\cite{SeifertROPP2012,PelitiBOOK2021}, aiming to extend thermodynamic notions and theorems to fluctuation-dominated microscopic systems, such as driven colloidal suspensions and individual motor proteins. Since we are no longer in the thermodynamic limit, it is not obvious whether state functions such as temperature or pressure carry over to the microscale or whether the (fluctuating) mechanical work satisfies any relations akin to the second law (Eq.~\eqref{eq:2nd_law}). The immense success of stochastic thermodynamics, summarized in part here and in the coming sections, is the fact that many functions, including mechanical work, can indeed be used still; \textit{e.g.}, the fluctuating work is just any externally applied forces along a stochastic displacement. Upon selecting the surrounding bath properties (temperature, pressure, etc.) and adopting the Shannon entropy as the microscopic analogue of a nonequilibrium thermodynamic entropy, the second law of thermodynamics and many additional useful new fluctuation relations can be derived. The aim of this section is to explore the correspondence between the thermodynamics of stochastic systems and optimal control.

To demonstrate this point, we will consider the paradigmatic example of an overdamped colloid submerged in a thermal bath, which an agent drives via an external force $F_t(x)$,

\begin{equation}
    \dot x_t=D\beta F_t(x_t)+\sqrt{2D}\eta_t,\label{eq:THERMOlangevin}
\end{equation}
where $D$ is the particle diffusivity and $\eta_t$ is standard white noise. Thus, $\varepsilon=2D$, the stochastic drift is $f_t(x)=D\beta F_t(x)$, and $F_t$ denotes a physical force. The time-dependent probability distribution to obtain a microstate, $p_t(x)$, satisfies the Fokker-Planck equation~\cite{SchussBOOK2010}\,---\,the continuity equation with the current velocity

\begin{equation}
    v_t(x)=D\beta F_t(x)-D\nabla\log p_t(x).\label{eq:FPEdrift}
\end{equation}
As preparation for later, we distinguish between two types of steady states: One is an equilibrium state (as in Sec.~\ref{sec:energy-based_modelling}), where the external driving force is constant in time and conservative, $F=-\nabla E$, yielding the Boltzmann factor as the equilibrium distribution, Eq.~\eqref{eq:Bfactor}, and a zero current, $v\to0$. The other is a nonequilibrium steady state, where the current need not be zero ($v\ne0$), but the flux is divergence-free $\nabla\cdot(pv)\to0$ so $\partial p/\partial t=0$.

Since we may not be close to equilibrium, we may not rely on Onsager's dissipation functional, and instead refer to the machinery of stochastic thermodynamics~\cite{SekimotoPTP1998,PelitiBOOK2021}. Given our precaution in Sec.~\ref{sec:energy-based_modelling}, we note that the Shannon entropy (Eq.~\eqref{eq:ShannonEnt}) is indeed an adequate choice for the thermodynamic entropy out of equilibrium in stochastic processes, $S[p_t]=-k_{\mathrm B}\int \dd x\,p_t(x)\log p_t(x)$, since it will provide insightful and thermodynamically consistent results shortly. The change in entropy over time is

\begin{equation}
    \frac{\dd S}{\dd t}=-k_\mathrm{B}\int\dd xv_t(x)\cdot\nabla p_t(x).\label{eq:dentropy}
\end{equation}
According to the first law of thermodynamics, in this setting, the heat exchanged with the bath is minus the external work\,---\,the average of the externally applied forces along a displacement,

\begin{equation}
    \dot Q=-\dot W=-\int\dd xp_t(x)v_t(x)\cdot F_t(x).\label{eq:heatwork}
\end{equation}
Thus, the entropy production is given by

\begin{equation}
\begin{aligned}
    \dot\Sigma_t&=\frac{\dd S}{\dd t}-\frac{\dot Q}T =\frac{k_\mathrm{B}}D\int\dd{x} p_t(x)|v_t(x)|^2.
\end{aligned}
\label{eq:EPRlangevin}
\end{equation}
Indeed, it is zero at equilibrium ($v=0$), otherwise positive, and functionally consistent with Onsager's Eq.~\eqref{eq:EPR}. 

Upon further comparison with the Benamou-Brenier approach, Eq.~\eqref{eq:bb}, we obtain an optimal-transport-inspired bound on the entropy production~\cite{AurellJSP2012}. The mere fact that one started and terminated the process at particular distributions during a finite time $\tau$ means that one must have produced entropy which is bounded by the Wasserstein metric between the distributions

\begin{equation}
    \int_0^\tau\dd t\dot\Sigma_t\geq \frac{k_\mathrm{B}}{D\tau} W_2^2(p_0,p_\tau).\label{eq:W2EPR}
\end{equation}
Since $W_2^2$ is itself nonnegative, Eq.~\eqref{eq:W2EPR} is a nontrivial bound on the total dissipated heat $T\Sigma$ that is tighter than the second law of thermodynamics, Eq.~\eqref{eq:2nd_law}. 

The minimization of Eq.~\eqref{eq:W2L} necessitates a gradient current velocity, $v_t=-\nabla\chi_t$. Therefore, unless $F_t$ is conservative ($F_t=k_\mathrm{B}T\nabla[\log p_t-\chi_t/D]$), there is no chance to saturate the inequality in Eq.~\eqref{eq:W2EPR}, let alone approach $T\dot\Sigma=0$. 
Indeed, unless $F_t$ is given by the optimal value of $\chi_t$, even the Onsager least-dissipation principle will dissipate more heat than that prescribed by the Wasserstein-metric bound. To illustrate a related consequence of the above thermodynamic interpretations of Wasserstein geometry and its connection to the kinetics of stochastic processes, we outline in Sec.~\ref{sec:JKO} the Jordan-Kinderlehrer-Otto scheme~\cite{JKOSIAM1998} in the language of optimal transport.

Interestingly, since $W_2^2$ is a property of the two distributions (but not time; \textit{cf.} Eq.~\eqref{eq:w2_static}), the bound of Eq.~\eqref{eq:W2EPR} is inversely proportional to the elapsed time. Indeed, the second law only saturates for quasistatic transformations; otherwise, Eq.~\eqref{eq:W2EPR} provides a lower bound on the minimal heat that must have been dissipated to enable a finite-time process. Conversely, Eq.~\eqref{eq:W2EPR} can be interpreted as a thermodynamic speed limit on how quickly one may transform from $p_0(x)$ to $p_\tau(x)$, giving rise to the lower bound $\tau\geq k_\mathrm{B} W_2^2/(D\Sigma)$. 

To conclude, the Benamou-Brenier optimal transport problem sets a bound on the thermodynamic cost of very general, finite-time, nonequilibrium transformations~\cite{AurellJSP2012}. There are additional, fascinating extensions, stemming from combining optimal transport with stochastic mesoscopic systems. For instance, the above statements were shown to apply to discrete-state Markov systems as well~\cite{VanVuPRX23}. Upon introducing feedback via measurements, the dissipated heat bound can be made tighter in terms of the acquired information~\cite{TaghvaeiIEEE2022}. It is important to stress that although the above are nontrivial connections for far-from-equilibrium systems, these connections only apply in cases where the Einstein relation is valid~\cite{AurellJSP2012,VanVuPRX23}. 
Otherwise, the Wasserstein metric only yields a bound on the entropy production in athermal and active systems (which is still given by Eq.~\eqref{eq:EPRlangevin}~\cite{SeifertROPP2012,SorkinJSTAT2025}), but not on the dissipated heat, as the connection between the latter two breaks in the absence of a scalar temperature~\cite{SorkinPRL2024}.

Is it possible to saturate the inequalities provided by the second law of thermodynamics? We present here two strategies\,---\,one involving the expression of entropy production as a thermodynamic metric in protocol space~\cite{SivakPRL2012,Rotskoff.Crooks2015}, and the other involving fluctuation properties of stochastic processes~\cite{Jarzynski1997,Crooks1999}.

\subsubsection{Geometric interpretation of entropy production}\label{sec:geometry_quasistatic}

In the previous sections we found nontrivial bounds on the entropy production using optimal transport. Here, our aim is more practical\,---\,is it possible to engineer efficient thermodynamic transformations? By efficient we mean that the entropy production $\Sigma$ is minimal so, according to Eq.~\eqref{eq:2nd_law}, the extractable work is closest to the upper bound dictated by free energy. More concretely, while we did not specify $F_t(x)$ in Sec.~\ref{sec:far_from_equilibrium}, here we consider specifically a conservative force, $F_t(x)=-\nabla E_t(x)$. We furthermore note that the time dependence of the potential, in practice, arises from a ``control knob'' that the experimentalist varies over time (\textit{e.g.}, increasing the electric power of an optical trap, thereby increasing the radiation-pressure force on a colloid). These knobs are often called protocols, which we denote by $u_t$. Finding the most efficient thermodynamic transformation, therefore, entails optimizing the entropy production over the protocols $u_t$. For the purpose of this section, the protocol $u_t$ can be thought of as the controls in Sec.~\ref{sec:mechanics_control}.

In trying to find the optimization framework for the entropy production, we mention several noteworthy properties of entropy production, Eq.~\eqref{eq:EPRlangevin}: First, clearly the quasi-static transformations are most efficient\,---\,since, close to equilibrium $v(x)\to0$ everywhere in $x$, the dissipation $\Sigma\to0$, necessarily from above due to the quadratic form. Even more so, we see that work extraction in the quasi-static limit is indeed possible. Qualitatively, the slower the time scale over which the process occurs, $\tau\to\infty$, then $v_t(x)\sim\tau^{-1}$. Since changes in free energy, entropy (Eq.~\eqref{eq:dentropy}), work, and heat (Eq.~\eqref{eq:heatwork}) are proportional to $v$ but involve the integration over time, order-$\int_0^\tau\dd tv\sim1$ energy extraction is possible while the dissipation is $\sim\int_0^\tau\dd tv^2\sim1/\tau\to0$. Are all quasi-static transformations the same, or are some more efficient than others?
For instance, due to critical slowing down, it may be counterproductive to move throughout Ising-model phase space through the critical point~\cite{Rotskoff.Crooks2015}.

We restrict our attention to exploring the space of quasi-static transformations, as their entropy production is smallest. Since $v_t(x)$ is an involved function, requiring the knowledge of the instantaneous distribution in addition to the force, we seek to make the connection of quasistaticity to the protocol $u_t$ more concrete, as follows. For the process to be quasistatic, $p_t(x)$ must equilibrate on a faster timescale than the protocol evolves; namely all eigenmodes of the Fokker-Planck operator converge to equilibrium faster than the protocol varies. We first define the deviation $\delta p_t(x)$ of the instantaneous distribution $p_t(x)$ from the equilibrium Boltzmann factor corresponding to the instantaneous protocol value $p_{u_t}^\mathrm{eq}(x)$:

\begin{equation}
\begin{aligned}
    p_t(x)&:=p_{u_t}^\mathrm{eq}(x)+\delta p_t(x),\\
    p_u^\mathrm{eq}(x)&=\frac1{Z_u}\exp\left[-\frac{E_u(x)}{k_\mathrm{B}T}\right],\\
    Z_u&=\int \dd x \exp\left[-\frac{E_u(x)}{k_\mathrm{B}T}\right].
\end{aligned}
\end{equation}
By inserting this solution into the Fokker-Planck equation,

\begin{equation}
\begin{aligned}
    \frac{\partial p_t(x)}{\partial t}&=-\mathcal{L}_{u_t}p_t(x),\\
    \mathcal{L}_u&=-D\nabla\cdot\left[\frac{\nabla E_{u}(x)}{k_\mathrm{B}T}+\nabla\right],
\end{aligned}
\end{equation}
where $\mathcal{L}_up_u^\mathrm{eq}=0$, we find the exact expression

\begin{equation}
    \delta p_t(x)=-\int_{-\infty}^t\dd se_\mathrm{T}^{-\int_s^t\dd s'\mathcal{L}_{u_{s'}}}\dot u_s\cdot\frac{\partial p_{u_s}^\mathrm{eq}(x)}{\partial u},\label{eq:p1exact}
\end{equation}
where $e_\mathrm{T}$ is a time-ordered exponential. We will now use the assumption of a quasistatic protocol: Since all modes of $\LL_u$ converge prior to $u_t$ changing considerably, $e_\mathrm{T}^{-\int_s^t\dd s'\mathcal{L}_{u_{s'}}}$ suppresses integrand contributions in Eq.~\eqref{eq:p1exact} where $u_s$ departed much from $u_t$. Thus, we approximately replace all protocol values $u_s$ with $u_t$,

\begin{equation}
    \delta p_t(x)\simeq-\int_{-\infty}^t\dd se^{-(t-s)\mathcal{L}_{u_t}}\dot u_t\cdot\frac{\partial p_{u_t}^\mathrm{eq}(x)}{\partial u}.\label{eq:p1quasistatic}
\end{equation}
In preparation for later, we define the protocol score (a vector for a multicomponent protocol):
\begin{equation}
\begin{aligned}
    \Theta_u(x)&:=\frac{1}{p_u^\mathrm{eq}(x)}\frac{\partial p_u^\mathrm{eq}(x)}{\partial u}\\
    &=- \frac{1}{k_{\text{B}}T}\left(\frac{\partial E_u(x)}{\partial u}-\left\langle\frac{\partial E_u(x)}{\partial u}\right\rangle^\mathrm{eq}_u\right)
\end{aligned}
\end{equation}
where $\langle a(x)\rangle^\mathrm{eq}_u=\int \dd xp^\mathrm{eq}_u(x)a(x)$. 

Recall that the work is, by definition, the energy involved in shifting the potential in time~\cite{Jarzynski1997}, 

\begin{equation}
    W=\int_0^1\dd t\int \dd xp_t(x)\dot u_t\cdot\frac{\partial E_{u_t}(x)}{\partial u}.\label{eq:Work_Jarz}
\end{equation}
and the change in free energy is $\Delta \mathcal F^{\mathrm{eq}}=-k_\mathrm{B}T[\log Z_{u_1}-\log Z_{u_0}]$. Using Eq.~\eqref{eq:2nd_law}, we find the quasi-static expression for the entropy production:

\begin{equation}
\begin{aligned}
    \Sigma&=\frac1T\int_0^1\dd t\int \dd x\delta p_t(x)\dot u_t\cdot\frac{\partial E_{u_t}(x)}{\partial u}\\
    &=k_\mathrm{B}\int_0^1\dd t\dot u_t\cdot\zeta_{u_t}\cdot \dot u_t,
\end{aligned}
\label{eq:EPR_RieGeom}
\end{equation}
where the friction coefficient can be expressed as the integrated autocorrelation function of the thermodynamic force

\begin{equation}
\begin{aligned}
    \zeta_{u}&=\int_0^\infty\dd s \int \dd x\,p_u^\mathrm{eq}(x)\Theta_u(x)e^{-s\mathcal{L}_u^\dagger}\Theta_u^{\mathsf T}(x)\\
    &=\int_0^\infty \dd s\langle\Theta_u(x_s)\Theta_u^{\mathsf T}(x_0)\rangle_u^\mathrm{eq},
\end{aligned}
\label{eq:zeta_RieGeom}
\end{equation}
where the average is in the equilibrium problem, Eq.~\eqref{eq:THERMOlangevin}, where the protocol is fixed (with $x_0\sim p_u^\mathrm{eq}$). 

Thus, Eq.~\eqref{eq:EPR_RieGeom} is an exact expression for the entropy production in terms of a Riemannian geometry in the protocol space~\cite{SivakPRL2012}. (This contrasts with Eq.~\eqref{eq:W2EPR}, which provides an optimal-transport-inspired bound on the entropy production.) The thermodynamic metric, $\zeta_u$, describes the ``friction'' one would experience in a given position in phase space $u$. We now concretely see how entropy production decreases quadratically with slower driving $\dot u$. However, we may now further ask questions such as how to optimally transition between states quasistatically, \textit{e.g.}, how to vary temperature and magnetic field to flip the magnetization in the zero-temperature Ising model~\cite{Rotskoff.Crooks2015}. These trajectories would correspond to finding the geodesics on the curved surface described by $\zeta_u$, \textit{e.g.}, using geometric minimum-action methods~\cite{RotskoffPRE2017} or by backpropagation~\cite{EngelPRX2023}. Eq.~\eqref{eq:EPR_RieGeom} then allows us to investigate the efficiency and geometry of near-equilibrium thermodynamic transformations.

\subsubsection{Thermodynamic fluctuation theorems}\label{sec:thermodynamic_fluctuation_theorems}

While the above has been a detailed, geometric treatise on close-to-equilibrium thermodynamics, one can find additional, surprising statistical characterizations of the work distributions far from equilibrium. Eq.~\eqref{eq:Work_Jarz} provides the mean work, and in the previous sections we derived second-law-like bounds $T\Sigma=W-\Delta\mathcal F^{\mathrm{eq}}\geq0$. Here we seek additional statistical properties of the work along a stochastic trajectory $\{x_t\}$,

\begin{equation}
    w_t[x(\cdot)]=\int_0^t\dd s\,\dot u_s\cdot\frac{\partial E_{u_s}(x_s)}{\partial u},\qquad W=\langle w_1\rangle_P,
\end{equation}
where $P$ is the forward path distribution and $w_1$ is the work over the complete unit-time protocol. The trajectories start and end at equilibria ($x_0\sim p_{u_0}^\mathrm{eq}$ and $x_1\sim p_{u_1}^\mathrm{eq}$). 
For that purpose, we will be computing its moment-generating function (MGF), $\langle e^{\lambda w_1}\rangle$. We will show that it satisfies two useful equalities: the Jarzynski equality~\cite{Jarzynski1997,JarzynskiPRL1997},

\begin{equation}
    \langle e^{-w_1/(k_\mathrm{B}T)}\rangle = e^{-\Delta \mathcal F^{\mathrm{eq}}/(k_\mathrm{B}T)},\label{eq:Jarzynski}
\end{equation}
and the Crooks fluctuation theorem~\cite{Crooks1999}.

\begin{equation}
    \frac{p(w_1)}{\bar p(-w_1)}=e^{(w_1-\Delta \mathcal F^{\mathrm{eq}})/(k_\mathrm{B}T)},\label{eq:Crooks}
\end{equation}
where $\bar P$ is the reverse path distribution and $\bar p$ is its distribution of work along a trajectory that is being orchestrated by the reversed protocol, $\bar u_t:= u_{1-t}$, which starts from the corresponding (reversed) equilibrium initial condition, $p_{\bar u_0}^\mathrm{eq}=p_{u_1}^\mathrm{eq}$.

The Jarzynski equality has been proven using the Feynman-Kac formula in  Ref.~\cite{HummerPNAS01}. Inspired by their approach, we present a new method for deriving the Crooks fluctuation theorem, involving a Feynman-Kac calculation of the forward- and reverse-process MGFs. While Eq.~\eqref{eq:Jarzynski} vividly involves the work MGF at $\lambda=-1/(k_\mathrm{B}T)$, we convert Eq.~\eqref{eq:Crooks} to involve the MGFs by multiplying it by $\bar p(-w_1)e^{\lambda w_1}$ and integrating over $w_1$, to find:

\begin{equation}
    \langle e^{\lambda w_1}\rangle_P=e^{-\Delta \mathcal F^{\mathrm{eq}}/(k_\mathrm{B}T)}\langle e^{-[\lambda+1/(k_\mathrm{B}T)]w_1}\rangle_{\bar P}.\label{eq:CrooksMGF}
\end{equation}
If the Crooks fluctuation theorem holds, Eq.~\eqref{eq:CrooksMGF} readily reproduces the Jarzynski relation, Eq.~\eqref{eq:Jarzynski}, by inserting $\lambda=-1/(k_\mathrm{B}T)$. Therefore, we aim to derive the Crooks fluctuation theorem, involving both the forward- and backward-running controls, $u_t$ and $\bar u_t=u_{1-t}$.

Define the work-weighted endpoint density (similar to a moment-generating function) to appear at $x$ at time $t$ along the $u_t$-controlled trajectories:

\begin{equation}
    M_t(x,\lambda)
    =\left\langle \exp\left[\lambda \int_0^t\dd s\dot u_s\cdot\frac{\partial E_{u_s}(x_s)}{\partial u}\right]\delta(x-x_t)\right\rangle_P.
\label{eq:Mdef}
\end{equation}
Using It\^o's lemma~\cite{SchussBOOK2010}, one can show that it follows the initial-value partial-differential equation:

\begin{equation}
\begin{gathered}
    \begin{aligned}
    \frac{\partial M_t(x,\lambda)}{\partial t}&=-\mathcal{L}_{u_t}M_t(x,\lambda)+\lambda\dot u_t\cdot\frac{\partial E_{u_t}(x)}{\partial u}M_t(x,\lambda),
    \\
    M_0(x,\lambda)&=p_{u_0}^\mathrm{eq}(x).
    \end{aligned}
\end{gathered}
\label{eq:Mpde}
\end{equation}
We define the initially conditioned MGF along the reversed-protocol, $\bar u_t\equiv u_{1-t}$-controlled trajectories:

\begin{equation}
    \bar M_t(x,\lambda)\quad=\left\langle \exp\left[\lambda \int_t^1\dd s\dot{\bar u}_s\cdot\frac{\partial E_{\bar u_s}(x_s)}{\partial \bar u}\right]\Bigg|x_t=x\right\rangle_{\bar P}.
\label{eq:barMdef}
\end{equation}
Using the Feynman-Kac formula, it satisfies the terminal-value partial-differential equation:

\begin{equation}
\begin{gathered}
    \begin{aligned}
    \frac{\partial \bar M_t(x,\lambda)}{\partial t}&=\mathcal{L}_{\bar u_t}^\dagger \bar M_t(x,\lambda)-\lambda\dot{\bar u}_t\cdot\frac{\partial E_{\bar u_t}(x)}{\partial \bar u}\bar M_t(x,\lambda),\\
    \bar M_1(x,\lambda)&=1.
    \end{aligned}
\end{gathered}
\label{eq:barMpde}
\end{equation}
In order to prove the Jarzynski and Crooks fluctuation theorems, we should find the connection between $M$ and $\bar M$, which will involve a time-reversal of the latter.

A useful, intermediate MGF we define is $\widetilde M_t(x,\lambda):=\bar M_{1-t}(x,-\lambda-\beta)$. Based on Eq.~\eqref{eq:barMpde}, one can show that $\widetilde M$ satisfies the following equation:

\begin{equation}
\begin{aligned}
    \frac{\partial \widetilde M_t(x,\lambda)}{\partial t}=&-\mathcal{L}_{u_t}^\dagger \widetilde M_t(x,\lambda)\\
    &+\left(\lambda+\frac1{k_\mathrm{B}T}\right)\dot u_t\cdot\frac{\partial E_{u_t}(x)}{\partial u}\widetilde M_t(x,\lambda),
\end{aligned}
\end{equation}
where the forward-time protocol has been restored, $\bar u_{1-t}=u_t$, and $\dd/\dd (1-t)=-\dd/\dd t$. Note the property $e^{-E_u(x)/(k_\mathrm{B}T)}\mathcal L_u^\dagger e^{E_u(x)/(k_\mathrm{B}T)}=\mathcal L_u$. With this in mind, we find that 

\begin{equation}
    M_t(x,\lambda)=\frac1{Z_{u_0}}e^{-E_{u_t}(x)/(k_\mathrm{B}T)}\widetilde M_t(x,\lambda),\label{eq:CrooksEquivPrep}
\end{equation}
as by direct substitution, we see that the right-hand side of Eq.~\eqref{eq:CrooksEquivPrep} indeed satisfies Eq.~\eqref{eq:Mpde} along with its initial condition. Upon integrating Eq.~\eqref{eq:CrooksEquivPrep} over $x$, evaluated at $t=1$, and recalling the definitions of the forward and backward MGFs, Eqs.~\eqref{eq:Mdef} and~\eqref{eq:barMdef}, we recover the Crooks fluctuation theorem, Eq.~\eqref{eq:CrooksMGF},

\begin{align}
    \begin{aligned}
    \langle e^{\lambda w_1}\rangle_P&=\int \dd xM_1(x,\lambda)\\
    &=\frac{Z_{u_1}}{Z_{u_0}}\int \dd xp_{u_1}^\mathrm{eq}(x)\bar M_0(x,-\lambda-\beta)\\
    &=\frac{Z_{u_1}}{Z_{u_0}}\langle e^{-[\lambda+1/(k_\mathrm{B}T)]w_1}\rangle_{\bar P}.
    \end{aligned}
\end{align}
Upon replacing the partition functions with the free energy, $Z_u=e^{-\mathcal F_u^{\mathrm{eq}}/(k_\mathrm{B}T)}$, and inverting the Laplace transform, we find the original Crooks fluctuation theorem, Eq.~\eqref{eq:Crooks}.

It is remarkable that for any equilibrium-to-equilibrium transformation, independently of how fast or sub-optimal the controls were, how strong the stochasticity is in the microscopic system, and how far the system has been out of equilibrium during the transformation, equalities such as Eqs.~\eqref{eq:Jarzynski} and~\eqref{eq:Crooks} hold without any approximation. As a corollary, by the Jensen inequality, they reproduce the second law of thermodynamics, $\mathcal F_{u_1}^{\mathrm{eq}}-\mathcal F_{u_0}^{\mathrm{eq}}\leq\langle w_1\rangle$, which is an inequality for the average work involved in a protocol-driven transformation of the stochastic system. Previous sections involved near-equilibrium expansions or optimal-transport-inspired bounds. Fluctuation relations such as these are routinely used to estimate free energy differences in microscopic transformations~\cite{HummerPNAS01,CollinNature2005}. In Sec.~\ref{sec:AISjarz}, we show that the Jarzynski relation can also have utility for importance sampling.

\section{Sampling and control}
\label{sec:sampling}

Sampling from complex, high-dimensional probability distributions is a fundamental task in scientific computing, statistics, and machine learning.
The sampling problem asks for the production of independent samples distributed according to a target probability distribution $p_1$.
In statistical mechanics, for instance, samples from the Boltzmann distribution allow estimation of physical observables, while in Bayesian inference, samples from the posterior distribution allow predictions based on observed data. 
This chapter describes the close connection between sampling, inference, control theory, and non-equilibrium thermodynamics, which forms the basis for modern machine-learning methods in reinforcement learning (Sec.~\ref{sec:RL}) and generative modeling (Sec.~\ref{sec:flows_diffusions}).
These links arise from two directions. 
On the one hand, sampling provides a general-purpose method for solving inference and control problems.
On the other hand, control theory and non-equilibrium thermodynamics give rise to practical sampling algorithms.
Indeed, such algorithms typically use a controlled, dynamic process that transforms samples from a simple \emph{base distribution} $p_0$ into the more complex target $p_1$. Sampling thus becomes a control and transport problem of ``guiding'' base samples into regions of high target probability.

Sec.~\ref{sec:sampling_review} begins by briefly reviewing the use cases and challenges of sampling in high dimension. Many classic algorithms generate samples using equilibrium physics. Sec.~\ref{sec:AISjarz} introduces \emph{Annealed Importance Sampling} (AIS), which allows harnessing \emph{non-equilibrium} processes to generate samples, and can greatly outperform classical methods. AIS is closely connected to non-equilibrium thermodynamics (Sec.~\ref{sec:OTthermo}).
Finally, Sec.~\ref{sec:sampling_as_control} casts sampling as a control problem by optimizing the statistical efficiency of the sample-generation process. Vice versa, control problems can be solved by sampling trajectories and keeping count of the resulting reward (Sec.~\ref{sec:feynman-kac-formula-doobs-h-transform}).

\subsection{What is sampling and why is it hard?}\label{sec:sampling_review}

We begin by fixing the setting. Our goal is to draw $N$ independent samples  $x_n\sim p_1, n=1,...,N$ from a complex target distribution $p_1$.
We assume that we can evaluate an unnormalized density $\widetilde p_1(x)$ for the target, with $p_1(x)=\widetilde p_1(x)/Z_1$. Throughout this chapter, $p$ denotes a normalized probability density and a tilde denotes its unnormalized form.
Below, we use dimensionless ``energies'' and write $E_i(x)=-\log\widetilde p_i(x)$, so $p_i=e^{-E_i}/Z_i$ and $-\log p_i=E_i+\log Z_i$.
As we explain next, the sampling task naturally arises in statistical mechanics and Bayesian inference. 

\subsubsection{The Monte Carlo method}\label{sec:monte_carlo}

A key problem in equilibrium statistical mechanics is to calculate expectation values of physical observables, denoted $a(x)$ for ``average'', for example, the magnetization of a piece of metal~\cite{Mezard.Montanari2009}. 
This requires summing (or integrating) over all $d$-dimensional configurations $x$, weighted by the Boltzmann measure:

\begin{align}
\begin{aligned}
    \left\langle a\right\rangle &= \int a(x) p_1(x)\,\dd^d x,\\
    p_1(x) &= \exp\Big(-\beta E(x)\Big) / Z
\end{aligned}
\end{align}
where $\beta=(k_{\mathrm B}T)^{-1}$, $E$ is the energy, and $Z$ is the normalization constant.
Computing this integral in large dimensions requires summing a number of configurations exponential in $d$ (the number of system degrees of freedom). This is generally intractable. A similar problem occurs in Bayesian inference~\cite{Gelman.etal2013}, already encountered in Sec.~\ref{sec:varational_basis_for_inference}. Here, the target distribution $p_1$ is the \emph{posterior} of a statistical model.
By Bayes' rule, the posterior for parameters or hidden variables $x$, conditioned on observations $A$, reads: 

\begin{align}
\begin{aligned}
   p(x | A) &= p(A| x) p(x) / p(A)\\
   &= \exp\Big(\log p(A| x) + \log p^0(x) \Big) / Z
\end{aligned}
\label{NC_eq:posterior}
\end{align}
The target is now $p_1(x)=p(x| A)$.
The final form emphasizes that the prior $p(x)$, the statistical model $p(A| x)$, and the observations $A$ define an effective ``energy landscape'' for the parameters $x$. The $x$-independent $p(A)$ acts like a normalization constant.
Like in statistical mechanics, in complex, high-dimensional models, directly computing expectation values, like the parameter means or normalized posterior probabilities, is intractable.

The \emph{Monte Carlo} method relies on a set of $N$ (independent) samples from the target distribution $p_1$ to estimate expectation values:

\begin{align}
\begin{gathered}
\langle a\rangle_{1}\approx\frac1N\sum_{n=1}^N a(x_n)
    =:\hat a_N,\\
    x_n\sim p_1,\quad n=1,\ldots,N.
\end{gathered}
\label{NC_eq:Monte_Carlo}
\end{align}
We use $\hat{\cdot}$ for estimators, and simplify subscripts $\langle a\rangle_{1} := \langle a\rangle_{p_1}$ where unambiguous.
Since it uses a finite number of samples, the estimator Eq.~\eqref{NC_eq:Monte_Carlo} is not perfect (throughout, we use $\hat{\cdot}$ for estimators). It is correct on average, but has a nonzero mean squared error

\begin{align}
 \left\langle (\hat{a}_n -\langle a\rangle)\right\rangle = 0, \quad \left\langle (\hat{a}_n-\langle a\rangle)^2\right\rangle 
    =\frac{\operatorname{Var}(a)}{N}.
\label{NC_eq:MC_error}
\end{align}
Estimating strongly fluctuating quantities requires many samples. 

Computing statistical-mechanical observables and Bayesian inference thus become sampling tasks: Given an energy or an unnormalized probability density, draw independent samples to estimate observables or normalized probabilities.

\subsubsection{Probability distributions in high dimension}\label{sec:high_dimension}

\begin{figure}[t]
    \centering
    \includegraphics[width=0.6\linewidth]{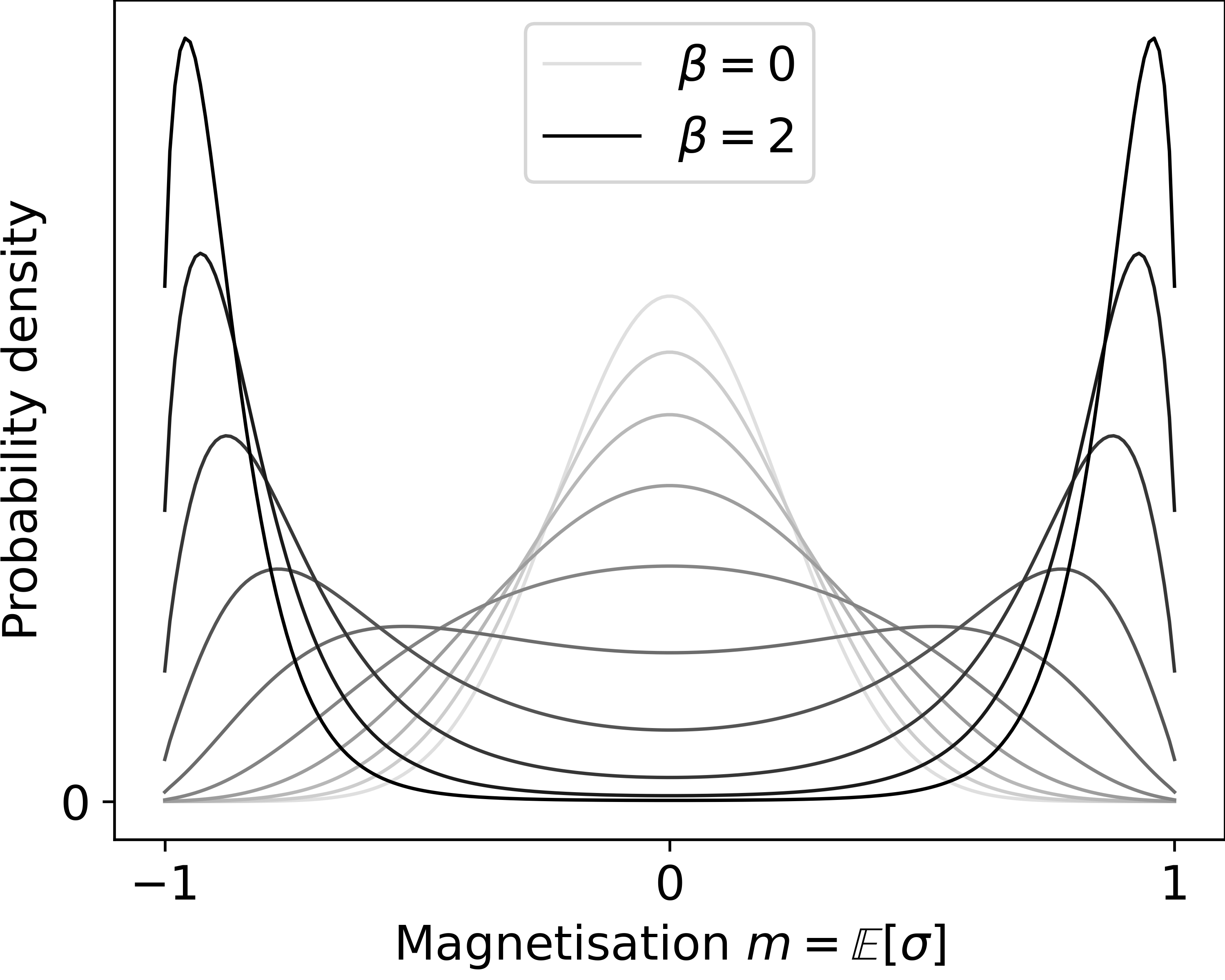}
    
    \includegraphics[width=0.6\linewidth]{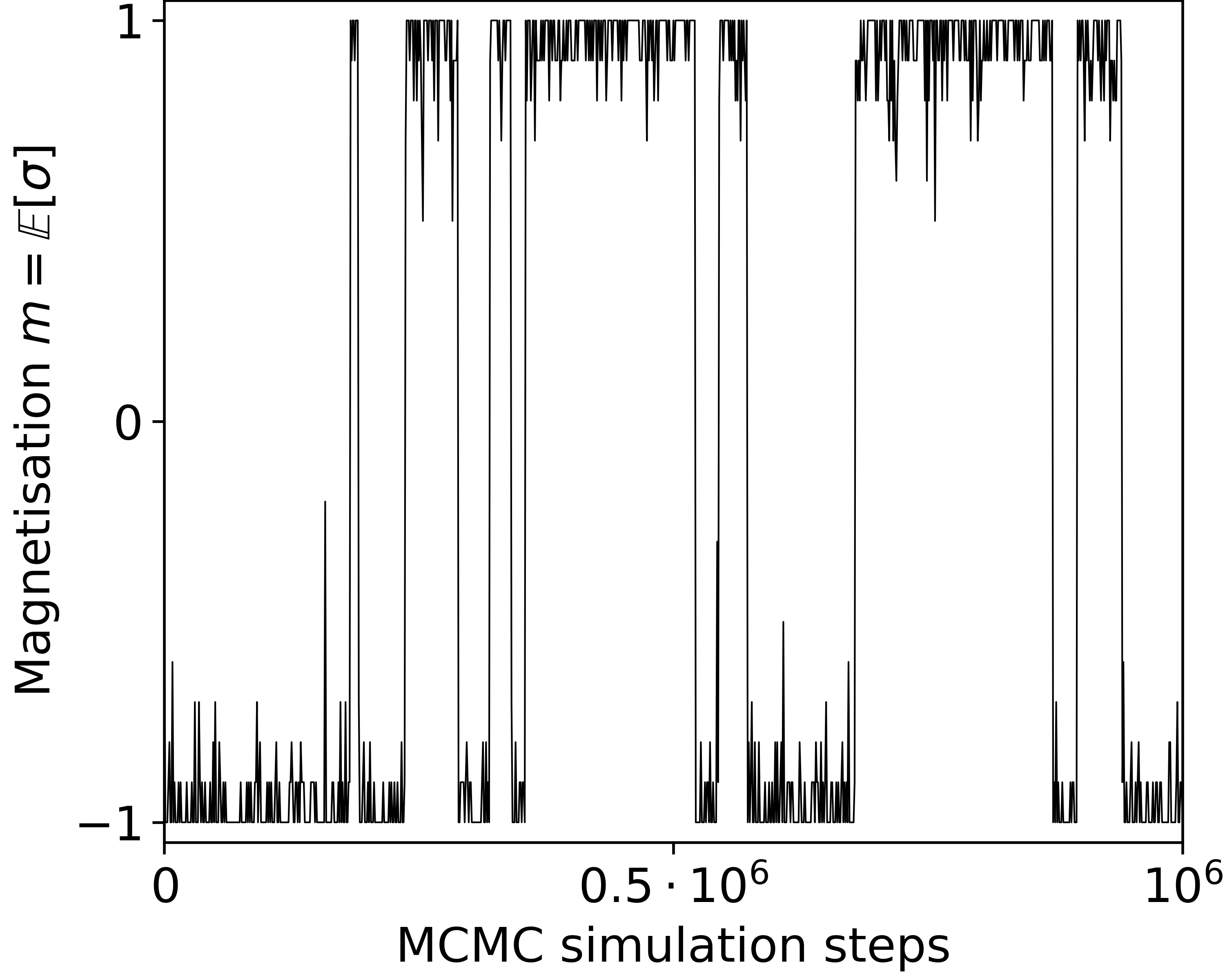}
    \caption{
    (Left) Boltzmann distribution of the magnetization at different temperatures in the fully connected Ising model with energy $E = d^{-1}\sum_{i\neq j}\sigma_i\sigma_j$ where the $\sigma_i \in \{-1, 1\}$ are binary ``spins''. The magnetization is the average over its spins, $m(x)=d^{-1}\sum_i\sigma_i$.
    At high temperature, the distribution is simple and unimodal, but at low temperature, it has two disconnected modes.
    (Right) Magnetization of MCMC samples with $d=20$ spins in the low-temperature ferromagnetic phase $\beta=2$. 
    The MCMC sampler carries out a random walk over spin configurations; jumps are barrier-crossing events between positive and negative magnetization.
    Convergence is very slow: it takes $\sim 10^6$ steps to escape a local free energy minimum.}
    \label{NC_fig:MCMC}
\end{figure}

Sampling in high dimension $d\gg 1$ is challenging.
First, complicated 
probability distributions often feature multiple, disconnected modes. An example from physics is the Ising model in the ordered phase (Fig.~\ref{NC_fig:MCMC}, left). In the high-dimensional (thermodynamic) limit, the free energy barrier between the two magnetization states scales like $d$ and thus becomes very large~\cite{Mezard.Montanari2009}. 
In Bayesian statistics, different modes in the posterior are distinct ``explanations'' of the observed data. Capturing all modes is therefore essential. 

\begin{figure}[t]
    \centering
    \includegraphics[width=0.5\linewidth]{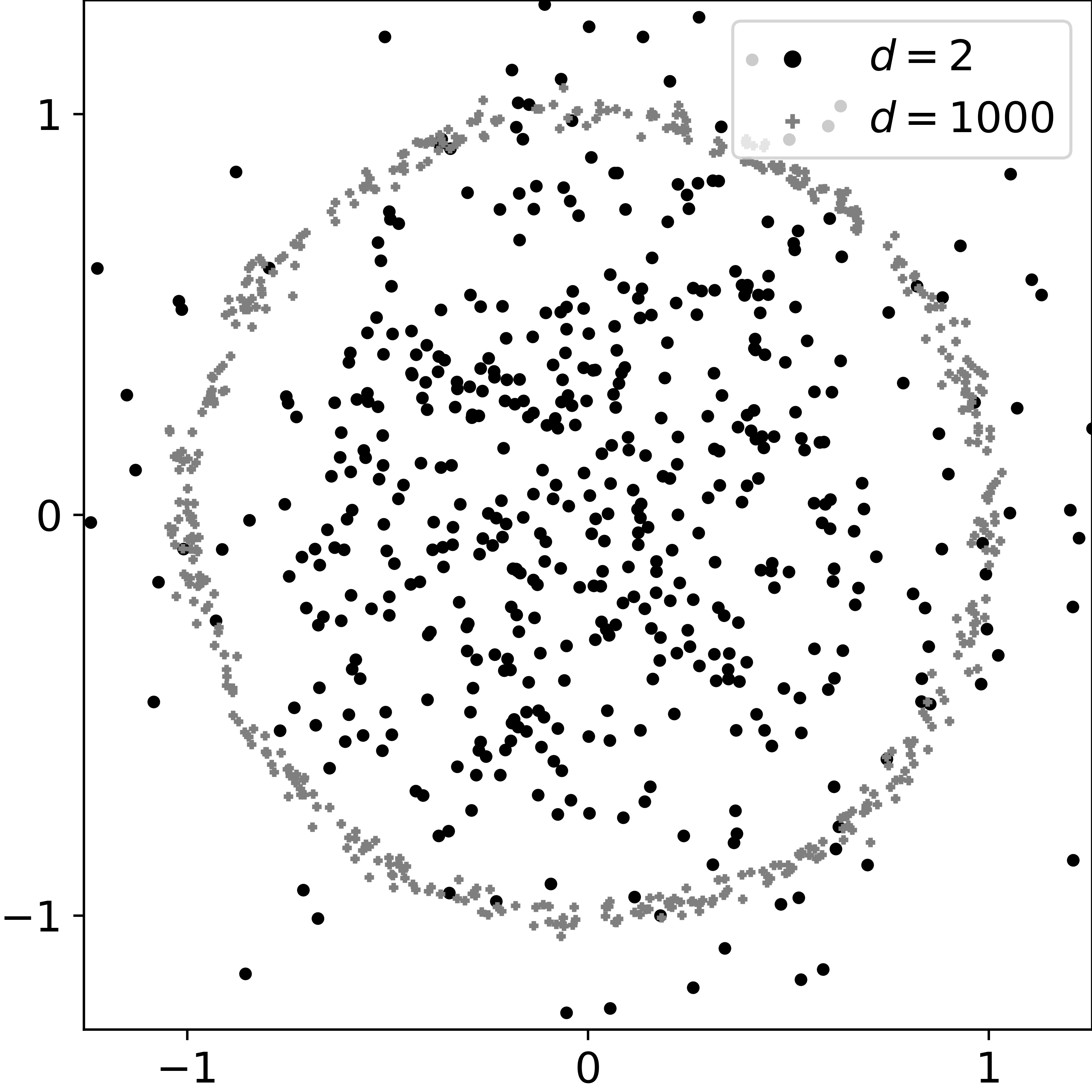}
    \caption{Samples from a Gaussian with zero mean and variance $\sigma^2 =1/d$ in dimension $d=2$ and $d=1000$, plotted in (projected) polar coordinates $r = \sqrt{\sum_i x_i^2}, \, \theta = \arctan(x_1/x_2)$. For large $d$, samples concentrate on a sphere of radius $1$ around the mean.
    } 
    \label{NC_fig:Gaussian_high_d}
\end{figure}

Second, the high-probability regions that sampling must locate and explore become  needles in a high-dimensional haystack~\cite{Betancourt2018}. 
This issue is often called the \emph{curse of dimensionality}.
More precisely,
high-dimensional probability measures tend to concentrate into very narrow sets (Fig.~\ref{NC_fig:Gaussian_high_d}, right), which sampling algorithms must explore. This phenomenon, a generalization of the central limit theorem, is called \emph{concentration of measure}~\cite{Vershynin2026}.
In statistical mechanics, it underlies the equivalence of micro- and macrocanonical ensembles in the thermodynamic limit $d\rightarrow\infty$: large energy fluctuations become unlikely.

How fluctuations grow with the number of dimensions is also of great importance in sampling.
As a rule of thumb, estimating an extensive quantity like the energy is feasible. 
Let us consider a scalar observable $a$ that is approximately Gaussian and for which the mean and standard deviation are of the same order. 
If the magnitude is extensive, $|a|\sim d$,
the estimator Eq.~\eqref{NC_eq:Monte_Carlo} requires $N \gtrsim d$ samples for an accurate estimate. 
However, estimating the exponential of an extensive quantity $e^a$ is very hard.
Since $\mathrm{Var}[e^a]\sim e^{2\mathrm{Var}[a]} \sim e^{2d}$, it requires $N\gtrsim e^{d}$ samples and is thus infeasible. 
Controlling this variance is a key question in sampling algorithm design.

\subsubsection{Basic sampling algorithms: importance and Langevin sampling}\label{sec:basic_sampling}

Most sampling algorithms begin with a \emph{base distribution} $p_0$ from which one can sample easily (often, the standard $d$-dimensional Gaussian  $\mathcal{N}(0, \mathbb{I})$)~\cite{Krauth2006}. 
The game is then to convert samples from $p_0$ into samples of the \emph{target distribution} $p_1$.
The most straightforward implementation of this idea, \emph{transform sampling}, applies a map to base samples. This idea underlies modern generative
models and will be discussed in Application Sec.~\ref{sec:flows_diffusions}.

\emph{Importance sampling} is based on reweighting.
Assuming that the unnormalized density $\widetilde p_1$ is known, one uses the base distribution $p_0$ to produce a set of \emph{proposals} $x_n \sim p_0$ (in this context, $p_0$ is often called the \emph{proposal} distribution).
Weights $\omega_n$ compensate for the difference between $p_0$ and $p_1$:

\begin{align}
\begin{aligned}
    \langle a\rangle_{1}
    &=\left\langle a\,\omega\right\rangle_{0}
     =\frac{\langle a\,\widetilde\omega\rangle_{0}}
            {\langle\widetilde\omega\rangle_{0}}
    \approx\frac{\sum_{n=1}^N\widetilde\omega_n a(x_n)}
                  {\sum_{n=1}^N\widetilde\omega_n},\\
    \omega(x)&=\frac{p_1(x)}{p_0(x)},\qquad
    \widetilde\omega_n=\frac{\widetilde p_1(x_n)}{\widetilde p_0(x_n)},
    \qquad x_n\sim p_0.
\end{aligned}
\label{NC_eq:importance_sampling}
\end{align}
Crucially, the sample estimate only requires evaluating the unnormalized probability densities, since the normalization constants $Z_0, Z_1$ drop out.

Reweighting converts proposals into target samples.
The closer $p_0$ is to $p_1$, the more efficiently the method works; if $p_0$ and $p_1$ are very different, most proposals land in regions of vanishing $p_1$-probability. This means that the finite-sample estimate will be dominated by a few samples with the largest weights.
This phenomenon can be quantified by the \emph{effective sample size}, the number $N_{\mathrm{eff}}$ of samples from $p_1$ that yield the same error as $N$ importance samples from $p_0$, often approximated as~\cite{Gelman.etal2013} 

\begin{align}
N_{\mathrm{eff}}\approx
\left(\sum_{n=1}^N\widetilde\omega_n\right)^2 / \left(
     {\sum_{n=1}^N\widetilde\omega_n^2} \right).
\label{NC_eq:effective_sample_size}
\end{align}
The effective sample size is thus set by the $2^\mathrm{nd}$ moment of the weights. Let us see how this arises.
By Eq.~\eqref{NC_eq:MC_error}, $N_{\mathrm{eff}}$ samples from $p_1$ yield an error $\mathrm{Var}[a]/N_{\mathrm{eff}}$. Instead of the ``bare'' observable $a$, importance sampling estimates the product $a\cdot \omega$, where $\omega = p_1/p_0 = \exp(\log p_1 - \log p_0)$ is the weight. The error of the importance sampling estimator Eq.~\eqref{NC_eq:importance_sampling} is therefore $\mathrm{Var}[a\cdot \omega]/N \sim(\mathrm{Var}[a] \cdot \mathrm{Var}[\omega])/N$. Comparing the two yields $N_{\mathrm{eff}} = N / \mathrm{Var}[e^{\log p_1 -\log p_0}]$. 
In high dimensions, importance sampling thus becomes ineffective. The ratio $p_1/ p_0$ generally fluctuates wildly since probability measures concentrate in narrow regions~\cite{ChatterjeeDiaconis2018}. The ``energies'' $E_{0,1} = -\log\widetilde p_{0,1}$ are usually extensive, $E_{0,1} \sim d$, and so $\omega = p_1/p_0 \sim e^d$. Indeed, constraining $p_1/ p_0$ is a key part of importance-sampling-based machine learning algorithms like PPO~\cite{schulman2017proximal}, which we will discuss in Sec.~\ref{sec:RL}.

\emph{Markov chain Monte Carlo} (MCMC) uses an easy-to-simulate stochastic process that converges to the target distribution $p_1$ over time $t$. This generates samples incrementally and allows simulating complex, high-dimensional distributions. 
Its simplest variant, Langevin sampling, uses Brownian motion in a potential $E_1 = -\log\widetilde p_1$:

\begin{align}\label{NC_eq:Langevin}
    \dot x_t = - \nabla E_1  +  \sqrt{\varepsilon}\,\eta_t, \quad x_0 \sim p_0
\end{align}
Here $\eta_t$ is the standard Gaussian white noise. We denote the marginal distribution of $x_t$, the sampler density, by $q_t(x)$.
The stationary distribution of this process is precisely $p_1=\exp(-E_1)/Z_1$, so after waiting long enough, $x_t$ can generate target samples.
MCMC succeeds or fails based on how rapidly $q_t$ equilibrates to the target distribution. 
However, for complex, high-dimensional distributions, convergence can be painfully slow. For multi-modal distributions, Langevin dynamics can get stuck in a single energy minimum and thus perform poorly (Fig.~\ref{NC_fig:MCMC}, right).
For instance, for a mixture of Gaussians with well-separated peaks $\mu_1, \mu_2$, trajectories must cross an energy barrier $\Delta E \sim |\mu_1-\mu_2|^2/\sigma^2$, which takes a time $\sim \exp(\Delta E)$.

While Langevin sampling is conceptually based on equilibrium physics, the algorithms we discuss next use \emph{non-equilibrium} physics. They transform the base into the target distribution over a fixed, finite time frame, akin to the thermodynamic protocols encountered in Sec.~\ref{sec:OTthermo}. The target distribution is no longer a stationary state. Intuitively, this allows proceeding at ``full speed'' where equilibrium approaches must ``slow down'' as the target is approached.

\subsection{Annealed importance sampling}\label{sec:AISjarz}

The method of \emph{simulated annealing}, originally introduced in statistical physics~\cite{Kirkpatrick.etal1983,Mezard.Montanari2009}, overcomes this issue and can be seen as an antecedent of modern diffusion methods.
Simulated annealing was originally an algorithm to optimize an energy $E(x)$.
Because greedy optimization methods like gradient descent, $\dot x_t = -\nabla E$, can get stuck in local minima, one introduces noise (a finite temperature) to allow escape: $\dot x_t = - \nabla E + \sqrt{\varepsilon} \, \eta_t$. However, if $\varepsilon$ is too large, the state $x_t$ spends much time away from the energy minimum due to ``thermal'' fluctuations. Simulated annealing splits the difference by starting with a high temperature and gradually lowering it over time. The initial high-temperature distribution is supposed to act as a ``bridge'' connecting different energy minima.

\subsubsection{Sampling with time-varying energies}

By reparametrization, a changing noise level can be absorbed into a changing energy.
In this section, we rescale so that $\varepsilon=2$ to simplify notation. One thus generates samples by:

\begin{align}\label{NC_eq:time_dependent}
   \dot x_t = -\nabla E_t  + \sqrt{2}\,\eta_t, \quad x_0 \sim p_0
\end{align}
The process generates proposal samples $x_t$, with actual marginal density $q_t(x)$. For independent runs, $x_t^{(n)}$ denotes the state in run $n$.
By convention, time runs from $t=0$ to $t=1$. 
Thus, simulated annealing departs from conventional Langevin dynamics: samples are generated over finite time and using a time-dependent energy $E_t$.
One simple strategy is linear interpolation between the energies:
\begin{align}
\begin{gathered}
    E_t = (1-t) E_0 + t E_1,\\
    \text{where} \quad p_0 = e^{-E_0}/Z_0, \;\; p_1 = e^{-E_1}/Z_1
\end{gathered}
\label{NC_eq:energy_interpolation}
\end{align}
Eq.~\eqref{NC_eq:time_dependent} can be thought of as a thermodynamic protocol (Sec.~\ref{sec:OTthermo}). The ``instantaneous equilibrium'' distribution corresponding to Eq.~\eqref{NC_eq:energy_interpolation} is $p_t^{\mathrm{eq}} = e^{-E_t}/Z_t$, with $p_0^{\mathrm{eq}}=p_0$ and $p_1^{\mathrm{eq}}=p_1$.

However, this strategy is not yet perfect. The distribution $q_1$ of the trajectory endpoints  $x_1$ from Eq.~\eqref{NC_eq:time_dependent} is not the same as the equilibrium distribution $p_1$, our target.
Physically, the transformation induced by the time-dependent energy is non-adiabatic.
Mathematically, the Fokker-Planck equation for $q_t$ reads $\partial_t q_t = \nabla\cdot(q_t\nabla E_t) +\nabla^2 q_t$. The $p_t^{\mathrm{eq}}$ cannot be a solution since they are instantaneous equilibria, $\nabla\cdot(p_t^{\mathrm{eq}}\nabla E_t) + \nabla^2 p_t^{\mathrm{eq}} = 0 \neq \partial_t p_t^{\mathrm{eq}}$~\cite{Coste2026}.
Therefore, $x_1$ cannot be used directly as a sample for $p_1$.

\emph{Annealed importance sampling}~\cite{Neal1998} (AIS) corrects the annealing process Eq.~\eqref{NC_eq:time_dependent} to produce samples from $p_1$. 
Similarly to Girsanov's theorem (Sec.~\ref{sec:Girsanov}), AIS computes how the marginal distribution of $x_t$ changes due to the additional force $\nabla E_t - \nabla E_0 =  t\nabla(E_1-E_0)$ from the time-dependent energy.
AIS computes a weight $\omega_t$ for each realization $x_t$ that compensates for the difference 
between $q_t$ and $p_t^{\mathrm{eq}}$ via importance sampling Eq.~\eqref{NC_eq:importance_sampling}, i.e. $\omega_t = p_t^{\mathrm{eq}}(x_t) / q_t(x_t)$. 
As we will derive shortly, this weight can be calculated ``step-by-step'' by integrating along the trajectory.

\begin{align}
\begin{aligned}
    \frac{\dd}{\dd t}\log\omega_t &=(\partial_t\log p_t^{\mathrm{eq}})(x_t),\\
    \omega_t=\frac{Z_0}{Z_t}e^{-w_t},&\qquad
    w_t=\int_0^t(\partial_sE_s)(x_s)\,\dd s.
\end{aligned}
\label{NC_eq:AIS_weights}
\end{align}
where $w_t$ is the work  performed on the trajectory particle due to the changing energy. In practice, the common factor $Z_0/Z_t$ need not be computed: when estimating averages, one simply normalizes by the total weight of all AIS samples. A special case of AIS is \emph{thermodynamic integration}~\cite{FRENKEL2002167}, in which the energy $E_t$ is changed quasi-statically.

\begin{sloppypar}
In non-equilibrium thermodynamics, Eq.~\eqref{NC_eq:AIS_weights} is known as \emph{Jarzynski's equality}~\cite{Jarzynski1997} (Sec.~\ref{sec:thermodynamic_fluctuation_theorems} deduced it from Crooks' fluctuation theorem, another non-equilibrium thermodynamics result).
Jarzynski's equality relates the free energy difference $\mathcal F_1^{\mathrm{eq}}-\mathcal F_0^{\mathrm{eq}}=-\log Z_1+\log Z_0$ to the work $w_t$ on individual trajectories.
It notably allows calculating the free energy difference between two states. This result from thermodynamics was, in fact, an important inspiration for AIS~\cite{Neal1998}.
\end{sloppypar}

\subsubsection{Proof of the annealed importance sampling identity}\label{sec:AIS_proof}

Jarzynski's equality was proven in Sec.~\ref{sec:thermodynamic_fluctuation_theorems} using the Feynman-Kac formula. Here, we provide an alternative derivation of Eq.~\eqref{NC_eq:AIS_weights}, effectively using a path-integral approach~\cite{Jarzynski1997,Neal1998}.
The biggest challenge lies in notation: we already distinguished the instantaneous equilibria $p_t^{\mathrm{eq}}$ from the distribution $q_t$ of the non-equilibrium process Eq.~\eqref{NC_eq:time_dependent}. In addition, we now introduce the time-reversed version of Eq.~\eqref{NC_eq:time_dependent}, in which the energy is changed from $E_1$ back to $E_0$:

\begin{align}
   \dot{\bar{x}}_t = -\nabla E_{1-t}  + \sqrt{2}\,\eta_t, \quad \bar{x}_0 \sim p_1    
\end{align}
Note that the reverse process is initialized at $p_1$, our target: $\bar q_0=p_1$. We denote its marginal density by $\bar q_t$ and its path law by $\bar P$.
We now show how $p_t^{\mathrm{eq}}$ (equilibrium), $q_t$ (forward process), and $\bar q_t$ (reverse process) are related.

We discretize time into small intervals, $t= 0, dt, 2dt, ..., 1$.
Formally, the probability of a trajectory equals the product of the (infinitesimal) transition probabilities:

\begin{align}
	P[x(\cdot)] = p_0(x_0) \prod_t k_t(x_{t+dt}| x_t)
\end{align}
In the discretization, we alternate between taking trajectory steps $x_t \mapsto x_t + (-\nabla E_t +\sqrt{2}\,\eta_t  )dt$ and updating the energy, $E_t\mapsto E_{t+dt}$. The $k_t(y| x)$ are one-step transition matrices. They are the \emph{equilibrium transition probabilities} for an energy ``frozen'' at $E_t$, and, in particular, obey \emph{detailed balance}.
The transition probabilities of the reverse process are therefore\footnote{
For the SDE Eq.~\eqref{NC_eq:time_dependent}, one can explicitly compute the (Gaussian) transition kernel to confirm Eq.~\eqref{NC_eq:detailed_balance}.
Alternatively, Eq.~\eqref{NC_eq:detailed_balance} can serve as the definition of the reverse process (well-defined as $p_t^{\mathrm{eq}}$ is the instantaneous equilibrium).}:

\begin{align}
	\label{NC_eq:detailed_balance}
	k_t(x_{t} | x_{t+dt})  &= k_t(x_{t+dt}| x_t) \frac{p_t^{\mathrm{eq}}(x_t)}{p_t^{\mathrm{eq}}(x_{t+dt})}
\end{align}
Hence, the difference in a trajectory $x_t$'s log-probability under the forward and reverse processes reads:

\begin{align}
    &\log\frac{\bar P[x(1-\cdot)]}{P[x(\cdot)]} =\sum_t\dd t\,\frac{\log p_{t+dt}^{\mathrm{eq}}(x_t)-\log p_t^{\mathrm{eq}}(x_t)}{\dd t} \nonumber \\
    &\quad=\int\dd t\,(\partial_t\log p_t^{\mathrm{eq}})(x_t)
      =\log\omega_1.
\label{NC_eq:discretized_transitions}
\end{align}
The reverse process starts at $\bar q_0 = p_1$, so the weight $\omega_1$ corrects for the difference between $q_1$ and $p_1$. After reweighting with $\omega$, the trajectory endpoints $x_1^{(n)}$ can be used as importance samples.

\begin{figure}[t]
    \centering
    \includegraphics[width=0.6\linewidth]{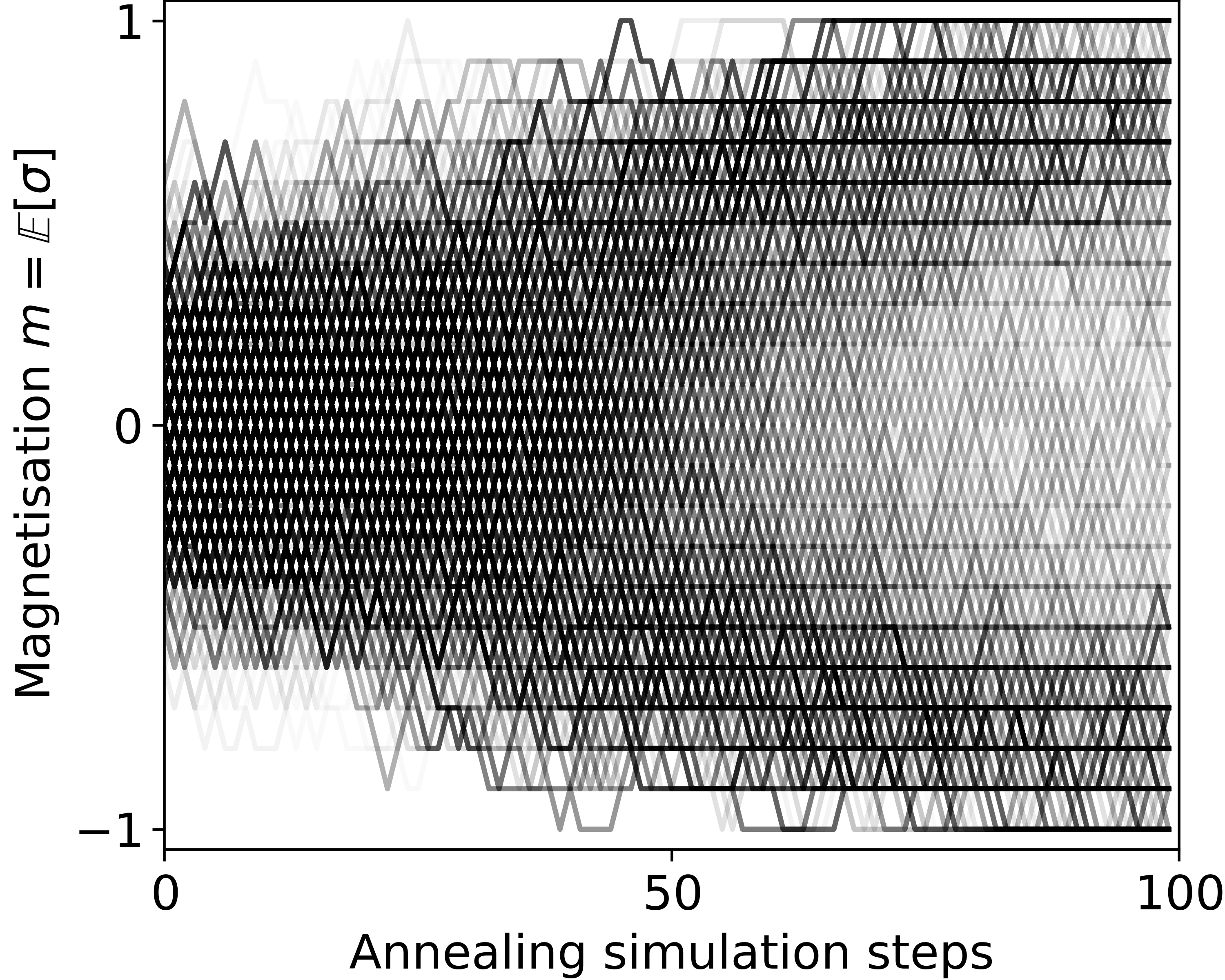}

    \includegraphics[width=0.6\linewidth]{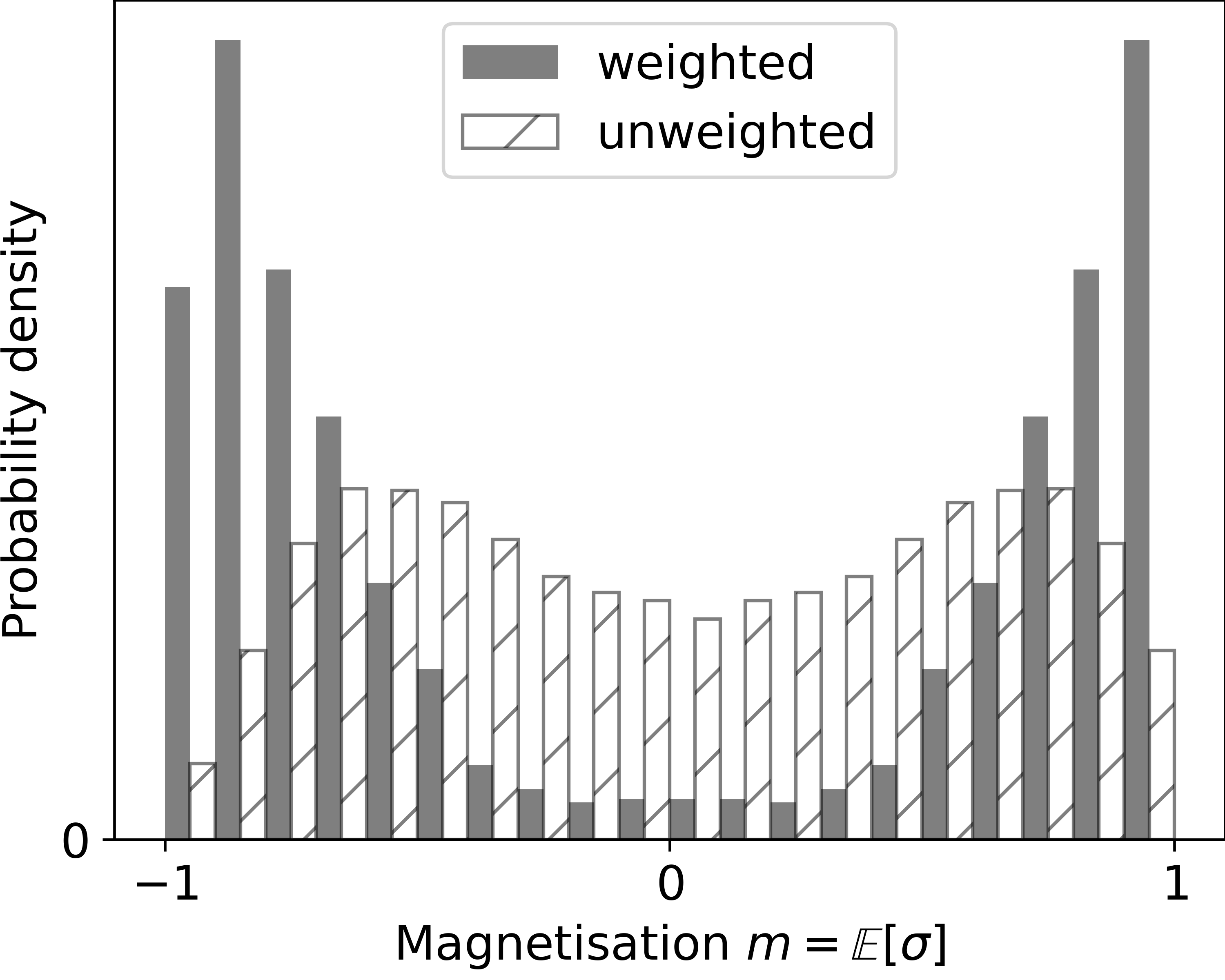}
    \caption{
    (Left) Simulated annealing trajectories for the fully connected Ising model with $d=20$ spins, going from $\beta=0$ to $\beta=2$ in 100 steps.
    (Right) Distribution of magnetization at $\beta=2$ based on annealing runs. Without reweighting, the samples are not distributed correctly.
    }
    \label{NC_fig:annealing}
\end{figure}

In summary, AIS is an importance sampling scheme that uses time-dependent energy landscapes to generate samples from a target distribution.
Because AIS leverages a non-equilibrium process, it can greatly outperform, say, Langevin sampling, when convergence to equilibrium is slow, for example due to energy barriers.
The AIS importance weights are easy to calculate numerically\footnote{Using the discrete Eq.~\eqref{NC_eq:discretized_transitions}. Note that there is no ``discretization error'': AIS sampling is mathematically exact also for discrete-time Markov chains defined by a set of transition matrices.}.
An example is shown in Fig.~\ref{NC_fig:annealing} for the Ising model in the spontaneously magnetized phase. While Langevin sampling took $10^6$ steps to escape a local free energy minimum, AIS can generate independent samples in $100$ steps by cooling down from the unmagnetized phase.
While AIS is guaranteed to deliver unbiased importance samples, its practical performance greatly depends on the annealing schedule, i.e., on the choice of the time-dependent energy $E_t$. We discuss this question next.

\subsection{Sampling as a control problem}\label{sec:sampling_as_control}

We now understand how to generate target samples using the AIS method. 
Does that mean that the sampling problem is ``solved''? 
Further, the AIS scheme works for any time-dependent energy landscape. In which sense is one choice better than another?
Just like thermodynamic protocols and the transport plans of Sec.~\ref{sec:dynamic_OT} can be optimized to minimize energy dissipation or transport costs, we will now discuss how sampling protocols can be optimized for statistical efficiency. As we noted above, importance sampling works best when the proposal and the target distributions are close. Otherwise, many samples land in regions of vanishing probability and are thus ``lost''.
By Eq.~\eqref{NC_eq:effective_sample_size}, the performance of importance sampling is quantified by the variance of the weights $\omega = \exp(\log p_1 - \log p_0)$. Our goal is thus to minimize weight variance to maximize statistical efficiency.

In the thermodynamic language of the Jarzynski equality (Sec.~\ref{sec:thermodynamic_fluctuation_theorems}), minimal weight variance means that the work for moving the system between the two equilibria with free energies $\mathcal F_{0,1}^{\mathrm{eq}}=-\log Z_{0,1}$ is minimal.
To see this, recall Eq.~\eqref{NC_eq:AIS_weights}: the unnormalized AIS weights are $\widetilde\omega_t = \exp(-w_t)$, where $w_t$ is the work performed on a trajectory $x_t$. The Jarzynski equality Eq.~\eqref{eq:Jarzynski} states that the free energy difference is $\mathcal F_1^{\mathrm{eq}}-\mathcal F_0^{\mathrm{eq}} = -\log \left\langle e^{-w_1}\right\rangle \leq \left\langle w_1\right\rangle$. This bound is saturated when the work $w_1$ on all trajectories is equal, and thus the weight variance is zero
(we caution that minimal work coincides with minimal variance, but away from the minimum, the two objectives differ).
Physically, an efficient protocol is close to adiabatic: it ensures that samples at time $t$ rapidly thermalize to the next energy landscape $E_{t+ dt}$, and avoids, for instance, regions of phase transitions~\cite{Rotskoff.Crooks2015}.

In an ideal importance sampler, $\mathrm{Var}[\omega]=0$ and thus $\omega(x)=\mathrm{const}$. That is, the ideal proposals $x$ are already distributed according to the target $p_1$.
To get around this somewhat circular conclusion, we will formulate
the design of a good proposal distribution as an optimization problem.
Indeed, importance sampling can be seen as an optimal control problem by considering the time-dependent energy gradient $-\nabla E_t$ in Eq.~\eqref{NC_eq:time_dependent} as a control force $u_t$. 

\subsubsection{Sampler design as a control problem }\label{sec:sampler_design}

To formalize sampler design as a Path Integral Stochastic Control (PISC) problem (Sec.~\ref{sec:pisc}), we consider a generalized setting. We aim to evaluate expectation values of the form $a \cdot e^{R_0}$ over trajectories of a stochastic process:

\begin{align}
    \label{NC_eq:control_sampling}
    \left\langle a[x(\cdot)] e^{R[x(\cdot)]}\right\rangle, \quad \dot x_t = f_t(x_t) +  \sqrt{\varepsilon}\,\eta_t,
\end{align}
The brackets $[\cdot]$ indicate that $a$ and $R$ may now be functionals of the entire trajectory $x(\cdot)$.

To map Eq.~\eqref{NC_eq:control_sampling} to the importance sampling setting, choose:

\begin{align}
    &R[x(\cdot)]=\log\frac{p_1(x_1)}{p_0(x_1)}=E_0(x_1)-E_1(x_1)+\log\frac{Z_0}{Z_1}, \nonumber \\
    &f_t=-\nabla E_0,\; \varepsilon=2, \;
    a[x(\cdot)]=a(x_1),\; x_0\sim p_0.
\end{align}
The stochastic process produces Langevin samples from the base distribution; $R$ is the log importance weight, so $e^{R}$ reweights them to the target; the expectation on the left of Eq.~\eqref{NC_eq:control_sampling} is then $\langle a\rangle_{p_1}$.
As discussed below Eq.~\eqref{NC_eq:effective_sample_size}, in high dimensions, the importance weight $e^{R} \sim e^d$ fluctuates wildly. This makes a finite-sample estimate of Eq.~\eqref{NC_eq:control_sampling} unreliable.

To remedy this, we add a control force $u_t$ to drive trajectories toward higher rewards, and then reweigh the trajectories to account for the change in distribution~\cite{Kappen2016,Theodorou2010},

\begin{align}
\dot x_t=f_t(x_t)+u_t(x_t)+\sqrt\varepsilon\,\eta_t.
\label{NC_eq:controlled}
\end{align}
We denote by $P^0[x(\cdot)]$ the probability of a path without control, and by $P^u[x(\cdot)]$ the probability with control.
To compensate for the modified dynamics, we reweigh:

\begin{align}
\begin{aligned}
    &\left\langle a[x(\cdot)] e^{ R[x(\cdot)]}\right\rangle_{0}=\left\langle a[x(\cdot)] e^{ R[x(\cdot)]} \frac{P^0[x(\cdot)]}{P^u[x(\cdot)]}\right\rangle_{u}
\end{aligned}
\label{NC_eq:control_sampling_importance}
\end{align}
We now need a suitable reward to choose the control $u$. We will aim to maximize $R$, balanced against a quadratic control cost:

\begin{align}
u^*=\arg\min_u\left\langle
    \int_0^1\frac{\|u_t(x_t)\|^2}{2\varepsilon}\,\dd t
    -R[x(\cdot)]\right\rangle_{u}.
\label{NC_eq:sampling_objective}
\end{align}
We thus arrive at a PISC problem. Recall from Eq.~\eqref{eq:optimal-conditional-path-measure} that the optimal path measure is a tilted version of the uncontrolled measure:

\begin{align}
    P^{*}[x(\cdot) | x_0=x]  = P^0[x(\cdot)| x_0=x] \cdot e^{R[x(\cdot)]}/ Z
\end{align}
\begin{sloppypar}
Indeed, Girsanov's theorem from Sec.~\ref{sec:Girsanov} shows that $\left\langle\int_0^1\|u_t(x_t)\|^2/(2\varepsilon)\,\dd t\right\rangle_{u}=\dkl(P^u\Vert P^0)$. The control cost is thus an entropy, and as we saw in Sec.~\ref{sec:MEM}, entropy maximization generally leads to solutions that are exponential tilts. Plugging the optimal control into Eq.~\eqref{NC_eq:control_sampling_importance}:
\end{sloppypar}

\begin{align}
\begin{aligned}
    &\left\langle a[x(\cdot)]e^{R[x(\cdot)]} \mid x_0=x \right\rangle_{0}\\
    &\quad=\left\langle a[x(\cdot)]e^{R[x(\cdot)]}
       \frac{P^0[x(\cdot)]}{P^{*}[x(\cdot)]} \mid x_0=x \right\rangle_{u^*}\\
    &\quad=Z\left\langle a[x(\cdot)] \mid x_0=x \right\rangle_{u^*}.
\end{aligned}
\label{eq:exp_a_optctrl}
\end{align}
The weights $e^{R}$ in the expectation are now constant (conditional on the initial condition $x_0$ -- unconditional expectation values must additionally average over the easy-to-sample $x_0\sim p_0$).
In the case of importance sampling, we can compute $\langle a\rangle_{p_1}$ purely from samples $x_t, x_0\sim p_0$ of the Langevin equation Eq.~\eqref{NC_eq:controlled}, with the control force found by optimizing Eq.~\eqref{NC_eq:sampling_objective}. Since no reweighting is necessary, the effective sample size is maximal, $N_{\mathrm{eff}} = N$: every proposal sample is worth one true sample.

\subsubsection{Non-equilibrium systems and transition path sampling}

The formulation Eq.~\eqref{NC_eq:control_sampling} also applies to 
non-equilibrium systems, like energetically driven chemical reactions~\cite{Singh.etal2025,DasLimmer2019}.
Instead of a Boltzmann distribution, the state $x_t$ of the system at time $t$ follows a non-equilibrium Langevin equation, combining deterministic forces and random fluctuations.
The dynamics in Eq.~\eqref{NC_eq:control_sampling} thus become physically meaningful, rather than a sampling device.
Evaluating expectation values over physical trajectories is of great importance, for example, to estimate chemical reaction rates.
In this setting, so-called \emph{rare events} can have outsize importance: the crossing of a potential barrier determines whether a reaction occurs or not.
Mathematically, such events can be encoded in the reward $R$ in Eq.~\eqref{NC_eq:control_sampling}. 
One faces the same challenge as above: strong fluctuations in $e^{R}$ lead to poor finite-sample estimates (the target event may simply not occur in a finite sample). 
\emph{Variational Path Sampling}~\cite{Singh.etal2025} addresses it precisely as discussed in Sec.~\ref{sec:sampler_design}. One introduces a control $u_t$ to encourage the target rare events, and reweighs trajectories to compensate for the fictitious control when computing physical observables (Eq.~\eqref{NC_eq:control_sampling_importance}).

\subsubsection{Control as a sampling problem}

In summary, the design of an optimal importance sampler is a control problem, where the reward is the ``target'' probability of the trajectories. This result enables systematic optimization of annealing samplers. But, in some sense, we simply kicked the can down the road: how do we solve the resulting control problem?
The key to resolving this conundrum is that, vice versa, control problems can be solved by sampling.
Indeed, we saw in Sec.~\ref{sec:inference_as_control} that control problems can be seen as probabilistic inference problems. Since sampling provides a general-purpose method to solve the latter, it should also be applicable to the former.
This suggests an iterative approach in which the sampler is used to solve a control problem, whose solution in turn improves the sampler. 
This duality between sampling and control will play an important role in Sec.~\ref{sec:RL} on reinforcement learning.

Let us make this precise. The notations ``$\psi$'' and ``$R$'' in Eq.~\eqref{NC_eq:control_sampling} are no coincidence.
Conditioned on the trajectory's initial  condition, the weighted expectation in Eq.~\eqref{NC_eq:control_sampling}
is the Cole--Hopf transform $\psi_s=e^{V_s}$ of the all-important value function of a control problem with accumulated reward $R[x(\cdot)]$ (Sec.~\ref{sec:feynman-kac-formula-doobs-h-transform}).
We write $R_s[x(\cdot)]$  for the accumulated reward from time $s$ onward, including the terminal reward and excluding control cost.
Recall that the value function obeys the HJB equation (the recursive formulation of optimality), which can be Cole-Hopf transformed into a Schrödinger equation (in imaginary time). The Feynman-Kac formula Eq.~\eqref{eq:psi-as-path-integral} shows that the solution to this Schrödinger equation can be obtained by a Feynman path integral over stochastic trajectories:

\begin{align}
\begin{gathered}
\psi_s(x)=\left\langle e^{R_s[x(\cdot)]}| x_s=x\right\rangle_{P^0},\\
\dot x_t=f_t(x_t)+\sqrt\varepsilon\,\eta_t.
\end{gathered}
\label{NC_eq:feynman_kac}
\end{align}
The stochastic trajectories are now initialized at $x_s = x$ instead of at random. Up to this detail, the Feynman-Kac formula is precisely Eq.~\eqref{NC_eq:control_sampling} with $a=1$, $f_t$ as the dynamics in the absence of control, and $R_s$ as the reward of the control problem. 
Intuitively, Eq.~\eqref{NC_eq:feynman_kac} computes the value function by ``undirected exploration'' (no control applied) and taking inventory of the rewards encountered. In the language of this chapter, the Feynman-Kac formula is a sampling-based approach to control problems: applying the Monte Carlo method to Eq.~\eqref{NC_eq:feynman_kac} yields an estimate of the value function from a finite number of stochastic trajectories.
However, Monte Carlo estimation faces the same challenge as in importance sampling. In high dimensions, $e^{R_s} \sim e^d$ fluctuates wildly. This makes a finite-sample estimate of Eq.~\eqref{NC_eq:feynman_kac} unreliable. Uncontrolled trajectories are very unlikely to gather high reward $R_s$, while it is precisely high-reward trajectories that dominate Eq.~\eqref{NC_eq:feynman_kac}.

Sec.~\ref{sec:sampler_design} explains the solution: add a control force to bias trajectories towards higher rewards.
This suggests an iterative procedure, starting with a guess for the optimal control $u_t^{(0)}$:
\begin{enumerate}
    \item Generate importance samples from the $u_t^{(k)}$-controlled process (Eq.~\eqref{NC_eq:control_sampling_importance})
    \item Estimate $\log\widehat{\psi}^{\,(k)}$ by Monte Carlo, using the reweighted trajectories (Eq.~\eqref{NC_eq:feynman_kac})
    \item Recompute the optimal control from the value function by optimizing the expected reward at each timestep ($u^{(k+1)}_t(x)=\varepsilon\nabla\log\widehat{\psi}_t^{\,(k)}(x)$, Eq.~\eqref{eq:PISC_optimal_control})
\end{enumerate}
and repeat until convergence. This procedure is closely related to the policy iteration algorithm for control problems~\cite{sutton1998reinforcement}, which iterates \emph{evaluation} of the current control's value function (steps 1-2) with \emph{improvement} of the control using the value function (step 3). Here, evaluation happens through Monte Carlo estimation.
In Sec.~\ref{sec:RL} on \emph{reinforcement learning} (RL), we will see that policy iteration forms the basis for state-of-the-art methods like the soft actor-critic algorithm~\cite{haarnoja2018soft}, and that controlling the variance of Monte Carlo estimates is a crucial question in RL algorithm design~\cite{sutton1998reinforcement}.
RL generalizes control theory to the case where the dynamics $f_t$ and reward $r_t$ are not known in advance but must be learned through exploration. RL is often formulated in a somewhat different language, but makes use of the same principles that connect control, inference, and sampling problems.

\section{Applications}\label{sec:application}

The Applications chapter will cover three topics.  Sec.~\ref{sec:RL} on reinforcement learning builds on control theory (Sec.~\ref{sec:mechanics_control}), maximum-entropy methods (Sec.~\ref{sec:path_inference}), and sampling (Chapter~\ref{sec:sampling});  Sec.~\ref{sec:WGFApps} on Wasserstein gradient flows draws on transport (Chapter~\ref{sec:OT}) and non-equilibrium thermodynamics (Sec.~\ref{sec:OTthermo}); and Sec.~\ref{sec:flows_diffusions} on generative models combines ideas from transport (Chapter~\ref{sec:OT}) and sampling (Chapter~\ref{sec:sampling}).
A general theme is moving from settings where the dynamics, rewards, or distributional constraints are known, to settings where they have to be \emph{learned} from data. The Applications chapter is thus centered on ideas from machine learning.

\subsection{Reinforcement learning}\label{sec:RL}

\begin{figure*}[t]
    \centering
    \includegraphics[width=0.7\linewidth]{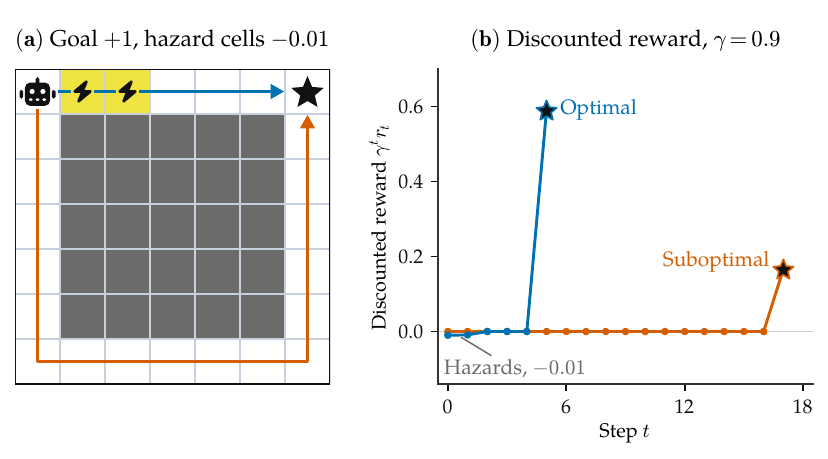}
    \caption{\textbf{The explore-exploit problem in reinforcement learning.} The robot wishes to reach a goal, and is given a reward of $1$ at the goal (star), a reward of $0$ in empty cells, and a slightly negative reward $-0.01$ in hazard cells (electricity). Furthermore, at any point in time, the robot might break down with a large independent probability $10\%$. Thus, if the robot wants to maximize its expected reward, it should heavily prioritize shorter routes to the goal. Assume that the robot's initial state is fixed to the top left of the warehouse. At initialization, the robot's control is random and chooses these two actions with equal probability. From this point, the agent can choose one of two controls: go right or go down. These two controls lead to two paths towards the goal, a suboptimal path that is long and circuitous with ample opportunity to break down but with no penalties, and an optimal path that routes directly to the goal with a small incurred penalty (e.g. an activation energy barrier). }
\label{fig:explore-exploit}
\end{figure*}

Before introducing the reinforcement learning (RL) problem, we begin with a simple example to build intuition and illustrate what makes the problem hard.
Take a simple problem setting of a robot in a new warehouse that it has never seen before (Fig.~\ref{fig:explore-exploit}), where its goal is to reach a target as soon as possible. Importantly, the robot has \emph{no prior knowledge about the warehouse}, and can only make decisions based on its collected experience and observed rewards.
Let's say that the robot executes random actions until it reaches the goal. This initial path is likely suboptimal and circuitous. Should the robot continue taking this path or explore to find a potentially shorter route? Thus, if the robot wishes to find the optimal route, it must  balance between collecting information about its environment and refining the solutions found so far. This balance, known as the exploitation-exploration problem, is the core difficulty of the RL problem.

The RL formalism and methods build on the stochastic optimal control (SOC) framework. As discussed in Sec.~\ref{sec:mechanics_control}, SOC aims to find the optimal control $u_t(x)$ that maximizes the expected cumulative reward in a stochastic environment.
Like the SOC problem, the RL problem also aims to solve for optimal controls that maximize a cumulative reward. However, general RL aims to solve for optimal actions from the perspective of a decision-making entity in a new environment where both environmental statistics and rewards are unknown~\cite{sutton1998reinforcement}, like the robot in a new warehouse. This entity is known as an \emph{agent}.
Recall that, given the dynamics of the environment, a reward function, and boundary conditions, the Hamilton–Jacobi–Bellman (HJB) equation determines the value function and hence the optimal control. Its underlying recursion decomposes the value into the immediate reward and the expected value at the next state. This recursion forms the theoretical backbone of modern RL methods.
In RL, finding the optimal control not only involves solving HJB but also requires the agent to accumulate experience such that finding the optimal controls is tractable. For instance, while completely random controls are very suboptimal if an agent already has access to environment dynamics and reward functions (SOC setting), such a strategy may actually be somewhat sensible in a new environment (RL setting), where random controls can help the agent explore and build a better model of the world; more generally, the agent must carefully balance exploitation (e.g. greedily navigating to states which have already been shown to lead to high total rewards) and exploration. The desiderata to maximize reward and gather data to improve controls in the future often compete, in what is known as the exploitation-exploration problem. This balancing act is a function of the restricted, harder RL problem setting and makes the SOC and the RL problem fundamentally different, despite the shared overall objective of finding optimal controls. 

The central tension between exploration and exploitation is illustrated by the so-called \emph{two-armed bandit problem}. An agent repeatedly chooses between two slot machines---the ``two arms of a bandit''---
each offering one unit of reward with a fixed but unknown probability and zero otherwise. 
Only the outcome of the chosen arm is observed, and the aim is to maximize the accumulated reward. Playing the arm with the larger observed average reward exploits existing information, whereas exploring the other arm ensures that the arm that has been deemed optimal with finite observations is indeed optimal. 
The bandit problem admits a solution to the exploration-exploitation tradeoff in a minimal setting, known as the \emph{Lai--Robbins bound}~\cite{LaiRobbins1985}  (in the asymptotic limit of a large number of plays). This bound shows that for strategies that learn efficiently across possible winning probabilities, the expected reward lost relative to knowing the better arm from the outset grows at least logarithmically with the number of plays. 
The intuition comes from the fact that large deviations in the observed reward can make a sub-optimal arm appear optimal. Sanov's theorem, Eq.~\eqref{eq:MeasureEmpiP}, tells us that the probability of the better arm producing observations that make it seem sub-optimal decreases exponentially with the number of times it is sampled, at a rate determined by the KL divergence. 
Such misleading data could cause the agent to neglect the better arm, with a loss that grows with the duration of play. 
Suppressing these rare but costly errors therefore requires continued exploration. The competition between exponentially decreasing probabilities and accumulating losses explains the logarithmic scale in the bound. Thus, successful strategies for exploration and exploitation are ones that increasingly favor the apparently better arms (or, more generally, actions) while continuing to test alternatives that the data have not yet ruled out.

\subsubsection{The Markov decision process}\label{sec:mdp}
Now that we have some intuition about the structure and fundamental challenges in the RL problem, we introduce necessary terminology to discuss key algorithms.
\emph{States} $x$ are the RL vocabulary for coordinates in a particular \emph{state space} $\mathcal{X}$. For example, in the state space of the Cartesian plane, one possible state an agent can occupy is the origin.
An \emph{environment} is a space of states that an agent can navigate subject to transition dynamics $k(\cdot | x, u)$, where the transitions are a (generally stochastic) function of the current state and current control. A \emph{reward function} $r(x, u)$ is a scalar function that takes as input a state and control and outputs a reward.
Reinforcement learning is defined on discrete time steps $n$; we will use the subscripts $x_{n}$ and $u_{n}$ to denote the states and controls at decision time $t_{n}$. For finite-horizon problems, we include the timestep $n$ in the state $x$ throughout this section; policies and value functions therefore need no separate time subscript.
An agent is a decision-making entity that executes controls in the world based on its observed transitions and rewards~\cite{sutton1998reinforcement}. The controls an agent executes are known as \emph{actions} $u$ which lie within an \emph{action space} $\mathcal U$. Reactive controls are also referred to as a \emph{policy} in RL, where the policy $\pi(u| x)$ is a distribution over controls conditioned on the current state $x$.

If the transition dynamics are Markovian -- that is, if the statistics of the next state depend only on the current state and action and not any prior states and actions -- the transition dynamics, reward function, state space, and action space define what is called a Markov Decision Process (MDP). If the transition dynamics are non-Markovian and rely on additional history, this problem is referred to as a Partially Observed Markov Decision Process, or a POMDP~\cite{Kaelbling1998planning}. We do not consider POMDPs here, but note that effective approaches include augmenting model-free MDP algorithms with recurrent neural networks to incorporate memory~\cite{Ni2022recurrent}.

For the finite-horizon formulas, let $n=0,\ldots,N$ label decision times $t_{n}=n\Delta t$. The agent takes an action and receives a reward at the final step $N$. A policy induces a joint state-action path law $P^\pi[x_{0:N},u_{0:N}]$, where $x_{0:N}=(x_0,\ldots,x_{N})$ and similarly for $u_{0:N}$.

\paragraph{The Objective.}
Let $\pi$ be an arbitrary policy.
Denote the path measure of this policy as $P^\pi$. Then, for some reward function $r(x,u)$, the goal of RL is to maximize the objective $\mathcal J[\pi]$:

\[
    \mathcal J[\pi] \triangleq \EE_{P^\pi}
    \left[\sum_{n=0}^{N}r(x_{n},u_{n})\right],
    \qquad \max_\pi\mathcal J[\pi].
\]
In RL, it is standard to consider a cumulative reward modified by a \emph{discount factor} $0 < \gamma < 1$ that is taken to be close to $1$. The discount factor weighs the reward and incentivizes accumulating reward sooner, rather than later:

\[
    \mathcal J[\pi] \triangleq \EE_{P^\pi}
    \left[\sum_{n=0}^{N}\gamma^{n} r(x_{n},u_{n})\right],
    \qquad \max_\pi\mathcal J[\pi].
\]
Another way to understand the discount factor is that it imposes an independent $1-\gamma$ probability at every timestep that an agent permanently transitions to a ``death'' state with no reward (i.e. an absorbing state with no reward).

\paragraph{Value Functions}
For the value definitions and Bellman equations in this subsection, we use the undiscounted convention $\gamma=1$.
To solve the MDP, RL algorithms invoke theoretical tools from SOC. These methods rely on \emph{value estimation}, where value denotes the expected cumulative future reward conditioned on a starting state $x$ at step $n$, for a particular policy $\pi$:

\begin{equation}
\begin{aligned}
V^{\pi}(x)
 &=\EE_{P^\pi}\!\left[\sum_{j=n}^{N}r(x_j,u_j)\,\middle|\,x_{n}=x\right].
\end{aligned}
\end{equation}
This value function is also known as a \emph{critic} that critiques the ``goodness'' of a particular policy $\pi$ at state $x$.
For $n<N$, the value function satisfies the Bellman equation for policy $\pi$:

\begin{equation}
 V^{\pi}(x)
 =
 \EE_{u\sim\pi(\cdot| x)}\!\left[
 r(x,u)
 + \EE_{x'\sim k(\cdot| x,u)}\!\left[
 V^{\pi}(x')
 \right]
 \right].
\label{eq:bellman-policy-value}
\end{equation}
Here $x'$ denotes the next state. The \emph{optimal value function} is

\begin{equation}
\begin{aligned}
V^{*}(x)&=\max_\pi V^{\pi}(x)\\
 &=\max_\pi\EE_{P^\pi}\!\left[\sum_{j=n}^{N}r(x_j,u_j)\,\middle|\,x_{n}=x\right],
\end{aligned}
\end{equation}
and measures the maximal cumulative expected reward achievable by a policy from a particular starting point. Like the value function in the continuous stochastic control setting, this discretized optimal value function similarly follows a recursive relationship:

\begin{equation}
 V^{*}(x)
 =
 \max_u \left[
 r(x,u)
 + \EE_{x'\sim k(\cdot| x,u)}\!\left[
 V^{*}(x')
 \right]
 \right].
\end{equation}
This equation is known as the Bellman optimality equation, the discrete-time analogue of the Hamilton-Jacobi-Bellman equation.
The \emph{Q function} critiques the ``goodness'' of a particular policy $\pi$ for some state and control:

\begin{equation}
 Q^{\pi}(x, u)
 =
 \EE_{P^\pi}\!\left[
 \sum_{j=n}^{N} r(x_j,u_j)
 \,\middle|\, x_{n}=x, u_{n} = u
 \right].
\end{equation}
In other words, the Q function measures the cumulative expected reward of a policy when fixing the control at step $n$ to $u$.
For $n<N$, the Q function satisfies the Bellman equation for policy $\pi$:

\begin{align}
\begin{aligned}
Q^{\pi}&(x,u)=r(x,u)\\
 &\quad+\EE_{x'\sim k(\cdot| x,u)}\!\left[
 \EE_{u'\sim\pi(\cdot| x')}\!\left[Q^{\pi}(x',u')\right]\right].
\end{aligned}
\label{eq:bellman-policy-q}
\end{align}
Here $x'$ and $u'$ denote the next state and action, respectively. The Q function also shows how to improve the policy $\pi$     by scoring the values of all actions $u$. The simplest scheme for \emph{policy improvement} is to choose, at each state $x$, an action that maximizes $Q^\pi(x,u)$. With exact Q values, the resulting policy has expected cumulative reward at least as high as that of the original policy.
However, clearly, this greedy scheme can lead to failed exploration if policy improvement is performed using an estimate of the true Q function from prior experience. Later in this section, we discuss alternative policy formulations that naturally incentivize exploration.

The \emph{optimal Q function} definition directly follows:

\begin{equation}
\begin{aligned}
&Q^{*}(x,u)=\max_\pi Q^{\pi}(x,u)\\
 &=\max_\pi\EE_{P^\pi}\!\left[
 \sum_{j=n}^{N}r(x_j,u_j)\,\middle|\,x_{n}=x,\;u_{n}=u\right].
\end{aligned}
\end{equation}
Just like the previously introduced objects, the optimal Q function also satisfies a Bellman optimality equation:

\begin{equation}
 Q^{*}(x,u)
 =
 r(x,u)
 + \EE_{x'\sim k(\cdot| x,u)}\!\left[
 \max_{u'} Q^{*}(x',u')
 \right].
\end{equation}
The starting point of RL methods is based on these Bellman equations, which are the discretized analogues of the Hamilton-Jacobi-Bellman equation.

\subsubsection{An overview of online RL algorithms}\label{sec:online_rl}

We explicitly consider the RL setting where an agent has access to the environment, and can actively choose transitions to gather new data. This problem setting is referred to as online RL.
Key methods in this category include Soft Actor Critic (SAC)~\cite{haarnoja2018soft} and Twin-Delayed Deep Deterministic Policy Gradient (TD3)~\cite{fujimoto2018addressing}. In this section, we focus on building up to the SAC method, which, in addition to being an established method for continuous control on which newer algorithms build~\cite{Bhatt2024crossq}, is also a Maximum Entropy RL (MaxEnt RL) method with close connections to maximum entropy inference principles discussed throughout this review. To get there, we start from basic, early precursors to SAC that do not require deep learning at all, then go through the practical considerations that ultimately culminate in SAC.

\paragraph{Tabular Q Learning.}
A starting point to solve the RL optimization problem is to utilize the theoretical tools from SOC. Recall from Sec.~\ref{sec:mechanics_control} that recursion provides a PDE called the Hamilton-Jacobi-Bellman equation. This PDE is a functional of the environment dynamics and the true reward function, where the solution is the optimal value function. Then, one route to find optimal controls is to first find the optimal value function, now with the Bellman Equation instead of the HJB equation, and extract the corresponding optimal control. However, the RL problem setting makes this route difficult: without access to the environment dynamics or the reward, we cannot directly solve the Bellman equation for the true optimal value function. One possible solution is to, instead, utilize some of the intuition from \emph{sampling}. In RL, the agent actively collects experience that it must use to find optimal controls. When the agent executes a control at timestep $n$, the agent observes a transition with reward $r_{n}$ from state $x_{n}$ to the next state $x_{n+1}$, which is a sample from the true environment dynamics and reward.

Recall the iterative connection between sampling and control developed in Sec.~\ref{sec:sampling_as_control}: sampled trajectories update the value estimates, which guide changes to the controls, which change the next set of sampled trajectories. One route to update the value estimates and policy is to directly average the realized cumulative rewards of the current policy, improve the policy using these estimates, and then evaluate the updated policy. Evaluation and improvement thus iterate and build on each other, because changing the policy changes the expected future rewards, which is referred to as policy iteration. In the most naive approach, we can roll out $M$ independent trajectories each starting from state $x$ with initial control $u$ at time $t_{n}$, then follow policy $\pi$ for all subsequent controls. We now switch to a discounted reward formulation, with $0<\gamma<1$, so that rewards received later count less. Averaging the realized cumulative rewards of these rollouts gives:

\begin{equation}
 \widehat Q^{\pi}(x,u)
 =\frac{1}{M}\sum_{i=1}^{M}
 \sum_{j=n}^{N}\gamma^{j-n}r(x_{j}^{(i)},u_{j}^{(i)}).
\end{equation}
Then, a simple way to find a good policy based on these value estimates is to take the argmax over these estimated Q values:

\begin{equation}
 u \leftarrow \argmax_{u} \widehat Q^{\pi}(x,u).
\end{equation}
However, evaluating different actions from complete rollouts can have high variance. Another approach uses a running estimate of value to replace the remainder of a trajectory~\cite{Sutton1988}. Instead of sampling and summing all realized rewards after a particular state, we can directly use a running value estimate for the future returns from the next state. This substitution can reduce sampling variance at the cost of potentially introducing error from utilizing an incorrect running value estimate. The observed immediate reward $r$ and the remaining estimated value from the next state $x'$ provide a new prediction for the current Q value, referred to as the target $y$:
\begin{equation}
\begin{aligned}
y&=r+\gamma\max_{u'}\widehat Q^*(x^{\prime},u'),\\
 &\qquad\text{for observation }(x,u,r,x^{\prime}).
\end{aligned}
\end{equation}
We can use these targets to update the current Q estimate, using a learning rate of $\xi$:

\begin{equation}
\begin{aligned}
\widehat Q^{*}(x,u)
 &\leftarrow\widehat Q^{*}(x,u)+\xi\left[y-\widehat Q^{*}(x,u)\right].
\end{aligned}
\label{eq:q-learning-update}
\end{equation}
Repeatedly updating the current Q estimate toward these target predictions converges to the optimal Q function \emph{under the assumption that every state and action is sampled infinitely often}~\cite{sutton1998reinforcement,watkins1992q}.
Conceptually, this update can be thought of as a ``backup'' of value from the future value estimate to the present estimate. Because the future value is ``easier'' to estimate (e.g. requires aggregating rewards over fewer timesteps) and is trivial at the endpoint of a trajectory, where it becomes exactly the expected reward, this backup moves value backwards in time from the easier subproblem  to harder subproblems. This intuition is also why we can refer to Hamilton-Jacobi-Bellman as a continuous form of dynamic programming and the Bellman Equation as discretized dynamic programming. Practically, a table holds these Q values, and the Bellman Backup indexes into this table and updates the value entries. As such, this algorithm is known as tabular Q-learning. 

\paragraph{Deep Q-Networks.}
The practical limitations of tabular Q Learning motivate the sequence of methods (Deep Q-Networks, Soft Q-Learning, and Soft Actor-Critic) we discuss below. First and foremost, storing and updating a separate Q value for every state and action in a table becomes infeasible in a large state space. Indeed, the tabular setup does not transfer to the continuous setting. Deep Q-networks (DQN) address this restriction by approximating the Q function with a neural network, allowing experience at one state to inform predictions at others~\cite{mnih2015human}. We denote learned estimates of $Q^\pi$ and $Q^*$ by $Q_\theta^\pi$ and $Q_\theta^*$, where $\theta$ denotes the network weights.

DQN fits a neural-network estimate $Q_\theta^*$ of the optimal Q function using all of an agent's prior observations -- including transitions and observed rewards -- aggregated within a \emph{replay buffer} $\mathcal{D}$. Rather than do tabular updates of the value functions following the Bellman equation, DQN directly minimizes the ``error'' between the value function and the recursed form of the value function over this replay buffer of all past experiences, which is amenable to neural network parameterization:

\begin{align*}
 L(\theta)&=\frac12\EE_{(x,u,r,x')\sim\mathcal D}\Big[\\
 &\quad\left(
 Q_\theta^*(x,u)
 \quad-(r+\gamma\max_{u'}Q_{\bar\theta}^*(x',u'))
 \right)^2\Big],\\
 \theta&\leftarrow\theta-\xi\nabla_\theta L(\theta).
\end{align*}
The notation $\bar{\theta}$ means that gradients cannot flow through the parameters in the target. The reason for this ``stop gradient'' is to force value to flow backwards in time: the value at time $t$ should not be updated with the rewards that occurred before time $t$.
While tabular Q learning provably converges, DQN does not provably converge without additional assumptions due to function approximation error~\cite{mnih2015human,Baird1995residual}.

\subsubsection{MaxEnt RL \& Soft Q-learning}\label{sec:maxent_rl}

We now return to the undiscounted convention $\gamma=1$ for the MaxEnt derivation and the two algorithms below. Throughout the remainder of the RL section, $Q^\pi$ and $V^\pi$ denote entropy-regularized (``soft'') values. The entropy coefficient $\alpha$ has the same units as the per-step reward; it is distinct from the convexity parameter used in Wasserstein geometry.

Learning values alone does not ensure that the agent explores, as is shown in Fig.~\ref{fig:explore-exploit}. Indeed, the tabular Q-learning algorithm can only guarantee convergence if the policies visit every state infinitely often. In large state spaces, this condition can be impossible to satisfy even with random actions. Furthermore, always choosing the action with the largest estimated Q value can ignore routes with activation energy. In RL, a common heuristic for exploration is to take a random action a small, epsilon proportion of the time. This strategy is called epsilon-greedy exploration where, typically, the probability epsilon goes to $0$ over the course of training~\cite{sutton1998reinforcement,mnih2015human}. This exploration mechanism can be thought of as a heuristic version of simulated annealing. Later works employ \emph{Maximum-entropy RL} (MaxEnt RL) to control exploration in a more principled way, incorporating the policy Shannon entropy $H[\pi(\cdot| x_{n})]$ in the objective itself:

\begin{equation}
\begin{aligned}
    \mathcal J[\pi] \triangleq
    \EE_{P^\pi}
    \qty[
        \sum_{n=0}^{N}
        \qty[r(x_{n}, u_{n})
        +
        \alpha H[\pi(\cdot| x_{n})]]
    ]
\end{aligned}
\label{eq:max-ent-intro}
\end{equation}
where $\alpha$ is a temperature that controls the strength of the entropy term.
This approach encourages the policy to spread out actions while exponentially weighting those with high expected reward. Here, we discuss Soft Q-learning, one of the early examples of MaxEnt RL that utilizes entropy as a principled mechanism to encourage exploration~\cite{haarnoja2017reinforcement}.

The crux of MaxEnt RL is the idea that the policy should induce the maximum-entropy path measure, where the energy of the measure is the expected cumulative reward.
Crucially, MaxEnt RL has close connections to control as inference principles (Sec.~\ref{sec:path_inference}). In fact, one can derive the soft Q-learning objective to learn optimal policies directly from a maximum-entropy inference problem over path measures~\cite{levine2018reinforcement}. We lay out this derivation here, connecting MaxEnt RL to the approaches and perspectives discussed previously in the review.

The inference-as-control problem begins with a definition of a path measure induced by a particular control policy. If this path measure exhibits mean observables (for example, the Schr\"odinger Bridge problem), and these moments are all that is known about the realized trajectories, Sanov's theorem states that the most likely path measure is the maximum entropy distribution that satisfies these moments. Each of these observables corresponds to a Lagrange Multiplier $r$ that enforces this constraint, and the resulting max-ent path measure is:
\begin{equation}
\begin{aligned}
    P^{u^*}[x(\cdot)]
    \propto
    P^0[x(\cdot)]
    \exp\qty(
        \sum_m
        r_m
        a_m[x(\cdot)]
    ),
\end{aligned}
\end{equation}
Here $a_m[x(\cdot)]$ is the $m$-th observable measured along a trajectory. The Lagrange multiplier $r_m$ enforces a contraint on the averaged observable, as in Sec.~\ref{sec:path_inference}. Thus, these approaches give controls $u$ that give a max-ent realization of these observables.

In the control setting, we instead \emph{wish} for the path measure to satisfy particular observations -- for example, that the path measure realizes a high expected cumulative $r$ for some fixed reward function. The rewards $r$ no longer directly result from enforcing specific observables (e.g. termination within a particular set). Therefore, despite MaxEnt RL optimizing path measures that look identical to the control-as-inference approach, maximum entropy is a \emph{postulate} of the MaxEnt RL approach -- there is not a fundamental physical principle like Sanov's theorem for fixed observables that underlies the MaxEnt RL formulation.

To derive the MaxEnt RL objective, we now express the continuous-time path measures above using the discrete-time MDP notation introduced in Sec.~\ref{sec:mdp}.
As an example, the controlled stochastic dynamics of Sec.~\ref{sec:mechanics_control}
\[
\dot{x}_t = f_t(x_t) + u_t(x_t) + \sqrt{\varepsilon}\,\eta_t
\]
discretized by an Euler--Maruyama step of duration $\Delta t$ result in the Gaussian transition kernel:
\begin{equation}
\begin{aligned}
&
k(x^{\prime}|x,u)\\
 &=\mathcal N\qty(x';\;z+\qty[f_{t_{n}}(x)+u]\Delta t,\;\varepsilon\Delta t\,I).
\end{aligned}
\label{eq:mdp-kernel-langevin}
\end{equation}
where $\mathcal N$ denotes the Gaussian density for the physical coordinates, where $I$ is the identity matrix. Then, an agent executes controls and transitions according to this discretized transition kernel. When an agent visits the pair $(x,u)$, the agent receives a reward $r(x,u)$. Together, the kernel (or transition dynamics), the reward, and the state and action spaces define the finite-horizon Markov decision process (MDP), thus showing how the continuous control problem becomes the MDP. \footnote{We assume the problem is fully observed, so the dynamics are Markovian. We consider the objective where $\gamma=1$ (undiscounted), noting that discounting can be recovered by augmenting the MDP with stochastic transitions to a zero-reward, permanent ``death'' state.} The \emph{policy} $\pi(u_{n}| x_{n})$ denotes the agent's probability of choosing action $u_{n}$ in state $x_{n}$ at step $n$ and plays the role of the control law
$u_t(x)$. Iteratively combining the transition kernel and the policy through time gives the discrete-time analog of the controlled path measure $P^u[x(\cdot)]$, which now induces a measure over discretized trajectories $\{x_{n}, u_{n}\}$:
\begin{equation}
\begin{aligned}
&P^\pi[x_{0:N},u_{0:N}] \\
 &=p_0(x_0)\prod_{n=0}^{N}\pi(u_{n}| x_{n}) \prod_{n=0}^{N-1}k(x_{n+1}| x_{n},u_{n}),
\end{aligned}
\label{eq:policy-path-measure}
\end{equation}
and the analog of the reference measure $P^0[x(\cdot)]$ is the measure of the passive dynamics with uniform action distribution $\pi^0(u_{n}| x_{n})=c_0$:
\begin{equation}
\begin{aligned}
&P^0[x_{0:N},u_{0:N}] \\
 &=p_0(x_0)\prod_{n=0}^{N}\pi^0(u_{n}| x_{n})\prod_{n=0}^{N-1}k(x_{n+1}| x_{n},u_{n}).
\end{aligned}
\label{eq:discrete-reference-measure}
\end{equation}
Say that we wish to find controls that lead to large rewards. One way to do so is to construct a target measure that upweights high-reward trajectories, then regress the control-induced path measure towards this target measure for the training objective. Following the maximum entropy postulate in the MaxEnt RL approach, one possible choice for this target measure is (a possibly unrealizable) maximum-entropy, reward-maximizing path measure with respect to a base measure $P^0[x_{0:N},u_{0:N}]$:

\begin{align}
\begin{aligned}
 &P^*[x_{0:N},u_{0:N}] \triangleq\\
 &\argmax_P\left[\EE_P\left[\frac1\alpha\sum_{n=0}^{N}r(x_{n},u_{n})\right]-\dkl(P\Vert P^0)\right] \\
 &= \frac1Z\,P^0[x_{0:N},u_{0:N}]\exp\qty(\frac1\alpha\sum_{n=0}^{N}r(x_{n},u_{n})).
\end{aligned}
\end{align}
Here the maximization over $P$ is over normalized path measures. $P^*$ is the reward-tilted target measure and $\alpha$ is a temperature that controls the strength of the tilt. Importantly, this $P^*$ is the \emph{theoretically optimal} max-ent target distribution and need not be achievable by a particular policy. In practice, the rewards are only observed, as the agent does not have a priori access to the true reward function following standard RL assumptions.
We seek a policy whose path measure approximates this target, where the limited policies can only determine the executed controls, not the dynamics of the system. The optimal policy for this inference objective therefore produces the closest realizable path measure, which is the minimizer of $\dkl\qty(P^\pi \Vert P^*)$ \footnote{The choice of reverse KL is intentional, as it requires sampling from $P^\pi$ (something we know how to do) rather than $P^*$ (which we do not know how to sample from a priori). }:

\begin{align}
    \pi^* \triangleq \arg \min_{\pi} \dkl\qty(P^\pi \Vert P^*)
\end{align}
Note that the path distribution $P^{\pi^*}$ is the closest realizable distribution and need not equal $P^*$. Thus, we have derived an inference objective that should provide high-reward, maximum-entropy controls. What is the relationship between this inference objective and the MaxEnt RL objective? It turns out that in discrete time, the transition kernels cancel in the ratio of the two measures, which ``simplifies'' this inference into the MaxEnt RL objective. The result is shown in Eq.~\eqref{eq:long_RL},

\begin{figure*}[t] 
  \centering
\begin{equation}
\begin{aligned}
&\dkl\qty(P^\pi\Vert P^*)\!=\!\int\prod_{n=0}^{N}\dd{x_{n}}\dd{u_{n}}\,P^\pi[x_{0:N},u_{0:N}]\!\!\log\frac{
 \textstyle p_0(x_0)\prod_{n=0}^{N}\pi(u_{n}| x_{n})
 \textstyle \prod_{n=0}^{N-1}k(x_{n+1}| x_{n},u_{n})
 }{
 \textstyle \frac1Z\,p_0(x_0)\prod_{n=0}^{N}\qty[\pi^0(u_{n}| x_{n})e^{r(x_{n},u_{n})/\alpha}]
 \textstyle \prod_{n=0}^{N-1}k(x_{n+1}| x_{n},u_{n})
 }\\
 &=\int\prod_{n=0}^{N}\dd{x_{n}}\dd{u_{n}}\,P^\pi[x_{0:N},u_{0:N}]\log\frac{\prod_{n=0}^{N}\pi(u_{n}| x_{n})}{
 \frac1Z\prod_{n=0}^{N}\qty[\pi^0(u_{n}| x_{n})e^{r(x_{n},u_{n})/\alpha}]}\\
 &=\EE_{P^\pi}\qty[\sum_{n=0}^{N}\qty(\log\frac{\pi(u_{n}| x_{n})}{\pi^0(u_{n}| x_{n})}-\frac{r(x_{n},u_{n})}{\alpha})] +\log Z \\
 &=-\frac1\alpha\EE_{P^\pi}\qty[\sum_{n=0}^{N}\qty[r(x_{n},u_{n})+\alpha H[\pi(\cdot| x_{n})]]] +\log Z-(N+1)\log c_0.
\end{aligned}
\label{eq:long_RL}
\end{equation}
  \caption{Eq.~\eqref{eq:long_RL}. As the reader can see, RL-notation is no joke.}
  \label{fig:long_RL}
\end{figure*}

\noindent Here $H[\pi(\cdot| x_{n})] = -\EE_{u\sim\pi(\cdot| x_{n})}[\log\pi(u| x_{n})]$ denotes the Shannon entropy of the policy at state $x_{n}$ and step $n$ and the normalizing constant $Z$ is independent of $\pi$. Minimizing the KL divergence is therefore equivalent, up to the positive factor $1/\alpha$ and the additive constant $\log Z-(N+1)\log c_0$ (neither of which should change the fixed point), to maximizing the maximum-entropy RL objective:

\begin{equation}
\begin{aligned}
    \mathcal J[\pi] \triangleq
    \EE_{P^\pi}
    \qty[
        \sum_{n=0}^{N}
        \bigl[r(x_{n}, u_{n})
        +
        \alpha H[\pi(\cdot| x_{n})]\bigr]
    ]
    .
\end{aligned}
\label{eq:max-ent}
\end{equation}
Thus, as desired, we have theoretically derived the maximum-entropy RL objective from MaxEnt inference principles. As a note,
Eq.~\eqref{eq:max-ent} has the same free-energy structure as entropic transport in Eq.~\eqref{eq:static_maxent2}, with negative reward playing the role of cost and temperature $\alpha$ the role of $\varepsilon$.
Maximum-entropy RL introduces the entropy bonus in Eq.~\eqref{eq:max-ent} as a regularizer that aids robustness~\cite{eysenbach2022maximum} and exploration~\cite{haarnoja2018soft}. In the above derivation, we show that the entropy bonus emerges directly from the inference problem following Ref.~\cite{levine2018reinforcement}: the entropy bonus is the KL cost of deviating from the base dynamics, now without any assumption of Brownian noise. With Brownian noise, the same divergence reduces to the quadratic control cost of Eq.~\eqref{eq:girsanov-kl-section2}.

While the above derivation gives the MaxEnt RL objective, it is unclear how to practically solve this maximization.
In continuous time, we solved the KL control problem of Eq.~\eqref{eq:pre-pisc} by dynamic programming, arriving at the HJB equation linearized using the Cole--Hopf transformation $V^*_t(x) = \log \psi_t(x)$ (Eq.~\eqref{eq:cole-hopf-hjb-psi}). The same transformation in discrete time leads to soft Q-learning~\cite{haarnoja2017reinforcement}, the MaxEnt RL variant of Q-learning~\cite{watkins1992q}. The central object is the soft Q function, the action-conditioned, maximum-entropy counterpart of the value function with the state and the first action both held fixed:
\begin{widetext}
\begin{equation}
\begin{aligned}
&Q^\pi(x,u)=r(x,u) +\EE_{P^\pi}\left[
 \sum_{j=n+1}^{N}
 \bigl(r(x_{j},u_{j})
 -\alpha\log\pi(u_{j}|x_{j})\bigr) \middle|\,x_{n}=x,\;
 u_{n}=u\right].
\end{aligned}
\label{eq:soft-q-def}
\end{equation}
The soft Q and soft value functions are related by
\begin{equation}
\begin{aligned}
&V^\pi(x)=\EE_{u\sim\pi(\cdot|x)}\Bigl[ Q^\pi(x,u)-\alpha\log\pi(u|x)\Bigr].
\end{aligned}
\end{equation}
Here $Q^\pi$ excludes the entropy contribution of the fixed current action, while $V^\pi$ includes the current entropy contribution.
By construction, $Q^\pi$ satisfies the ``soft'' Bellman Eq.~\eqref{eq:soft-bellman},
\begin{equation}
\begin{aligned}
&Q^\pi(x,u)=r(x,u)+\EE_{x'\sim k(\cdot|x,u)}\Biggl[
\EE_{u'\sim\pi(\cdot|x')}\Bigl[
 Q^\pi(x^{\prime},u^{\prime})-\alpha\log\pi(u^{\prime}|x^{\prime})\Bigr]\Biggr].
\end{aligned}
\label{eq:soft-bellman}
\end{equation}
\end{widetext}

\noindent with the boundary condition $Q^{\pi}(x,u) = r(x,u)$ at the final step $N$. Then, the optimal policy maximizes the conditional expectation on the right-hand side of Eq.~\eqref{eq:soft-bellman}, the expected Q value plus the policy entropy, independently at the next states $x^{\prime}$. Setting the variation with respect to $\pi(u^{\prime}| x^{\prime})$ to zero gives a Boltzmann distribution over actions,\footnote{Greedily maximizing this quantity at each state is a form of soft policy iteration. Ref.~\cite{haarnoja2018soft} proves that it globally optimizes Eq.~\eqref{eq:max-ent}.}

\begin{equation}
\begin{aligned}
\pi^{*}(u| x)
=\frac{e^{Q^{*}(x,u)/\alpha}
 }{\sum_{v} e^{Q^{*}(x,v)/\alpha}
 }.
\end{aligned}
\label{eq:soft-q-policy}
\end{equation}
Substituting $\pi^*$ into Eq.~\eqref{eq:soft-bellman} then reveals an alternative form of the soft Bellman equation, assuming that the policy $\pi$ is the true Boltzmann distribution:

\begin{equation}
\begin{aligned}
Q^{*}(x,u)&=r(x,u)\\
 &\quad+\EE_{x'\sim k(\cdot| x,u)}\!\left[
 \alpha\log\sum_{u^{\prime}}e^{Q^{*}(x^{\prime},u^{\prime})/\alpha}\right].
\end{aligned}
\label{eq:soft-bellman-2}
\end{equation}
Thus, if the policy $\pi^*$ is known exactly, fixing $(x,u)$ leaves the environment transition $k(x^{\prime}| x,u)$ as the only remaining source of randomness in this backup.

Eq.~\eqref{eq:soft-bellman-2} is the discrete counterpart of the max-ent HJB equation after the Cole--Hopf transformation, providing a consistency equation whose right-hand side is a conditional expectation under the path measure. In continuous time, we solved the consistency equation with explicit knowledge of the dynamics and the reward. In RL, neither is known, but every observed transition $(x,u,r,x^{\prime})$ provides a one-sample Monte Carlo estimate of the right-hand side, giving an update rule:

\begin{equation}
\begin{aligned}
y&\mathrel{=}r+\alpha\,\log\,\sum_{u'}\,e^{\widehat Q^*(x',u')/\alpha},\\
\widehat Q^*(x,u)&\gets \widehat Q^*(x,u)+\xi\bigl[y-\widehat Q^*(x,u)\bigr].
\end{aligned}
\label{eq:soft-q-update}
\end{equation}
Here, $\widehat Q^*$ is the learned estimate of $Q^*$ and $\xi$ is the learning rate.
This update rule is the soft Q-learning algorithm (Algorithm~\ref{alg:soft-q-learning}), where the log-sum-exp is a softened maximum. As the temperature $\alpha \rightarrow 0$, the softened maximum in the log-sum-exp becomes a hard maximum, recovering standard tabular Q-learning~\cite{watkins1992q}.

Soft Q-learning~\cite{haarnoja2017reinforcement} has formed the basis for many popular RL algorithms, in which a neural network $Q_\theta^*$ approximates the soft Q function using the update rule in Eq.~\eqref{eq:soft-q-update}. Implementation details, such as the use of replay buffers and target networks, are important for getting these algorithms to work in practice.

\begin{algorithm}[H]
\caption{Soft Q-learning}
\label{alg:soft-q-learning}
\begin{algorithmic}
\State Initialize $Q_\theta^*$ and replay buffer $\mathcal D$;
       fix temperature $\alpha>0$ and learning rate $\xi$
\For{each iteration}
    \State \textbf{Collect experience:} Run
           $\pi_\theta(u| x)\propto e^{Q_\theta^*(x,u)/\alpha}$,
           and store $(x,u,r,x')$ in $\mathcal D$
    \State \textbf{Sample past experience:}
           $(x,u,r,x')\sim\mathcal D$
    \State \textbf{Q target:}
           $y=r+\alpha\log \sum_{u'} e^{Q_\theta^*(x',u')/\alpha}$
    \State \textbf{Update Q:}
           $\theta\leftarrow\theta-\xi\nabla_\theta
           \bigl(Q_\theta^*(x,u)-y\bigr)^2$,
           holding $y$ fixed
\EndFor
\end{algorithmic}
\end{algorithm}

\subsubsection{Soft actor-critic}\label{sec:sac}

Each update must still evaluate the log-sum-exp over every action, which is cheap for a few discrete actions. With continuous actions, the sum becomes an integral, and computing the Q-conditioned partition function $Z_Q(x) = \int \dd{u}\, e^{Q(x, u)/\alpha}$ can become intractable and high-variance as the action dimension grows due to the curse of dimensionality. This curse makes it difficult to evaluate the normalized \emph{Boltzmann policy} $\pi_Q(u | x) = e^{Q(x, u)/\alpha}/Z_Q(x)$ defined by the current critic $Q$, or the soft value $\alpha\log Z_Q(x^{\prime})$ on the right-hand side of the backup in Eq.~\eqref{eq:soft-bellman-2}.

This leads to Soft Actor-Critic (SAC), which learns a policy together with the value estimate~\cite{haarnoja2018soft}. The iteration in Algorithm~\ref{alg:sac} alternates between updating the critic's estimate of value under the current policy and improving the policy using that estimate, with entropy included in both steps. The parameterized actor $\pi_\theta$ aims to fit the true Boltzmann target $\pi_Q$ defined by the current critic $Q$. Simultaneously, the learned critic $Q_\theta^\pi$ aims to fit the value induced by the current policy $\pi_\theta$.

The unknown reward function and transition dynamics are only available through samples. Each observed transition supplies $r$ and $x^{\prime}$. Thus, the integral in $Z_Q(x^{\prime})$ can nevertheless remain intractable even when the current critic $Q$ is directly evaluable. To address this integration problem, consider the Boltzmann policy $\pi_Q$ defined above. Suppose we could draw the next action $u^{\prime}$ from $\pi_Q(u^{\prime} | x^{\prime})$ and evaluate its log density. A single draw would then replace the log-sum-exp, because substituting $\log \pi_Q(u^{\prime} | x^{\prime}) = Q(x^{\prime}, u^{\prime})/\alpha - \log Z_Q(x^{\prime})$ gives

\begin{equation}
\begin{aligned}
&Q(x^{\prime},u^{\prime})-\alpha\log\pi_Q(u^{\prime}| x^{\prime})\\
 &=Q(x^{\prime},u^{\prime})\\
 &\quad-\alpha\qty[\frac{Q(x^{\prime},u^{\prime})}{\alpha}-\log Z_Q(x^{\prime})]\\
 &=\alpha\log Z_Q(x^{\prime})
\end{aligned}
\label{eq:sac-zero-variance}
\end{equation}
for every sampled $u^{\prime}$. Thus, with an exact $\pi_Q (u^{\prime} | x^{\prime})$, this difference is independent of the sampled action $u^{\prime}$ and provides a zero-variance estimate of the soft value function.
In practice, we do not have access to the true Boltzmann $\pi_Q$, so the estimator is not variance-free.
Soft actor-critic~\cite{haarnoja2018soft,Haarnoja2018applications} keeps the combination $Q(x^{\prime}, u^{\prime}) - \alpha\log \pi_{\phi}(u^{\prime}| x^{\prime})$ as the estimate of the soft value. Evaluating the normalized density $\pi_Q$ itself requires the same intractable partition function $Z_Q(x)$. Soft actor-critic therefore draws $u^{\prime}$ from a second network, a policy $\pi_\phi(u| x)$ with a tractable form, trained to stand in for $\pi_Q$. This substitution turns Eq.~\eqref{eq:soft-bellman-2} into the practical target used in the critic update in Algorithm~\ref{alg:sac}

\begin{equation}
\begin{aligned}
y&=r\,+\,Q_\theta^\pi(x^{\prime},u^{\prime})\,-\,\alpha\log\pi_\phi(u^{\prime}| x^{\prime}),\\
 &\quad\qquad\qquad\qquad\qquad\quad \text{where} \; u^{\prime}\sim\pi_\phi(\cdot| x^{\prime}).
\end{aligned}
\label{eq:sac-critic}
\end{equation}
where, like Soft Q-learning, the learned Q function $Q_\theta^\pi$ is directly regressed towards the target. Then,
training $\pi_\phi$ to maximize
\begin{equation}
\begin{aligned}
\max_\phi\;\EE_{u\sim\pi_\phi(\cdot| x)}\left[
Q_\theta^\pi(x,u)-\alpha\log\pi_\phi(u| x)\right].
\end{aligned}
\label{eq:sac-actor}
\end{equation}
is equivalent to minimizing $\dkl\qty(\pi_\phi \Vert \pi_Q)$, since the actor objective equals $-\alpha\dkl\qty(\pi_\phi \Vert \pi_Q)+\alpha\log Z_Q(x_k)$ at fixed $x_k$. Thus, for a \emph{fixed} critic $Q$, the policy objective corresponds to projecting the learned policy $\pi_\phi$ onto the Boltzmann policy $\pi_Q$.
However, in practice, the critic estimate $Q_\theta^\pi$ can be noisy. Then, targets built from a maximum, or a soft maximum, of the estimates inherit an upward \emph{maximization bias}. Practical implementations reduce overestimation by training two Q networks and using the smaller of the two targets, adding back in some pessimism~\cite{haarnoja2018soft}.

\begin{algorithm}[H]
\caption{Soft Actor-Critic}
\label{alg:sac}
\begin{algorithmic}
\State Initialize critic $Q_\theta^\pi$, actor $\pi_\phi$,
       and replay buffer $\mathcal D$; fix temperature $\alpha>0$
\For{each iteration}
    \State \textbf{Collect experience:} Run $\pi_\phi$,
           observe $(x,u,r,x')$, and store in $\mathcal D$
    \State \textbf{Sample past experience:}
           $(x,u,r,x')\sim\mathcal D$
    \State Sample $u'\sim\pi_\phi(\cdot| x')$
    \State \textbf{Critic target:}
           $y=r+Q_\theta^\pi(x',u')-\alpha\log\pi_\phi(u'| x')$
    \State \textbf{Update critic:}
           $\theta\leftarrow\theta-\xi\nabla_\theta
           \bigl(Q_\theta^\pi(x,u)-y\bigr)^2$, holding $y$ fixed
    \State \textbf{Update actor:}
    \begin{align*}
            \phi \leftarrow\phi+\xi\nabla_\phi
           &\EE_{v\sim\pi_\phi(\cdot| x)}
           [ \\
           &\qquad Q_\theta^\pi(x,v)-\alpha\log\pi_\phi(v| x)]
    \end{align*}
\EndFor
\end{algorithmic}
\end{algorithm}

Solving a fixed-point equation by stochastic approximation raises the question of convergence. When the algorithm stores one Q value per state-action pair, the update converges to the optimal Q function~\cite{watkins1992q, agarwal2019reinforcement}, and similar guarantees exist for simple function classes such as linear MDPs~\cite{doi:10.1287/moor.2022.1309, agarwal2019reinforcement}.
In continuous or high-dimensional settings, however, the algorithm cannot enumerate every state-action pair and must instead approximate $Q^*$ with a parametric function such as a neural network. Unfortunately, the convergence guarantee does not generally survive function approximation~\cite{Baird1995residual}, and the failure is visible in practice. A standard benchmark is mountain car~\cite{Moore90efficientmemory-based}, an underpowered car on a sinusoidal hill that must accelerate back and forth to build enough energy to reach the top. Even in mountain car, learned RL solutions show a large gap from the known optimal control solution~\cite{huberchebyshev}.
Thus, modern RL methods learn controls from samples alone and approximate intractable objects, at the cost of the exactness guarantees of the earlier control methods.

\subsection{Wasserstein gradient flows}
\label{sec:WGFApps}

In this section, we examine three classes of applied problems for which Wasserstein geometry and Wasserstein gradient flows provide elegant solutions. The first problem is the numerical integration of Fokker-Planck equations, in which the Wasserstein geometry yields a powerful integration algorithm known as the \emph{Jordan-Kinderlehrer-Otto (JKO)} scheme. Second, we discuss modeling the dynamics of many interacting particles, particularly the regime known as the \emph{mean-field limit} in which each particle effectively interacts with a continuous density. The collective evolution of the particle density often takes the form of Wasserstein gradient flow, and this perspective has yielded insights into the training of wide neural networks, in which the parameters can be thought of as particles. Finally, we discuss the general problem of \emph{variational inference (VI)}, in which a complex distribution is approximated by a simpler one that can be used for sampling. The problem of finding the optimal parameters can be viewed as an instance of Wasserstein gradient flow, and this perspective yields physical intuition and stable algorithms. As we will see, variational inference as discussed in this section forms an ``equilibrium'' methodology, whose ``non-equilibrium'' counterparts are powerful, contemporary algorithms discussed in Sec.~\ref{sec:flows_diffusions} on flows and diffusion.

\subsubsection{The Jordan-Kinderlehrer-Otto (JKO) scheme}\label{sec:JKO}

The JKO scheme is a canonical example of using Wasserstein geometry to tackle numerical challenges, resulting in algorithms that are often more efficient, stable, and interpretable. Consider the general problem of numerically solving Fokker-Planck equations (FPE) driven by a conservative flow and constant diffusivity, which take the form:

\begin{equation}
    \frac{\partial \rho_t(x)}{\partial t}=\nabla\cdot\left[\rho_t(x)\nabla \Phi(x)\right]+\nabla^2\rho_t(x) 
    \label{eq:JKO_FPE}
\end{equation}
The unit diffusion coefficient corresponds to $\varepsilon=2$ in the convention of Chapter~\ref{sec:variational_structure}. It may be tempting to numerically solve Eq.~\eqref{eq:JKO_FPE} by an explicit finite step update:

\begin{align}
\begin{aligned}
 \rho^{(k+1)}(x)=\rho^{(k)}(x)+\Delta t&\Bigl\{\nabla\cdot\left[\rho^{(k)}(x)\nabla\Phi(x)\right]\\
 &+\nabla^2\rho^{(k)}(x)\Bigr\}
\end{aligned}
\label{eq:naive}
\end{align}
where $\rho^{(k)}(x) = \rho_{k\Delta t}(x)$ is the numerical discretization and $\Delta t$ is the step size. While conceptually straightforward, this direct approach and those like it encounter certain challenges. First, under Eq.~\eqref{eq:naive}, $\rho^{(k)}$ is not guaranteed to be either positive or normalized, which is particularly problematic if $\rho^{(k)}$ represents a probability distribution. Second, one must explicitly compute derivatives of $\rho^{(k)}(x)$, which is costly in high dimensions and can introduce instabilities if $\rho^{(k)}(x)$ naturally concentrates onto singular manifolds such as points or lines. Third, the algorithm only converges for sufficiently small $\Delta t$. Fourth, the continuous-time FPE takes the form of a Wasserstein gradient flow $v=-\Wgrad\mathcal F$, yet under Eq.~\eqref{eq:naive}, $\mathcal{F}$ need not monotonically decrease at finite $\Delta t$.

Jordan, Kinderlehrer, and Otto proposed an integration scheme that overcomes these shortcomings~\cite{JKOSIAM1998}. It has two ingredients. First, the algorithm is variational: it breaks the numerical integration into a sequence of optimization problems. For intuition, consider integrating an ODE $\dot x = -\nabla \Phi(x)$. A variational approach takes the form

\begin{align}
x^{(k+1)}&=\arg\min_x\left\{\frac{|x-x^{(k)}|^2}{2\Delta t}+\Phi(x)\right\}\\
 &\begin{aligned}
 \approx\arg\min_x\Bigl\{&\frac{|x-x^{(k)}|^2}{2\Delta t}+\Phi(x^{(k)})\\
 &+\left(x-x^{(k)}\right)\cdot\nabla\Phi(x^{(k)})\Bigr\}
 \end{aligned}\\
 &=x^{(k)}-\Delta t\,\nabla\Phi(x^{(k)})
\end{align}
which recovers the expected Euler update as $\Delta t \to 0$. As another example, the diffusion equation $\partial_t \rho  = \nabla^2 \rho$, i.e. Eq.~\eqref{eq:JKO_FPE} with $\Phi=0$, can readily be put in variational form:

\begin{align}
\hspace{-0.55em}
 \rho^{(k+1)} = \arg\min_\rho \left( \frac{\|\rho-\rho^{(k)}\|_{L^2}^2}{ 2 \Delta t} + \int | \nabla \rho(x)|^2 \dd{x} \right) \label{eq:L2}
\end{align}
where $\|f\|_{L^2}^2=\int\dd{x}|f(x)|^2$ is the squared $L^2$ norm of $f$.
While Eq.~\eqref{eq:L2} is certainly variational, it still suffers some of the practical and conceptual limitations of Eq.~\eqref{eq:naive}. Practically, one still needs to compute derivatives of $\rho$, and Eq.~\eqref{eq:L2} can only capture Eq.~\eqref{eq:JKO_FPE} with $\nabla \Phi=0$, i.e., there is no way to amend the action of Eq.~\eqref{eq:L2} to account for the drift term $\nabla \cdot (\rho \nabla \Phi)$. Conceptually, there is no reason to think of $\rho$ as a probability distribution in Eq.~\eqref{eq:L2}: both the $L_2$ norm and the functional in the second term make sense even when $\rho$ is non-normalized or negative. In this sense, Eq.~\eqref{eq:L2} emphasizes a particular interpretation of the diffusion equation — it punishes gradients. But we also know that diffusion has a probabilistic interpretation: it promotes randomness, thereby increasing entropy. Is there a version of Eq.~\eqref{eq:L2} that emphasizes this interpretation?

These shortcomings are addressed by the second ingredient of the JKO scheme. Namely, one replaces the $L_2$ norm in Eq.~\eqref{eq:L2} with one that is adapted to probability: the Wasserstein distance. In doing so, one also replaces the gradient penalty with a thermodynamically meaningful quantity: the free energy. The resulting update scheme takes the form

\begin{equation}
    \rho^{(k+1)}=\arg\min_\rho\left(\frac{W_2^2(\rho^{(k)},\rho)}{2\Delta t}+\mathcal{F}[\rho]\right)
    \label{eq:JKO_scheme}
\end{equation}
Eq.~\eqref{eq:JKO_scheme} has several advantages. First, it accommodates both drift and diffusion within a single variational principle, and it has a clear probabilistic interpretation: descend $\FF$ as fast as possible while not changing $\rho$ too much in the Wasserstein sense. This change in conceptual approach also resolves the practical issues arising in Eq.~\eqref{eq:naive}: $\rho^{(k)}$ is guaranteed to remain a valid probability distribution for all $k$, the integration is stable, and $\FF[\rho^{(k)}]$ is guaranteed to decrease monotonically for all $\Delta t$. Moreover, Eq.~\eqref{eq:JKO_scheme} has the advantage of being derivative-free — the gradients and divergences of Eqs.~\eqref{eq:naive}--\eqref{eq:L2} are replaced by an optimal transport problem~\cite{CarrilloFoCM2022}. In particular, the JKO scheme can be brought to the Benamou-Brenier form~\cite{MartinKRM2010} (see Sec.~\ref{sec:eot}), which allows one to rigorously take the continuum ($\Delta t\to0$) limit, as well as to introduce efficient optimization methods~\cite{CarrilloFoCM2022} such as the Sinkhorn algorithm or point-mass approximations.
The clear geometric motivation behind Eq.~\eqref{eq:JKO_scheme} also results in better interpretability; for instance, one can write down explicit formulas for the implicit bias (i.e., effective modifications of $\FF$) introduced at finite $\Delta t$~\cite{halmos2026implicit}.

As a practical matter, the JKO scheme tends to shine in the following contexts. In very high dimensions, explicit spatial derivatives or spectral methods become computationally expensive, whereas solving the optimal transport problem at each time step remains comparatively efficient using particle-based or other methods. Another context where JKO is helpful is when $\rho(x)$ has finite or irregular support, which is difficult to accommodate with spectral or other grid-based methods but is naturally handled by optimal transport optimization. Finally, JKO handles pattern-forming PDEs well, particularly when mass tends to aggregate onto singular points, lines, or manifolds. In short, JKO offers a practical advantage in high dimensions, with irregular boundaries, or in mass-concentrating situations.

\subsubsection{The mean-field limit of interacting particles}
\label{sec:mf}

Now we study how WGF helps us understand the dynamics of the interaction of many identical particles, particularly the so-called \emph{mean-field} limit, in which the dynamics of the individual particles can be replaced by the dynamics of a coarse-grained density. This limit is of fundamental interest in mathematical physics (e.g., how do continuum theories like fluid mechanics rigorously emerge from interacting particles?), and in practical applications, such as the training of wide neural networks, in which the neurons are viewed as interacting particles. 
In certain situations, the coarse-grained density obeys WGF, which is typically simpler to analyze than a full many-body system and offers geometric intuition that underlies convergence theorems. 
For example, this approach allows one to prove under what conditions training a wide two-layer NN converges to global minima of its loss.  

The mathematical setting of interest is a collection of $N_{\mathrm p}$ interacting particles with coordinates $x_t^{(1)},\ldots,x_t^{(N_{\mathrm p})}$, obeying the following Langevin dynamics:

\begin{align}
\begin{aligned}
 \dd x_t^{(i)}&=-\nabla_i\Phi^{(N_{\mathrm p})}(x_t^{(1)},\dots,x_t^{(N_{\mathrm p})})\dd t\\
 &\quad+\sqrt\varepsilon\dd W_t^{(i)}
\end{aligned}
\label{eq:micro}
\end{align}
where $\dd W_t^{(i)}$ are independent Wiener processes and $\nabla_i$ means that the gradient is evaluated with respect to argument $i$. To permit a mean-field limit, we assume that the particles are ``identical'' in the sense that their potential is only a function of their unlabeled density. That is, we can write $\Phi^{(N_{\mathrm p})} (x_t^{(1)}, \dots, x_t^{(N_{\mathrm p})}) = \VV[\widehat\rho_t]$ where 

\begin{align}
\widehat\rho_t(x) = \frac1{N_{\mathrm p}} \sum_{i=1}^{N_{\mathrm p}} \delta (x - {x_t^{(i)}})
\end{align}
is the empirical probability measure, normalized to one. We reserve $\rho_t$ for the continuous mean-field density and $N_{\mathrm p}$ is the particle count. A common class of such interactions consists of single- and two-particle potentials:

\begin{align}
&\begin{aligned}
 &\Phi^{(N_{\mathrm p})}(x^{(1)},\dots x^{(N_{\mathrm p})})=\sum_i\Phi(x^{(i)})+\frac1{2N_{\mathrm p}}\sum_{i,j}U(x^{(i)},x^{(j)}) \nonumber
\end{aligned}\\
&\begin{aligned}
 &=N_{\mathrm p}\int\Phi(x)\widehat\rho(x)\,\dd{x}\frac{N_{\mathrm p}}2\int U(x,y)\widehat\rho(x)\widehat\rho(y)\,\dd{x}\,\dd y \nonumber
\end{aligned}\\
&\equiv\VV[\widehat\rho]
\end{align}
where we assume $U(x,x)=0$ to exclude self interactions. Now let's assume that the initial positions of the particles are drawn independently from a continuous density $\rho_0$, namely $x_0^{(i)} \overset{\mathrm{iid}}{\sim} \rho_0$.
The mean-field limit boils down to the following approximation: instead of tracking all the individual particles, one tracks the evolution of the continuous density. 
More precisely, one assumes that at a later time $t$, the particles are once again independently and identically distributed, but now according to an updated density $\rho_t$. 
A representative particle $x_t$, for which $\rho_t$ is the distribution, evolves according to

\begin{align}
\dd x_t  = - \Wgrad\VV[\rho_t](x_t) \dd t + \sqrt{\varepsilon} \dd W_t \label{eq:mv}
\end{align}
where $\Wgrad \VV = \nabla (\delta \VV/\delta \rho)$ is the Wasserstein gradient introduced in Sec.~\ref{sec:WGF}. Eq.~\eqref{eq:mv} is often referred to as a \emph{McKean-Vlasov process}~\cite{McKean1966class}, and is simply a rewriting of Eq.~\eqref{eq:micro} in which the particle label has been dropped, and the potential is expressed in terms of the continuous density $\rho_t$ instead of the empirical density $\widehat\rho_t$. 
Qualitatively, Eq.~\eqref{eq:mv} mathematizes the intuition that as $N_{\mathrm p} \to \infty$, the typical particle $x_t$ effectively interacts with an average continuous field $\rho_t$ of particles. 
The corresponding Fokker-Planck equation for $\rho_t$ is given by:

\begin{align}
\begin{aligned}
 \partial_t\rho_t&=\nabla\cdot\bigl(\rho_t\Wgrad\FF[\rho_t]\bigr)\\
 &=\nabla\cdot\bigl(\rho_t\Wgrad\VV[\rho_t]\bigr)+\frac\varepsilon2\nabla^2\rho_t.
\end{aligned}
\label{eq:mve}
\end{align}
where $\FF = \VV - \frac{\varepsilon}{2} H$ and $H[\rho] = -\int \rho \log \rho \dd{x} $ is the entropy.  Eq.~\eqref{eq:mve} is sometimes referred to as a \emph{McKean-Vlasov equation}\footnote{Historically, McKean refers to the stochastic process $\varepsilon>0$ and  Vlasov refers to the deterministic process $\varepsilon=0$}, and it precisely takes the form of WGF. Whether and for what values of $N_{\mathrm p}$ and $t$ the mean-field approximation is a good approximation is a field of mathematical research known as \emph{propagation of chaos}~\cite{sznitman1991topics}.

As a practical example, the WGF structure of Eq.~\eqref{eq:mve} has been useful in analyzing the training dynamics of large machine learning (ML) models such as neural networks (NNs)~\cite{Chizat2018global, Rotskoff2022trainability, Mei2018mean, Sirignano2020mean}. Large ML models are often trained by defining a loss function and performing versions of gradient descent, e.g., Stochastic Gradient Descent (SGD) or Adaptive Moment Estimation (Adam), in their parameters. For a large but finite number of parameters, the loss landscape typically features many saddles, large manifolds of degenerate minima, and different minima-manifolds separated by shallow barriers. These features make the convergence of gradient descent difficult to analyze.  However, when many of the parameters enter the loss identically, such as in a very wide NN, an $N_{\mathrm p} \to \infty$ limit can be taken, allowing one to analyze the evolution of the density $\rho_t$ of the parameters over training time. The density $\rho_t$ generally obeys WGF and the functional $\FF$ that it minimizes is generally much simpler than the loss landscape governing the evolution of a finite number $N_{\mathrm p}$ of parameters. This simplification allows convergence results to be obtained. It should be noted that there is an older collection of results, sometimes referred to as \emph{Universal Approximation Theorems}~\cite{Cybenko1989approximation,Barron1993universal,Park1991universal}, that ask which classes of functions can be approximated by a given ML architecture. The question at stake here is perhaps even more relevant: under what conditions are these functions actually reached via training?

To illustrate typical ingredients of such an approach, consider a two-layer neural network 

\begin{align}
f^{N_{\mathrm p}}(x; \Theta_t) = \frac{1}{N_{\mathrm p}}\sum_{i=1}^{N_{\mathrm p}} w_t^{(i)} \varphi(x; z_t^{(i)}) 
\label{eq:model}
\end{align}
In Eq.~\eqref{eq:model}, $f^{N_{\mathrm p}}( x ; \Theta_t)$ is the model to be trained. Here $\varphi(x;z_t^{(i)})$ is a nonlinear function, referred to as a ``feature'', with weight $w_t^{(i)}$ and internal parameters $z_t^{(i)}$. All the parameters of $f^{N_{\mathrm p}}$ may be collected in a single vector $\Theta_t = (\theta_t^{(1)}, \dots, \theta_t^{(N_{\mathrm p})})$ where $\theta_t^{(i)} = (w_t^{(i)}, z_t^{(i)})$. The parameters are given a time index $t$ because they will evolve over training. The function $f^{N_{\mathrm p}}(x; \Theta_t)$ is trained to approximate a target function $y(x)$ with $x$ distributed according to a density $\nu(x)$. This is done by minimizing a loss function $L(\Theta_t)$, a common choice for which is the mean-squared error (MSE):

\begin{align}
\begin{aligned}
 L(\Theta_t)&=\frac12\int\abs{y(x)-f^{N_{\mathrm p}}(x;\Theta_t)}^2\nu(x)\,\dd{x}\\
 &=\frac12\EE_\nu\!\left[\abs{y-f_t^{N_{\mathrm p}}}^2\right]
\end{aligned}
\end{align}
where $f_t^{N_{\mathrm p}} = f^{N_{\mathrm p}}(\cdot ; \Theta_t)$ for brevity. The loss is minimized by using some form of gradient descent, typically with stochasticity. For instance  

\begin{align}
\dd \theta_t^{(i)} = - N_{\mathrm p}\nabla_i L \dd t + \sqrt{\varepsilon} \dd W_t^{(i)} \label{eq:tsde}
\end{align}
The factor $N_{\mathrm p}$ compensates for the $1/N_{\mathrm p}$ in the network output: differentiating $L$ with respect to one neuron's parameters produces a factor $1/N_{\mathrm p}$. Multiplying the learning rate by $N_{\mathrm p}$ therefore keeps each neuron's drift of order one as the network grows.
In practice, more practical forms of gradient descent are used, for instance stochastic gradient descent in which stochasticity does not come from a Wiener process $\dd W_t^{(i)}$, but from using finite sample-size Monte Carlo estimates of the loss. 
Regardless of the precise method, the loss $L$ is a complicated, nonconvex function of the parameters $\Theta_t$, so proving (or understanding) anything about training may at first seem to be a daunting task.

To gain intuition, we can think of each neuron $i$ as being a particle with coordinate $\theta_t^{(i)}$, and (noisy) gradient descent, e.g. Eq.~\eqref{eq:tsde}, as introducing interactions between the particles. For the two-layer neural network, the key insight is that Eq.~\eqref{eq:model} can be written as a function of neuron density only:  

\begin{align}
f^{N_{\mathrm p}}(x; \Theta_t) = \int a(x;\theta) \widehat\rho_t(\theta) \dd \theta 
\end{align}
where $\theta = (w,z)$, $a(x;\theta) = w \varphi(x;z)$, and  $\widehat\rho_t(\theta) = \frac{1}{N_{\mathrm p}} \sum_{i=1}^{N_{\mathrm p}} \delta( \theta- \theta_t^{(i)})$ is the empirical distribution of the parameters. Because $f^{N_{\mathrm p}}$ only depends on the density of parameters, the loss $L(\Theta_t)$ can also be expressed in terms of the density alone. For example, the MSE becomes: 

\begin{align}
\begin{aligned}
L(\Theta_t)&=\LL[\widehat\rho_t]=\int\Phi(\theta)\widehat\rho_t(\theta)\dd\theta\\
 &\quad+\frac12\int U(\theta,\theta')\widehat\rho_t(\theta)\widehat\rho_t(\theta')\dd\theta\dd\theta'+C_y
\end{aligned}
\label{eq:mfloss}
\end{align}
where $\Phi(\theta)=-\EE_\nu[y(x)a(x;\theta)]$ and $U(\theta,\theta')=\EE_\nu[a(x;\theta)a(x;\theta')]$ are single- and two-particle potentials, and $C_y=\tfrac12\EE_\nu[y(x)^2]$ is independent of $\theta$. The fact that the interactions only depend on the density $\widehat\rho(\theta)$ makes the convergence of training dynamics amenable to mean-field-based proof techniques. 

The exact convergence statements proven in the literature~\cite{Chizat2018global, Rotskoff2022trainability, Mei2018mean, Sirignano2020mean} require some mathematical formalism to state precisely, and depend on the exact technical hypotheses placed on the features $\varphi$, the loss $L$, and the training algorithm. There are, however, some shared ingredients: one typically assumes that the network is ``initialized'' by drawing  initial parameters, $\theta_0^{(i)}$, independently from a well-behaved distribution $\rho_0$, typically obeying some precise regularity conditions.  One is then interested in the evolution of the coarse-grained distribution $\rho_t$, which obeys a McKean-Vlasov equation, whose precise form depends on the exact architecture, loss, or version of gradient descent under consideration. Typically, the McKean-Vlasov equation can be interpreted in terms of WGF of a functional $\FF$, $\partial_t\rho_t=\nabla\cdot(\rho_t\Wgrad\FF[\rho_t])$, where $\FF$ consists of the loss $\LL$ plus entropic terms arising from stochasticity. The dynamical problem of training the neural network is then mathematically transformed into studying the comparatively simple geometric and topological properties of $\FF$. 
While the functional $\FF$ is not necessarily geodesically convex in the Wasserstein sense, it often permits enough structure to make mathematical statements about convergence to global minima for $\rho_0$ obeying suitable conditions. 
The conditions placed on $\rho_0$ for convergence are of some practical use, because they can be interpreted as instructions on reasonable ways to initialize the NN for training. Intuitively, WGF is the right mathematical framework for analyzing these problems because of the underlying physical picture: the parameters $\theta_t^{(i)}$ are particles, but these particles cannot jump, or be created or destroyed. Therefore, continuous motion of a density, i.e. WGF, naturally emerges. 

To give a sense of the results proven, Ref.~\cite{Mei2018mean} considers a quadratic loss and training via either stochastic gradient descent or noisy stochastic gradient descent; for the latter case, they prove that any absolutely continuous $\rho_t$ converges to a unique global minimum of $\FF$ that controllably approximates the MSE. They also prove that $f_t^{N_{\mathrm p}}$ converges to the $N_{\mathrm p}\to \infty$ limit with bounds on the rate of convergence as a function of $N_{\mathrm p}$ and training time $t$. Ref.~\cite{Rotskoff2022trainability} considers models of the form of Eq.~\eqref{eq:model} with technical hypotheses on the forms of the nonlinearity and data distribution, and identifies convergence to a global minimum, and studies the size of fluctuations for finite $N_{\mathrm p}$ and $t$. Ref.~\cite{Chizat2018global} provides convergence proofs for exact gradient descent for a class of architectures and loss functions defined by the property that the functional $\FF$ they give rise to obeys certain homogeneity conditions. 

A couple of additional general remarks are in order. First, the mean-field approach based on a single density of neurons works for a two-layer neural network, but deeper networks require additional information about the connections between layers. Mean-field methods have been extended to these settings, with different assumptions and conclusions~\cite{nguyen_rigorous_2023,sirignano2021meanfieldanalysisdeep}. Second, there is a common source of confusion: one observes that $\LL$ in Eq.~\eqref{eq:mfloss} is manifestly convex in the parameter density $\rho$, i.e. $\LL[\lambda\rho'+(1-\lambda)\rho]\leq\lambda\LL[\rho']+(1-\lambda)\LL[\rho]$ for any two densities $\rho$, $\rho'$ and $\lambda\in[0,1]$. While true, this mixture-interpolation notion of convexity is not the useful one for training based on gradient descent. Under gradient descent, the parameters flow along Wasserstein gradients, not weighted mixtures, and they certainly do not teleport (as is sometimes required for interpolation).  Therefore, convexity with respect to Wasserstein geodesics is the useful mathematical framework to use, and one cannot directly conclude that $\rho_t$ converges to a unique global minimum under training due to interpolation convexity. 
Finally, one might wonder: why not freeze the internal parameters $z_t^{(i)}$ and only optimize the feature weights $w_t^{(i)}$'s? After all, the MSE loss $L(\Theta_t)$ is explicitly convex in $w_t^{(i)}$. This approach, often referred to as \emph{kernel regression}, is certainly feasible, but it suffers from two drawbacks. First, a much larger $N_{\mathrm p}$ is required for kernel regression to approximate the $N_{\mathrm p}=\infty$ global minimum~\cite{Chizat2018global}. Second, kernel regression exhibits a different implicit bias: when both the $w$'s and the $z$'s are dynamic, it is observed that the network learns comparatively sparse distributions for the $w_t^{(i)}$'s, and the selected features $\varphi(x;z_t^{(i)})$ have been updated via $z_t^{(i)}$, and thus tend to be informative about the learned properties of the data. When only the $z_t^{(i)}$'s are frozen, the learned distribution over the $w$'s tends to be broader and, by construction, no individual feature $\varphi(x;z_t^{(i)})$ reveals properties about the data. Whether ingredients from the mean-field approach can be extended to other architectures remains an open question.

\subsubsection{Variational inference}
\label{sec:variational_inference}

Our final example of applications of WGF is to \emph{variational inference} (VI)~\cite{Blei2017variational}.  
VI addresses the following challenge: one is confronted with a probability density $p_1=\widetilde p_1/Z_1$ that is difficult to sample from and whose unnormalized form $\widetilde p_1$ may be all that is available; however, we assume that we have access to a family of distributions $p_\theta$ defined by a finite number of parameters $\theta$, and that the $p_\theta$ are easy to normalize and sample from, e.g. the family of Gaussians. Our goal is to find the optimal parameters $\theta^*$ such that $p_{\theta^*}$ best approximates $p_1$. 
This arises, for example, in Bayesian settings, in which one has a prior $p^0(x)$  and observations $A$, and one would like to sample from the posterior $p_1(x)=p(x| A)=p(A| x)p^0(x)/p^0(A)$, where the normalization constant $p^0(A)=\int p(A| x)p^0(x)\,\dd x$ is often intractable. It is also thematically related to variational methods in quantum mechanics or statistical physics, in which the ground state of a model may be approximated by an ansatz featuring only a few parameters. 

For the connection to WGF, we are interested in the setting where ``best approximation'' is defined as minimizing the Kullback-Leibler divergence 

\begin{align}
\theta^* = \underset{ \theta}{\argmin} \, \dkl( p_\theta\Vert p_1 ) \label{eq:viobj}
\end{align}
As noted earlier in the review, the gradient of $\dkl(p_\theta\Vert p_1)$ with respect to $\theta$ is independent of $Z_1=\int\widetilde p_1\,\dd x$, which is often infeasible to compute. 
One tempting option to minimize Eq.~\eqref{eq:viobj} is to simply perform gradient descent in the parameters, i.e. 
\begin{align}
\dot \theta = - \nabla_\theta L(\theta)\label{eq:naive2}
\end{align}
with $L(\theta)= \dkl(p_\theta\Vert p_1)$.
However, in practice, this naive parameter-centric approach is often problematic. The loss  $L(\theta)$ may be poorly conditioned and nonconvex. Moreover, certain families of probability distributions are only well defined on a subset of parameters; for instance, a Gaussian distribution is only well defined if its covariance matrix is positive-definite. Unfortunately, Eq.~\eqref{eq:naive2} may lead to violations of these constraints. Finally, if one were to reparameterize the same family of probability distributions, Eq.~\eqref{eq:naive2} would yield a different trajectory through probability space---so Eq.~\eqref{eq:naive2} lacks parameterization invariance. 

Wasserstein geometry is useful in addressing these issues. For convenience, write $p_1(x)=e^{-\Phi(x)}/Z_1$, with dimensionless potential $\Phi=-\log\widetilde p_1$, and write 

\begin{align}
\begin{aligned}
 \LL[p_\theta]&=\dkl(p_\theta\Vert p_1) =\int\Phi(x)p_\theta(x)\,\dd x\\
 &\quad+\int p_\theta(x)\log p_\theta(x)\,\dd x+\log Z_1.
\end{aligned}
\label{eq:lossp}
\end{align}
It seems that one approach to minimizing $\dkl(p_\theta\Vert  p_1)$ would be to perform Wasserstein gradient flow down the functional $\LL[p_\theta]$, namely update $p_\theta$ according to $\nabla \cdot p_\theta \Wgrad\LL[p_\theta]$. Yet, there is a fundamental problem: there is no guarantee that one can find a trajectory of parameters $\theta_t$ such that $\dot\theta\cdot\partial_\theta p_\theta=\nabla\cdot(p_\theta\Wgrad\LL[p_\theta])$. In other words, the naive WGF may induce a flow field that flows out of the subspace of probability distributions defined by the parameters $\theta$. 

The resolution is to project the Wasserstein descent velocity onto the tangent space of the parametric family. Recall from Sec.~\ref{sec:WGF} that a density tangent is represented by a minimum-energy velocity $v=-\nabla\chi$, so $\partial_t p_t=-\nabla\cdot(p_t v_t)$. The inner product of two such velocities at density $p$ is

\begin{align}
 \langle v,v'\rangle_p=\int p(x)v(x)\cdot v'(x)\,\dd x.
 \label{eq:ddk}
\end{align}
Consider a parametric family $\mathcal M=\{p_\theta:\theta\in\Omega\}$, where $\Omega$ is a parameter domain of dimension $d_\theta$. In local coordinates $\theta=(\theta_1,\ldots,\theta_{d_\theta})$, define velocities $v_i=-\nabla\chi_i$ by $\partial_{\theta_i}p_\theta=-\nabla\cdot(p_\theta v_i)$. They induce the metric $g_{ij}=\langle v_i,v_j\rangle_{p_\theta}$. The projection of a velocity field $v$ is

\begin{align}
 \operatorname{proj}_{\mathcal M}v
 =\sum_{i,j}v_i(g^{-1})_{ij}\langle v_j,v\rangle_{p_\theta}.
\end{align}
The projected descent velocity is $-\operatorname{proj}_{\mathcal M}\Wgrad\LL[p_\theta]$, and its density evolution is

\begin{align}
 \partial_t p_{\theta_t}
 =\nabla\cdot\left(p_{\theta_t}\operatorname{proj}_{\mathcal M}\Wgrad\LL[p_{\theta_t}]\right).
 \label{eq:wgfproj}
\end{align}
Writing $L(\theta)=\LL[p_\theta]$, the same evolution in parameter coordinates is

\begin{align}
 \dot\theta_i=-\sum_{j=1}^{d_\theta}(g^{-1})_{ij}\partial_{\theta_j}L(\theta).
 \label{eq:projdiscrete}
\end{align}
Comparing Eq.~\eqref{eq:projdiscrete} to Eq.~\eqref{eq:naive2}, the metric induced by Wasserstein geometry essentially acts to redirect the gradients with respect to $\theta$. 
Moreover, one can show that Eq.~\eqref{eq:projdiscrete} is equivalent to a parametric version of the JKO style update introduced in Sec.~\ref{sec:JKO}

\begin{align}
\theta_{t+\Delta t} = \arg \min_\theta \left\{ \frac{W_2^2(p_{\theta_t},p_\theta)}{2\Delta t}  + \LL[p_\theta]\right\}
\end{align}
in the limit $\Delta t \to 0$. 

So what does all this formalism buy us? 
The minimization algorithm in Eq.~\eqref{eq:wgfproj}, or equivalently Eq.~\eqref{eq:projdiscrete}, comes with elegant convergence guarantees. Recall from Sec.~\ref{sec:WGF} that the functional $\LL$ can have a notion of convexity (namely $\alpha$-geodesic convexity) completely analogous to convexity in finite-dimensional optimization, and that this notion of convexity ensures that WGF converges to its global minimum. Recall also from Sec.~\ref{sec:WGF} that $\mathcal{M}$ is a geodesically convex subset of probability distributions if it contains the entire Wasserstein geodesic between any two of its elements. One can show that whenever $\mathcal{M}$ is geodesically convex and  $\LL$ is $\alpha$-geodesically convex, the projected WGF in Eq.~\eqref{eq:wgfproj} is guaranteed to reach its unique minimum on $\mathcal{M}$, and do so exponentially fast. Especially important in the context of VI is that the loss $\LL$ in Eq.~\eqref{eq:lossp} is in fact $\alpha$-geodesically convex whenever $p_1$ is \emph{strongly log-concave}, which means that it is written as $p_1 \propto e^{-\Phi} $ for some $\alpha$-convex $\Phi$ (see Sec.~\ref{sec:WGF} for the definition of $\alpha$-convex). So in short, one can use the structure of Wasserstein geometry to turn the problematic optimization in Eq.~\eqref{eq:naive2} into a well-behaved one. 

Let's explore two pedagogical examples of parametric subspaces. First,  take $\mathcal{M}$ to be the set of all non-degenerate Gaussians on $\RR^d$, which goes by the name \emph{Bures-Wasserstein} space $\bw(\RR^d)$~\cite{Takatsu2011}. Each distribution in this family can be represented by the mean $m \in \RR^d$ and covariance matrix $\Sigma \in \mathbb S_{++}^d$, where $\mathbb S_{++}^d$ is the set of all symmetric, positive-definite $d\times d$ matrices. 
To see that the optimization problem is nontrivial, one can check that $L(m,\Sigma)$ defined by Eq.~\eqref{eq:lossp} is not necessarily convex in the parameters $(m, \Sigma)$ even when $\Phi(x)$ is $\alpha$-convex in $x$---for example, consider $\Phi(x) = x^4$ in $d=1$ dimension. Now let's apply the machinery of WGF and see what the dynamical optimization equations for $m$ and $\Sigma$ look like. 
First, let's check that $\bw(\RR^d)$ is geodesically convex. One can show that the Wasserstein geodesic between any two Gaussians $p_0 = \NN(m_0, \Sigma_0)$ and $p_1 = \NN(m_1, \Sigma_1)$ is given by the transport map 

\begin{align}
\begin{aligned}
 x_t&=(1-t)x_0 +t\,\Bigl[\Sigma_0^{-1/2}\left(\Sigma_0^{1/2}\Sigma_1\Sigma_0^{1/2}\right)^{1/2}\\
 &\qquad\Sigma_0^{-1/2}(x_0-m_0)+m_1\Bigr]
\end{aligned}
\label{eq:affinemap}
\end{align}
Notably, Eq.~\eqref{eq:affinemap} is an affine transformation for all $t$. Recall that any affine transformation of a normal distribution is a normal distribution, so the $p_t \in \bw(\RR^d)$  for all $t$, confirming that $\bw(\RR^d)$ is geodesically convex. Next, we would like to find an explicit expression for the projected Wasserstein gradient $\BWgrad \FF = \mathrm{proj}_{\mathrm{BW}(\RR^d)} \Wgrad \FF$ on a functional $\FF$. Because all Gaussians are connected by affine maps, and all affine maps take Gaussians to Gaussians, it follows that the tangent space to $\bw(\RR^d)$ corresponds to all affine maps. Using this fact and a bit of algebra, one finds that:

\begin{align}
 (\BWgrad\FF[p])(x)
 =\left\langle\nabla^2\frac{\delta\FF}{\delta p}\right\rangle_p(x-m)
  +\left\langle\nabla\frac{\delta\FF}{\delta p}\right\rangle_p.
\end{align}
Here $\nabla^2$ in a matrix expression denotes the Hessian; the Laplacian is its trace. The parameter dynamics are

\begin{align}
 \dot m_t&=-\left\langle\nabla\frac{\delta\FF}{\delta p}\right\rangle_{p_t},\label{eq:bwgrad1}\\
 \dot\Sigma_t&=-\left\langle\nabla^2\frac{\delta\FF}{\delta p}\right\rangle_{p_t}\Sigma_t
 -\Sigma_t\left\langle\nabla^2\frac{\delta\FF}{\delta p}\right\rangle_{p_t}.\label{eq:bwgrad2}
\end{align}
For the special case of VI, we take $\FF = \dkl(p\Vert p_1)$ and the flow equations become

\begin{align}
 \dot m_t&=-\langle\nabla\Phi\rangle_{p_t},\label{eq:gvi1}\\
 \dot\Sigma_t&=-\langle\nabla^2\Phi\rangle_{p_t}\Sigma_t
 -\Sigma_t\langle\nabla^2\Phi\rangle_{p_t}+2I.\label{eq:gvi2}
\end{align}
Eqs.~\eqref{eq:gvi1}--\eqref{eq:gvi2} can be implemented straightforwardly because the expectations are taken with respect to the Gaussian measure $p_t = \NN (m_t, \Sigma_t)$. In the naive approach, Eq.~\eqref{eq:naive2}, of directly optimizing with respect to $m$ and $\Sigma$, the equations for the mean $\dot m_t$ are unchanged, while Eq.~\eqref{eq:gvi2} for $\dot \Sigma_t$ differs in its dependence on $\Sigma_t$. With the standard Frobenius inner product for covariance matrices, ordinary gradient descent gives: 

\begin{align}
    \dot\Sigma_t=-\frac12\langle\nabla^2\Phi\rangle_{p_t}+\frac12\Sigma_t^{-1},
\end{align}
which can create situations in which $\Sigma$ becomes singular (loses its positive definiteness) during optimization.  
The geometric perspective of WGF avoids this pathology, makes the convex nature of the problem more transparent, and leads to other generalizations and principled convergence guarantees~\cite{lambert2022variational, Dio2023forward}. 

As a second, theoretically rich, example, suppose $\mathcal{M}$ is the set of all discrete measures comprising $N_{\mathrm p}$ point masses: $\widehat p_t(x) = \frac{1}{N_{\mathrm p}} \sum_{i=1}^{N_{\mathrm p}} \delta(x-x_t^{(i)})$~\cite{Liu2016stein,Liu2017stein}. Here, the parameters $x_t^{(i)}$ can be thought of as point particles, whose positions need to be optimally situated to approximate the target distribution $p_1$. Upon inspection, this is exactly the setting of interacting particles considered in Sec.~\ref{sec:mf}. Therefore, using the loss in Eq.~\eqref{eq:lossp}, the particles update according to 

\begin{align}
    \dd x_t^{(i)} = - \nabla \Phi(x_t^{(i)}) \dd t +\sqrt2\,\dd W_t^{(i)} \label{eq:langevin} 
\end{align}
where the noise convention is $\varepsilon=2$, as in Chapter~\ref{sec:sampling}. This is simply traditional MCMC sampling (see Chapter~\ref{sec:sampling}). However, there is a conceptually interesting point: Unlike Eqs.~\eqref{eq:gvi1}--\eqref{eq:gvi2}, or more generally, Eq.~\eqref{eq:projdiscrete}, the evolution in Eq.~\eqref{eq:langevin} is stochastic, not deterministic. This is ultimately connected to the fact that the discrete measure $\widehat p_t$ is singular, so gradients of the entropic contribution to Eq.~\eqref{eq:lossp} are ambiguous. While stochastic updates are not inherently problematic, this has led to a conceptual question: are there purely deterministic equations for the $x_t^{(i)}$ that would nevertheless give rise to an equivalent approximation in the $N_{\mathrm p} \to \infty$ limit? In the language of Sec.~\ref{sec:mf}, is there a deterministic version of Eq.~\eqref{eq:langevin} that gives rise to the same mean-field limit in Eq.~\eqref{eq:mve}? An intuitive attempt would be to replace the delta functions $\delta(x- x_t^{(i)})$ with a smoother function $k_h(x-x_t^{(i)})$ of width $h$, which gives the evolution equations

\begin{align}
\begin{aligned}
 \dot x_t^{(i)}&=-\nabla\Phi(x_t^{(i)})\\
 &\quad-\left.\nabla_x\log\left[\frac1{N_{\mathrm p}}\sum_j k_h(x-x_t^{(j)})\right]\right|_{x=x_t^{(i)}}
\end{aligned}
\end{align}
where the density gradient differentiates the evaluation argument $x$ with particle centers held fixed, before setting $x=x_t^{(i)}$. The continuum limit $N_{\mathrm p} \to \infty$ yields

\begin{align}
\partial_t p_t=\nabla\cdot\left[p_t\nabla\bigl(\Phi+\log(k_h\star p_t)\bigr)\right] \label{eq:kern1}
\end{align}
where $(k_h \star p_t)(x) = \int k_h(x-y) p_t(y) \dd y$ denotes convolution. Notably, this approach is not entirely straightforward because $p_t$ in Eq.~\eqref{eq:kern1} does not converge to $p_1$ unless $h=0$, which raises a subtle question of limits for $N_{\mathrm p} \to \infty$ and $h \to 0$. This subtlety motivates an alternative, more recent, approach known as \emph{Stein variational gradient descent}~\cite{Liu2016stein,Liu2017stein}. The basic idea is to put the regularization in a different location:  

\begin{align}
\partial_t p_t = \nabla \cdot \left[ p_t \mathcal{K}^{p_t}_h \left(\nabla \log \frac{p_t}{p_1} \right) \right] \label{eq:stein} 
\end{align}
In Eq.~\eqref{eq:stein}, $\mathcal{K}^{p}_h$ implements convolution weighted by $p$. Namely,  given $f(x)$, the smoothed version is 

\begin{align}
\begin{aligned}
 \mathcal{K}^{p}_h(f)(x)&=(k_h\star(pf))(x)\\
 &=\int f(y)p(y)k_h(x-y)\dd y
\end{aligned}
\end{align}
Interestingly, the smoothing operator $\mathcal{K}_h^p$ can also be interpreted as a projection of the Wasserstein gradient $\Wgrad \FF =\nabla\log(p_t/p_1)$~\cite{Liu2017stein}; however, this projection is distinct from $\mathrm{proj}_\mathcal{M}$ in Eq.~\eqref{eq:wgfproj}; it instead projects the velocity onto the  reproducing kernel Hilbert space associated with the kernel $k_h$, see e.g.~\cite{Liu2017stein} for more detail.
The discretized version of Eq.~\eqref{eq:stein} takes the form:

\begin{align}
\begin{aligned}
 \dot x_t^{(i)}&=-\frac1{N_{\mathrm p}}\sum_{j=1}^{N_{\mathrm p}}\Bigl[
 k_h(x_t^{(i)}-x_t^{(j)})\nabla\Phi(x_t^{(j)})\\
 &\qquad+\nabla k_h(x_t^{(i)}-x_t^{(j)})\Bigr]
\end{aligned}
\label{eq:implement}
\end{align}
where $\nabla\Phi(x_t^{(j)})$ differentiates the potential at particle $j$, while $\nabla k_h(z)$ differentiates the kernel argument $z=x_t^{(i)}-x_t^{(j)}$. Notice that Eq.~(\ref{eq:implement}) is straightforward to implement, and  Eq.~\eqref{eq:stein} genuinely permits $p_1$ as a stationary solution. Yet, questions surrounding its convergence remain a topic of investigation~\cite{Lu2019scaling,Korba2020nonasymptotic, Salim2022convergence,Das2023provably,Shi2023finite}. Nonetheless, Eq.~\eqref{eq:implement} has a clear physical interpretation: the first term captures gradient descent, while the second is a particle repulsion encouraging exploration of the full potential.

In addition to these two examples, several other parametric spaces for VI have received attention. 
For example, the treatment of Bures-Wasserstein space above can be extended to Gaussian mixtures by considering measures over $\bw(\RR^d)$ itself~\cite{lambert2022variational, Chen2019gaussian,Delon2020Wasserstein}\footnote{In the development of WGF above, we have assumed that the base space $\RR^d$ is equipped with the standard Euclidean metric. As would be appropriate for $\bw(\RR^d)$ one can consider generalizations to intrinsically curved spaces by treating $\nabla$ as a covariant derivative and using Riemannian dot products where appropriate.}. Another powerful approach is to consider a \emph{mean-field} approximation (not to be confused with the mean-field limit of interacting particles) in which $\mathcal{M} = \otimes_{i=1}^{d} \mathcal{W}_2(\RR) \subset \mathcal{W}_2 (\RR^d)$ 
is the space of products of one-dimensional distributions~\cite{Jiang2024algorithms,ChewiBOOK2025,Ghosh2024representations}. Across these examples, VI benefits from theoretical guarantees when $\Phi(x)$ is $\alpha$-convex. Much modern interest is in situations in which $\Phi(x)$ is not convex, but may be multi-modal and feature low-dimensional structure. Building models that are effective for sampling in this more advanced context motivates the next section on flow and diffusion (Sec.~\ref{sec:flows_diffusions}).

\subsection{Generative modeling with flows and diffusion}\label{sec:flows_diffusions}

In Chapter~\ref{sec:sampling} on Sampling, we explored the following problem: given a (potentially unnormalized) probability distribution $p(x)$, how to draw a set of $N$ independent samples $\{x_n\}_{n=1}^N$? 
In this section, we ask the inverse problem: given a set of samples $\{x_n\}_{n=1}^N$, how to infer the probability distribution $p(x)$ and generate a new independent sample $x_{N+1}$ from the estimated distribution? The former task is known as \emph{density estimation}, and the latter task is \emph{generative modeling}. Like many inverse problems, generative modeling is a somewhat ill-defined task: one aims to generate samples similar, but not \emph{identical} to those from the training set (an issue referred to as ``generalization versus memorization'').
Generative modeling is an important problem: In many contexts, like protein structures~\cite{Watson.etal2023} or natural images~\cite{Ho.etal2020}, training samples are abundant, but unlike in physics or chemistry, an explicit model for the data is not available.  
Over the last ten years, deep-learning-based generative models have proven astonishingly successful at producing samples from such complex, high-dimensional distributions. 
This section provides an overview of these new methods, their utility for scientific simulations, and their connection to the ideas around sampling, transport, and non-equilibrium physics discussed in Chapters~\ref{sec:variational_structure}--\ref{sec:sampling}. Note that we do not attempt a systematic literature review.

\begin{figure}[t]
    \centering
    \includegraphics[width=0.5\linewidth]{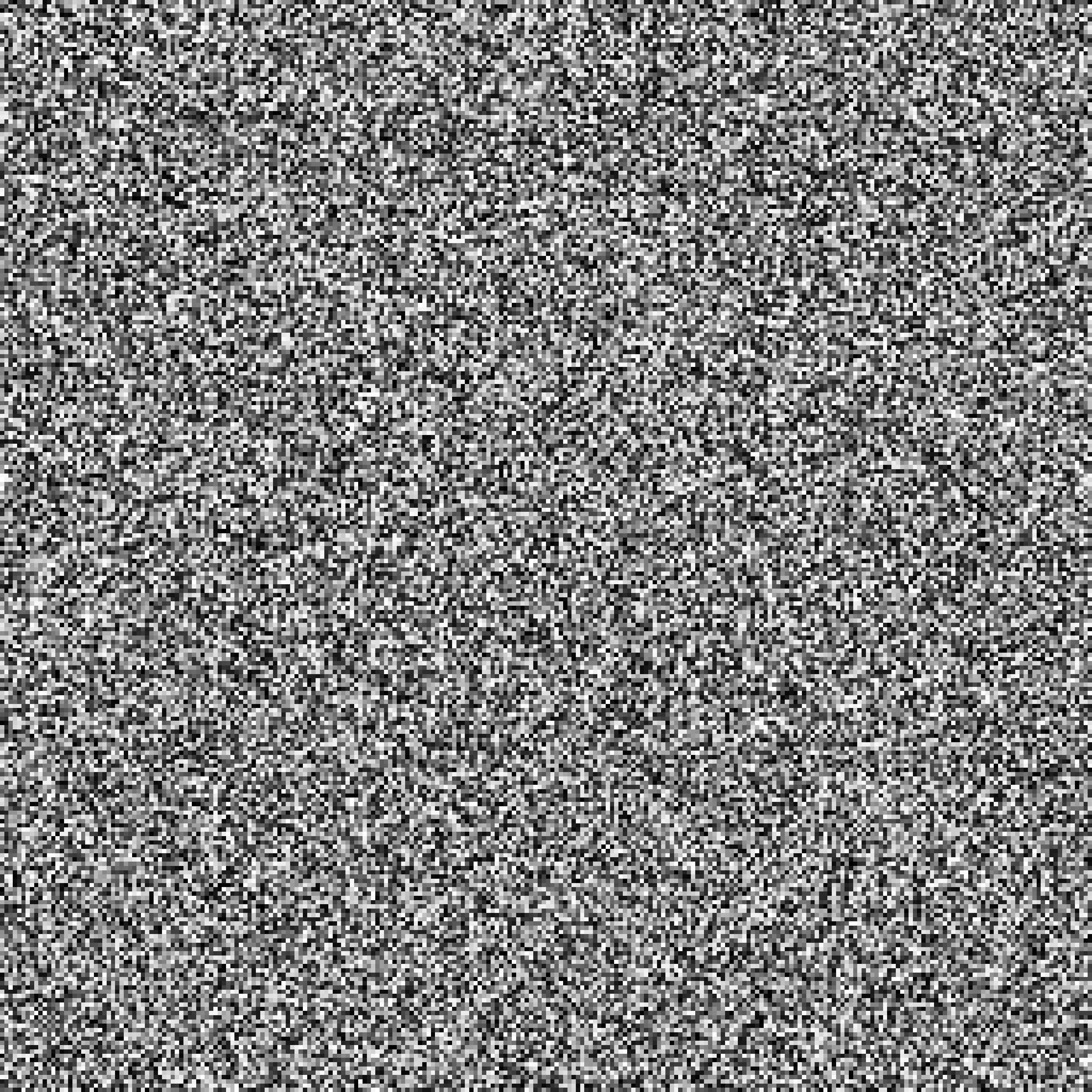}
    \caption{Not a cat.} 
    \label{NC_fig:cat}
\end{figure}

Like for sampling, the key difficulty in generative modeling is the high dimension $d$ of the sample space. A typical example is image generation, where $d\approx 200,000$ (for $256\times 256$-pixel color images). 
Learning high-dimensional probability distributions is difficult, since the training samples only cover a vanishing fraction of high-dimensional space (the ``curse of dimensionality''). Vice versa, a generative model needs to target this vanishing fraction to succeed. Put flippantly, randomly choosing pixel values has vanishing probability of producing an image of a cat (Fig.~\ref{NC_fig:cat}). In addition, complex probability distributions are typically \emph{multi-modal}.
As for sampling, modern designed models generate samples from a high-dimensional target distribution $p_1$ (e.g., the space of images) by dynamically transforming a simple base distribution $p_0$ (e.g., the standard Gaussian). 
We already encountered such transport problems in Chapter~\ref{sec:OT} on optimal transport. Like optimal transport, generative models are recipes for transforming probability distributions, now designed to make it easy to learn the transformation from training data.
\emph{Diffusion models}~\cite{Sohl-Dickstein.etal2015,Ho.etal2020} are perhaps the most well-known machine learning methods of this type.
They learn a stochastic process that generates samples from a target distribution defined by a set of training samples.
Diffusion models were originally inspired by the physical ideas of non-equilibrium thermodynamics and simulated annealing discussed in Chapter~\ref{sec:sampling}~\cite{Sohl-Dickstein.etal2015}.
This section approaches generative modeling using the (effectively equivalent) frameworks of \emph{stochastic interpolation}~\cite{Albergo.etal2025} and \emph{flow matching}~\cite{Lipman.etal2024}. We believe that this choice will make the core idea behind diffusion models easier to understand and generalize.

\subsubsection{Variational transform sampling}\label{sec:variational_transform}

Fundamentally, this section is based on the method of transform sampling, which applies a map $T$ to transform a sample $x \sim p_0$ into $y = T(x)$. 
Throughout this section, we use ``$y$'' to denote samples generated by a transform.
The map $T$ can also be thought of as a ``decoder'', and $x$ as a ``hidden variable''~\cite{Kingma.Welling2019}.
The probability distribution of $y$ is called the \emph{pushforward} $T_* p_0$.
By changing variables, the probability density of $T_* p_0$ is:

\begin{align}\label{NC_eq:change_of_variables}
    (T_* p_0)(y)  = p_0(T^{-1}(y)) \cdot |\det \nabla_y T^{-1}|
\end{align}
For example, the linear map $y = M \cdot x$ transforms a standard Gaussian into one with covariance matrix $\Sigma=MM^{\mathsf T}$. 
A second example is provided by one-dimensional distributions, where  the map $T$ is defined by the cumulative distribution functions of $p_0, \; p_1$ (Eq.~\eqref{eq:CDF_OT}).

The algorithms in this section combine transform sampling with \emph{variational inference} (VI, Sec.~\ref{sec:variational_inference}). VI approximates a target distribution $p_1$ by a member $q$ of a parametrized family. Here, the trial family is defined by a set of transformations $\widehat T$ via Eq.~\eqref{NC_eq:change_of_variables}. 
VI is an intuitive match for modern ML techniques like expressive neural networks and large-scale, gradient-based optimization.
To bring these techniques to bear, one needs a \emph{loss function} $\mathcal{L}$ to measure the difference between 
the estimated $q$ and the target distribution $p_1$. 
Here, our goal is to build loss functions that can be efficiently approximated using training samples, without explicit knowledge of $p_1$. 
Then, the transformation $\widehat T$ can be learned directly from data by minimizing $\mathcal{L}$.

\subsubsection{Flow matching}\label{sec:flow_matching}

The key idea of flow matching is to construct the transformation $\widehat T$ step-by-step from a velocity field $\widehat v$, parametrized by a neural network~\cite{Lipman2023flow}.
Flow matching generates target samples $y_1$ by integrating the ODE $\dot{y}_t=\widehat{v}_t(y_t)$ starting from a random initial condition $y_0 = x_0\sim p_0$ (say, an image of white noise). Flow matching defines a suitable \emph{loss function} that allows learning the velocity $\widehat{v}$ from training samples.
In contrast to diffusion models, the sample generation process is deterministic; Sec.~\ref{sec:flow_to_diffusion} will introduce stochasticity.

Both flow matching and diffusion models are based on \emph{mass transport}. Consider a time-dependent\footnote{
Note that in practice, explicit time-dependence of the velocity field is often unnecessary, and can even be detrimental to performance~\cite{Kadkhodaie.etal2026}.
In high dimensions, source $p_0$ and target $p_1$ do not overlap, so the velocity can change along trajectories without being explicitly time-dependent.} 
vector field $\widehat v_t(x)$ on $\mathbb{R}^d$.
Here and below, hats denote learned maps or fields. The marginal of the generated trajectories $y_t$ is $q_t$.
By convention, time $t$ ranges from $0$ to $1$, so that $q_{t=0} =p_0$ is the base, and we want to adjust the generative process so that $q_1$ matches the target distribution $p_{1}$ (reader beware: the reverse convention is also often used in the literature).
The \emph{flow map} $\widehat T_t(x)$ transports particles along the integral curves of $\widehat v_t$, and is defined by solutions to the ODE

\begin{align}
\begin{aligned}
 \dot y_t&=\widehat v_t(y_t),\quad y_{t=0}=x_0\sim p_0, \quad 
 \widehat T_t(x_0):=y_t.
\end{aligned}
\label{NC_eq:flow_sampling}
\end{align}
If the initial conditions are sampled $y_{t=0}\sim p_0$, the time-dependent distribution $q_t=(\widehat T_t)_*p_0$ is governed by conservation of mass~\cite{Villani2009}:

\begin{align}
    \label{NC_eq:mass_conservation}
    \partial_t q_t(x) + \nabla \cdot \left(q_t(x) \cdot \widehat v_t (x)\right) = 0
\end{align}
If a $q_t$ fulfills Eq.~\eqref{NC_eq:mass_conservation}, it is said to be generated by the flow $\widehat v$.
In fluid dynamics terms, Eq.~\eqref{NC_eq:flow_sampling} is the Lagrangian, and Eq.~\eqref{NC_eq:mass_conservation} the Eulerian formulation.

Note that many different velocity fields can generate the same $q_t$. Indeed, adding a \emph{divergence-free} vector field, $\widehat v_t\mapsto\widehat v_t+v_t'$, with $\nabla\cdot(q_t v_t')=0$, leaves Eq.~\eqref{NC_eq:mass_conservation} invariant. 
In Sec.~\ref{sec:eot}, we saw how optimal transport singles out one of these vector fields by minimizing a transport cost. 
On the other hand, not all time-dependent distributions can be generated by a flow. For example, direct interpolation between two probability densities, $p_t(x)=(1-t)p_0(x)+tp_1(x)$ with disjoint support requires ``mass teleportation'' and cannot be generated by any flow.

\begin{figure*}[t]
    \centering
    \includegraphics[width=0.8\linewidth,trim=7.56bp 39.36bp 8.04bp 59.16bp,clip]{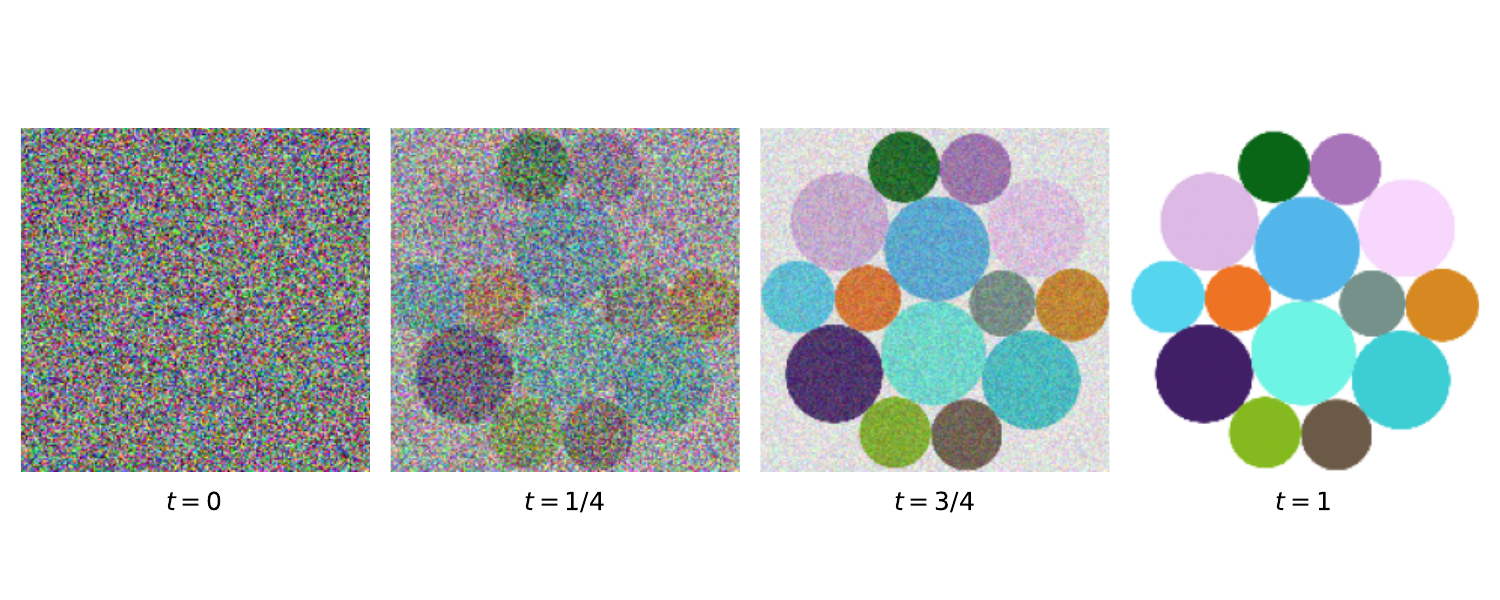}
    \caption{A stochastic interpolant adds progressively larger amounts of uncorrelated Gaussian noise to an initial image.
    }
    \label{NC_fig:noising}
\end{figure*}

Flow-generated distributions are very convenient: given $\widehat v$, one can generate samples of $q_1$ by integrating Eq.~\eqref{NC_eq:flow_sampling}.
How can one identify a velocity field $\widehat v$ so that $q_1 = (\widehat T_1)_*p_0$ matches the target distribution $p_1$?
A first idea is to directly optimize the transport map $\widehat T_1$, obtained by integrating Eq.~\eqref{NC_eq:flow_sampling}.
This approach is, however, very inefficient: evaluating the objective requires \emph{simulating} the entire trajectory.
Instead, one uses a \emph{stochastic interpolant}~\cite{Albergo.etal2025,Lipman.etal2024}, a stochastic process $x_t$ with $x_0\sim p_0$ and $x_1\sim p_1$.
Recall that in generative modeling, the target distribution $p_1$ is unknown. Instead, one has access to a set of training samples (for example, an image database). One constructs a stochastic interpolant from these training samples, for example:

\begin{align}\label{NC_eq:stochastic_interpolant}
    x_t  = (1-t)x_0 + t x_1, \quad x_0 \sim p_0, \; x_1 \sim p_1.
\end{align}
The endpoints $x_0, x_1$ are sampled independently\footnote{One can also sample $x_0, x_1$ dependently, for instance, to predict the next frame of a movie from the preceding one}.
For instance, $x_1$ could be an image sampled from the database, and $x_0$ could be Gaussian noise, so that $x_t$ is a sequence of increasingly noisy images (Fig.~\ref{NC_fig:noising}). In this sense, Eq.~\eqref{NC_eq:stochastic_interpolant} can be thought of as a version of the Brownian bridge from Sec.~\ref{sec:brownian_bridge}, conditioned to terminate in the target sample $x_1$.
The interpolant Eq.~\eqref{NC_eq:stochastic_interpolant} defines a time-dependent \emph{probability path} $x_t \sim p_t$ from $p_0$ to $p_1$.
The $p_t$ play a role similar to the instantaneous equilibria in annealed importance sampling (Sec.~\ref{sec:AISjarz}).
It is important to distinguish the interpolant $x_t$ from the generating process $y_t$; even if initialized at the same starting point $x_0\sim p_0$, the two processes lead to different trajectories.

\begin{figure}[t]
    \centering
    \includegraphics[width=0.49\linewidth]{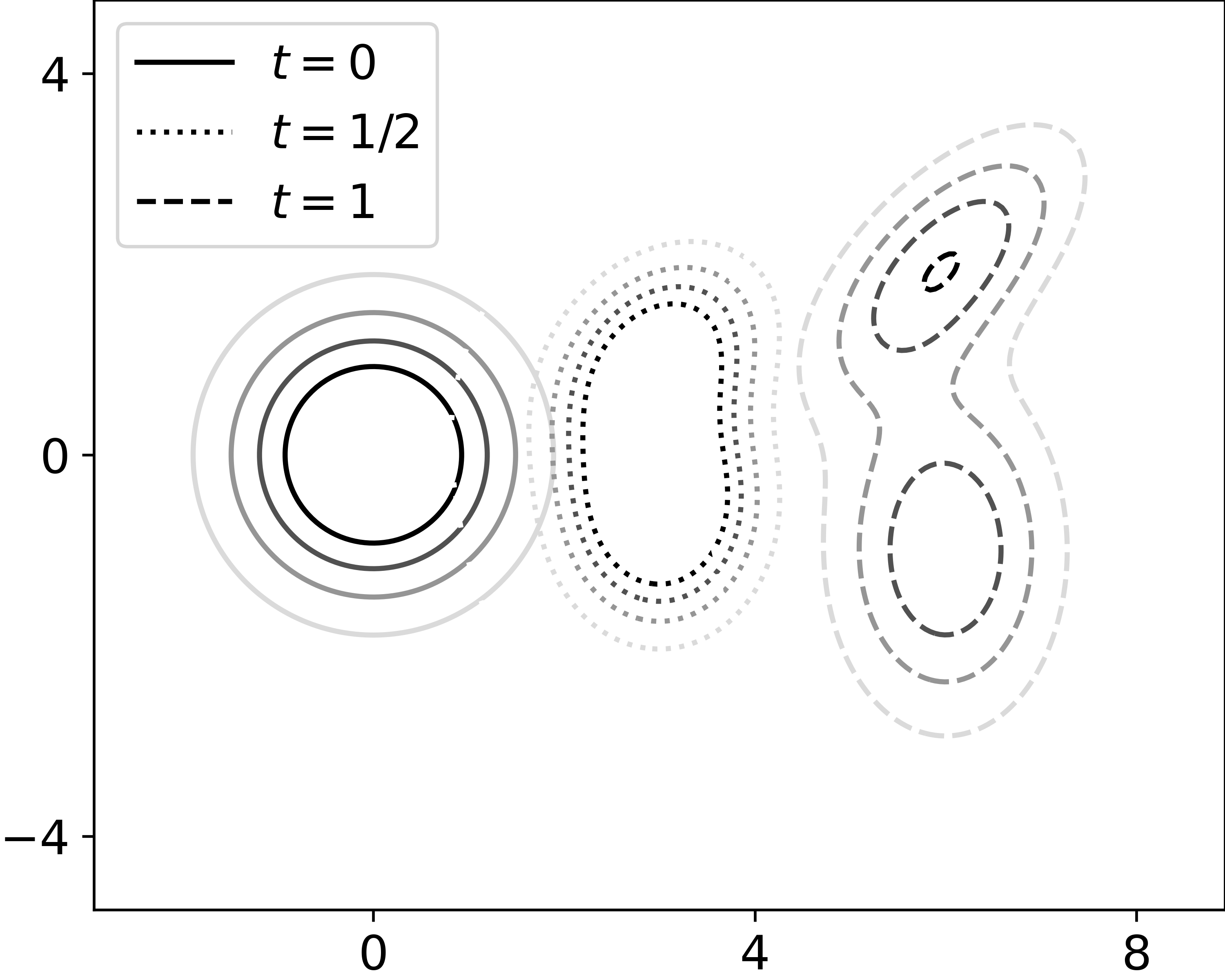}
    \includegraphics[width=0.49\linewidth]{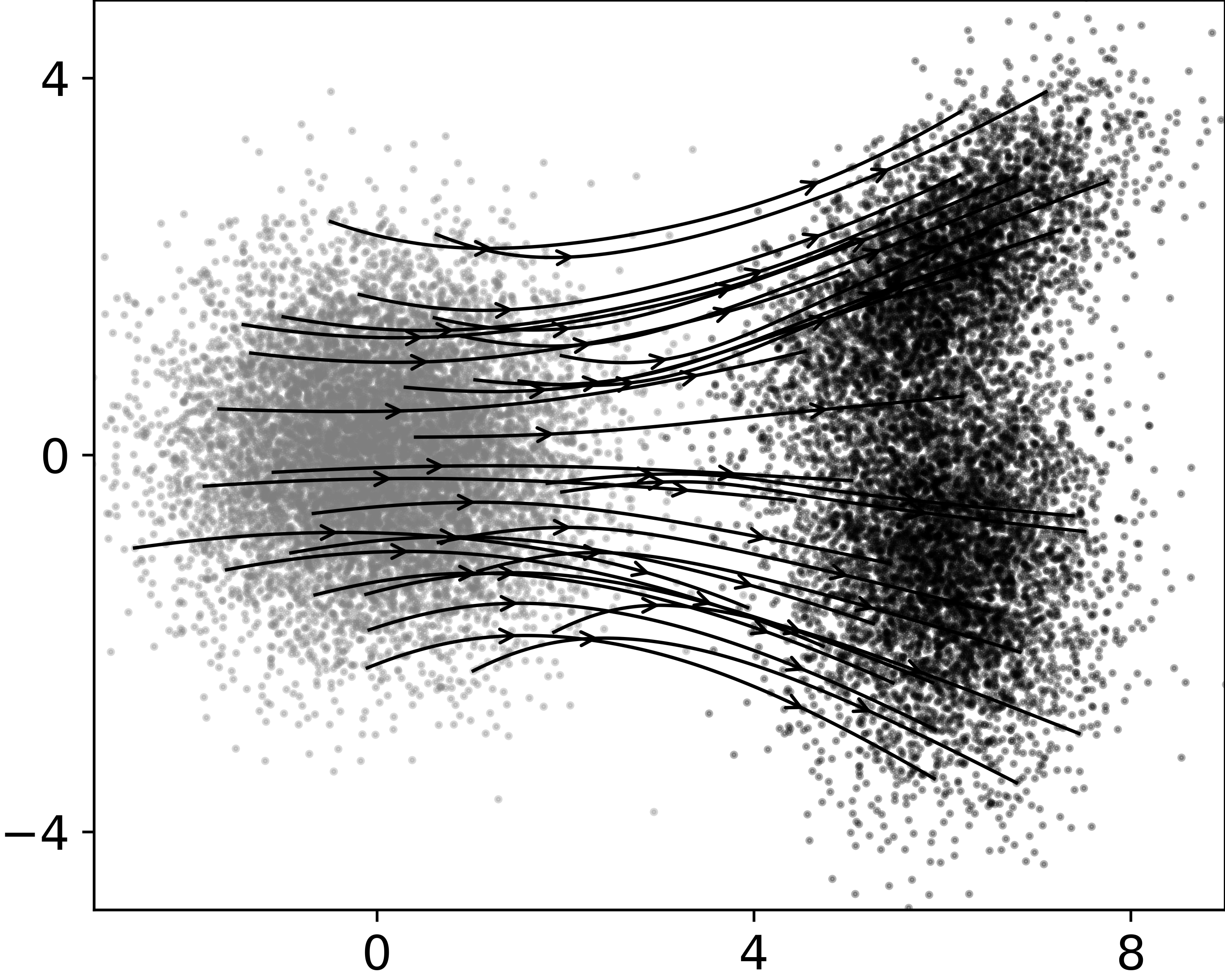}
    \caption{(Left) Stochastic interpolation using Eq.~\eqref{NC_eq:stochastic_interpolant} between $p_0 =\mathcal{N}(0,I)$, a standard Gaussian, and $p_1= \sum_j c_j \,\mathcal{N}(\mu_j,\Sigma_j)$, a Gaussian mixture with weights $c_j$ (in the 2d example shown, $c_1=0.4, c_2=0.6$).
    Since $x_t =(1-t)x_0 +  tx_1$ is a sum of Gaussian mixture variables, the intermediate distributions are also Gaussian mixtures, $p_t = \sum_j c_j \, \mathcal{N}\left(t\mu_j, (1-t)^2I+t^2\Sigma_j\right)$.
    The plot shows contour lines of $p_t$ for $t=0, 1/2, 1$. 
    (Right) The flow map can be calculated from the intermediate $p_t$'s using Eq.~\eqref{NC_eq:score_flow}. It transports samples from $p_0$ to samples from $p_1$ in unit time (black). Note that the integral curves are bent while the trajectories of the interpolant are straight lines.
    }
    \label{NC_fig:stochastic_interpolation}
\end{figure}

Crucially, the probability path $p_t$ defined by Eq.~\eqref{NC_eq:stochastic_interpolant} can be generated by a flow field (in the sense of mass conservation,  Eq.~\eqref{NC_eq:mass_conservation}). 
As we show shortly, the velocity $v$ of the interpolant $p_t$ is the average of the trajectories' speed:

\begin{align}
    \label{NC_eq:true_velocity}
    v_t(x) = \left\langle \dot{x}_t |x_t=x\right\rangle
\end{align}
Here and below, $\langle \cdot | \cdot \rangle$ denotes the conditional expectation. The average is over the randomly sampled endpoints $x_0, \, x_1$, which determine $x_t, \; \dot{x}_t$.
To derive Eq.~\eqref{NC_eq:true_velocity}, integrate $\partial_t p_t$ against a test function $\varphi(x)$, and use the law of total expectation:

\begin{align}
\begin{aligned}
 &\int\varphi(x)\partial_t p_t(x)d^dx
 =\frac{d}{dt}\left\langle\varphi(x_t)\right\rangle \\
 &=\left\langle\nabla\varphi\cdot\dot{x_t}\right\rangle
 =\left\langle\nabla\varphi\cdot\left\langle\dot{x_t}|x_t\right\rangle\right\rangle\\
 &=-\int\varphi(x)\nabla\cdot(p_t\,\left\langle\dot{x}_t|x_t=x\right\rangle)d^dx
\end{aligned}
\end{align}
Fig.~\ref{NC_fig:stochastic_interpolation} shows a toy example ($p_1$ is a mixture of Gaussians), where the true flow field $v$ can be computed analytically.

The stochastic interpolant defines the objective: match the estimated flow field $\widehat v$ to the true one $v$.
The resulting \emph{flow matching loss} reads, using the law of total expectation:

\begin{align}
\begin{aligned}
\label{NC_eq:flow_matching_loss}
 \mathcal L_\mathrm{FM}[\widehat v]
 &=\int_0^1\left\langle\left\|\widehat v_t(x_t)-v_t(x_t)\right\|^2\right\rangle dt\\
 &=\int_0^1\left\langle\left\|\widehat v_t(x_t)-\left\langle\dot x_t|x_t\right\rangle\right\|^2\right\rangle dt \\
 &=\int_0^1dt\;\left\langle\|\widehat v_t(x_t)\|^2-2\widehat v_t(x_t)\cdot\dot{x}_t\right\rangle\\
 &\quad+\mathrm{const.}   = \langle\|\widehat v_t(x_t)-\dot x_t\|^2\rangle+\mathrm{const.}
 \end{aligned}
\end{align}
The additive constant in Eq.~\eqref{NC_eq:flow_matching_loss} is independent of $\widehat v$, and thus does not affect training. Eq.~\eqref{NC_eq:flow_matching_loss} has three important properties. First, the unique minimizer is the true flow field. Thus, at the minimum,  $q_t$ matches the interpolant distribution $p_t$, and $y_{t=1}$ is a sample from $p_1$ (note that $y_t$ and $x_t$ have the same marginals, but the trajectory laws can differ).
Second, since $x_t=(1-t)x_0+tx_1$, the expectation value is easily estimated using independent samples $x_0\sim p_0, \; x_1\sim p_1$ from the training set. Third, the loss is \emph{simulation-free}: it does not require simulating the entire trajectory $y_t$.
Thus, one can efficiently train the neural network $\widehat v$ to generate target samples by stochastic gradient descent on $\mathcal{L}_\mathrm{FM}$. In summary, the overall flow-matching workflow reads:
\begin{enumerate}
    \itemsep-.3em 
    \item Design a stochastic interpolant $x_t$ whose distribution $p_t$ bridges the simple base distribution $p_0$ and the target $p_1$.
    \item Using interpolant samples generated from a training set, train a velocity field $\widehat v$ by minimizing $\|\widehat v_t(x_t)-\dot x_t\|^2$.
    \item Generate samples by integrating the ODE $\dot{y}_t=\widehat v_t(y_t)$, with $y_{t=0}=x_0\sim p_0$.
\end{enumerate}

One can see that flow matching bears a close resemblance to dynamic optimal transport (OT). Both aim to transform a base into a target distribution via ``local'', mass-conserving dynamics, Eq.~\eqref{NC_eq:mass_conservation}.
Like in optimal transport, in Eq.~\eqref{NC_eq:stochastic_interpolant}, samples move along straight lines between endpoints. However, the choice of the endpoints is precisely the opposite: OT couples the endpoints deterministically via the transport map $T$, $x_1 = T(x_0)$, while in Eq.~\eqref{NC_eq:stochastic_interpolant}, the endpoints are independent.
Therefore, OT and stochastic interpolation are not identical.
Sampling from the OT process is computationally hard, since it requires finding a very specific flow field that additionally minimizes a transport cost.
In flow matching, the stochastic interpolant is instead designed to be easy to sample. These samples serve as a device to find a flow field $\widehat v$ that takes $p_0$ to $p_1$.

Flow matching can be viewed not only as data-driven transport, but also as an iterative denoising process. This perspective sheds light on how flow matching is able to learn \emph{multi-modal} distributions. Multi-modality was already a challenge in the context of sampling from a known distribution (Chapter~\ref{sec:sampling}).
Since $x_0$ is typically Gaussian white noise, the stochastic interpolant is a sequence of increasingly noisy images (Fig.~\ref{NC_fig:noising}).
One can think of the velocity field as a \emph{denoiser}, which estimates $x_{t+dt}$ from the (infinitesimally) more noisy version $x_t$ as $\left\langle x_{t+dt}|x_t\right\rangle \approx  x_t + v_t(x_t)dt$.
The flow-matching loss Eq.~\eqref{NC_eq:flow_matching_loss} measures the squared error of the denoiser.
Intuitively, flow matching thus breaks down a difficult sampling task (generating an image) into a sequence of simpler tasks (incrementally denoising an image). 

This step-wise approach is essential to handle multi-modality.
Indeed, suppose one were to add a large, finite amount of noise to a target sample and attempt to denoise it in one step. The denoiser $D_t$ at interpolation time $t$ has mean squared error $\langle\|D_t(x_t)-x_1\|^2\rangle$ (smaller $t$ corresponds to more noise).
Because the noise removes details, several different reconstructions are compatible with the noisy sample.
The optimal denoiser (with minimum least-squares error) will return the average of all reconstructions, $D_t^*(x)=\langle x_1| x_t=x\rangle$.
If the distribution is multi-modal, this average is a blurred mix of different modes that does not resemble any sample of the target distribution.
However, for an \emph{infinitesimal} amount of noise, the reconstruction problem becomes uni-modal.
This is the principle that flow matching exploits to learn multi-modal distributions.

\subsubsection{The score function}

\label{sec:score_function}

Let us take a closer look at the distributions $p_t$ for a linear interpolant between a standard Gaussian and an arbitrary target. If $x_0$ is standard Gaussian and independent of $x_1$, the conditional density is also Gaussian, $p_t(\cdot| x_1)=\mathcal N(tx_1,(1-t)^2I)$. Taking the gradient, 
one obtains $\nabla\log p_t(x_t| x_1)=-x_0/(1-t)$.  Averaging over endpoints using the law of total expectation gives~\cite{Vincent2011}: 

\begin{align}
\begin{aligned}
 \nabla\log p_t(x)
 &=\left\langle\nabla\log p_t(x| x_1)| x_t=x\right\rangle\\
 &=-\frac{\langle x_0| x_t=x\rangle}{1-t}.
\end{aligned}
 \label{NC_eq:score_vs_denoiser}
\end{align}
The quantity $\nabla\log p_t$ is called the \emph{score}. 
We already encountered the score in Sec.~\ref{sec:feynman-kac-formula-doobs-h-transform} in the context of optimal control, where it served as a control force that drives a particle towards a target. Here, the score plays a similar role. Eq.~\eqref{NC_eq:score_vs_denoiser} shows that the score can separate the noise $x_0$ from the current $x_t$, and thus allows reconstructing the target $x_1$. After a little algebra:

\begin{align}
\begin{aligned}
 v_t(x)&=\langle\dot x_t| x_t=x\rangle
       =\frac{x-\langle x_0| x_t=x\rangle}{t}\\
       &=\frac{x+(1-t)\nabla\log p_t(x)}{t},\qquad 0<t<1.
\end{aligned}
\label{NC_eq:score_flow}
\end{align}
Flow matching with an initially Gaussian distribution is therefore closely related to \emph{score matching}~\cite{Hyvarinen2005}, a method that uses the score to fit statistical models. Flow and diffusion models can be viewed as a combination of simulated annealing (Sec.~\ref{sec:AISjarz}) and score matching~\cite{SongErmon2019}.
Eq.~\eqref{NC_eq:score_flow} demonstrates that the velocity field is a potential gradient: $\log p_t$ acts as a time-dependent energy landscape that guides a sample from the initial to the target distribution (see the example in Fig.~\ref{NC_fig:stochastic_interpolation}). 
Note that the relation between score and flow only holds for linear interpolation where one endpoint is Gaussian; in general, there is no simple relation between $p_t$ and $v$. 

\subsubsection{From flow matching to diffusion}\label{sec:flow_to_diffusion}

In flow matching, sampling is a deterministic process, defined by the ODE Eq.~\eqref{NC_eq:flow_sampling}.
Diffusion models instead use a \emph{stochastic sampling} process, which, however, has the same marginals.
To see how, add $\nabla \log p_t$ to and subtract it from the mass conservation equation for the interpolant (target) density $p_t$:

\begin{align}
\begin{aligned}
 \partial_t p_t&=-\nabla\cdot\Bigl(\bigl(v_t+\tfrac\varepsilon2\nabla\log q_t -\tfrac\varepsilon2\nabla\log p_t\bigr)q_t\Bigr)\\
 &=-\nabla\cdot\left((\widehat v_t+\tfrac\varepsilon2\nabla\log p_t)p_t\right)+\tfrac\varepsilon2\nabla^2p_t
\end{aligned}
\label{NC_eq:Fokker-Planck}
\end{align}
Eq.~\eqref{NC_eq:mass_conservation} is thus reformulated as a \emph{Fokker-Planck} equation with diffusion, compensated by an additional drift. This seemingly tautological manipulation allows defining a new sampling process. The Fokker-Planck Eq.~\eqref{NC_eq:Fokker-Planck} gives a Langevin equation for trajectories:

\begin{align}
\begin{aligned}
 \dot y_t&=v_t(y_t)+\frac\varepsilon2\nabla\log p_t(y_t)+\sqrt\varepsilon\,\eta_t\\
 &=\frac1t\Bigl(y_t+\left(1+\frac{\varepsilon-2}{2}t\right)\nabla\log p_t(y_t)\Bigr) + \sqrt\varepsilon\,\eta_t.
\end{aligned}
\label{NC_eq:stochastic_sampling}
\end{align}
In the second step, we used the relation between score and velocity.
Because the density obeys Eq.~\eqref{NC_eq:Fokker-Planck}, the stochastic process Eq.~\eqref{NC_eq:stochastic_sampling} still generates samples from $p_1$. 
Effectively, diffusion models add an additional stochastic term to the interpolant Eq.~\eqref{NC_eq:stochastic_interpolant}, such that there is stochasticity at intermediate $t$ even after fixing the endpoints.
Stochastic sampling has practical advantages: it can improve stability so that even an imperfectly learned score produces reasonable samples~\cite{Albergo.etal2025}.

With the above results, flow matching can be compared to the original formulation of denoising diffusion models~\cite{Song.etal2021}.
Here, a forward diffusion process gradually adds noise to a target sample $x_1$. This forward process is the pendant of the stochastic interpolant and generates samples from which the score function is learned.
Given the score function, target samples are generated by Eq.~\eqref{NC_eq:stochastic_sampling}. 
In this context, Eq.~\eqref{NC_eq:stochastic_sampling} is called the reverse process, since it is the time-reversal of the forward diffusion process. 

To see this, reparametrize time as $s=-\log t$, so $s=0$ is the target distribution and $s=\infty$ is standard Gaussian noise. The forward process is  initialized at the target distribution:

\begin{align}
    \label{NC_eq:forward}
   \dot{x}_s = - x_s +  \sqrt{2} \,\eta_s , \quad  x_{s=0}\sim p_1
\end{align}
Since we set $\varepsilon=2$, as $s\rightarrow\infty$, the distribution $p_s(x)$ converges to a standard Gaussian (the base distribution). 
(Note that the intermediate distributions $p_s$ are not equivalent to those of the stochastic interpolant Eq.~\eqref{NC_eq:stochastic_interpolant}.)
The forward process obeys the Fokker-Planck equation:

\begin{align}
\begin{aligned}
 \partial_s p_s(x)&=\nabla\cdot(xp_s)+\tfrac12\nabla^2p_s\\
 &=\nabla\cdot(p_s(x+\tfrac12\nabla\log p_s))
\end{aligned}
\end{align}
To reverse the probability flow, one flips the sign of the effective flow field $(x + \tfrac{\varepsilon}{2}\nabla \log p_s) \mapsto - (x + \tfrac{\varepsilon}{2}\nabla \log p_s)$. One thus reverses time, starting from a large final time $S\gg 1$.
The reverse process, which generates target samples starting from noise, reads~\cite{Anderson1982,Song.etal2021}

\begin{align}
    \label{NC_eq:reverse_diffusion}
   \dot y_s = y_s+ \tfrac{\varepsilon}{2} \nabla\log p_{S-s}(y_s) +  \sqrt{\varepsilon}\, \eta_s
\end{align}
This idea of time-reversing a process that starts at the target distribution $p_1$ is precisely what underlies the proof of annealed importance sampling and the Jarzynski identity in Sec.~\ref{sec:AIS_proof}.

Finally, Eq.~\eqref{NC_eq:stochastic_sampling} also allows revisiting the distinction between flow matching and Langevin sampling with energy $E= -\log p_1$. Flow matching uses a non-equilibrium process to transform probability distributions in finite time, and the target $p_1$ is not a stationary state. The potential is time-dependent, and the intermediate-time distributions $p_t$ are explicitly distinct from the target and instead act as a bridge between $p_0$ and $p_1$. By contrast, Langevin dynamics only converge to its equilibrium distribution $p_1$ as $t\rightarrow\infty$.

\subsubsection{One-step generation and normalizing flows}\label{sec:normalizing_flows}

Once a flow-matching model is trained, generating samples requires integrating the flow field $\widehat v$ (or an SDE for a diffusion model). This can be slow and makes the generation process difficult to fine-tune for downstream applications. Consistency models~\cite{Song.etal2023} and rectifying flows~\cite{Liu.etal2022} distill the flow field $\widehat v$ into a neural-network model for the flow map $\widehat T_1$, which allows generating samples in one step. Such one-step models parallel the transition from Benamou-Brenier dynamic optimal transport (Sec.~\ref{sec:eot}, Eq.~\eqref{eq:bb}) to the static Monge-Kantorovich problem (Sec.~\ref{sec:eot}). 

In this section, we will discuss another one-step generative method, \emph{normalizing flows}~\cite{Rezende.Mohamed2015}, one of the inspirations for  flow matching models. Despite their name, normalizing flows are not flows, but generate samples using a single-step transform.
While flow matching excels at learning generative models from training data, 
normalizing flows can also learn from an energy function or unnormalized probability density. This makes them very useful for improving sampling algorithms in statistical mechanics simulations and Bayesian statistics (Chapter~\ref{sec:sampling}). In this setting, no training samples are available: producing them is the whole point.

As before, the goal is to transform a base $p_0$ into a target $p_1$ distribution via Eq.~\eqref{NC_eq:change_of_variables}.
Samples are generated as $y = \widehat T(x_0), \; x_0 \sim p_0$ where the base distribution $p_0$ is often Gaussian (hence, ``normalizing'' flow).
In contrast to flow matching, which constructs $\widehat T$ iteratively, normalizing flows directly parametrize $\widehat T$ by a neural network.
However, not any neural network will do: Eq.~\eqref{NC_eq:change_of_variables} shows that the inverse $\widehat T^{-1}$ and the Jacobian determinant $\det \nabla_x\widehat T^{-1}$ are needed to compute the transformed density $q = \widehat T_* p_0$.
One thus uses neural network architectures designed specifically to make these quantities easy to evaluate. A minimal example is the following neural network layer for a 2-component vector %
$x=\begin{pmatrix} x^{(1)} & x^{(2)} \end{pmatrix}$~\cite{Dinh2017realnvp}
:

\begin{align}
\begin{gathered}
 \begin{pmatrix}x^{(1)}&x^{(2)}\end{pmatrix}\mapsto\\
 \begin{pmatrix}\widehat a(x^{(2)})\cdot x^{(1)}+\widehat b(x^{(2)})&x^{(2)}\end{pmatrix}
\end{gathered}
\label{NC_eq:normalizing_layer}
\end{align}
where $\widehat a$ and $\widehat b$ are neural networks for scale and shift. Since Eq.~\eqref{NC_eq:normalizing_layer} is an affine function in $x^{(1)}$ and copies $x^{(2)}$, it is easy to invert.
More complex networks can be built by stacking layers.

Next, we need a procedure to adjust $q=\widehat T_*p_0.
$ to match the target distribution $p_1$, paralleling the minimization of the flow matching loss in Eq.~\eqref{NC_eq:flow_matching_loss}.  
One possibility is to use a training dataset $\{x_n\}$ of samples from $p_1$. In this case, one maximizes the log-likelihood $\sum_{n=1}^N\log q(x_n)$ of the training data.
In the setting of Bayesian inference or statistical mechanics simulations, one has access to the unnormalized density $\widetilde p_1$ instead of training samples.
In Sec.~\ref{sec:variational_inference}, we saw that variational inference uses $p_1$ to define a \emph{data-free} objective:

\begin{figure}[t]
    \centering
    \includegraphics[width=0.8\linewidth,trim=6.60bp 28.56bp 6.84bp 62.28bp,clip]{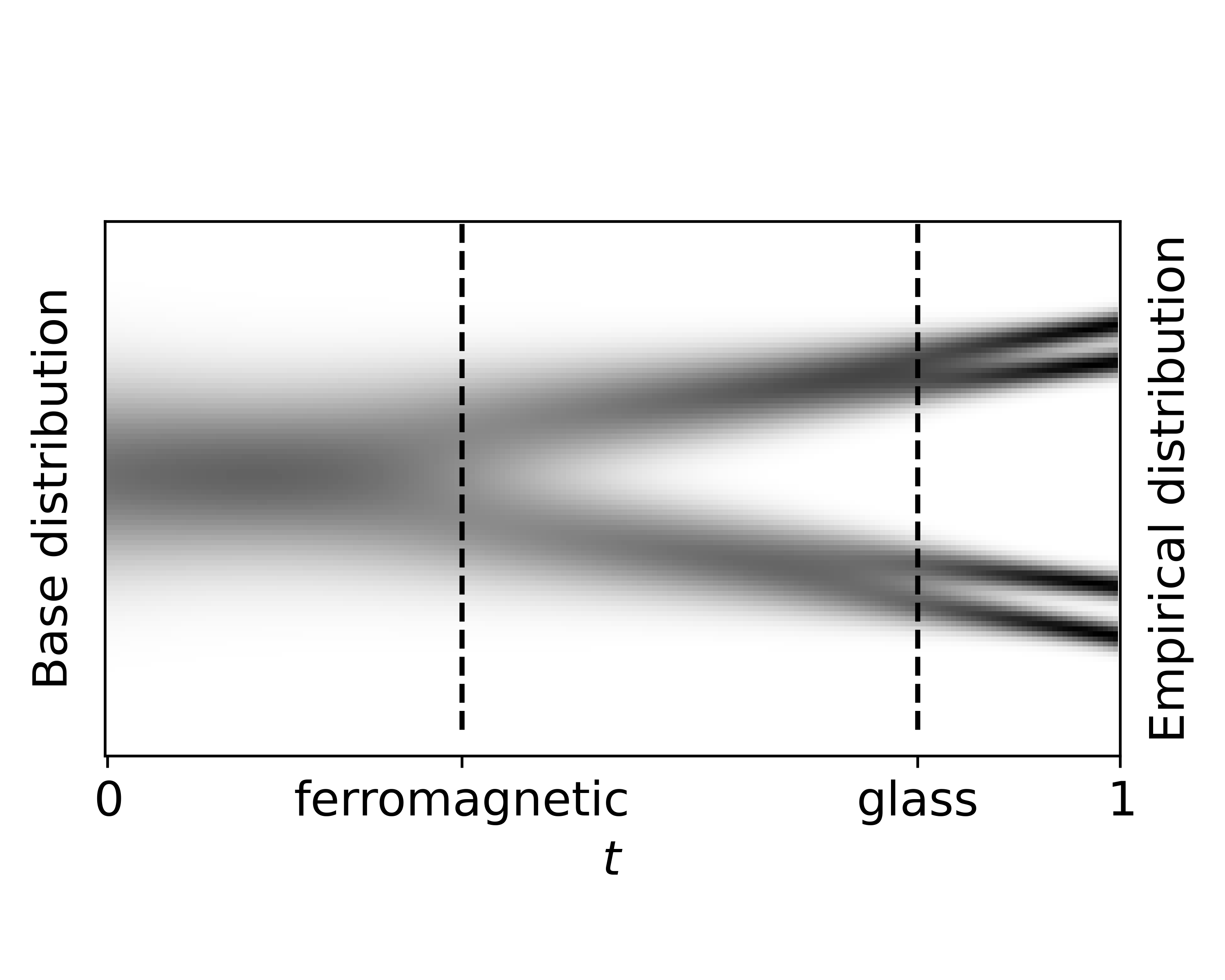}
    \caption{
    Memorization-generalization transition for a 1D mixture of Gaussians: interpolating distributions $p_t$ between a Gaussian base and the empirical distribution of a Gaussian mixture ($N=4$ samples). 
    At the ferromagnetic transition, $p_t$ splits into the distinct modes of the target distribution. At the glass transition, trajectories become trapped near a training sample from the empirical distribution.
    }
    \label{NC_fig:memorization}
\end{figure}

\begin{align}
\begin{aligned}
 \dkl(q\Vert p_1)&=\langle\log q-\log p_1\rangle_q\\
 &\approx\frac1N\sum_{n=1}^N\left[\log q(x^{(n)})-\log p_1(x^{(n)})\right],\\
 x^{(n)}&=\widehat T(x_0^{(n)}),\qquad x_0^{(n)}\sim p_0\end{aligned}
\label{NC_eq:data_free}
\end{align}
Here, $D_\mathrm{KL}$ is the Kullback-Leibler (KL) divergence, a measure of the distance between two probability distributions encountered several times in previous chapters. Because of the logarithm, the normalization constant $Z_1$ drops out when optimizing Eq.~\eqref{NC_eq:data_free}.
The expectation value in Eq.~\eqref{NC_eq:data_free} is approximated using samples from $q$, so no target distribution samples are required (hence, ``data-free''). Now, $\widehat T$ can be trained by minimizing $ \dkl(q\Vert p_1)$ via gradient descent~\cite{NoeScience2019}.
The data-free objective has certain drawbacks: it can encourage \emph{mode collapse}, where the map $\widehat T$ always returns samples from a single mode and fails to explore the full distribution $p_1$~\cite{Gabrie.etal2022}. This is a well-known effect of the KL divergence. It strongly penalizes any probability density outside of the support of $p_1$ due to the term $\log(q/p_1)$, but failing to sample from a part of $p_1$ receives a much milder penalty.

Once trained, one can generate samples $\widehat T(x_0)$. However, if the variational approximation is not perfect, these samples will be biased (not exactly distributed $\sim p_1$). This is a serious issue in scientific contexts.
Fortunately, one can combine a normalizing flow with another sampling technique like importance sampling from Chapter~\ref{sec:sampling} to correct the bias~\cite{Gabrie.etal2022}. As long as $q$ covers the support of $p_1$, the result is guaranteed to be unbiased.
This achieves the ``best of both worlds'': the conventional sampling algorithm provides mathematical guarantees, and the power of machine learning allows rapid identification of high-probability regions in $d$-dimensional space.
Normalizing flows have been successfully used to accelerate statistical physics simulations, for instance in lattice QCD~\cite{Kanwar.etal2020}.
More generally, gradient-based optimization can systematically tune parameters of sampling algorithms that were traditionally tuned by hand, like the proposal distribution for an importance sampler.

\subsubsection{Generalizations and ongoing work}\label{sec:diffusion_outlook}

In summary, machine learning has made striking progress on the classic problem of sampling from complex, high-dimensional distributions. These techniques transform a simple base distribution to the target via a transform that is fit variationally to data. The greatest success has come from methods based on mass transport, like diffusion and flow matching, which construct the transform map step by step. In flow matching, one first designs a dynamical process that interpolates between the base (often, uniform noise) and the target, and then learns the corresponding probability flow. 
In the domain of image and video generation, flows and diffusions are the state of the art and have eclipsed previous approaches, like Variational Autoencoders~\cite{Kingma.Welling2019} and Generative Adversarial Networks~\cite{Goodfellow.etal2014}. 
On the other hand, for language and other ``sequential datasets'', \emph{autoregressive} models, which iteratively predict the next sequence element or ``token'', dominate.

Flow matching can be generalized in several directions. First, the design of the stochastic interpolant offers significant flexibility. Eq.~\eqref{NC_eq:stochastic_interpolant}, linear interpolation, is only the simplest possibility.
Second, there is an equal amount of freedom in choosing a sampling algorithm (stochastic or deterministic sampling), which can lead to important numerical efficiency gains. Lastly, different neural network architectures can be used to approximate the velocity field or score. This leads to a large zoo of methods~\cite{Albergo.etal2025,Lipman.etal2024}.

Next, in generative modeling, one is rarely interested in generating ``random'' samples from the target distribution $p_1$ (in contrast to statistical mechanics). After all, a random sample could simply be drawn from the training set.
Instead, one is interested in a sample that fulfills some particular task: an image that matches a given caption, or a protein with a particular functionality.
\emph{Guided diffusion} uses additional ``guidance'' terms to nudge the sampling process $y_t$ towards a user-defined aim~\cite{DhariwalNichol2021}. 
The population distribution $p_1$ of images/proteins/... serves as a conduit, a convenient fiction, towards the ultimate goal of a task-directed model.
Guided diffusion is a form of sampling with a control $u_t$ which we discussed above. A running reward can be supplied by an ML model (like a classifier) to score whether the trajectory $\mathbf{x}_t$ progresses towards a desired sample like a particular type of image, and a terminal reward  can score the final sample. 

Finally, flow matching can also be generalized from continuous distributions on $\mathbb{R}^d$, such as images, to discrete ones, such as text data. In this case, a sample is a sequence $x=\{x_1,...,x_d\}, x_i\in \mathcal{A}$ from a finite alphabet $\mathcal{A}$. The stochastic interpolant becomes a Markov chain that converts a sequence into a sample from a simple base distribution $p_0$. For example, \emph{masked diffusion}~\cite{Sahoo.etal2024} continuously masks out sequence elements (so $p_0$ has only one element, the fully masked sequence). The goal is now to learn a Markov process that reverses the masking. The transition rates of this process play the role of the velocity field in the continuous setting, and one minimizes a loss similar to Eq.~\eqref{NC_eq:flow_matching_loss}. Alternatively, the discrete data can be embedded into a continuous space. 
For language modeling, discrete diffusion has recently become competitive with the next-token-prediction paradigm that underlies most large language models~\cite{Sahoo.etal2024}. In contrast to the latter, diffusion models can generate many tokens at once, and thus leverage more computing power.

\paragraph*{Statistical mechanics of memorization and generalization.}

Flow matching and diffusion models are optimized by using a set of training samples, but so far, we have not paid any attention to the limited size of this training set. The models are fit not to the true target $p_1$, but to the \emph{empirical distribution} $\widehat p_{\mathrm{data}}(x)=N^{-1}\sum_{n=1}^N\delta(x-x_n)$ of the finite training set. Therefore, the ``optimal'' flow map under Eq.~\eqref{NC_eq:flow_matching_loss} reproduces the empirical distribution, and generation always returns a sample from the training set.
This evidently undesirable phenomenon is called \emph{memorization}; the model fails to generalize to new samples.
In practice, memorization is avoided by stopping the training process early, and systematic approaches to avoid memorization are a topic of current research.
The phenomenon of \emph{benign overfitting}, where a model perfectly interpolates the training data without impeding performance, is thus absent in flow models. A model with zero training flow matching loss always memorizes.

The dynamics of diffusion models starting from pure noise bear a close resemblance to a statistical mechanics system that is slowly cooled from an initially high temperature.
Flow-matching time is thus analogous to temperature. Indeed, flow matching and diffusion models can be fruitfully analyzed using statistical mechanics, which has shed light on the \emph{generalization-memorization} transition.
Refs.~\cite{Raya.Ambrogioni2023,Biroli.etal2024} argue that flow matching dynamics undergo a series of symmetry-breaking phase transitions as a function of time $t$.
At a first ``ferromagnetic'' transition, samples $x_t$ commit to one of the modes of the target distribution, generating a novel sample. However, as the dynamics continue, a second ``glass'' transition traps $x_t$ near a sample from the training dataset (Fig.~\ref{NC_fig:memorization}).

In Chapter~\ref{sec:sampling}, we saw how an analogy with physical annealing gave rise to powerful algorithms for sampling from a known distribution. An interesting avenue for future research is whether the analogy between annealing and flow matching could be used to design ``better'', more statistically efficient generative models, now learned from training data.

\section*{Contributions and acknowledgments}

\StartHideFromToC

\subsection*{Contributions}
\noindent E.B.: Chapters~\ref{sec:variational_structure} and~\ref{sec:OT} \\
N.C.: Chapters~\ref{sec:variational_structure} and~\ref{sec:sampling}, and Sec.~\ref{sec:flows_diffusions} \\
B.E.: Sec.~\ref{sec:RL} \\
C.J.: Chapter~\ref{sec:variational_structure} and Sec.~\ref{sec:RL} \\
G.R.: Chapters~\ref{sec:variational_structure} and~\ref{sec:OT} \\
C.S.: Chapter~\ref{sec:OT} and Sec.~\ref{sec:WGFApps} \\
B.S.: Chapters~\ref{sec:variational_structure} and~\ref{sec:thermodynamics} \\
Overlapping sections were written collaboratively by the authors listed above. All authors reviewed and revised the manuscript.  
\subsection*{Acknowledgments}

\StopHideFromToC

E.B. acknowledges support from the Fannie and John Hertz Foundation Fellowship.
E.B. and C.J. acknowledge this material is based upon work supported by the National Science Foundation Graduate Research Fellowship Program under Grant No. DGE-2444107. 
Any opinions, findings, and conclusions or recommendations expressed in this material are those of the authors and do not necessarily reflect the views of the National Science Foundation.
N.C. was supported in part by Princeton University through the Princeton Center for Theoretical Science and the Center for the Physics of Biological Function, as well as by the National Institute of General Medical Sciences (NIGMS) of the National Institutes of Health (NIH) under award numbers R01GM082938 and R35GM156427. The content is solely the responsibility of the authors and does not necessarily represent the official views of the National Institutes of Health. Part of this work was performed at the Kavli Institute for Theoretical Physics (KITP), supported by NSF grant PHY-2309135 and  Gordon and Betty Moore Foundation Grant No. 2919.02.
N.C. and B.S. acknowledge useful discussions at the Beg Rohu Summer School of Physics 2026 and thank all summer school lecturers.
C.S. was supported in part by Princeton University through the Center for the Physics of Biological Function. This work was performed in part at the Aspen Center for Physics, which is supported by National Science Foundation grant PHY-2210452.
B.S. was supported by the Princeton Center for Theoretical Science and in part by the Center for the Physics of Biological Function at Princeton University.
G.R. was partially supported by a joint research agreement between NTT Research Inc and Princeton University. We acknowledge the use of generative AI for editorial purposes. 

\bibliography{references}

@incollection{oksendal2003stochastic,
  title={Stochastic differential equations},
  author={{\O}ksendal, Bernt},
  booktitle={Stochastic differential equations: an introduction with applications},
  pages={38--50},
  year={2003},
  publisher={Springer}
}

@article{levine2018reinforcement,
  title={Reinforcement learning and control as probabilistic inference: Tutorial and review},
  author={Levine, Sergey},
  journal={arXiv preprint arXiv:1805.00909},
  year={2018}
}

@article{todorov2006linearly,
  title={Linearly-solvable Markov decision problems},
  author={Todorov, Emanuel},
  journal={Advances in neural information processing systems},
  volume={19},
  year={2006}
}

@book{sutton1998reinforcement,
  title={Reinforcement learning: An introduction},
  author={Sutton, Richard S and Barto, Andrew G and others},
  volume={1},
  number={1},
  year={1998},
  publisher={MIT press Cambridge}
}

@article{CasasVazquez2003,
Author = {Casas-Vazquez, J. and Jou, D.},
Title = {{Temperature in non-equilibrium states: A review of open problems and current proposals}},
Journal = {{Rep. Prog. Phys.}},
Year = {{2003}},
Volume = {{66}},
Number = {{11}},
Pages = {{1937}},
Month = {{NOV}},
DOI = {10.1088/0034-4885/66/11/R03},
}

@article{solon2015pressure,
  title={Pressure is not a state function for generic active fluids},
  author={Solon, A. P. and Fily, Y. and Baskaran, A. and Cates, M. E. and Kafri, Y. and Kardar, M. and Tailleur, J.},
  journal={Nat. Phys.},
  volume={11},
  number={8},
  pages={673--678},
  year={2015},
  publisher={Nature Publishing Group},
  doi={10.1038/nphys3377}
}

@book{KardarBOOK2007,
    author = {Kardar, M.},
    title = {Statistical Physics of Particles},
    publisher = {Cambridge University Press},
    year = {2007}
}

@article{JaynesPR1957,
  title = {Information Theory and Statistical Mechanics},
  author = {Jaynes, E. T.},
  journal = {Phys. Rev.},
  volume = {106},
  issue = {4},
  pages = {620--630},
  numpages = {0},
  year = {1957},
  month = {May},
  publisher = {American Physical Society},
  doi = {10.1103/PhysRev.106.620},
  url = {https://link.aps.org/doi/10.1103/PhysRev.106.620}
}

@Article{JaynesIEEE1982,
  author={Jaynes, E.T.},
  journal={Proc. IEEE}, 
  title={On the rationale of maximum-entropy methods}, 
  year={1982},
  volume={70},
  number={9},
  pages={939-952},
  doi={10.1109/PROC.1982.12425}
}

@article{SorkinPRL2024,
  title = {Second Law of Thermodynamics without Einstein Relation},
  author = {Sorkin, Benjamin and Diamant, Haim and Ariel, Gil and Markovich, Tomer},
  journal = {Phys. Rev. Lett.},
  volume = {133},
  issue = {26},
  pages = {267101},
  numpages = {8},
  year = {2024},
  month = {Dec},
  publisher = {American Physical Society},
  doi = {10.1103/PhysRevLett.133.267101},
  url = {https://link.aps.org/doi/10.1103/PhysRevLett.133.267101}
}

@article{Shannon1948,
  author={Shannon, C. E.},
  journal={Bell Syst. Tech. J.}, 
  title={A mathematical theory of communication}, 
  year={1948},
  volume={27},
  number={3},
  pages={379-423},
  doi={10.1002/j.1538-7305.1948.tb01338.x}}

@book{cover1999elements,
  title={Elements of information theory},
  year={1999},
  publisher={Wiley},
  author = {Cover, T. M. and Thomas, J.}
}

@book{HansenBOOK2003,
    author = {Hansen, J. P. and McDonald, I.R.},
    title = {Theory of Simple Liquids},
    publisher = {Academic Press},
    edition = {3},
    year = {2003}
}

@article{SeifertROPP2012,
doi = {10.1088/0034-4885/75/12/126001},
url = {https://doi.org/10.1088/0034-4885/75/12/126001},
year = {2012},
month = {nov},
publisher = {IOP Publishing},
volume = {75},
number = {12},
pages = {126001},
author = {Seifert, Udo},
title = {Stochastic thermodynamics, fluctuation theorems and molecular machines},
journal = {Rep. Prog. Phys.}
}

@book{PelitiBOOK2021,
    author = {Peliti, L. and Pigolotti, S.},
    title = {Stochastic Thermodynamics},
    subtitle={An Introduction},
    publisher = {Princeton University Press},
    year = {2021}
}

@article{LebowitzPHYSA1999,
title = {Microscopic origins of irreversible macroscopic behavior},
journal = {Phys. A: Stat. Mech. Appl.},
volume = {263},
number = {1},
pages = {516-527},
year = {1999},
issn = {0378-4371},
doi = {https://doi.org/10.1016/S0378-4371(98)00514-7},
url = {https://www.sciencedirect.com/science/article/pii/S0378437198005147},
author = {Joel L. Lebowitz}
}

@article{LevinPR2014,
title = {Nonequilibrium statistical mechanics of systems with long-range interactions},
journal = {Phys. Rep.},
volume = {535},
number = {1},
pages = {1-60},
year = {2014},
issn = {0370-1573},
doi = {https://doi.org/10.1016/j.physrep.2013.10.001},
url = {https://www.sciencedirect.com/science/article/pii/S0370157313003761},
author = {Yan Levin and Renato Pakter and Felipe B. Rizzato and Tarcísio N. Teles and Fernanda P.C. Benetti}
}

@article{CavagnaPRE2014,
  title = {Dynamical maximum entropy approach to flocking},
  author = {Cavagna, Andrea and Giardina, Irene and Ginelli, Francesco and Mora, Thierry and Piovani, Duccio and Tavarone, Raffaele and Walczak, Aleksandra M.},
  journal = {Phys. Rev. E},
  volume = {89},
  issue = {4},
  pages = {042707},
  numpages = {10},
  year = {2014},
  month = {Apr},
  publisher = {American Physical Society},
  doi = {10.1103/PhysRevE.89.042707},
  url = {https://link.aps.org/doi/10.1103/PhysRevE.89.042707}
}

@Article{SorkinSM2023,
author ="Sorkin, Benjamin and Be’er, Avraham and Diamant, Haim and Ariel, Gil",
title  ="Detecting and characterizing phase transitions in active matter using entropy",
journal  ="Soft Matter",
year  ="2023",
volume  ="19",
issue  ="27",
pages  ="5118-5126",
doi  ="10.1039/D3SM00482A",
url  ="http://dx.doi.org/10.1039/D3SM00482A"
}

@article{CavagnaARCMP2014,
   author = "Cavagna, Andrea and Giardina, Irene",
   title = "Bird Flocks as Condensed Matter", 
   journal= "Annu. Rev. Condens. Matter Phys.",
   year = "2014",
   volume = "5",
   number = "Volume 5, 2014",
   pages = "183-207",
   doi = "https://doi.org/10.1146/annurev-conmatphys-031113-133834",
   url = "https://www.annualreviews.org/content/journals/10.1146/annurev-conmatphys-031113-133834"
}

@article{MartinianiPRX19,
  title = {Quantifying Hidden Order out of Equilibrium},
  author = {Martiniani, Stefano and Chaikin, Paul M. and Levine, Dov},
  journal = {Phys. Rev. X},
  volume = {9},
  issue = {1},
  pages = {011031},
  numpages = {13},
  year = {2019},
  month = {Feb},
  publisher = {American Physical Society},
  doi = {10.1103/PhysRevX.9.011031},
  url = {https://link.aps.org/doi/10.1103/PhysRevX.9.011031}
}

@article{AvineryPRL19,
  title = {Universal and Accessible Entropy Estimation Using a Compression Algorithm},
  author = {Avinery, Ram and Kornreich, Micha and Beck, Roy},
  journal = {Phys. Rev. Lett.},
  volume = {123},
  issue = {17},
  pages = {178102},
  numpages = {5},
  year = {2019},
  month = {Oct},
  publisher = {American Physical Society},
  doi = {10.1103/PhysRevLett.123.178102},
  url = {https://link.aps.org/doi/10.1103/PhysRevLett.123.178102}
}

@article{SorkinPRL2023,
  title = {Universal Relation between Entropy and Kinetics},
  author = {Sorkin, Benjamin and Diamant, Haim and Ariel, Gil},
  journal = {Phys. Rev. Lett.},
  volume = {131},
  issue = {14},
  pages = {147101},
  numpages = {6},
  year = {2023},
  month = {Oct},
  publisher = {American Physical Society},
  doi = {10.1103/PhysRevLett.131.147101},
  url = {https://link.aps.org/doi/10.1103/PhysRevLett.131.147101}
}

@article{SorkinPRE2023,
  title = {Resolving entropy contributions in nonequilibrium transitions},
  author = {Sorkin, Benjamin and Ricouvier, Joshua and Diamant, Haim and Ariel, Gil},
  journal = {Phys. Rev. E},
  volume = {107},
  issue = {1},
  pages = {014138},
  numpages = {12},
  year = {2023},
  month = {Jan},
  publisher = {American Physical Society},
  doi = {10.1103/PhysRevE.107.014138},
  url = {https://link.aps.org/doi/10.1103/PhysRevE.107.014138}
}

@article{MeshulanRMP2025,
  title = {Statistical mechanics for networks of real neurons},
  author = {Meshulam, Leenoy and Bialek, William},
  journal = {Rev. Mod. Phys.},
  volume = {97},
  issue = {4},
  pages = {045002},
  numpages = {66},
  year = {2025},
  month = {Nov},
  publisher = {American Physical Society},
  doi = {10.1103/jcrn-3nrc},
  url = {https://link.aps.org/doi/10.1103/jcrn-3nrc}
}

@article{TouchettePR2009,
title = {The large deviation approach to statistical mechanics},
journal = {Phys. Rep.},
volume = {478},
number = {1},
pages = {1-69},
year = {2009},
issn = {0370-1573},
doi = {https://doi.org/10.1016/j.physrep.2009.05.002},
url = {https://www.sciencedirect.com/science/article/pii/S0370157309001410},
author = {Hugo Touchette}
}

@book{WainwrightBOOK2008,
  title={Graphical models, exponential families, and variational inference},
  author={Wainwright, Martin J and Jordan, Michael Irwin and others},
  year={2008},
  publisher={Now Publishers}
}

@ARTICLE{ZhuNC1997,
  author={Zhu, Song Chun and Wu, Ying Nian and Mumford, David},
  journal={Neural Comput.}, 
  title={Minimax Entropy Principle and Its Application to Texture Modeling}, 
  year={1997},
  volume={9},
  number={8},
  pages={1627-1660},
  doi={10.1162/neco.1997.9.8.1627}}

@article{WeigtPNAS2009,
author = {Martin Weigt  and Robert A. White  and Hendrik Szurmant  and James A. Hoch  and Terence Hwa },
title = {Identification of direct residue contacts in protein–protein interaction by message passing},
journal = {Proc. Natl. Acad. Sci. U.S.A.},
volume = {106},
number = {1},
pages = {67-72},
year = {2009},
doi = {10.1073/pnas.0805923106},
URL = {https://www.pnas.org/doi/abs/10.1073/pnas.0805923106}
}

@article{OnsagerPR1931i,
  title = {Reciprocal Relations in Irreversible Processes. {I}.},
  author = {Onsager, Lars},
  journal = {Phys. Rev.},
  volume = {37},
  issue = {4},
  pages = {405--426},
  numpages = {0},
  year = {1931},
  month = {Feb},
  publisher = {American Physical Society},
  doi = {10.1103/PhysRev.37.405},
  url = {https://link.aps.org/doi/10.1103/PhysRev.37.405}
}

@article{OnsagerPR1931ii,
  title = {Reciprocal Relations in Irreversible Processes. {II}.},
  author = {Onsager, Lars},
  journal = {Phys. Rev.},
  volume = {38},
  issue = {12},
  pages = {2265--2279},
  numpages = {0},
  year = {1931},
  month = {Dec},
  publisher = {American Physical Society},
  doi = {10.1103/PhysRev.38.2265},
  url = {https://link.aps.org/doi/10.1103/PhysRev.38.2265}
}

@book{deGrootBOOK84,
  title={Non-Equilibrium Thermodynamics},
  author={De Groot, Sybren Ruurds and Mazur, Peter},
  year={1984},
  publisher={Dover Publications}
}

@article{DoiJPCM2011,
doi = {10.1088/0953-8984/23/28/284118},
url = {https://doi.org/10.1088/0953-8984/23/28/284118},
year = {2011},
month = {jun},
publisher = {},
volume = {23},
number = {28},
pages = {284118},
author = {Doi, Masao},
title = {Onsager’s variational principle in soft matter},
journal = {J. Phys.: Condens. Matter},
}

@book{ZwanzigBOOK2001,
    author = {Zwanzig, R.},
    title = {Nonequilibrium Statistical Mechanics},
    publisher = {Oxford University Press},
    year = {2001}
}

@book{ChewiBOOK2025,
    author = {Chewi, Sinho and Niles-Weed, Jonathan and Rigollet, Philippe},
    title = {Statistical Optimal Transport},
    publisher = {Springer},
    year = {2025}
}

@article{DyreJCP18,
    author = {Dyre, Jeppe C.},
    title = {Perspective: Excess-entropy scaling},
    journal = {J. Chem. Phys.},
    volume = {149},
    number = {21},
    pages = {210901},
    year = {2018},
    month = {12},
    issn = {0021-9606},
    doi = {10.1063/1.5055064},
    url = {https://doi.org/10.1063/1.5055064},
}

@article{BenamouNUMATH2000,
  title={A computational fluid mechanics solution to the {M}onge-{K}antorovich mass transfer problem},
  author={Benamou, Jean-David and Brenier, Yann},
  journal={Numer. Math.},
  volume={84},
  number={3},
  pages={375--393},
  year={2000},
  publisher={Springer-Verlag Berlin/Heidelberg},
  url={https://doi.org/10.1007/s002110050002}
}

@article{VanVuPRX23,
  title = {Thermodynamic Unification of Optimal Transport: Thermodynamic Uncertainty Relation, Minimum Dissipation, and Thermodynamic Speed Limits},
  author = {Van Vu, Tan and Saito, Keiji},
  journal = {Phys. Rev. X},
  volume = {13},
  issue = {1},
  pages = {011013},
  numpages = {45},
  year = {2023},
  month = {Feb},
  publisher = {American Physical Society},
  doi = {10.1103/PhysRevX.13.011013},
  url = {https://link.aps.org/doi/10.1103/PhysRevX.13.011013}
}

@book{SchussBOOK2010,
  title     = {Theory and Applications of Stochastic Processes},
  subtitle  = {An Analytical Approach},
  author    = {Schuss, Z},
  year      = {2010},
  publisher = {Springer},
  address   = {New York}
}

@article{SekimotoPTP1998,
    author = {Sekimoto, K.},
    title = {Langevin Equation and Thermodynamics},
    journal = {Prog. Theor. Phys.},
    volume = {130},
    pages = {17-27},
    year = {1998},
    month = {01},
    issn = {0375-9687},
    doi = {10.1143/PTPS.130.17},
    url = {https://doi.org/10.1143/PTPS.130.17}
}

@article{AurellJSP2012,
author = {Aurell, Erik and Gawedzki, Krzysztof and Mejia-Monasterio, Carlos and Mohayaee, Roya and Muratore-Ginanneschi, Paolo},
  title   = {Refined Second Law of Thermodynamics for Fast Random Processes},
  journal = {J. Stat. Phys.},
  volume  = {147},
  number  = {3},
  pages   = {487--505},
  year    = {2012},
  doi     = {10.1007/s10955-012-0478-x}
}

@ARTICLE{TaghvaeiIEEE2022,
  author={Taghvaei, Amirhossein and Miangolarra, Olga Movilla and Fu, Rui and Chen, Yongxin and Georgiou, Tryphon T.},
  journal={IEEE Control Sys. Lett.}, 
  title={On the Relation Between Information and Power in Stochastic Thermodynamic Engines}, 
  year={2022},
  volume={6},
  number={},
  pages={434-439},
  doi={10.1109/LCSYS.2021.3078716}}

@article{SorkinJSTAT2025,
doi = {10.1088/1742-5468/ad99c8},
url = {https://dx.doi.org/10.1088/1742-5468/ad99c8},
year = {2025},
publisher = {IOP Publishing and SISSA},
pages = {013208},
author = {Sorkin, Benjamin and Ariel, Gil and Markovich, Tomer},
title = {Consistent expansion of the {L}angevin propagator with application to entropy production},
journal = {J. Stat. Mech.: Theor. Exp.}
}

@article{JKOSIAM1998,
author = {Jordan, Richard and Kinderlehrer, David and Otto, Felix},
title = {The Variational Formulation of the {F}okker--{P}lanck Equation},
journal = {SIAM J. Math. Analys.},
volume = {29},
number = {1},
pages = {1-17},
year = {1998},
doi = {10.1137/S0036141096303359}
}

@articleInfo{MartinKRM2010,
title = {A mixed finite element method for nonlinear diffusion equations},
journal = {Kinet. Relat. Model.},
volume = {3},
number = {1},
pages = {59-83},
year = {2010},
issn = {1937-5093},
doi = {10.3934/krm.2010.3.59},
url = {https://www.aimsciences.org/article/id/b169a0d4-beb2-4c5b-88b4-61be422a9570},
author = {Martin Burger and José A. Carrillo and Marie-Therese Wolfram}
}

@article{CarrilloFoCM2022,
	author = {Carrillo, Jos{\'e}A. and Craig, Katy and Wang, Li and Wei, Chaozhen},
	date = {2022/04/01},
	doi = {10.1007/s10208-021-09503-1},
	id = {Carrillo2022},
	isbn = {1615-3383},
	journal = {F. Comput. Math.},
	number = {2},
	pages = {389--443},
	title = {Primal Dual Methods for {W}asserstein Gradient Flows},
	url = {https://doi.org/10.1007/s10208-021-09503-1},
	volume = {22},
	year = {2022}
}

@article{Crooks1999,
  title = {Entropy production fluctuation theorem and the nonequilibrium work relation for free energy differences},
  author = {Crooks, Gavin E.},
  journal = {Phys. Rev. E},
  volume = {60},
  issue = {3},
  pages = {2721--2726},
  numpages = {0},
  year = {1999},
  month = {Sep},
  publisher = {American Physical Society},
  doi = {10.1103/PhysRevE.60.2721},
  url = {https://link.aps.org/doi/10.1103/PhysRevE.60.2721}
}

@article{RotskoffPRE2017,
  title = {Geometric approach to optimal nonequilibrium control: Minimizing dissipation in nanomagnetic spin systems},
  author = {Rotskoff, Grant M. and Crooks, Gavin E. and Vanden-Eijnden, Eric},
  journal = {Phys. Rev. E},
  volume = {95},
  issue = {1},
  pages = {012148},
  numpages = {7},
  year = {2017},
  month = {Jan},
  publisher = {American Physical Society},
  doi = {10.1103/PhysRevE.95.012148},
  url = {https://link.aps.org/doi/10.1103/PhysRevE.95.012148}
}

@article{HummerPNAS01,
author = {Gerhard Hummer  and Attila Szabo },
title = {Free energy reconstruction from nonequilibrium single-molecule
 pulling experiments},
journal = {Proc. Natl. Acad. Sci. U.S.A.},
volume = {98},
number = {7},
pages = {3658-3661},
year = {2001},
doi = {10.1073/pnas.071034098},
URL = {https://www.pnas.org/doi/abs/10.1073/pnas.071034098}
}

@article{CollinNature2005,
	author = {Collin, D. and Ritort, F. and Jarzynski, C. and Smith, S. B. and Tinoco, I. and Bustamante, C.},
	doi = {10.1038/nature04061},
	id = {Collin2005},
	isbn = {1476-4687},
	journal = {Nature},
	number = {7056},
	pages = {231--234},
	title = {Verification of the {C}rooks fluctuation theorem and recovery of {RNA} folding free energies},
	url = {https://doi.org/10.1038/nature04061},
	volume = {437},
	year = {2005}
}

@misc{halmos2026implicit,
      title={Implicit Bias of the JKO Scheme}, 
      author={Peter Halmos and Boris Hanin},
      year={2026},
      eprint={2511.14827},
      archivePrefix={arXiv},
      primaryClass={stat.ML},
      url={https://arxiv.org/abs/2511.14827}, 
}

@article{Rotskoff2022trainability,
author = {Rotskoff, Grant and Vanden-Eijnden, Eric},
title = {Trainability and Accuracy of Artificial Neural Networks: An Interacting Particle System Approach},
journal = {Communications on Pure and Applied Mathematics},
volume = {75},
number = {9},
pages = {1889-1935},
doi = {https://doi.org/10.1002/cpa.22074},
url = {https://onlinelibrary.wiley.com/doi/abs/10.1002/cpa.22074},
eprint = {https://onlinelibrary.wiley.com/doi/pdf/10.1002/cpa.22074},
year = {2022}
}

@inproceedings{Chizat2018global,
author = {Chizat, L\'{e}na\"{\i}c and Bach, Francis},
title = {On the global convergence of gradient descent for over-parameterized models using optimal transport},
year = {2018},
publisher = {Curran Associates Inc.},
address = {Red Hook, NY, USA},
booktitle = {Proceedings of the 32nd International Conference on Neural Information Processing Systems},
pages = {3040–3050},
numpages = {11},
location = {Montr\'{e}al, Canada},
series = {NIPS'18}
}

@article{
Mei2018mean,
author = {Song Mei  and Andrea Montanari  and Phan-Minh Nguyen },
title = {A mean field view of the landscape of two-layer neural networks},
journal = {Proceedings of the National Academy of Sciences},
volume = {115},
number = {33},
pages = {E7665-E7671},
year = {2018},
doi = {10.1073/pnas.1806579115},
URL = {https://www.pnas.org/doi/abs/10.1073/pnas.1806579115},
eprint = {https://www.pnas.org/doi/pdf/10.1073/pnas.1806579115}}

@inproceedings{Liu2017stein,
author = {Liu, Qiang},
title = {Stein variational gradient descent as gradient flow},
year = {2017},
isbn = {9781510860964},
publisher = {Curran Associates Inc.},
address = {Red Hook, NY, USA},
booktitle = {Proceedings of the 31st International Conference on Neural Information Processing Systems},
pages = {3118–3126},
numpages = {9},
location = {Long Beach, California, USA},
series = {NIPS'17}
}

@inproceedings{Liu2016stein,
author = {Liu, Qiang and Wang, Dilin},
title = {Stein variational Gradient descent: a general purpose Bayesian inference algorithm},
year = {2016},
isbn = {9781510838819},
publisher = {Curran Associates Inc.},
address = {Red Hook, NY, USA},
booktitle = {Proceedings of the 30th International Conference on Neural Information Processing Systems},
pages = {2378–2386},
numpages = {9},
location = {Barcelona, Spain},
series = {NIPS'16}
}

@article{Ghosh2024representations,
url = {https://doi.org/10.1515/jaa-2024-0110},
title = {On representations of mean-field variational inference},
title = {},
author = {Soumyadip Ghosh and Yingdong Lu and Tomasz Nowicki and Edith Zhang},
journal = {Journal of Applied Analysis},
doi = {doi:10.1515/jaa-2024-0110},
year = {},
lastchecked = {2026-03-09}
}

@inproceedings{sznitman1991topics,
	address = {Berlin, Heidelberg},
	title = {Topics in propagation of chaos},
	isbn = {978-3-540-46319-1},
	booktitle = {Ecole d'{Eté} de {Probabilités} de {Saint}-{Flour} {XIX} — 1989},
	publisher = {Springer Berlin Heidelberg},
	author = {Sznitman, Alain-Sol},
	editor = {Hennequin, Paul-Louis},
	year = {1991},
	pages = {165--251},
}

@inproceedings{
lambert2022variational,
title={Variational inference via Wasserstein gradient flows},
author={Marc Lambert and Sinho Chewi and Francis Bach and Silv{\`e}re Bonnabel and Philippe Rigollet},
booktitle={Advances in Neural Information Processing Systems},
editor={Alice H. Oh and Alekh Agarwal and Danielle Belgrave and Kyunghyun Cho},
year={2022},
url={https://openreview.net/forum?id=K2PTuvVTF1L}
}

@article{Sirignano2020mean,
author = {Sirignano, Justin and Spiliopoulos, Konstantinos},
title = {Mean Field Analysis of Neural Networks: A Law of Large Numbers},
journal = {SIAM Journal on Applied Mathematics},
volume = {80},
number = {2},
pages = {725-752},
year = {2020},
doi = {10.1137/18M1192184},
URL = { 
    
        https://doi.org/10.1137/18M1192184
    
    

}
}

@article{Benamou2002monge,
author = {Benamou, J.-D. and Brenier, Y. and Guittet, K.},
title = {The Monge–Kantorovitch mass transfer and its computational fluid mechanics formulation},
journal = {International Journal for Numerical Methods in Fluids},
volume = {40},
number = {1-2},
pages = {21-30},
doi = {https://doi.org/10.1002/fld.264},
url = {https://onlinelibrary.wiley.com/doi/abs/10.1002/fld.264},
eprint = {https://onlinelibrary.wiley.com/doi/pdf/10.1002/fld.264},
year = {2002}
}

@book{Villani2003topics,
  title={Topics in Optimal Transportation},
  author={Villani, C. and American Mathematical Society},
  isbn={9781470418045},
  lccn={2003040350},
  series={Graduate studies in mathematics},
  url={https://books.google.com/books?id=MyPjjgEACAAJ},
  year={2003},
  publisher={American Mathematical Society}
}

@book{Ambrosio2008gradient,
  langid = {english},
  location = {{Basel}},
  title = {Gradient Flows in Metric Spaces and in the Space of Probability Measures},
  edition = {2. ed},
  isbn = {978-3-7643-8722-8 978-3-7643-8721-1},
  pagetotal = {334},
  series = {Lectures in Mathematics {{ETH Zürich}}},
  publisher = {{Birkhäuser}},
  date = {2008},
  author = {Ambrosio, Luigi and Gigli, Nicola and Savaré, Giuseppe},
  note = {OCLC: 254181287}
}

@inproceedings{Dio2023forward,
author = {Diao, Michael and Balasubramanian, Krishnakumar and Chewi, Sinho and Salim, Adil},
title = {Forward-backward Gaussian variational inference via JKO in the Bures–Wasserstein space},
year = {2023},
publisher = {JMLR.org},
booktitle = {Proceedings of the 40th International Conference on Machine Learning},
articleno = {316},
numpages = {32},
location = {Honolulu, Hawaii, USA},
series = {ICML'23}
}

@InProceedings{Jiang2024algorithms,
  title = 	 {Algorithms for mean-field variational inference via polyhedral optimization in the {W}asserstein space},
  author =       {Jiang, Yiheng and Chewi, Sinho and Pooladian, Aram-Alexandre},
  booktitle = 	 {Proceedings of Thirty Seventh Conference on Learning Theory},
  pages = 	 {2720--2721},
  year = 	 {2024},
  editor = 	 {Agrawal, Shipra and Roth, Aaron},
  volume = 	 {247},
  series = 	 {Proceedings of Machine Learning Research},
  month = 	 {30 Jun--03 Jul},
  publisher =    {PMLR},
  url = 	 {https://proceedings.mlr.press/v247/jiang24a.html}
}

@article{
McKean1966class,
author = {H. P. McKean },
title = {A CLASS OF MARKOV PROCESSES ASSOCIATED WITH NONLINEAR PARABOLIC EQUATIONS},
journal = {Proceedings of the National Academy of Sciences},
volume = {56},
number = {6},
pages = {1907-1911},
year = {1966},
doi = {10.1073/pnas.56.6.1907},
URL = {https://www.pnas.org/doi/abs/10.1073/pnas.56.6.1907}}

@article{Cybenko1989approximation,
	title = {Approximation by superpositions of a sigmoidal function},
	volume = {2},
	issn = {1435-568X},
	url = {https://doi.org/10.1007/BF02551274},
	doi = {10.1007/BF02551274},
	number = {4},
	journal = {Mathematics of Control, Signals and Systems},
	author = {Cybenko, G.},
	month = dec,
	year = {1989},
	pages = {303--314},
}

@article{Barron1993universal,
author = {Barron, A. R.},
title = {Universal approximation bounds for superpositions of a sigmoidal function},
year = {1993},
issue_date = {May 1993},
publisher = {IEEE Press},
volume = {39},
number = {3},
issn = {0018-9448},
url = {https://doi.org/10.1109/18.256500},
doi = {10.1109/18.256500},
journal = {IEEE Trans. Inf. Theor.},
month = may,
pages = {930–945},
numpages = {16}
}

@article{Park1991universal,
    author = {Park, J. and Sandberg, I. W.},
    title = {Universal Approximation Using Radial-Basis-Function Networks},
    journal = {Neural Computation},
    volume = {3},
    number = {2},
    pages = {246-257},
    year = {1991},
    month = {06},
    issn = {0899-7667},
    doi = {10.1162/neco.1991.3.2.246},
    url = {https://doi.org/10.1162/neco.1991.3.2.246},}

@ARTICLE{Chen2019gaussian,
  author={Chen, Yongxin and Georgiou, Tryphon T. and Tannenbaum, Allen},
  journal={IEEE Access}, 
  title={Optimal Transport for Gaussian Mixture Models}, 
  year={2019},
  volume={7},
  number={},
  pages={6269-6278},
  doi={10.1109/ACCESS.2018.2889838}}

@article{Delon2020Wasserstein,
author = {Delon, Julie and Desolneux, Agn\`{e}s},
title = {A Wasserstein-Type Distance in the Space of Gaussian Mixture Models},
journal = {SIAM Journal on Imaging Sciences},
volume = {13},
number = {2},
pages = {936-970},
year = {2020},
doi = {10.1137/19M1301047},

URL = { 
        https://doi.org/10.1137/19M1301047
}
}

@article{Lu2019scaling,
author = {Lu, Jianfeng and Lu, Yulong and Nolen, James},
title = {Scaling Limit of the Stein Variational Gradient Descent: The Mean Field Regime},
journal = {SIAM Journal on Mathematical Analysis},
volume = {51},
number = {2},
pages = {648-671},
year = {2019},
doi = {10.1137/18M1187611},

URL = { 
    
        https://doi.org/10.1137/18M1187611
}
}

@inproceedings{Korba2020nonasymptotic,
  title={A Non-Asymptotic Analysis for Stein Variational Gradient Descent},
  author={Korba, Anna and Salim, Adil and Arbel, Michael and Luise, Giulia and Gretton, Arthur},
  booktitle={Advances in Neural Information Processing Systems},
  volume={33},
  pages={4672--4682},
  year={2020}
}

@InProceedings{Salim2022convergence,
  title = 	 {A Convergence Theory for {SVGD} in the Population Limit under Talagrand’s Inequality T1},
  author =       {Salim, Adil and Sun, Lukang and Richtarik, Peter},
  booktitle = 	 {Proceedings of the 39th International Conference on Machine Learning},
  pages = 	 {19139--19152},
  year = 	 {2022},
  editor = 	 {Chaudhuri, Kamalika and Jegelka, Stefanie and Song, Le and Szepesvari, Csaba and Niu, Gang and Sabato, Sivan},
  volume = 	 {162},
  series = 	 {Proceedings of Machine Learning Research},
  month = 	 {17--23 Jul},
  publisher =    {PMLR},
  url = 	 {https://proceedings.mlr.press/v162/salim22a.html}
}

@inproceedings{Das2023provably,
 author = {Das, Aniket and Nagaraj, Dheeraj},
 booktitle = {Advances in Neural Information Processing Systems},
 editor = {A. Oh and T. Naumann and A. Globerson and K. Saenko and M. Hardt and S. Levine},
 pages = {49748--49760},
 publisher = {Curran Associates, Inc.},
 title = {Provably Fast Finite Particle Variants of SVGD via Virtual Particle Stochastic Approximation},
 url = {https://proceedings.neurips.cc/paper_files/paper/2023/file/9bf1962c5b65a243ee243bb03ff2c506-Paper-Conference.pdf},
 volume = {36},
 year = {2023}
}

@inproceedings{Shi2023finite,
author = {Shi, Jiaxin and Mackey, Lester},
title = {A finite-particle convergence rate for stein variational gradient descent},
year = {2023},
publisher = {Curran Associates Inc.},
address = {Red Hook, NY, USA},
booktitle = {Proceedings of the 37th International Conference on Neural Information Processing Systems},
articleno = {1166},
numpages = {14},
location = {New Orleans, LA, USA},
series = {NIPS '23}
}

@article{Kappen2005PRL,
  author    = {Kappen, Hilbert J.},
  title     = {Linear Theory for Control of Nonlinear Stochastic Systems},
  journal   = {Physical Review Letters},
  volume    = {95},
  number    = {20},
  pages     = {200201},
  year      = {2005},
  doi       = {10.1103/PhysRevLett.95.200201},
}

@article{Kappen2005JSTAT,
  author    = {Kappen, Hilbert J.},
  title     = {Path Integrals and Symmetry Breaking for Optimal Control Theory},
  journal   = {Journal of Statistical Mechanics: Theory and Experiment},
  volume    = {2005},
  number    = {11},
  pages     = {P11011},
  year      = {2005},
  doi       = {10.1088/1742-5468/2005/11/P11011},
}

@article{Todorov2009,
  author    = {Todorov, Emanuel},
  title     = {Efficient Computation of Optimal Actions},
  journal   = {Proceedings of the National Academy of Sciences},
  volume    = {106},
  number    = {28},
  pages     = {11478--11483},
  year      = {2009},
  doi       = {10.1073/pnas.0710743106},
}

@article{Theodorou2010,
  author    = {Theodorou, Evangelos and Buchli, Jonas and Schaal, Stefan},
  title     = {A Generalized Path Integral Control Approach to Reinforcement Learning},
  journal   = {Journal of Machine Learning Research},
  volume    = {11},
  number    = {104},
  pages     = {3137--3181},
  year      = {2010},
}

@article{Kappen2016,
  author    = {Kappen, Hilbert J. and Ruiz, Hans Christian},
  title     = {Adaptive Importance Sampling for Control and Inference},
  journal   = {Journal of Statistical Physics},
  volume    = {162},
  number    = {5},
  pages     = {1244--1266},
  year      = {2016},
  doi       = {10.1007/s10955-016-1446-7},
}

@inproceedings{Theodorou2012,
  author    = {Theodorou, Evangelos and Todorov, Emanuel},
  title     = {Relative Entropy and Free Energy Dualities: Connections to Path Integral and {KL} Control},
  booktitle = {IEEE 51st Annual Conference on Decision and Control (CDC)},
  pages     = {1466--1473},
  year      = {2012},
  doi       = {10.1109/CDC.2012.6426381},
}

@article{Schrodinger1931,
  author    = {Schr\"{o}dinger, Erwin},
  title     = {\"{U}ber die {U}mkehrung der {N}aturgesetze},
  journal   = {Sitzungsberichte der Preussischen Akademie der Wissenschaften, Physikalisch-Mathematische Klasse},
  volume    = {8--9},
  pages     = {144--153},
  year      = {1931},
}

@article{Schrodinger1932,
  author    = {Schr\"{o}dinger, Erwin},
  title     = {Sur la Th\'{e}orie Relativiste de l'\'{E}lectron et l'Interpr\'{e}tation de la M\'{e}canique Quantique},
  journal   = {Annales de l'Institut Henri Poincar\'{e}},
  volume    = {2},
  number    = {4},
  pages     = {269--310},
  year      = {1932},
}

@article{Leonard2014,
  author    = {L\'{e}onard, Christian},
  title     = {A Survey of the {S}chr\"{o}dinger Problem and Some of Its Connections with Optimal Transport},
  journal   = {Discrete and Continuous Dynamical Systems},
  volume    = {34},
  number    = {4},
  pages     = {1533--1574},
  year      = {2014},
  doi       = {10.3934/dcds.2014.34.1533},
}

@article{Chen2021,
  author    = {Chen, Yongxin and Georgiou, Tryphon T. and Pavon, Michele},
  title     = {Stochastic Control Liaisons: {R}ichard {S}inkhorn Meets {G}aspard {M}onge on a {S}chr\"{o}dinger Bridge},
  journal   = {SIAM Review},
  volume    = {63},
  number    = {2},
  pages     = {249--313},
  year      = {2021},
  doi       = {10.1137/20M1339982},
}

@article{Chen2016,
  author    = {Chen, Yongxin and Georgiou, Tryphon T. and Pavon, Michele},
  title     = {On the Relation between Optimal Transport and {S}chr\"{o}dinger Bridges: A Stochastic Control Viewpoint},
  journal   = {Journal of Optimization Theory and Applications},
  volume    = {169},
  number    = {2},
  pages     = {671--691},
  year      = {2016},
  doi       = {10.1007/s10957-015-0803-z},
}

@inproceedings{Cuturi2013,
  author    = {Cuturi, Marco},
  title     = {Sinkhorn Distances: Lightspeed Computation of Optimal Transport},
  booktitle = {Advances in Neural Information Processing Systems 26},
  pages     = {2292--2300},
  year      = {2013},
}

@book{Fleming2006,
  author    = {Fleming, Wendell H. and Soner, H. Mete},
  title     = {Controlled {M}arkov Processes and Viscosity Solutions},
  edition   = {2nd},
  publisher = {Springer},
  address   = {New York},
  year      = {2006},
  series    = {Stochastic Modelling and Applied Probability},
  volume    = {25},
}

@misc{Albergo.etal2025,
  title = {Stochastic {{Interpolants}}: {{A Unifying Framework}} for {{Flows}} and {{Diffusions}}},
  shorttitle = {Stochastic {{Interpolants}}},
  author = {Albergo, Michael S. and Boffi, Nicholas M. and {Vanden-Eijnden}, Eric},
  year = 2025,
  month = oct,
  number = {arXiv:2303.08797},
  eprint = {2303.08797},
  primaryclass = {cs},
  publisher = {arXiv},
  doi = {10.48550/arXiv.2303.08797},
  urldate = {2026-03-13},
  archiveprefix = {arXiv}
}

@misc{Betancourt2018,
  title = {A {{Conceptual Introduction}} to {{Hamiltonian Monte Carlo}}},
  author = {Betancourt, Michael},
  year = 2018,
  month = jul,
  number = {arXiv:1701.02434},
  eprint = {1701.02434},
  primaryclass = {stat},
  publisher = {arXiv},
  doi = {10.48550/arXiv.1701.02434},
  urldate = {2026-03-13},
  archiveprefix = {arXiv}
}

@article{Biroli.etal2024,
  title = {Dynamical Regimes of Diffusion Models},
  author = {Biroli, Giulio and Bonnaire, Tony and De Bortoli, Valentin and M{\'e}zard, Marc},
  year = 2024,
  month = nov,
  journal = {Nature Communications},
  volume = {15},
  number = {1},
  pages = {9957},
  issn = {2041-1723},
  doi = {10.1038/s41467-024-54281-3},
  urldate = {2026-03-13},
  langid = {english}
}

@misc{Coste2026,
  title = {Importance Sampling: {{The Jarzynski}} Connection},
  author = {Coste, Simon},
  year = 2026,
  urldate = {2026-03-14}
}

@article{Gabrie.etal2022,
  title = {Adaptive {{Monte Carlo}} Augmented with Normalizing Flows},
  author = {Gabri{\'e}, Marylou and Rotskoff, Grant M. and {Vanden-Eijnden}, Eric},
  year = 2022,
  month = mar,
  journal = {Proceedings of the National Academy of Sciences},
  volume = {119},
  number = {10},
  pages = {e2109420119},
  issn = {0027-8424, 1091-6490},
  doi = {10.1073/pnas.2109420119},
  urldate = {2026-03-13},
  langid = {english}
}

@book{Gelman.etal2013,
  title = {Bayesian {{Data Analysis}}},
  author = {Gelman, Andrew and Carlin, John B. and Stern, Hal S. and Dunson, David B. and Vehtari, Aki and Rubin, Donald B.},
  year = 2013,
  month = nov,
  edition = {0},
  publisher = {{Chapman and Hall/CRC}},
  doi = {10.1201/b16018},
  urldate = {2026-03-13},
  isbn = {978-0-429-11307-9},
  langid = {english}
}

@misc{Ho.etal2020,
  title = {Denoising {{Diffusion Probabilistic Models}}},
  author = {Ho, Jonathan and Jain, Ajay and Abbeel, Pieter},
  year = 2020,
  month = dec,
  number = {arXiv:2006.11239},
  eprint = {2006.11239},
  primaryclass = {cs},
  publisher = {arXiv},
  doi = {10.48550/arXiv.2006.11239},
  urldate = {2026-04-20},
  archiveprefix = {arXiv}
}

@article{Hyvarinen2005,
  title = {Estimation of Non-Normalized Statistical Models by Score Matching},
  author = {Hyv{\"a}rinen, Aapo},
  year = 2005,
  journal = {Journal of Machine Learning Research},
  volume = {6},
  number = {24},
  pages = {695--709}
}

@article{Jarzynski1997,
  title = {Equilibrium Free-Energy Differences from Nonequilibrium Measurements: {{A}} Master-Equation Approach},
  shorttitle = {Equilibrium Free-Energy Differences from Nonequilibrium Measurements},
  author = {Jarzynski, C.},
  year = 1997,
  month = nov,
  journal = {Physical Review E},
  volume = {56},
  number = {5},
  pages = {5018--5035},
  issn = {1063-651X, 1095-3787},
  doi = {10.1103/PhysRevE.56.5018},
  urldate = {2026-03-13},
  copyright = {http://link.aps.org/licenses/aps-default-license},
  langid = {english}
}

@article{SivakPRL2012,
  title = {Thermodynamic Metrics and Optimal Paths},
  author = {Sivak, David A. and Crooks, Gavin E.},
  journal = {Phys. Rev. Lett.},
  volume = {108},
  issue = {19},
  pages = {190602},
  numpages = {5},
  year = {2012},
  month = {May},
  publisher = {American Physical Society},
  doi = {10.1103/PhysRevLett.108.190602},
  url = {https://link.aps.org/doi/10.1103/PhysRevLett.108.190602}
}

@article{EngelPRX2023,
  title = {Optimal Control of Nonequilibrium Systems through Automatic Differentiation},
  author = {Engel, Megan C. and Smith, Jamie A. and Brenner, Michael P.},
  journal = {Phys. Rev. X},
  volume = {13},
  issue = {4},
  pages = {041032},
  numpages = {15},
  year = {2023},
  month = {Nov},
  publisher = {American Physical Society},
  doi = {10.1103/PhysRevX.13.041032},
  url = {https://link.aps.org/doi/10.1103/PhysRevX.13.041032}
}

@misc{Kadkhodaie.etal2026,
  title = {Blind Denoising Diffusion Models and the Blessings of Dimensionality},
  author = {Kadkhodaie, Zahra and Pooladian, Aram-Alexandre and Chewi, Sinho and Simoncelli, Eero},
  year = 2026,
  month = feb,
  number = {arXiv:2602.09639},
  eprint = {2602.09639},
  primaryclass = {cs},
  publisher = {arXiv},
  doi = {10.48550/arXiv.2602.09639},
  urldate = {2026-05-08},
  archiveprefix = {arXiv}
}

@misc{blumenthal2026selforganized,
      title={Self-organized robustness in mean-field interacting systems}, 
      author={Emmy Blumenthal and Gautam Reddy},
      year={2026},
      eprint={2606.27626},
      archivePrefix={arXiv},
      primaryClass={physics.bio-ph},
      url={https://arxiv.org/abs/2606.27626}, 
}

@article{Kirkpatrick.etal1983,
  title = {Optimization by {{Simulated Annealing}}},
  author = {Kirkpatrick, S. and Gelatt, C. D. and Vecchi, M. P.},
  year = 1983,
  month = may,
  journal = {Science},
  volume = {220},
  number = {4598},
  pages = {671--680},
  issn = {0036-8075, 1095-9203},
  doi = {10.1126/science.220.4598.671},
  urldate = {2026-03-13},
  langid = {english}
}

@book{Krauth2006,
  title = {Statistical {{Mechanics}}: {{Algorithms}} and {{Computations}}},
  shorttitle = {Statistical {{Mechanics}}},
  author = {Krauth, Werner},
  year = 2006,
  month = sep,
  publisher = {Oxford University PressOxford},
  doi = {10.1093/oso/9780198515357.001.0001},
  urldate = {2026-03-13},
  isbn = {978-0-19-851535-7 978-1-383-02272-8},
  langid = {english}
}

@misc{Lipman.etal2024,
  title = {Flow {{Matching Guide}} and {{Code}}},
  author = {Lipman, Yaron and Havasi, Marton and Holderrieth, Peter and Shaul, Neta and Le, Matt and Karrer, Brian and Chen, Ricky T. Q. and {Lopez-Paz}, David and {Ben-Hamu}, Heli and Gat, Itai},
  year = 2024,
  month = dec,
  number = {arXiv:2412.06264},
  eprint = {2412.06264},
  primaryclass = {cs},
  publisher = {arXiv},
  doi = {10.48550/arXiv.2412.06264},
  urldate = {2026-03-13},
  archiveprefix = {arXiv}
}

@book{Mezard.Montanari2009,
  title = {Information, {{Physics}}, and {{Computation}}},
  author = {M{\'e}zard, Marc and Montanari, Andrea},
  year = 2009,
  month = jan,
  edition = {1},
  publisher = {Oxford University PressOxford},
  doi = {10.1093/acprof:oso/9780198570837.001.0001},
  urldate = {2026-03-13},
  isbn = {978-0-19-857083-7 978-0-19-171875-5},
  langid = {english}
}

@misc{Neal1998,
  title = {Annealed {{Importance Sampling}}},
  author = {Neal, Radford M.},
  year = 1998,
  month = sep,
  number = {arXiv:physics/9803008},
  eprint = {physics/9803008},
  publisher = {arXiv},
  doi = {10.48550/arXiv.physics/9803008},
  urldate = {2026-03-13},
  archiveprefix = {arXiv}
}

@inproceedings{Rezende.Mohamed2015,
  title = {Variational Inference with Normalizing Flows},
  booktitle = {Proceedings of the 32nd International Conference on Machine Learning},
  author = {Rezende, Danilo and Mohamed, Shakir},
  editor = {Bach, Francis and Blei, David},
  year = 2015,
  month = jul,
  series = {Proceedings of Machine Learning Research},
  volume = {37},
  pages = {1530--1538},
  publisher = {PMLR},
  address = {Lille, France}
}

@misc{Sahoo.etal2024,
  title = {Simple and {{Effective Masked Diffusion Language Models}}},
  author = {Sahoo, Subham Sekhar and Arriola, Marianne and Schiff, Yair and Gokaslan, Aaron and Marroquin, Edgar and Chiu, Justin T. and Rush, Alexander and Kuleshov, Volodymyr},
  year = 2024,
  month = nov,
  number = {arXiv:2406.07524},
  eprint = {2406.07524},
  primaryclass = {cs},
  publisher = {arXiv},
  doi = {10.48550/arXiv.2406.07524},
  urldate = {2026-03-13},
  archiveprefix = {arXiv}
}

@inproceedings{Sohl-Dickstein.etal2015,
  title = {Deep Unsupervised Learning Using Nonequilibrium Thermodynamics},
  booktitle = {Proceedings of the 32nd International Conference on Machine Learning},
  author = {{Sohl-Dickstein}, Jascha and Weiss, Eric and Maheswaranathan, Niru and Ganguli, Surya},
  editor = {Bach, Francis and Blei, David},
  year = 2015,
  month = jul,
  series = {Proceedings of Machine Learning Research},
  volume = {37},
  pages = {2256--2265},
  publisher = {PMLR},
  address = {Lille, France}
}

@misc{Song.etal2021,
  title = {Score-{{Based Generative Modeling}} through {{Stochastic Differential Equations}}},
  author = {Song, Yang and {Sohl-Dickstein}, Jascha and Kingma, Diederik P. and Kumar, Abhishek and Ermon, Stefano and Poole, Ben},
  year = 2021,
  month = feb,
  number = {arXiv:2011.13456},
  eprint = {2011.13456},
  primaryclass = {cs},
  publisher = {arXiv},
  doi = {10.48550/arXiv.2011.13456},
  urldate = {2026-04-14},
  archiveprefix = {arXiv}
}

@book{Vershynin2026,
  title = {High-{{Dimensional Probability}}: {{An Introduction}} with {{Applications}} in {{Data Science}}},
  shorttitle = {High-{{Dimensional Probability}}},
  author = {Vershynin, Roman},
  year = 2026,
  month = feb,
  edition = {2},
  publisher = {Cambridge University Press},
  doi = {10.1017/9781009490672},
  urldate = {2026-03-13},
  copyright = {https://www.cambridge.org/core/terms},
  isbn = {978-1-009-49067-2 978-1-009-49064-1}
}

@book{Villani2009,
  title = {Optimal {{Transport}}},
  author = {Villani, C{\'e}dric},
  editor = {Berger, M. and Eckmann, B. and De La Harpe, P. and Hirzebruch, F. and Hitchin, N. and H{\"o}rmander, L. and Kupiainen, A. and Lebeau, G. and Ratner, M. and Serre, D. and Sinai, {\relax Ya}. G. and Sloane, N. J. A. and Vershik, A. M. and Waldschmidt, M.},
  year = 2009,
  series = {Grundlehren Der Mathematischen {{Wissenschaften}}},
  volume = {338},
  publisher = {Springer Berlin Heidelberg},
  address = {Berlin, Heidelberg},
  doi = {10.1007/978-3-540-71050-9},
  urldate = {2026-03-13},
  copyright = {http://www.springer.com/tdm},
  isbn = {978-3-540-71049-3 978-3-540-71050-9}
}

@article{schulman2017proximal,
  title={Proximal Policy Optimization Algorithms},
  author={Schulman, John and Wolski, Filip and Dhariwal, Prafulla and Radford, Alec and Klimov, Oleg},
  journal={arXiv preprint arXiv:1707.06347},
  year={2017},
  url={https://arxiv.org/abs/1707.06347}
}

@inproceedings{fujimoto2018addressing,
  title={Addressing Function Approximation Error in Actor-Critic Methods},
  author={Fujimoto, Scott and van Hoof, Herke and Meger, David},
  booktitle={Proceedings of the 35th International Conference on Machine Learning},
  pages={1587--1596},
  volume={80},
  series={Proceedings of Machine Learning Research},
  year={2018},
  publisher={PMLR},
  url={https://proceedings.mlr.press/v80/fujimoto18a.html}
}

@article{mnih2015human,
  title={Human-level control through deep reinforcement learning},
  author={Mnih, Volodymyr and Kavukcuoglu, Koray and Silver, David and Rusu, Andrei A. and Veness, Joel and Bellemare, Marc G. and Graves, Alex and Riedmiller, Martin and Fidjeland, Andreas K. and Ostrovski, Georg and Petersen, Stig and Beattie, Charles and Sadik, Amir and Antonoglou, Ioannis and King, Helen and Kumaran, Dharshan and Wierstra, Daan and Legg, Shane and Hassabis, Demis},
  journal={Nature},
  volume={518},
  number={7540},
  pages={529--533},
  year={2015},
  doi={10.1038/nature14236}
}

@inproceedings{haarnoja2017reinforcement,
  title={Reinforcement learning with deep energy-based policies},
  author={Haarnoja, Tuomas and Tang, Haoran and Abbeel, Pieter and Levine, Sergey},
  booktitle={International conference on machine learning},
  pages={1352--1361},
  year={2017},
  organization={PMLR}
}

@inproceedings{haarnoja2018soft,
  title={Soft actor-critic: Off-policy maximum entropy deep reinforcement learning with a stochastic actor},
  author={Haarnoja, Tuomas and Zhou, Aurick and Abbeel, Pieter and Levine, Sergey},
  booktitle={International conference on machine learning},
  pages={1861--1870},
  year={2018},
  organization={Pmlr}
}

@article{Kingma.Welling2019,
  title = {An {{Introduction}} to {{Variational Autoencoders}}},
  author = {Kingma, Diederik P. and Welling, Max},
  year = 2019,
  publisher = {arXiv},
  doi = {10.48550/ARXIV.1906.02691},
  urldate = {2026-06-30},
  journal = "",
  copyright = {arXiv.org perpetual, non-exclusive license}
}

@article{Rotskoff.Crooks2015,
  title = {Optimal Control in Nonequilibrium Systems: {{Dynamic Riemannian}} Geometry of the {{Ising}} Model},
  shorttitle = {Optimal Control in Nonequilibrium Systems},
  author = {Rotskoff, Grant M. and Crooks, Gavin E.},
  year = 2015,
  month = dec,
  journal = {Physical Review E},
  volume = {92},
  number = {6},
  pages = {060102},
  issn = {1539-3755, 1550-2376},
  doi = {10.1103/PhysRevE.92.060102},
  urldate = {2026-06-30},
  copyright = {http://link.aps.org/licenses/aps-default-license},
  langid = {english}
}

@article{Singh.etal2025,
  title = {Variational {{Path Sampling}} of {{Rare Dynamical Events}}},
  author = {Singh, Aditya N. and Das, Avishek and Limmer, David T.},
  year = 2025,
  month = apr,
  journal = {Annual Review of Physical Chemistry},
  volume = {76},
  number = {1},
  pages = {639--662},
  issn = {0066-426X, 1545-1593},
  doi = {10.1146/annurev-physchem-083122-115001},
  urldate = {2026-06-30},
  copyright = {http://creativecommons.org/licenses/by/4.0/},
  langid = {english}
}

@misc{Song.etal2023,
  title = {Consistency {{Models}}},
  author = {Song, Yang and Dhariwal, Prafulla and Chen, Mark and Sutskever, Ilya},
  year = 2023,
  publisher = {arXiv},
  doi = {10.48550/ARXIV.2303.01469},
  urldate = {2026-07-01},
  copyright = {arXiv.org perpetual, non-exclusive license}
}

@misc{Liu.etal2022,
  title = {Flow {{Straight}} and {{Fast}}: {{Learning}} to {{Generate}} and {{Transfer Data}} with {{Rectified Flow}}},
  shorttitle = {Flow {{Straight}} and {{Fast}}},
  author = {Liu, Xingchao and Gong, Chengyue and Liu, Qiang},
  year = 2022,
  publisher = {arXiv},
  doi = {10.48550/ARXIV.2209.03003},
  urldate = {2026-07-01},
  copyright = {arXiv.org perpetual, non-exclusive license}
}

@article{Kanwar.etal2020,
  title = {Equivariant {{Flow-Based Sampling}} for {{Lattice Gauge Theory}}},
  author = {Kanwar, Gurtej and Albergo, Michael S. and Boyda, Denis and Cranmer, Kyle and Hackett, Daniel C. and Racani{\`e}re, S{\'e}bastien and Rezende, Danilo Jimenez and Shanahan, Phiala E.},
  year = 2020,
  month = sep,
  journal = {Physical Review Letters},
  volume = {125},
  number = {12},
  pages = {121601},
  issn = {0031-9007, 1079-7114},
  doi = {10.1103/PhysRevLett.125.121601},
  urldate = {2026-07-03},
  langid = {english}
}

@inproceedings{eysenbach2022maximum,
  title = {Maximum Entropy {RL} (Provably) Solves Some Robust {RL} Problems},
  author = {Eysenbach, Benjamin and Levine, Sergey},
  booktitle = {10th International Conference on Learning Representations, ICLR 2022},
  year = {2022}
}

@article{watkins1992q,
  title = {Q-learning},
  author = {Watkins, Christopher JCH and Dayan, Peter},
  journal = {Machine learning},
  volume = {8},
  number = {3},
  pages = {279--292},
  year = {1992},
  publisher = {Springer}
}

@article{doi:10.1287/moor.2022.1309,
  author = {Jin, Chi and Yang, Zhuoran and Wang, Zhaoran and Jordan, Michael I.},
  journal = {Mathematics of Operations Research},
  number = {3},
  pages = {1496-1521},
  title = {Provably Efficient Reinforcement Learning with Linear Function Approximation},
  volume = {48},
  year = {2023}
}

@article{agarwal2019reinforcement,
  title = {Reinforcement learning: Theory and algorithms},
  author = {Agarwal, Alekh and Jiang, Nan and Kakade, Sham M and Sun, Wen},
  journal = {CS Dept., UW Seattle, Seattle, WA, USA, Tech. Rep},
  volume = {32},
  pages = {96},
  year = {2019}
}

@techreport{Moore90efficientmemory-based,
  author = {Andrew William Moore},
  title = {Efficient Memory-based Learning for Robot Control},
  institution = {University of Cambridge},
  year = {1990}
}

@inproceedings{huberchebyshev,
  title = {Chebyshev Policies and the Mountain Car Problem: Reinforcement Learning for Low-Dimensional Control Tasks},
  author = {Huber, Stefan and Unger, Hannes and Sch{\"a}fer, Georg and Rehrl, Jakob},
  booktitle = {Forty-third International Conference on Machine Learning},
  year = {2026}
}

@incollection{FRENKEL2002167,
title = {Chapter 7 - Free Energy Calculations},
booktitle = {Understanding Molecular Simulation (Second Edition)},
publisher = {Academic Press},
edition = {Second Edition},
address = {San Diego},
pages = {167-200},
year = {2002},
isbn = {978-0-12-267351-1},
doi = {https://doi.org/10.1016/B978-012267351-1/50009-2},
url = {https://www.sciencedirect.com/science/article/pii/B9780122673511500092},
author = {Daan Frenkel and Berend Smit}
}

@article{Bellman1954,
  author  = {Bellman, Richard},
  title   = {The Theory of Dynamic Programming},
  journal = {Bulletin of the American Mathematical Society},
  year    = {1954},
  volume  = {60},
  number  = {6},
  pages   = {503--515},
  doi     = {10.1090/S0002-9904-1954-09848-8}
}

@book{Pontryagin1962,
  author    = {Pontryagin, L. S. and Boltyanskii, V. G. and
               Gamkrelidze, R. V. and Mishchenko, E. F.},
  title     = {The Mathematical Theory of Optimal Processes},
  publisher = {Interscience Publishers},
  address   = {New York},
  year      = {1962},
  note      = {Translated from the Russian by K. N. Trirogoff;
               English edition edited by L. W. Neustadt}
}

@article{Dreyfus2002,
  author  = {Dreyfus, Stuart},
  title   = {{Richard Bellman} on the Birth of Dynamic Programming},
  journal = {Operations Research},
  year    = {2002},
  volume  = {50},
  number  = {1},
  pages   = {48--51},
  doi     = {10.1287/opre.50.1.48.17791}
}

@book{fleming1975deterministic,
  author    = {Fleming, Wendell H. and Rishel, Raymond W.},
  title     = {Deterministic and Stochastic Optimal Control},
  series    = {Applications of Mathematics},
  volume    = {1},
  publisher = {Springer-Verlag},
  address   = {New York},
  year      = {1975},
  isbn      = {978-0-387-90155-8},
  doi       = {10.1007/978-1-4612-6380-7}
}

@misc{evans2024control,
  author       = {Evans, Lawrence C.},
  title        = {An Introduction to Mathematical Optimal Control Theory},
  year         = {2024},
  howpublished = {Lecture notes, Department of Mathematics,
                  University of California, Berkeley},
  note         = {Spring 2024 version},
  url          = {https://math.berkeley.edu/~evans/control.course.pdf}
}

@article{Watson.etal2023,
  title = {De Novo Design of Protein Structure and Function with {{RFdiffusion}}},
  author = {Watson, Joseph L. and Juergens, David and Bennett, Nathaniel R. and Trippe, Brian L. and Yim, Jason and Eisenach, Helen E. and Ahern, Woody and Borst, Andrew J. and Ragotte, Robert J. and Milles, Lukas F. and Wicky, Basile I. M. and Hanikel, Nikita and Pellock, Samuel J. and Courbet, Alexis and Sheffler, William and Wang, Jue and Venkatesh, Preetham and Sappington, Isaac and Torres, Susana V{\'a}zquez and Lauko, Anna and De Bortoli, Valentin and Mathieu, Emile and Ovchinnikov, Sergey and Barzilay, Regina and Jaakkola, Tommi S. and DiMaio, Frank and Baek, Minkyung and Baker, David},
  year = 2023,
  month = aug,
  journal = {Nature},
  volume = {620},
  number = {7976},
  pages = {1089--1100},
  issn = {0028-0836, 1476-4687},
  doi = {10.1038/s41586-023-06415-8},
  urldate = {2026-08-08},
  langid = {english}
}

@misc{Goodfellow.etal2014,
  title = {Generative {{Adversarial Networks}}},
  author = {Goodfellow, Ian J. and {Pouget-Abadie}, Jean and Mirza, Mehdi and Xu, Bing and {Warde-Farley}, David and Ozair, Sherjil and Courville, Aaron and Bengio, Yoshua},
  year = 2014,
  publisher = {arXiv},
  doi = {10.48550/ARXIV.1406.2661},
  urldate = {2026-09-10},
  copyright = {arXiv.org perpetual, non-exclusive license}
}

@inproceedings{Raya.Ambrogioni2023,
  title = {Spontaneous Symmetry Breaking in Generative Diffusion Models},
  booktitle = {Advances in {{Neural Information Processing Systems}} 36},
  author = {Raya, Gabriel and Ambrogioni, Luca},
  year = 2023,
  pages = {66377--66389},
  publisher = {Neural Information Processing Systems Foundation, Inc. (NeurIPS)},
  address = {New Orleans, Louisiana, USA},
  doi = {10.52202/075280-2897},
  urldate = {2026-09-10},
  isbn = {978-1-7138-9911-2}
}

@article{Sinkhorn1967,
  author  = {Sinkhorn, Richard},
  title   = {Diagonal Equivalence to Matrices with Prescribed Row and Column Sums},
  journal = {The American Mathematical Monthly},
  year    = {1967},
  volume  = {74},
  number  = {4},
  pages   = {402--405},
  doi     = {10.2307/2314570}
}

@article{Brenier1991polar,
  author  = {Brenier, Yann},
  title   = {Polar factorization and monotone rearrangement of vector-valued functions},
  journal = {Communications on Pure and Applied Mathematics},
  volume  = {44},
  number  = {4},
  pages   = {375--417},
  year    = {1991},
  doi     = {10.1002/cpa.3160440402}
}

@article{LaiRobbins1985,
  author  = {Lai, Tze Leung and Robbins, Herbert},
  title   = {Asymptotically efficient adaptive allocation rules},
  journal = {Advances in Applied Mathematics},
  volume  = {6},
  number  = {1},
  pages   = {4--22},
  year    = {1985},
  doi     = {10.1016/0196-8858(85)90002-8}
}

@article{Kaelbling1998planning,
  author  = {Kaelbling, Leslie Pack and Littman, Michael L. and Cassandra, Anthony R.},
  title   = {Planning and acting in partially observable stochastic domains},
  journal = {Artificial Intelligence},
  volume  = {101},
  number  = {1--2},
  pages   = {99--134},
  year    = {1998},
  doi     = {10.1016/S0004-3702(98)00023-X}
}

@inproceedings{Baird1995residual,
  author    = {Baird, Leemon},
  title     = {Residual Algorithms: Reinforcement Learning with Function Approximation},
  booktitle = {Proceedings of the Twelfth International Conference on Machine Learning},
  pages     = {30--37},
  year      = {1995},
  publisher = {Morgan Kaufmann},
  doi       = {10.1016/B978-1-55860-377-6.50013-X}
}

@inproceedings{Ni2022recurrent,
  author    = {Ni, Tianwei and Eysenbach, Benjamin and Salakhutdinov, Ruslan},
  title     = {Recurrent Model-Free {RL} Can Be a Strong Baseline for Many {POMDP}s},
  booktitle = {Proceedings of the 39th International Conference on Machine Learning},
  volume    = {162},
  series    = {Proceedings of Machine Learning Research},
  pages     = {16691--16723},
  year      = {2022},
  publisher = {PMLR},
  url       = {https://proceedings.mlr.press/v162/ni22a.html}
}

@inproceedings{Bhatt2024crossq,
  author    = {Bhatt, A. and Palenicek, D. and Belousov, B. and Argus, M. and Amiranashvili, A. and Brox, T. and Peters, J.},
  title     = {{CrossQ}: Batch Normalization in Deep Reinforcement Learning for Greater Sample Efficiency and Simplicity},
  booktitle = {International Conference on Learning Representations},
  year      = {2024},
  url       = {https://lmbweb.informatik.uni-freiburg.de/Publications/2024/AAB24/}
}

@article{Sanov1957,
  author = {Sanov, I. N.},
  title = {On the probability of large deviations of random magnitudes},
  journal = {Matematicheskii Sbornik. Novaya Seriya},
  year = {1957},
  volume = {42(84)},
  number = {1},
  pages = {11--44},
  url = {https://www.mathnet.ru/eng/sm5043},
  note = {In Russian},
}

@article{PresseRMP2013,
  author = {Press{\'e}, Steve and Ghosh, Kingshuk and Lee, Julian and Dill, Ken A.},
  title = {Principles of maximum entropy and maximum caliber in statistical physics},
  journal = {Reviews of Modern Physics},
  year = {2013},
  volume = {85},
  number = {3},
  pages = {1115--1141},
  doi = {10.1103/RevModPhys.85.1115},
  url = {https://doi.org/10.1103/RevModPhys.85.1115},
}

@article{Girsanov1960,
  author = {Girsanov, I. V.},
  title = {On Transforming a Certain Class of Stochastic Processes by Absolutely Continuous Substitution of Measures},
  journal = {Theory of Probability \& Its Applications},
  year = {1960},
  volume = {5},
  number = {3},
  pages = {285--301},
  doi = {10.1137/1105027},
  url = {https://doi.org/10.1137/1105027},
}

@article{BoueDupuis1998,
  author = {Bou{\'e}, Michelle and Dupuis, Paul},
  title = {A variational representation for certain functionals of {Brownian} motion},
  journal = {The Annals of Probability},
  year = {1998},
  volume = {26},
  number = {4},
  doi = {10.1214/aop/1022855876},
  url = {https://doi.org/10.1214/aop/1022855876},
  pages = {1641--1659},
}

@article{ChetriteTouchette2015,
  author = {Chetrite, Rapha{\"e}l and Touchette, Hugo},
  title = {Nonequilibrium {Markov} Processes Conditioned on Large Deviations},
  journal = {Annales Henri Poincar{\'e}},
  year = {2015},
  volume = {16},
  number = {9},
  pages = {2005--2057},
  doi = {10.1007/s00023-014-0375-8},
  url = {https://doi.org/10.1007/s00023-014-0375-8},
}

@article{McCann1997,
  author = {McCann, Robert J.},
  title = {A Convexity Principle for Interacting Gases},
  journal = {Advances in Mathematics},
  year = {1997},
  volume = {128},
  number = {1},
  pages = {153--179},
  doi = {10.1006/aima.1997.1634},
  url = {https://doi.org/10.1006/aima.1997.1634},
}

@article{Otto2001,
  author = {Otto, Felix},
  title = {The geometry of dissipative evolution equations: The porous medium equation},
  journal = {Communications in Partial Differential Equations},
  year = {2001},
  volume = {26},
  number = {1--2},
  pages = {101--174},
  doi = {10.1081/PDE-100002243},
  url = {https://doi.org/10.1081/PDE-100002243},
}

@article{JarzynskiPRL1997,
  author = {Jarzynski, C.},
  title = {Nonequilibrium Equality for Free Energy Differences},
  journal = {Physical Review Letters},
  year = {1997},
  volume = {78},
  number = {14},
  pages = {2690--2693},
  doi = {10.1103/PhysRevLett.78.2690},
  url = {https://doi.org/10.1103/PhysRevLett.78.2690},
}

@article{ChatterjeeDiaconis2018,
  author = {Chatterjee, Sourav and Diaconis, Persi},
  title = {The sample size required in importance sampling},
  journal = {The Annals of Applied Probability},
  year = {2018},
  volume = {28},
  number = {2},
  doi = {10.1214/17-AAP1326},
  url = {https://doi.org/10.1214/17-AAP1326},
  pages = {1099--1135},
}

@article{DasLimmer2019,
  author = {Das, Avishek and Limmer, David T.},
  title = {Variational control forces for enhanced sampling of nonequilibrium molecular dynamics simulations},
  journal = {The Journal of Chemical Physics},
  year = {2019},
  volume = {151},
  number = {24},
  pages = {244123},
  doi = {10.1063/1.5128956},
  url = {https://doi.org/10.1063/1.5128956},
}

@article{Sutton1988,
  author = {Sutton, Richard S.},
  title = {Learning to predict by the methods of temporal differences},
  journal = {Machine Learning},
  year = {1988},
  volume = {3},
  number = {1},
  pages = {9--44},
  doi = {10.1007/BF00115009},
  url = {https://doi.org/10.1007/BF00115009},
}

@misc{Haarnoja2018applications,
  author = {Haarnoja, Tuomas and Zhou, Aurick and Hartikainen, Kristian and Tucker, George and Ha, Sehoon and Tan, Jie and Kumar, Vikash and Zhu, Henry and Gupta, Abhishek and Abbeel, Pieter and Levine, Sergey},
  title = {Soft Actor-Critic Algorithms and Applications},
  year = {2018},
  eprint = {1812.05905},
  archivePrefix = {arXiv},
  primaryClass = {cs.LG},
  url = {https://arxiv.org/abs/1812.05905},
}

@article{nguyen_rigorous_2023,
  author = {Nguyen, Phan-Minh and Pham, Huy Tuan},
  title = {A rigorous framework for the mean field limit of multilayer neural networks},
  journal = {Mathematical Statistics and Learning},
  year = {2023},
  volume = {6},
  number = {3},
  pages = {201--357},
  doi = {10.4171/msl/42},
  url = {https://doi.org/10.4171/msl/42},
}

@article{sirignano2021meanfieldanalysisdeep,
  author = {Sirignano, Justin and Spiliopoulos, Konstantinos},
  title = {Mean Field Analysis of Deep Neural Networks},
  journal = {Mathematics of Operations Research},
  year = {2022},
  volume = {47},
  number = {1},
  pages = {120--152},
  doi = {10.1287/moor.2020.1118},
  url = {https://doi.org/10.1287/moor.2020.1118},
}

@article{Blei2017variational,
  author = {Blei, David M. and Kucukelbir, Alp and McAuliffe, Jon D.},
  title = {Variational Inference: A Review for Statisticians},
  journal = {Journal of the American Statistical Association},
  year = {2017},
  volume = {112},
  number = {518},
  pages = {859--877},
  doi = {10.1080/01621459.2017.1285773},
  url = {https://doi.org/10.1080/01621459.2017.1285773},
}

@article{Takatsu2011,
  author = {Takatsu, Asuka},
  title = {{Wasserstein} geometry of {Gaussian} measures},
  journal = {Osaka Journal of Mathematics},
  year = {2011},
  volume = {48},
  number = {4},
  pages = {1005--1026},
  doi = {10.18910/4973},
  url = {https://ir.library.osaka-u.ac.jp/repo/ouka/all/4973/},
}

@inproceedings{Lipman2023flow,
  author = {Lipman, Yaron and Chen, Ricky T. Q. and Ben-Hamu, Heli and Nickel, Maximilian and Le, Matt},
  title = {Flow Matching for Generative Modeling},
  booktitle = {International Conference on Learning Representations},
  year = {2023},
  url = {https://openreview.net/forum?id=PqvMRDCJT9t},
  eprint = {2210.02747},
  archivePrefix = {arXiv},
}

@article{Vincent2011,
  author = {Vincent, Pascal},
  title = {A Connection Between Score Matching and Denoising Autoencoders},
  journal = {Neural Computation},
  year = {2011},
  volume = {23},
  number = {7},
  pages = {1661--1674},
  doi = {10.1162/NECO_a_00142},
  url = {https://doi.org/10.1162/NECO_a_00142},
}

@inproceedings{SongErmon2019,
  author = {Song, Yang and Ermon, Stefano},
  title = {Generative Modeling by Estimating Gradients of the Data Distribution},
  booktitle = {Advances in Neural Information Processing Systems},
  volume = {32},
  year = {2019},
  url = {https://papers.neurips.cc/paper_files/paper/2019/hash/3001ef257407d5a371a96dcd947c7d93-Abstract.html},
}

@article{Anderson1982,
  author = {Anderson, Brian D.O.},
  title = {Reverse-time diffusion equation models},
  journal = {Stochastic Processes and their Applications},
  year = {1982},
  volume = {12},
  number = {3},
  pages = {313--326},
  doi = {10.1016/0304-4149(82)90051-5},
  url = {https://doi.org/10.1016/0304-4149(82)90051-5},
}

@inproceedings{Dinh2017realnvp,
  author = {Dinh, Laurent and Sohl-Dickstein, Jascha and Bengio, Samy},
  title = {Density estimation using {Real NVP}},
  booktitle = {International Conference on Learning Representations},
  year = {2017},
  url = {https://openreview.net/forum?id=HkpbnH9lx},
  eprint = {1605.08803},
  archivePrefix = {arXiv},
}

@article{NoeScience2019,
  author = {No{\'e}, Frank and Olsson, Simon and K{\"o}hler, Jonas and Wu, Hao},
  title = {{Boltzmann} generators: Sampling equilibrium states of many-body systems with deep learning},
  journal = {Science},
  year = {2019},
  volume = {365},
  number = {6457},
  pages = {eaaw1147},
  doi = {10.1126/science.aaw1147},
  url = {https://doi.org/10.1126/science.aaw1147},
}

@inproceedings{DhariwalNichol2021,
  author = {Dhariwal, Prafulla and Nichol, Alexander},
  title = {Diffusion Models Beat {GANs} on Image Synthesis},
  booktitle = {Advances in Neural Information Processing Systems},
  volume = {34},
  year = {2021},
  url = {https://proceedings.neurips.cc/paper/2021/hash/49ad23d1ec9fa4bd8d77d02681df5cfa-Abstract.html},
}

\clearpage
\onecolumngrid
\section*{Tables of notation}
\label{sec:notation}

The final column links to the section where each symbol first appears. Each chapter has a separate table, and we list notation in the order in which each symbol was first introduced. Familiar symbols may have different meanings in clearly identified contexts, for example, mechanical action $S[x(\cdot)]$ and thermodynamic entropy $S[p]$. 
\begingroup
\renewcommand{\baselinestretch}{1}\small
\setlength{\tabcolsep}{5pt}
\renewcommand{\arraystretch}{1.1}
\setlength{\LTpre}{0pt}
\setlength{\LTpost}{0pt}
\setlength{\LTcapwidth}{\textwidth}
\begin{longtable}{@{}>{\raggedright\arraybackslash}p{0.27\textwidth}>{\raggedright\arraybackslash}p{0.59\textwidth}>{\raggedright\arraybackslash}p{0.09\textwidth}@{}}
\caption{Chapter~\ref{sec:variational_structure}: Mechanics, Control, and Inference.}\label{tab:notation-i}\\
\hline\hline
\textbf{Symbol} & \textbf{Meaning or convention} & \textbf{Section} \\
\hline
\endfirsthead
\multicolumn{3}{l}{\textbf{Chapter~\ref{sec:variational_structure} notation (continued)}}\\[6pt]
\hline\hline
\textbf{Symbol} & \textbf{Meaning or convention} & \textbf{Section} \\
\hline
\endhead
\hline
\multicolumn{3}{r}{\small Continued on the next page}\\
\endfoot
\hline\hline
\endlastfoot
$x,\ x_t,\ x(\cdot)$ & State argument, state at time $t$, and complete trajectory. In Chapter~\ref{sec:application}, $x_n$ denotes the state at decision step $n$ and $y_t$ distinguishes a generated path from an interpolant. & \ref{sec:mechanics_control} \\[3pt]
$u,\ u_t(x)$ & Control value and feedback control; the applied control is $u_t(x_t)$. In Chapter~\ref{sec:thermodynamics}, $u_t$ also denotes a prescribed experimental protocol. & \ref{sec:mechanics_control} \\[3pt]
$t,s$ & Time; the control horizon is $0\leq t\leq1$. & \ref{sec:mechanics_control} \\[3pt]
$g_t(x,u)$ & General controlled drift. & \ref{sec:mechanics_control} \\[3pt]
$\varepsilon$ & Noise covariance per unit time; amplitude $\sqrt\varepsilon$, diffusion coefficient $\varepsilon/2$. Also the entropic transport regularization in Sec.~\ref{sec:eot}. In Chapter~\ref{sec:thermodynamics}, $\varepsilon=2D$. & \ref{sec:mechanics_control} \\[3pt]
$\eta_t,\ W_t$ & Standard Gaussian white noise and standard Wiener process. & \ref{sec:mechanics_control} \\[3pt]
$r_t(x,u)$ & Running reward, including control costs. In Sec.~\ref{sec:RL}, $r(x,u)$ is a per-step reward; its running-reward limit includes a factor $\Delta t$. & \ref{sec:mechanics_control} \\[3pt]
$r_1(x)$ & Terminal reward, separate from running reward. In Sec.~\ref{sec:dynamic_OT} it enforces the final density constraint. & \ref{sec:mechanics_control} \\[3pt]
$V_t(x)$ & Optimal expected reward-to-go, including control costs. In Sec.~\ref{sec:RL}, $V^{\pi}$ evaluates a policy and $V^{*}$ is optimal. & \ref{sec:mechanics_control} \\[3pt]
$L_t(x,\dot x),\ S[x(\cdot)]$ & Mechanical Lagrangian and action. & \ref{sec:mechanics_control} \\[3pt]
$p_t$ & Costate; canonical momentum in the mechanics specialization. & \ref{sec:mechanics_control} \\[3pt]
$\mathcal H_t(x,p,u)$ & Control Hamiltonian before maximization over $u$. & \ref{sec:mechanics_control} \\[3pt]
$H_t(x,p)$ & Optimized Hamiltonian, $H_t=\max_u\mathcal H_t$. & \ref{sec:mechanics_control} \\[3pt]
$p(x),\ p_t(x)$ & Probability density and time-dependent marginal density. In Chapter~\ref{sec:sampling}, $p_0,p_1$ denote the base and target densities. & \ref{sec:varational_basis_for_inference}--\ref{sec:path_inference} \\[3pt]
$p^0,\ p^A$ & Reference density and density inferred from observations $A$. Superscripts distinguish these from time subscripts. & \ref{sec:varational_basis_for_inference} \\[3pt]
$\dkl(p\Vert q)$ & KL divergence, with averaging under its first argument; also used for path distributions. & \ref{sec:varational_basis_for_inference} \\[3pt]
$\langle\cdot\rangle_p$ & Expectation under the indicated density; conditioning is written inside the brackets. Also written $\mathbb E_p[\cdot]$ in Chapter~\ref{sec:application}. & \ref{sec:varational_basis_for_inference} \\[3pt]
$N$ & Number of independent samples; in Sec.~\ref{sec:RL}, $N$ instead denotes the final decision index and $M$ counts rollouts. & \ref{sec:varational_basis_for_inference} \\[3pt]
$\hat p$ & Empirical density estimated from samples; $\widehat p_{\mathrm{data}}$ denotes the training-set density in Sec.~\ref{sec:flows_diffusions}. Hats indicate estimates, not samples. & \ref{sec:varational_basis_for_inference} \\[3pt]
$I$ & Large-deviation rate function. & \ref{sec:varational_basis_for_inference} \\[3pt]
$a_m(x),\ a_m[x(\cdot)]$ & State observable and, in Sec.~\ref{sec:path_inference}, trajectory observable. & \ref{sec:varational_basis_for_inference}--\ref{sec:path_inference} \\[3pt]
$A_m$ & Prescribed mean of observable $a_m$. & \ref{sec:varational_basis_for_inference} \\[3pt]
$H[p]$ & Shannon entropy with natural logarithms. In Sec.~\ref{sec:RL}, $H[\pi(\cdot| x)]$ is the policy entropy at state $x$. & \ref{sec:varational_basis_for_inference} \\[3pt]
$M$ & Number of imposed observable constraints; in Sec.~\ref{sec:RL}, the number of independent rollouts. & \ref{sec:varational_basis_for_inference} \\[3pt]
$\lambda_m$ & Multiplier enforcing the mean of observable $a_m$, with the negative sign in the exponential tilt. & \ref{sec:varational_basis_for_inference} \\[3pt]
$Z$ & Normalization constant or partition function. In Sec.~\ref{sec:eot}, $Z(x_0)$ normalizes a transition kernel; in Sec.~\ref{sec:RL}, $Z_Q(x)=\int e^{Q(x,u)/\alpha}\,du$ normalizes the Boltzmann policy. & \ref{sec:varational_basis_for_inference} \\[3pt]
$\mathcal A,\ \mathbf1_{\mathcal A}$ & Conditioning set and its indicator. & \ref{sec:varational_basis_for_inference} \\[3pt]
$P^0,\ P^u,\ P^{u^*}$ & Reference, controlled, and optimally controlled path distributions. In Sec.~\ref{sec:RL}, $P^\pi$ is the joint state-action path law induced by a policy, and $P^{\pi^*}$ is its optimal realizable form. & \ref{sec:path_inference} \\[3pt]
$y$ & Prescribed terminal point in the Brownian-bridge examples. & \ref{sec:path_inference} \\[3pt]
$f_t(x)$ & Passive drift; additive control gives $g_t(x,u)=f_t(x)+u$. & \ref{sec:path_inference} \\[3pt]
$p_t^u(x)$ & Marginal density under $P^u$; denoted $\rho_t(x)$ in Chapter~\ref{sec:OT}. & \ref{sec:path_inference} \\[3pt]
$\langle\cdot\rangle_{P^u}$ & Expectation over the indicated path distribution. & \ref{sec:path_inference} \\[3pt]
$r_m$ & Reward coefficient enforcing a trajectory constraint, with $r_m=-\lambda_m$. & \ref{sec:path_inference} \\[3pt]
$r_t(x)$ & State reward, excluding the separately displayed quadratic control cost. & \ref{sec:path_inference} \\[3pt]
$\psi_t(x)$ & Exponentiated PISC value, $\psi_t=e^{V_t}$. Also denoted $Z_t(x)$ during its derivation; the backward factor in Sec.~\ref{sec:dynamic_OT}. & \ref{sec:pisc} \\[3pt]
$R_t[x(\cdot)]$ & Accumulated state and terminal rewards from $t$ onward, excluding control cost. & \ref{sec:pisc} \\[3pt]
$\nabla\log p_t(x)$ & Spatial score of a marginal density; distinct from the conditioning field $\nabla\log\psi_t$ and the protocol score $\Theta_u$ of Chapter~\ref{sec:thermodynamics}. & \ref{sec:pisc} \\[3pt]
$k_{t| s}(z| x)$ & Passive transition density from $x$ at $s$ to $z$ at $t>s$; $k=k_{1|0}$ in Sec.~\ref{sec:eot}. In Chapter~\ref{sec:sampling}, $k_t$ freezes the energy for one step; in Sec.~\ref{sec:RL}, $k(x\prime| x,u)$ is the environment transition kernel. & \ref{sec:pisc} \\[3pt]
\end{longtable}
\endgroup
\clearpage
\begingroup
\renewcommand{\baselinestretch}{1}\small
\setlength{\tabcolsep}{5pt}
\renewcommand{\arraystretch}{1.1}
\setlength{\LTpre}{0pt}
\setlength{\LTpost}{0pt}
\setlength{\LTcapwidth}{\textwidth}
\begin{longtable}{@{}>{\raggedright\arraybackslash}p{0.27\textwidth}>{\raggedright\arraybackslash}p{0.59\textwidth}>{\raggedright\arraybackslash}p{0.09\textwidth}@{}}
\caption{Chapter~\ref{sec:OT}: Transport and Wasserstein Geometry.}\label{tab:notation-ii}\\
\hline\hline
\textbf{Symbol} & \textbf{Meaning or convention} & \textbf{Section} \\
\hline
\endfirsthead
\multicolumn{3}{l}{\textbf{Chapter~\ref{sec:OT} notation (continued)}}\\[6pt]
\hline\hline
\textbf{Symbol} & \textbf{Meaning or convention} & \textbf{Section} \\
\hline
\endhead
\hline
\multicolumn{3}{r}{\small Continued on the next page}\\
\endfoot
\hline\hline
\endlastfoot
$\rho_t,\ \rho_0,\ \rho_1$ & Evolving density and prescribed endpoint densities. In Sec.~\ref{sec:WGFApps}, $\widehat\rho_t=N_{\mathrm p}^{-1}\sum_i\delta(x-x_t^{(i)})$ is the empirical particle density. & \ref{sec:dynamic_OT} \\[3pt]
$v_t(x)$ & Density-transport velocity: $\partial_t\rho_t+\nabla\cdot(\rho_t v_t)=0$. Generally distinct from $u_t$; $\widehat v_t$ is a learned estimate in Sec.~\ref{sec:flows_diffusions}. & \ref{sec:dynamic_OT} \\[3pt]
$\psi_t^\dagger(x)$ & Forward Schr\"odinger factor, with $\rho_t=\psi_t\psi_t^\dagger$. The dagger does not denote complex conjugation. & \ref{sec:dynamic_OT} \\[3pt]
$c(x_0,x_1)$ & Cost of transporting unit mass between two endpoints. & \ref{sec:eot} \\[3pt]
$T(x_0)$ & Transport map assigning a final position to an initial position. In Sec.~\ref{sec:flows_diffusions}, $T_t$ is a flow map and $\widehat T_t$ a learned map, not necessarily optimal. & \ref{sec:eot} \\[3pt]
$\pi(x_0,x_1),\ \pi^*$ & Endpoint coupling and optimal coupling, with marginals $\rho_0,\rho_1$. Noise dependence is displayed as $\pi_\varepsilon^*$ when taking a limit. & \ref{sec:eot} \\[3pt]
$\varphi_0,\ \varphi_1$ & Endpoint dual potentials for entropic transport; $\varphi_1=\varepsilon r_1$ with compatible normalization. Their limits are Kantorovich potentials. & \ref{sec:eot} \\[3pt]
$\zeta,\ \bar\varphi_1$ & Convex conjugate and shifted limiting endpoint potential: $\bar\varphi_1(x)=\|x\|^2/2-\varphi_1(x)$, $T=\nabla\zeta$. & \ref{sec:eot} \\[3pt]
$C_0,C_1$ & Cumulative distribution functions of the endpoint densities; $C_i^{-1}$ denotes the quantile function. & \ref{sec:eot} \\[3pt]
$\theta,\ p_\theta$ & Parameters and corresponding density in the geometric illustration; reused for the parametric VI family in Sec.~\ref{sec:WGFApps}. & \ref{sec:eot} \\[3pt]
$W_2(\rho_0,\rho_1)$ & 2-Wasserstein distance; its square is the minimum mean squared displacement. The cost $c=\|x_0-x_1\|^2/2$ has minimum $W_2^2/2$. & \ref{sec:eot} \\[3pt]
$F(x)$ & Ordinary scalar objective in the Euclidean gradient-descent analogy. & \ref{sec:WGF} \\[3pt]
$s(x),\ s_t(x)$ & Density tangent; along a curve, $s_t=\partial_t\rho_t$, with $\int s_t\,dx=0$. & \ref{sec:WGF} \\[3pt]
$\mathcal W_2$ & Space of probability measures with finite second moment, equipped with $W_2$. & \ref{sec:WGF} \\[3pt]
$\chi(x)$ & Potential for the minimum-energy transport velocity, with $v=-\nabla\chi$. In Sec.~\ref{sec:WGFApps}, $v_i=-\nabla\chi_i$ represents the parameter tangent $\partial_{\theta_i}p_\theta$. & \ref{sec:WGF} \\[3pt]
$\|s\|_\rho,\ \langle s,s'\rangle_\rho$ & Wasserstein tangent norm and inner product at density $\rho$. & \ref{sec:WGF} \\[3pt]
$\mathcal F[\rho]$ & Functional of a density, such as free energy. & \ref{sec:WGF} \\[3pt]
$\nabla_{\mathrm W}\mathcal F[\rho]$ & Wasserstein gradient in its velocity representation; the descent velocity is its negative. & \ref{sec:WGF} \\[3pt]
$\delta\mathcal F/\delta\rho$ & Functional derivative; $\nabla_{\mathrm W}\mathcal F=\nabla(\delta\mathcal F/\delta\rho)$. & \ref{sec:WGF} \\[3pt]
$\Phi(x)$ & Dimensionless potential in the illustrative free energy with unit diffusion coefficient; reused in Sec.~\ref{sec:WGFApps}, with $p_1=e^{-\Phi}/Z_1$ for VI. & \ref{sec:WGF} \\[3pt]
$\alpha$ & Convexity or geodesic-convexity parameter; positive values indicate strong convexity. & \ref{sec:WGF} \\[3pt]
\end{longtable}
\endgroup
\clearpage
\begingroup
\renewcommand{\baselinestretch}{1}\small
\setlength{\tabcolsep}{5pt}
\renewcommand{\arraystretch}{1.1}
\setlength{\LTpre}{0pt}
\setlength{\LTpost}{0pt}
\setlength{\LTcapwidth}{\textwidth}
\begin{longtable}{@{}>{\raggedright\arraybackslash}p{0.27\textwidth}>{\raggedright\arraybackslash}p{0.59\textwidth}>{\raggedright\arraybackslash}p{0.09\textwidth}@{}}
\caption{Chapter~\ref{sec:thermodynamics}: Thermodynamics.}\label{tab:notation-iii}\\
\hline\hline
\textbf{Symbol} & \textbf{Meaning or convention} & \textbf{Section} \\
\hline
\endfirsthead
\multicolumn{3}{l}{\textbf{Chapter~\ref{sec:thermodynamics} notation (continued)}}\\[6pt]
\hline\hline
\textbf{Symbol} & \textbf{Meaning or convention} & \textbf{Section} \\
\hline
\endhead
\hline
\multicolumn{3}{r}{\small Continued on the next page}\\
\endfoot
\hline\hline
\endlastfoot
$S[p]$ & Thermodynamic entropy, $S[p]=k_{\mathrm B}H[p]$ in the statistical description used here; distinct from mechanical action $S[x(\cdot)]$. & \ref{sec:energy-based_modelling} \\[3pt]
$T,\ k_{\mathrm B},\ \beta$ & Bath temperature, Boltzmann constant, and inverse thermal energy $\beta=(k_{\mathrm B}T)^{-1}$. & \ref{sec:energy-based_modelling} \\[3pt]
$p^{\mathrm{eq}},\ p_u^{\mathrm{eq}}$ & Equilibrium density, including dependence on a fixed protocol $u$. In Chapter~\ref{sec:sampling}, $p_t^{\mathrm{eq}}=e^{-E_t}/Z_t$ is the prescribed equilibrium bridge. & \ref{sec:energy-based_modelling} \\[3pt]
$E(x),\ E_u(x),\ E_t(x)$ & Microstate energy; $E_t=E_{u_t}$ for a protocol. General sampling algorithms in Chapter~\ref{sec:sampling} absorb $\beta$ into $E_t$ and use dimensionless energy. & \ref{sec:energy-based_modelling} \\[3pt]
$\mathcal F^{\mathrm{eq}},\ \mathcal F_u^{\mathrm{eq}}$ & Equilibrium free energy, $-k_{\mathrm B}T\log Z_u$; $-\log Z_u$ in Chapter~\ref{sec:sampling}'s thermal units. & \ref{sec:energy-based_modelling} \\[3pt]
$\mathcal E$ & Mean internal energy, $\langle E\rangle_p$. & \ref{sec:energy-based_modelling} \\[3pt]
$\Sigma,\ \dot\Sigma_t$ & Total entropy production and its rate; distinct from the change $\Delta S$ in system entropy. & \ref{sec:OTthermo} \\[3pt]
$Q,\ W$ & Mean heat received by the system and mean work done on it. & \ref{sec:OTthermo} \\[3pt]
$N_{\mathrm p}$ & Particle number; distinct from $N$ statistical samples. In Sec.~\ref{sec:WGFApps}, also the number of neurons viewed as parameter particles. & \ref{sec:OTthermo} \\[3pt]
$X_m,\ L_{mn}$ & Macroscopic variables ($m=1,\ldots,M$) and Onsager transport coefficients. & \ref{sec:onsager} \\[3pt]
$\mathcal R$ & Rayleighian, $\mathcal R=\tfrac12T\dot\Sigma+\dot{\mathcal F}$; distinct from accumulated reward $R_t$. & \ref{sec:onsager} \\[3pt]
$n_t(x)$ & Conserved particle number density, with $\int n_t\,dx=N_{\mathrm p}$. Its normalized form is $\rho_t=n_t/N_{\mathrm p}$. & \ref{sec:onsager} \\[3pt]
$\Gamma(n)$ & Dissipation or drag coefficient in the continuum Onsager description. & \ref{sec:onsager} \\[3pt]
$\eta,\ r$ & Fluid viscosity and colloid radius in the Stokes-drag example; $\eta$ here is distinct from white noise $\eta_t$. & \ref{sec:onsager} \\[3pt]
$\mu_t(x)$ & Chemical potential, $\delta\mathcal F/\delta n_t(x)$; for constant drag, relaxation has $\chi_t=\mu_t/\Gamma$. & \ref{sec:onsager} \\[3pt]
$D$ & Physical diffusion coefficient, $D=\varepsilon/2$. & \ref{sec:onsager} \\[3pt]
$F_t(x)$ & Physical force; the overdamped stochastic drift is $f_t=D\beta F_t$. & \ref{sec:far_from_equilibrium} \\[3pt]
$\tau$ & Process duration when physical elapsed time is displayed explicitly. & \ref{sec:far_from_equilibrium} \\[3pt]
$\delta p_t(x)$ & Deviation from instantaneous equilibrium: $p_t=p_{u_t}^{\mathrm{eq}}+\delta p_t$. & \ref{sec:geometry_quasistatic} \\[3pt]
$\mathcal L_u,\ \mathcal L_u^\dagger$ & Operators with convention $\partial_t p=-\mathcal L_u p$; $-\mathcal L_u^\dagger$ is the backward generator. The dagger denotes the operator adjoint. & \ref{sec:geometry_quasistatic} \\[3pt]
$\Theta_u(x)$ & Protocol score, $\partial_u\log p_u^{\mathrm{eq}}(x)$; a vector for a multicomponent protocol. & \ref{sec:geometry_quasistatic} \\[3pt]
$\zeta_u$ & Integrated equilibrium score-correlation tensor defining the protocol-space metric. Its energy-unit friction normalization is $k_{\mathrm B}T\zeta_u$. & \ref{sec:geometry_quasistatic} \\[3pt]
$w_t[x(\cdot)]$ & Work accumulated along one trajectory up to time $t$; $W=\langle w_1\rangle_P$ for the unit-time protocol. Dimensionless in Chapter~\ref{sec:sampling}'s sampling formulas. & \ref{sec:thermodynamic_fluctuation_theorems} \\[3pt]
$P,\ \bar u_t,\ \bar P$ & Forward path law, reversed protocol $\bar u_t=u_{1-t}$, and reverse path law. In Chapter~\ref{sec:sampling}, $\bar E_t=E_{1-t}$ and $x_t^{\mathrm{rev}}=x_{1-t}$ denotes a reversed realization. & \ref{sec:thermodynamic_fluctuation_theorems} \\[3pt]
$\lambda$ & Work-MGF transform variable in $e^{\lambda w}$; inverse-energy units when $w$ is physical work. & \ref{sec:thermodynamic_fluctuation_theorems} \\[3pt]
$M_t,\ \bar M_t,\ \widetilde M_t$ & Work-weighted endpoint density, reverse conditional MGF, and transformed reverse MGF used in the fluctuation-theorem proof. & \ref{sec:thermodynamic_fluctuation_theorems} \\[3pt]
\end{longtable}
\endgroup
\clearpage
\begingroup
\renewcommand{\baselinestretch}{1}\small
\setlength{\tabcolsep}{5pt}
\renewcommand{\arraystretch}{1.1}
\setlength{\LTpre}{0pt}
\setlength{\LTpost}{0pt}
\setlength{\LTcapwidth}{\textwidth}
\begin{longtable}{@{}>{\raggedright\arraybackslash}p{0.27\textwidth}>{\raggedright\arraybackslash}p{0.59\textwidth}>{\raggedright\arraybackslash}p{0.09\textwidth}@{}}
\caption{Chapter~\ref{sec:sampling}: Sampling and Control.}\label{tab:notation-iv}\\
\hline\hline
\textbf{Symbol} & \textbf{Meaning or convention} & \textbf{Section} \\
\hline
\endfirsthead
\multicolumn{3}{l}{\textbf{Chapter~\ref{sec:sampling} notation (continued)}}\\[6pt]
\hline\hline
\textbf{Symbol} & \textbf{Meaning or convention} & \textbf{Section} \\
\hline
\endhead
\hline
\multicolumn{3}{r}{\small Continued on the next page}\\
\endfoot
\hline\hline
\endlastfoot
$\widetilde p,\ p=\widetilde p/Z$ & Unnormalized density and the corresponding normalized probability density. & \ref{sec:sampling_review} \\[3pt]
$d$ & Dimension of configuration space or number of model degrees of freedom. & \ref{sec:monte_carlo} \\[3pt]
$\hat a_N$ & Monte Carlo estimate of $\langle a\rangle$ from $N$ independent samples $x_n$. For an ensemble of paths, $x_t^{(n)}$ labels time and run separately. & \ref{sec:monte_carlo} \\[3pt]
$\sigma_i,\ m(x)$ & Ising spin and configuration magnetization, $m(x)=d^{-1}\sum_i\sigma_i$. & \ref{sec:high_dimension} \\[3pt]
$\omega,\ \widetilde\omega$ & Normalized importance ratio and unnormalized weight. In AIS, $\widetilde\omega_t=e^{-w_t}$ and $\omega_t=(Z_0/Z_t)\widetilde\omega_t$. & \ref{sec:basic_sampling} \\[3pt]
$N_{\mathrm{eff}}$ & Effective sample size estimated from the weights, $(\sum_n\widetilde\omega_n)^2/\sum_n\widetilde\omega_n^2$. & \ref{sec:basic_sampling} \\[3pt]
$q_t,\ \bar q_t$ & Actual forward and reverse sampler densities, distinct from prescribed densities. In AIS, $q_0=p_0$, $\bar q_0=p_1$; in Sec.~\ref{sec:flows_diffusions}, $q_t=(\widehat T_t)_*p_0$ is the learned model's density. & \ref{sec:basic_sampling} \\[3pt]
\end{longtable}
\endgroup

\clearpage

\begingroup
\renewcommand{\baselinestretch}{1}\small
\setlength{\tabcolsep}{5pt}
\renewcommand{\arraystretch}{1.1}
\setlength{\LTpre}{0pt}
\setlength{\LTpost}{0pt}
\setlength{\LTcapwidth}{\textwidth}
\begin{longtable}{@{}>{\raggedright\arraybackslash}p{0.27\textwidth}>{\raggedright\arraybackslash}p{0.59\textwidth}>{\raggedright\arraybackslash}p{0.09\textwidth}@{}}
\caption{Chapter~\ref{sec:application}: Applications in Learning.}\label{tab:notation-v}\\
\hline\hline
\textbf{Symbol} & \textbf{Meaning or convention} & \textbf{Section} \\
\hline
\endfirsthead
\multicolumn{3}{l}{\textbf{Chapter~\ref{sec:application} notation (continued)}}\\[6pt]
\hline\hline
\textbf{Symbol} & \textbf{Meaning or convention} & \textbf{Section} \\
\hline
\endhead
\hline
\multicolumn{3}{r}{\small Continued on the next page}\\
\endfoot
\hline\hline
\endlastfoot
$\mathcal X,\ \mathcal U$ & RL state and action spaces. & \ref{sec:mdp} \\[3pt]
$\pi(u| x)$ & Policy: conditional action distribution; time is included in the state for finite-horizon problems. Later, $\pi_Q$ denotes the Boltzmann policy defined by a critic. & \ref{sec:mdp} \\[3pt]
 $n,\ N$ & Decision step and final decision index; $n=0,\ldots,N$ and $t_{n}=n\Delta t$. & \ref{sec:mdp} \\[3pt]
$Q^{\pi}(x,u),\ Q^{*}(x,u)$ & Exact policy-specific and optimal action values; hats denote learned tabular estimates. In MaxEnt RL, these denote entropy-regularized (``soft'') values. & \ref{sec:mdp} \\[3pt]
$\theta,\ Q_\theta^\pi,\ Q_\theta^*$ & Neural-network estimates of policy-specific and optimal values; $\theta$ denotes the network weights. & \ref{sec:online_rl} \\[3pt]
$i,\ M$ & Rollout sample index and number of independent rollouts, $i=1,\ldots,M$. & \ref{sec:online_rl} \\[3pt]
$\mathcal J[\pi]$ & Expected RL return; the text specifies discounting and any entropy bonus. & \ref{sec:online_rl} \\[3pt]
$\gamma$ & Reward discount factor. & \ref{sec:online_rl} \\[3pt]
$y$ & Sampled Bellman target used to update a critic. & \ref{sec:online_rl} \\[3pt]
$\xi$ & Learning rate. & \ref{sec:online_rl} \\[3pt]
$\mathcal D$ & Replay buffer of observed transitions and rewards. & \ref{sec:online_rl} \\[3pt]
$L(\theta),\ \mathcal L[\rho]$ & Loss as a function of parameters or a functional of a density, respectively. & \ref{sec:online_rl} \\[3pt]
$\bar\theta$ & Parameters held fixed when evaluating a critic target; target-network updates, if used, are specified separately. & \ref{sec:online_rl} \\[3pt]
$\alpha$ & RL entropy temperature in reward units; distinct from the convexity parameter in Wasserstein geometry. & \ref{sec:maxent_rl} \\[3pt]
$\pi^0(u| x),\ c_0$ & Reference action distribution and its constant value in the uniform case. & \ref{sec:maxent_rl} \\[3pt]
$P^*$ & Unrestricted reward-tilted target path law; generally distinct from $P^{\pi^*}$. & \ref{sec:maxent_rl} \\[3pt]
$\phi,\ \pi_\phi$ & Actor parameters and corresponding parameterized policy. & \ref{sec:sac} \\[3pt]
$\Phi^{(N_{\mathrm p})},\ \mathcal V[\rho]$ & Many-particle potential and its density-functional representation, with the normalization displayed in Sec.~\ref{sec:mf}. & \ref{sec:mf} \\[3pt]
$U(x,y)$ & Pair-interaction potential, also used for interactions between parameter particles. & \ref{sec:mf} \\[3pt]
$f^{N_{\mathrm p}}(x;\Theta_t)$ & Finite-width neural-network predictor; distinct from the passive drift $f_t$. & \ref{sec:mf} \\[3pt]
$\Theta_t,\ \theta_t^{(i)},\ w_t^{(i)},\ z_t^{(i)}$ & Full parameter collection, neuron parameters, feature weights and internal feature parameters; $\theta=(w,z)$. & \ref{sec:mf} \\[3pt]
$\varphi(x;z)$ & Nonlinear feature; its weighted form is the observable $a(x;\theta)=w\varphi(x;z)$. & \ref{sec:mf} \\[3pt]
$y(x),\ \nu(x)$ & Target regression function and input-data density; distinct from the Bellman target $y$. & \ref{sec:mf} \\[3pt]
$\mathcal M,\ \Omega,\ d_\theta$ & Parametric density family, parameter domain and parameter dimension. & \ref{sec:variational_inference} \\[3pt]
$g_{ij}$ & Parameter-space metric induced by Wasserstein geometry. & \ref{sec:variational_inference} \\[3pt]
$\operatorname{proj}_{\mathcal M}$ & Orthogonal projection of a velocity onto the tangent space of the parametric density family. & \ref{sec:variational_inference} \\[3pt]
$\mathrm{BW}(\mathbb R^d),\ \nabla_{\mathrm{BW}}$ & Gaussian Bures--Wasserstein space and its projected gradient. & \ref{sec:variational_inference} \\[3pt]
$m,\ \Sigma,\ \mathbb S_{++}^d$ & Gaussian mean, covariance and space of symmetric positive-definite covariance matrices. & \ref{sec:variational_inference} \\[3pt]
$h,\ k_h,\ \mathcal K_h^p$ & Smoothing bandwidth, spatial kernel and density-weighted smoothing operator. & \ref{sec:variational_inference} \\[3pt]
$T_*p$ & Pushforward of a distribution through a map $T$. & \ref{sec:variational_transform} \\[3pt]
$\hat{v}$ & Estimated probability flow field. & \ref{sec:flow_matching} \\[3pt]
$y_t$ & Sample-generating process in flow and diffusion models. & \ref{sec:flow_matching} \\[3pt]
$x_t$ & Stochastic interpolant defined from training samples. & \ref{sec:flow_matching} \\[3pt]
$\mathcal L_{\mathrm{FM}}[\widehat v]$ & Flow-matching loss. & \ref{sec:flow_matching} \\[3pt]
$c_j$ & Gaussian-mixture weights, with $c_j\geq0$ and $\sum_jc_j=1$. & \ref{sec:flow_matching} \\[3pt]
$\widehat a,\ \widehat b$ & Learned scale and shift functions in an affine normalizing-flow layer. & \ref{sec:normalizing_flows} \\[3pt]
\end{longtable}
\endgroup
\clearpage

\end{document}